\documentclass[twocolumn]{aastex62}

\shorttitle{SMBHB Candidates from PS1 MDS}
\shortauthors{T. Liu et al.}

\usepackage{epsfig}
\usepackage{natbib}
\usepackage{amsmath}
\usepackage{epstopdf}

\begin{document}

\title{Supermassive Black Hole Binary Candidates from the Pan-STARRS1 Medium Deep Survey}

\author[0000-0001-5766-4287]{T. Liu}
\affiliation{Center for Gravitation, Cosmology and Astrophysics, Department of Physics, University of Wisconsin--Milwaukee, P.O. Box 413, Milwaukee, WI 53201, USA}
\affiliation{Department of Astronomy, University of Maryland, College Park, MD 20742, USA}
\email{tingtliu@uwm.edu}

\author{S. Gezari} 
\affiliation{Department of Astronomy, University of Maryland, College Park, MD 20742, USA}

\author{M. Ayers} 
\affiliation{Department of Physics, Lewis \& Clark College, 0615 SW Palatine Hill Road, Portland, OR 97219, USA}
\affiliation{Maria Mitchell Observatory, 4 Vestal Street, Nantucket, MA 02554, USA}

\author{W. Burgett} 
\affiliation{GMTO Corporation, 465 North Halstead Stree, Suite 250, Pasadena, CA 91107, USA}

\author{K. Chambers}
\affiliation{Institute for Astronomy, University of Hawaii at Manoa, 2680 Woodlawn Drive, Honolulu, HI 96822, USA}

\author{K. Hodapp} 
\affiliation{Institute for Astronomy, University of Hawaii at Manoa, 2680 Woodlawn Drive, Honolulu, HI 96822, USA}

\author{M. E. Huber} 
\affiliation{Institute for Astronomy, University of Hawaii at Manoa, 2680 Woodlawn Drive, Honolulu, HI 96822, USA}

\author{R.-P. Kudritzki} 
\affiliation{Institute for Astronomy, University of Hawaii at Manoa, 2680 Woodlawn Drive, Honolulu, HI 96822, USA}

\author{N. Metcalfe} 
\affiliation{Department of Physics, University of Durham, South Road, Durham DH1 3LE, UK}

\author{J. Tonry} 
\affiliation{Institute for Astronomy, University of Hawaii at Manoa, 2680 Woodlawn Drive, Honolulu, HI 96822, USA}

\author{R. Wainscoat} 
\affiliation{Institute for Astronomy, University of Hawaii at Manoa, 2680 Woodlawn Drive, Honolulu, HI 96822, USA}

\author{C. Waters} 
\affiliation{Institute for Astronomy, University of Hawaii at Manoa, 2680 Woodlawn Drive, Honolulu, HI 96822, USA}


\begin{abstract}
We present a systematic search for periodically varying quasar and supermassive black hole binary (SMBHB) candidates in the Pan-STARRS1 Medium Deep Survey (MDS). From $\sim9000$ color-selected quasars in an $\sim50$ $\degr^{2}$ sky area, we initially identify $26$ candidates with more than $1.5$ cycles of variation. We extend the baseline of observations via our imaging campaign with the Discovery Channel Telescope and the Las Cumbres Observatory network and reevaluate the candidates using a more rigorous, maximum likelihood method. Using a range of statistical criteria and assuming the damped random walk model for normal quasar variability, we identify one statistically significant periodic candidate. We also investigate the capabilities of detecting SMBHBs with the Large Synoptic Survey Telescope using our study with MDS as a benchmark and explore any complementary multiwavelength evidence for SMBHBs in our sample.
\end{abstract}

\keywords{Quasars -- Supermassive black holes --- Surveys}


\section{Introduction}\label{sec:intro}

Supermassive black hole binaries (SMBHBs) are expected as a result of galaxy mergers occurring the universe (e.g., \citealt{Begelman1980}). As the supermassive black holes (SMBHs) in the centers of massive galaxies sink to the center of the merged system via dynamical friction, the pair of active SMBHs on a scale of $\sim$ a few kpc can be observable as a dual active galactic nucleus (AGN; e.g., \citealt{Comerford2015}). As its separation continues to shrink by ejecting stars in the ``loss cone,'' the pair becomes a gravitationally bound SMBHB at a subparsec separation. While spatially resolving close-separation SMBHBs has been achieved with very long baseline interferometry (e.g., \citealt{Rodriguez2006}), the direct imaging of SMBHBs at farther distances is beyond the capabilities of current, or even future, telescopes. An indirect method to search for SMBHBs is via spectroscopy, where the broad emission line from one black hole is shifted due to its radial velocity (e.g., \citealt{Eracleous2012, Runnoe2017}), or there is the presence of a double broad-line feature that is due to the broad-line region associated with each black hole (e.g., \citealt{Boroson2009}). 

Another indirect technique to search for SMBHBs is via their temporal variability signatures. (Magneto) hydrodynamical simulations of an SMBHB system (e.g., \citealt{MacFadyen2008, Noble2012, Shi2012, D'Orazio2013, Farris2014, Gold2014}) show that the binary tidal torque clears and maintains a low gas density cavity of a radius $\sim 2a$ (where $a$ is the binary separation) in the circumbinary disk, and material is ushered in through a pair of accretion streams. This distinct accretion pattern of a binary-disk system causes the accretion rate to strongly modulate on the order of the orbital frequency. Therefore, assuming the accretion rate directly translates to electromagnetic luminosity, these SMBHBs would manifest as AGNs or quasars that periodically vary on a timescale of months to years. More recently, \cite{D'Orazio2015} also proposed a relativistic Doppler-boosting model: the SMBHB system is viewed at a high inclination angle, and the emission dominated by the minidisk of the secondary black hole is Doppler-boosted as the secondary travels at a relativistic speed along the line of sight. In addition to optical variability, periodicity in the X-ray bands has also been predicted for SMBHBs at the inspiral stage due to gas being flung outward and hitting the cavity wall \citep{Tang2018}.

Observationally, there have been a number of systematic searches for periodically varying quasars in large optical time-domain surveys: \cite{Graham2015,Graham2015Nat}, using the Catalina Real-time Transient Survey (CRTS); \cite{Charisi2016}, using the Palomar Transient Factory (PTF); and \citet[hereafter L15 and L16, respectively]{Liu2015, Liu2016}, using the Pan-STARRS1 Medium Deep Survey (PS1 MDS). \cite{Graham2015} claimed $111$ SMBHB candidates from a search among $\sim 200,000$ spectroscopically confirmed quasars in the CRTS footprint, and \cite{Charisi2016} found $50$ SMBHB candidates from a sample of $\sim 35,000$ spectroscopic quasars in the PTF, $33$ of which remained significant after their reanalysis with extended data.

However, due to the stochastic nature of normal (i.e., single black hole) quasar variability, the search for a periodic signal is highly susceptible to red noise (i.e., increasing variability power on longer timescales) masquerading as periodicity over a small number of cycles \citep{Vaughan2016} and thus could produce a large number of false-positive detections in a systematic search. In fact, assuming the candidates reported by \cite{Graham2015} and \cite{Charisi2016} are all genuine SMBHBs with their claimed binary parameters, \cite{Sesana2018} concluded that the expected stochastic gravitational-wave background would exceed the current pulsar timing array (PTA) upper limit by a factor of a few to an order of magnitude. We addressed this issue of false positives due to red noise contamination in L16, where we tested the persistence of the periodic candidates with archival Sloan Digital Sky Survey (SDSS) Stripe 82 light curves and new monitoring data taken at the Discovery Channel Telescope (DCT) since 2015, extending the total length of the baseline to $N_{\rm cycle} >$ 5. We find three periodic candidates from $\sim1000$ color-selected quasars in one PS1 MD field, MD09, though none of them appear to be persistent over an extended baseline. Further, we have reanalyzed the best candidate from the CRTS SMBHB sample, PG 1302$-$102 \citep{Graham2015Nat}, by including new photometric data from the All-sky Automated Survey for Supernovae (ASAS-SN; \citealt{Shappee2014,Kochanek2017}), and we have shown that the detected periodicity does not persist, as expected for a true SMBHB \citep{Liu2018a}.

Here we expand our analysis in L16 to all $10$ fields in the PS1 MDS and extended the temporal baseline with monitoring programs with the DCT and the Las Cumbres Observatory (LCO) network telescopes (Section \ref{sec:data}). We systematically searched for periodically varying quasars over the PS1 MDS baseline and adopted a maximum likelihood method to put their periodicity to the test over the extended baseline, which was constructed by ``stitching'' new DCT and LCO observations to their PS1 light curves (Section \ref{sec:methods}). We will discuss the parent sample of $26$ candidates from PS1 MDS and the down-selected sample in Section \ref{sec:cand}. We also compare the cumulative SMBHB rate from our down-selected sample with previous work and look ahead to the era of the Large Synoptic Survey Telescope (LSST) using our study as a benchmark (Section \ref{sec:discuss}). We also explore the multiwavelength properties of the best SMBHB candidate from our sample. We summarize our results in Section \ref{sec:conclude}. We adopt the following cosmological parameters throughout this paper: $\Omega_{\rm m}$ = 0.3, $\Omega_{\rm \lambda}$ = 0.7, $H_{\rm 0}$ = 70 km s$^{-1}$Mpc$^{-1}$.


\section{Observations and Data}\label{sec:data}
\subsection{The PS1 MDS}\label{sec:mds}

The PS1 (\citealt{Kaiser2010,Chambers2016}) operated from 2009 to 2014 on the $1.8$ m PS1 telescope at the summit of Haleakala on Maui, Hawaii. About $25$\% of the survey time was dedicated to the MDS, a multifilter, high-cadence time-domain survey of $10$ circular fields (Table \ref{tab:mds}), each of which is $\sim$ 8 $\degr^2$ in size. The MDS observed in the $g_{\rm P1}$, $r_{\rm P1}$, $i_{\rm P1}$, $z_{\rm P1}$, and $y_{\rm P1}$\footnote{Although the $y_{\rm P1}$ filter was not used in our work.} filters on the AB photometric system \citep{Tonry2012photometry} and can reach a $5\sigma$ magnitude depth of $22.5$ mag in $g_{\rm P1}$, $r_{\rm P1}$, and $i_{\rm P1}$ and $22.0$ mag in the $z_{\rm P1}$ filter in a single exposure of $113$ s ($g_{\rm P1}$, $r_{\rm P1}$) or $240$ s ($i_{\rm P1}$, $z_{\rm P1}$). The data were processed by the PS1 image processing pipeline (IPP; \citealt{Magnier2006IPP}) and were made available to members of the PS1 Science Consortium through the PS1 Science Archive.


\begin{deluxetable}{lrr}
\tablecaption{MD Field Centers \label{tab:mds}}
\tablehead{
\colhead{MD Field} & \colhead{R.A. (J2000)} & \colhead{Decl. (J2000)}
}
\startdata
MD01 & 02:24:50 & --04:35:00 \\
MD02 & 03:32:24 & --28:08:00 \\
MD03 & 08:42:22 & +44:19:00 \\
MD04 & 10:00:00 & +02:12:00 \\ 
MD05 & 10:47:40 & +58:05:00 \\
MD06 & 12:20:00 & +47:07:00 \\
MD07 & 14:14:49 & +53:05:00 \\
MD08 & 16:11:09 & +54:57:00 \\
MD09 & 22:16:45 & +00:17:00 \\
MD10 & 23:29:15 & --00:26:00 \\
\enddata
\end{deluxetable}


Each nightly observation consisted of eight single exposures; although the subexposures can be combined to produce ``nightly stacks,'' we have used the single-exposure detections in this work, as well as in our previous work presented in L15 and L16. The telescope visited the field during the $6-8$ months that it was visible and rotated through the $g_{\rm P1}$, $r_{\rm P1}$, $i_{\rm P1}$, and $z_{\rm P1}$ filters every $3$ nights (observations in $g_{\rm P1}$, $r_{\rm P1}$ were carried out on the same night). Therefore, in the full MDS data set, most objects were observed $\sim 400$ times over the $\sim 4$ yr baseline.


\subsection{Extended Baseline Photometry}\label{sec:extended}

New imaging data presented in this work include those taken with the Large Monolithic Imager (LMI) in the $g_{\rm SDSS}$, $r_{\rm SDSS}$, $i_{\rm SDSS}$, and $z_{\rm SDSS}$ filters at DCT from 2015 May to 2017 November. In Table \ref{tab:extended}, we list the Modified Julian Dates (MJDs) on which the observations were carried out, as well as the filters that were used.

The images were reduced using standard \texttt{IRAF} routines and corrected for astrometry with \texttt{SCAMP} \citep{BertinSCAMP}. For the $z_{\rm SDSS}$-band images that are affected by fringe patterns, we also subtract a scaled master fringe pattern created via \texttt{create\_fringes} \citep{Snodgrass2013} from all $z_{\rm SDSS}$-band images taken on the same night and remove the fringes using the routine \texttt{remove\_fringes} \citep{Snodgrass2013}. We then coadd five subexposures in each filter (taken in a dither pattern to avoid bad pixels) with \texttt{SWARP} \citep{BertinSWARP} before performing aperture photometry using \texttt{SExtractor} \citep{BertinSE}. Following the method described in L16, we cross-match \texttt{SExtractor} detections with an SDSS catalog of point sources from DR12 \citep{SDSSDR12}, resulting in $\sim 200$ cross-matched pairs in LMI's $12'.3\times12'3$ field of view (FOV). We exclude bright, saturated detections ($m < 16$ mag), faint objects ($m > 22$ mag), outliers, and the target itself (which is variable) and obtain a linear transformation from the \texttt{SExtractor} instrumental magnitude to an SDSS magnitude. We then apply the transformation to the target and obtain a measurement of its magnitude on the SDSS photometric system. 


\begin{deluxetable*}{llll}
\tablecaption{Extended Baseline Monitoring of Candidates \label{tab:extended}}
\tablehead{
\colhead{PS1 Designation} & \colhead{Telescope/Instrument} & \colhead{MJD(s)} & \colhead{Filters} \\
\colhead{} & \colhead{} & \colhead{of follow-up observation(s)} & \colhead{}
}
\startdata
PSO J35.7068--4.23144 & DCT/LMI, \textbf{LCO/Spectral}  & 57,641, 57,682, \textbf{57,940, 57,993, 58,123} & $g$ $r$ $i$ $z$ \\
PSO J35.8704--4.0263 & DCT/LMI, \textbf{LCO/Spectral} & 57,641, 57,682, \textbf{57,939, 58,101} & $g$ $r$ $i$ $z$ \\
PSO J52.6172--27.6268 & \nodata & \nodata & \nodata \\
PSO J129.4288+43.8234 & DCT/LMI, \textbf{LCO/Spectral} & 57,682, \textbf{57,901, 58,208} & $g$ $r$ $i$ $z$ \\
PSO J130.9953+43.7685 & DCT/LMI & 57,682, 57,741, \textbf{58,147} & $g$ $r$ $i$ $z$ \\
PSO J131.1273+44.8582 & DCT/LMI & 57,682, \textbf{58,208} & $g$ $r$ $i$ $z$ \\
PSO J131.7789+45.0939 & DCT/LMI & 57,682, 57,741, 58,075 & $g$ $r$ $i$ $z$ \\
PSO J148.8485+1.8124 & DCT/LMI & 57,787, \textbf{58,126} & $r$ $i$ \\
PSO J149.4989+2.7827 & DCT/LMI & 57,787 & $g$ $r$ $i$ \\
PSO J149.2447+3.1393 & DCT/LMI & 57,369, 57,788, \textbf{58,148} & $g$ $r$ $i$ $z$ \\
PSO J149.9400+1.5090 & DCT/LMI & 57,788 & $g$ $r$ $i$ $z$ \\
PSO J149.6873+1.7192 & DCT/LMI & 57,788, 58,075, \textbf{58,148} & $g$ $r$ \\
PSO J150.9191+3.3880 & DCT/LMI, \textbf{LCO/Spectral} & 57,833, \textbf{57,845, 58,230} & $g$ $r$ $i$ $z$ \\
PSO J160.6037+56.9160 & DCT/LMI, \textbf{LCO/Spectral} & 57,741, \textbf{57,852, 58,122} & $g$ $r$ $i$ $z$ \\
PSO J161.2980+57.4038 & DCT/LMI & 57,741, \textbf{58,208} & $g$ $r$ $i$ $z$ \\
PSO J163.2331+58.8626 & DCT/LMI, \textbf{LCO/Spectral} & 57,741, \textbf{57,851, 58,123} & $g$ $r$ $i$ $z$ \\
PSO J185.8689+46.9752 & DCT/LMI, \textbf{LCO/Spectral} & 57,833, \textbf{57,858}, 58,075, \textbf{58,126, 58,269} & $g$ $r$ $i$ $z$ \\
PSO J213.9985+52.7527 & DCT/LMI & 57,833 & $g$ $r$ $i$ $z$ \\
PSO J214.9172+53.8166 & DCT/LMI, \textbf{LCO/Spectral} & 57,170, 57,284, 57,369, 57,522, \textbf{57,977} & $g$ $r$ $i$ $z$ \\
PSO J242.5040+55.4391  & DCT/LMI, \textbf{LCO/Spectral} & 57,522, 57,579, 57,641, \textbf{57,851, 57,977}, 58,012, \textbf{58,269} & $g$ $r$ $i$ $z$ \\
PSO J242.8039+54.0585 & DCT/LMI, \textbf{LCO/Spectral} & 57,522, 57,578, 57,642, \textbf{57,851, 57,977}, 58,012, \textbf{58,269} & $g$ $r$ $i$ $z$ \\
PSO J243.5676+54.9741 & DCT/LMI, \textbf{LCO/Spectral} & 57,522, 57,579, \textbf{57,851, 57,901, 57,976}, 58,012 & $g$ $r$ $i$ $z$ \\
PSO J333.0298+0.9687 & DCT/LMI, \textbf{LCO/Spectral} & 57,579, 57,641, \textbf{57,940, 57,990}, 58,012, 58,016 & $g$ $r$ $i$ $z$ \\
PSO J333.9832+1.0242 & DCT/LMI & 57,579, 57,641, \textbf{57,935}, 58,016, \textbf{58,269} & $g$ $r$ $i$ $z$ \\
PSO J334.2028+1.4075 & DCT/LMI & 57,170, 57,282, 57,284, 57,523, 57,579, 57,641, 57,682, & $g$ $r$ $i$ $z$ \\
 & & \textbf{57,935, 57,990}, 58,016 &  \\
PSO J351.5679--1.6795 & DCT/LMI, \textbf{LCO/Spectral} & 57,578, 57,641, 57,682, \textbf{57,940}, 58,016 & $g$ $r$ $i$ $z$ \\
\enddata
\tablecomments{Monitoring of the periodic candidates is being carried out in the SDSS $g$ $r$ $i$ $z$ filters on LMI at the DCT and the Spectral imager on the LCO network telescopes. To distinguish between the two telescopes, the MJDs of the observations on LCO/Spectral are in bold.}
\end{deluxetable*}


\begin{figure}
\centering
\epsfig{file=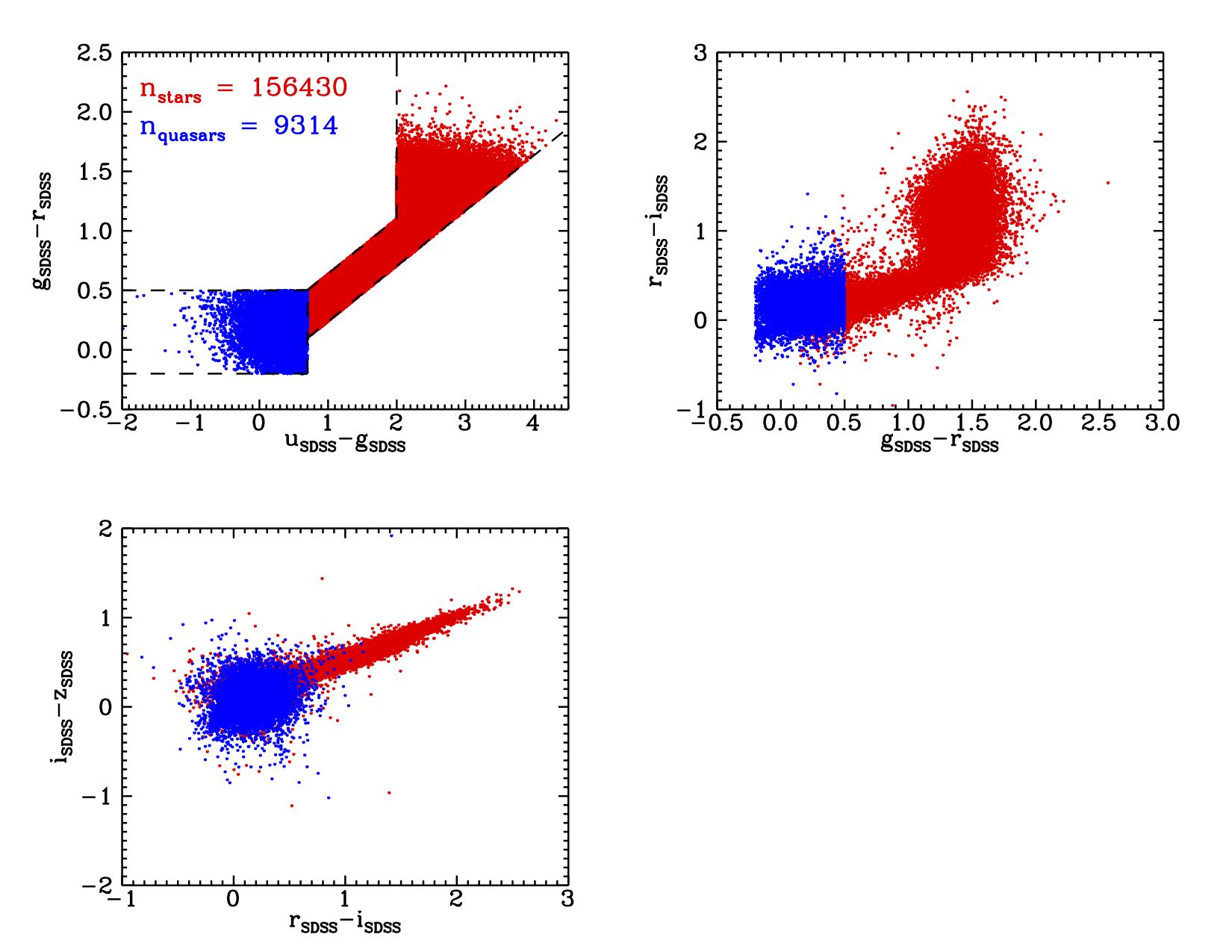,width=0.45\textwidth,clip=}
\caption{The CFHT $u$ and PS1 $griz$ magnitudes were first converted to the SDSS system, and quasars (blue dots) and stars (red dots) were selected by their $u_{\rm SDSS}-g_{\rm SDSS}$ and $g_{\rm SDSS}-r_{\rm SDSS}$ colors (dashed lines represent the color selection boxes).}
\label{fig:color-color}
\end{figure} 


To convert the SDSS magnitudes to the PS1 system, we adopt the same customized method in L16 that is suitable for quasar colors: we first calculate synthetic PS1 and SDSS magnitudes by convolving the (redshifted) composite quasar spectrum from \cite{VandenBerk2001} with the respective filter sensitivity. We then apply the PS1-SDSS magnitude offset to the LMI measurements to obtain their magnitudes on the PS1 system. 

We have also included data from our monitoring program with the LCO, a global network of telescopes in both hemispheres. The observations were carried out with the Spectral imager on the 2 m class telescopes at the Haleakala Observatory on Maui, Hawaii, and the Siding Spring Observatory in Australia between 2017 April and 2018 May (Project IDs: NOAO2017AB-013, NOAO2018A-004; PI: Liu) in the $g_{\rm SDSS}$, $r_{\rm SDSS}$, and $i_{\rm SDSS}$ filters (Table \ref{tab:extended}). The LCO images have been reduced by the \texttt{BANZAI} pipeline and are retrieved from the LCO Science Archive. Coadding of the subexposure and photometry on the coadded image are then run on the same custom-developed pipeline that we apply to LMI data. However, due to the smaller FOV of the 2 m class LCO telescope ($10'\times10'$) and shallower magnitude depth ($\sim 22$ mag), we instead obtain $\sim$$50$--$100$ SDSS cross-matched point sources on each image, and we avoid faint detections and potential saturated detections by excluding objects with $m > 21$ mag or $m<15$ mag when performing photometry. The same color correction for DCT/LMI is then applied to LCO/Spectral data before they are combined with PS1.


\section{Methods}\label{sec:methods}

\subsection{Color and Variability Selection of Quasars}\label{sec:sel_qso}

We first extract sources from the catalog from the PS1 Science Archive that meet the same criteria in L16 for MD09 data: (1) they are point sources (defined as deep stack mag$_{\rm psf}-$mag$_{\rm Kron}$$<0$) with good point-spread function (PSF) quality factors (\texttt{psfQF} $> 0.85$), (2) they have at least five detections, and (3) the same quality flags in L16 were applied to exclude bad or poor detections. The query returns $\sim 30,000$ sources from each MD field.

We then cross-match the PS1 sources with a catalog of deep stacked images in the Canada--France--Hawaii Telescope (CFHT) $u$ band and the PS1 $grizy$ bands (hereafter the PS1$\times$CFHT catalog; \citealt{Heinis2016AGN}) using a $1''$ radius. To extract point sources from the PS1$\times$CFHT catalog, we used the star/galaxy classification in the catalog that has been trained on a \emph{Hubble Space Telescope} Advanced Camera for Surveys sample of stars and galaxies \citep{Heinis2016SVM}. We then convert the $u_{\rm CFHT}$-, $g_{\rm P1}$-, and $r_{\rm P1}$-band magnitudes to the SDSS system, so that the quasar selection box in SDSS colors from \cite{Sesar2007} can be directly applied. This results in $\sim$ 9000 color-selected quasars in $\sim$ 50 $\degr^{2}$ of the total cross-matched sky area (Figure \ref{fig:color-color}). 


\begin{figure}[h] 
\centering
\epsfig{file=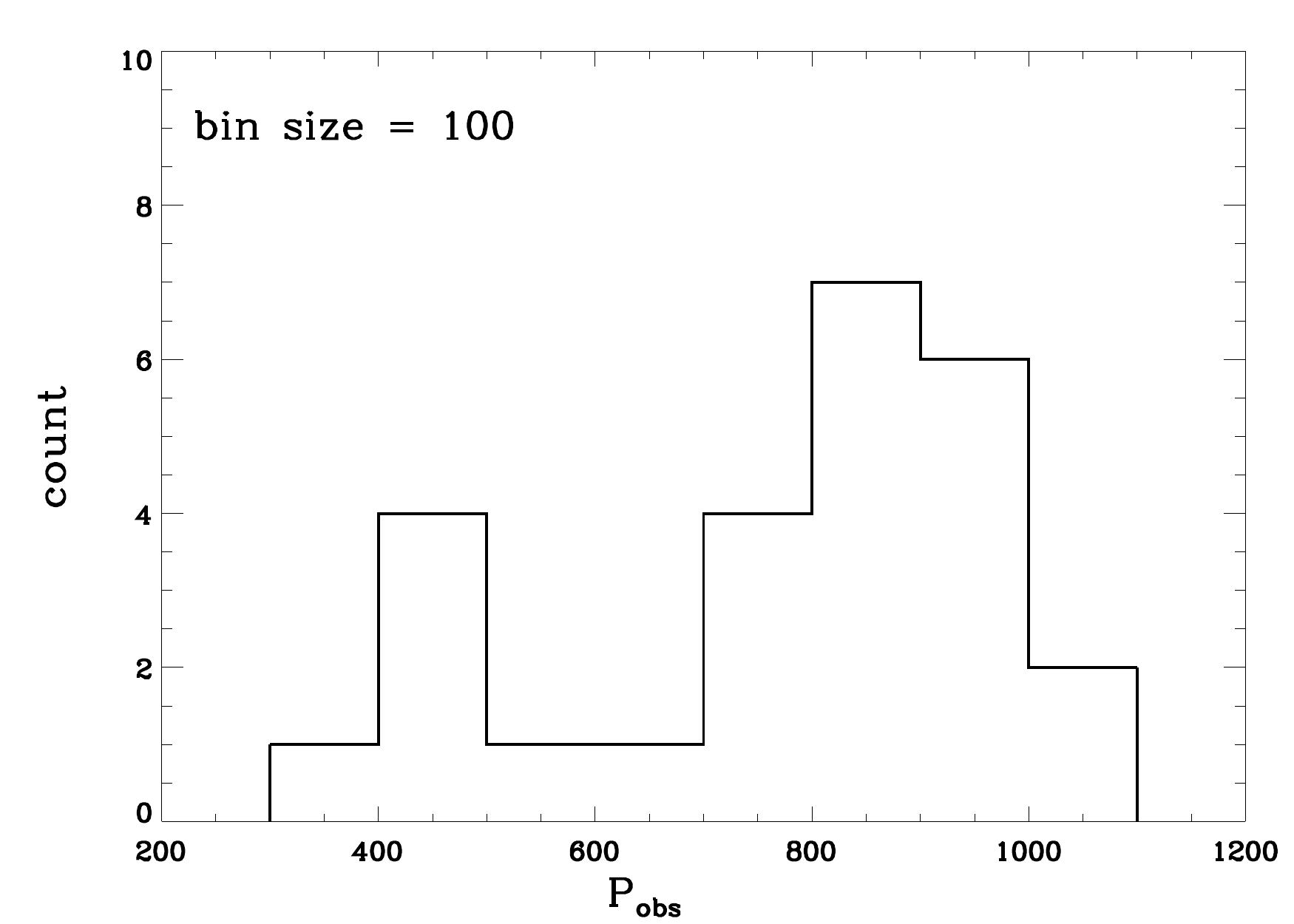,width=0.45\textwidth,clip=}
\caption{Distribution of $P_{\rm obs}$, the observed period determined by the LS periodogram.}
\label{fig:pobs_dist}
\end{figure} 


We then follow the method in L16 to select variable quasars: we construct an ensemble of objects within $\Delta$R.A. = $0\degr5$ and $\Delta$decl. = $0\degr5$ from each color-selected quasar. Then, in each filter, we compute the standard deviation $\sigma$ of the light curve for each object in the ensemble and iteratively exclude outliers by fitting a piecewise linear function to the $\sigma$--$m$ relation: $\sigma = \sigma(m)$. While most objects in the ensemble are stars and follow a tight $\sigma$--$m$ trend, intrinsic variable objects such as quasars have significantly larger $\sigma$ than stars of similar brightness and thus would appear as outliers from the trend. We identify the quasars with standard deviation $> 2 \sigma(m)$ in at least two filters as variables, and $\sim$1400 out of the $\sim$9000 color-selected quasars are identified as variable quasars.

We note that this fraction ($\sim 15$\%) of quasars being selected as variable is consistent with the anticorrelation of AGN variability amplitude with luminosity being processed through our pipeline (L16). We also note that optical colors (including the $u$ band) as a quasar selection technique is highly efficient ($\sim 98$\%) out to $z\sim2.7$ with $\sim 93$\% completeness, while combining color and multiband variability has an $\sim 97$\% efficiency and $\sim 97$\% completeness (e.g. \citealt{Peters2015}). As we will also show in our spectroscopic follow-up in Section \ref{sec:bhmass}, $100$\% of our candidates are spectroscopically confirmed as quasars. We will further discuss the effect of the incompleteness of the quasar sample on the detection rate in Section \ref{sec:rate}.


\begin{deluxetable*}{lrrrrrrrrrrr}
\tablecaption{MD Fields by the Numbers \label{tab:bythenumbers}}
\tablehead{
\colhead{Category} & \colhead{MD01} & \colhead{MD02} & \colhead{MD03} & \colhead{MD04} & \colhead{MD05} & \colhead{MD06} & \colhead{MD07} & \colhead{MD08} & \colhead{MD09} & \colhead{MD10} & \colhead{Full MDS}
}
\startdata
PS1 point sources & 30,109 & 28,845 & 31,350 & 32,661 & 29,517 & 34,112 & 29,031 & 38,194 & 40,488 & 28,455 & \nodata \\
PS1$\times$CFHT quasars & 983 & 1147 & 942 & 1030 & 1083 & 854 & 815 & 1013 & 670 & 777 & 9314 \\ 
PS1$\times$CFHT variable quasars & 109 & 112 & 202 & 200 & 163 & 115 & 120 & 138 & 104 & 106 & 1369 \\
Coherent periodogram peaks & 88 & 97 & 134 & 158 & 102 & 77 & 84 & 98 & 77 & 68 & \nodata \\
$\xi>$3.0 in at least one filter & 5 & 3 & 7 & 11 & 3 &1 & 3 & 5 & 6 & 3 & \nodata \\ 
$N_{\rm cycle}>$ 1.5 & 2 & 1 & 4 & 6 & 3 & 1 & 2 & 3 & 3 & 1 & 26 \\
\enddata
\end{deluxetable*}

\begin{deluxetable*}{lrrc}
\tablecaption{Period, Significance Factors, and Number of Cycles of Periodic Candidates \label{tab:P_xi}}
\tablehead{
\colhead{PS1 Designation} & \colhead{$P_{\rm LS}\pm \Delta P$ (day)} & \colhead{$\xi$ ($g r i z$)}  & \colhead{$N_{\rm cycle}$}
}
\startdata
PSO J35.7068--4.2314 & 427$\pm$4 & (3.6 3.1 3.6 2.2) & 3.6   \\
PSO J35.8704--4.0263& 829$\pm$23 & (3.5 3.8 3.5 2.0) & 1.9 \\
PSO J52.6172--27.6268 & 992$\pm$33 & (5.0 5.6 4.9 2.9) & 1.6  \\
PSO J129.4288+43.8234 & 313$\pm$5 & (2.6 3.2 1.9 1.8) & 4.9  \\
PSO J130.9953+43.7685 & 717$\pm$18 & (2.9 3.1 3.0 2.7) & 2.2   \\
PSO J131.1273+44.8582 & 843$\pm$31 & (3.5 3.5 3.0 2.1) & 1.8   \\
PSO J131.7789+45.0939 & 697$\pm$18 & (3.2 3.0 2.0 1.0) & 2.2   \\
PSO J148.8485+1.8124 & 816$\pm$5 & (3.7 4.0 2.9 1.4) & 1.9  \\
PSO J149.4989+2.7827 & 960$\pm$8 & (2.0 2.7 3.1 2.2) & 1.6  \\
PSO J149.2447+3.1393 & 810$\pm$8 & (4.0 3.1 2.0 1.2) & 1.9   \\
PSO J149.9400+1.5090 & 417$\pm$5 & (2.8 3.3 2.9 1.6) & 3.7    \\
PSO J149.6873+1.7192 & 820$\pm$5 & (2.8 4.3 4.5 3.3) & 1.9   \\
PSO J150.9191+3.3880 & 741$\pm$9 & (1.9 2.7 3.8 2.6) & 2.1 \\
PSO J160.6037+56.9160 & 988$\pm$17 & (3.0 2.0 1.6 1.2) & 1.6 \\
PSO J161.2980+57.4038 & 982$\pm$10 & (3.7 3.2 2.9 1.6) & 1.6   \\
PSO J163.2331+58.8626 & 1000$\pm$13 & (2.1 3.2 3.3 2.1) & 1.5   \\
PSO J185.8689+46.9752 & 958$\pm$19 & (3.3 2.9 2.1 1.6) & 1.6  \\
PSO J213.9985+52.7527 & 727$\pm$22 & (5.2 5.0 3.7 2.5) & 2.2    \\
PSO J214.9172+53.8166 & 1003$\pm$21 & (4.0 4.4 4.0 2.4) & 1.6   \\
PSO J242.5040+55.4391 & 862$\pm$24 & (2.9 3.5 2.8 2.0) & 1.8   \\
PSO J242.8039+54.0585 & 735$\pm$22 & (3.2 2.8 2.1 1.4) & 2.1   \\
PSO J243.5676+54.9741 & 984$\pm$17 & (3.2 2.6 1.2 0.4) & 1.6    \\
PSO J333.0298+0.9687 & 428$\pm$12 & (3.5 2.8 2.8 1.1) & 3.8    \\
PSO J333.9833+1.0242 & 466$\pm$11 & (3.9 2.6 2.2 1.3) & 3.5   \\
PSO J334.2028+1.4075 & 556$\pm$17 & (3.8 2.7 1.8 0.9) & 2.8    \\
PSO J351.5679--1.6795 & 805$\pm$6 & (1.9 2.0 3.2 2.5) & 1.9  \\
\enddata
\end{deluxetable*}


\subsection{Searching for Periodicity}\label{sec:sel_period}

To search for periodicity among the variable quasars, we compute the Lomb--Scargle (LS) periodogram \citep{Lomb1976,Scargle1982,Horne1986} and take advantage of the multifilter observations and their different sampling to determine a coherent periodic signal by a ``majority vote.'' We then define the best period as $P_{\rm LS} = \displaystyle\sum^{N}_{i}(P_{i})/N$, where $i = 1...N$ is the index of the filter in which a coherent period has been detected, and the uncertainty of $\bar{P}$ is determined from the uncertainty in each filter: $(\Delta P)^2 = (\sqrt{\sum{\delta P_{i}^2}}/N)^2+\sum{(P_{i}-\bar{P})^2}/(N-1)$, where the $\delta P$ in each filter is given by the uncertainty in the frequency $\delta \omega = 3 \pi \sigma/(2\sqrt{N_{0}}TA)$ \citep{Horne1986}.  We also calculate a signal-to-noise (S/N) $\xi=A_{0}^2/(2\sigma_{r}^2)$, where $\sigma_{r}$ is the standard deviation of the residual after a signal of amplitude $A_{0}$ is fitted to and subtracted from the data. We only select periodic candidates with high significance by requiring $\xi >$ 3 in at least one filter and require that the periodic variation has at least $1.5$ cycles over the $4$ yr PS1 baseline, where $N_{\rm cycle}$ is simply defined as [max(MJD)$-$min(MJD)]/$P_{\rm LS}$\footnote{Here the number of cycles gives a quantitative description of the periodic candidate and does not imply actual periodicity.}. The search results in $26$ periodic candidates from $10$ MD fields. We note that the significance $\xi$ is calculated against white noise and is only used as a preliminary cut, whereas the significance of the periodic signal against a background of colored noise is determined in Section \ref{sec:lkhd}. 

In Table \ref{tab:bythenumbers}, we break down the number from each step of the selection pipeline by the MD field, and the $P_{\rm LS}$, $\xi$, and $N_{\rm cycle}$ of the candidates are tabulated in Table \ref{tab:P_xi}. We note that only one candidate (PSO J129.4288+43.8234) has an observed period that is comparable to $1$ yr, indicating that our sample is not severely contaminated by the aliasing effect of the large seasonal gap. We note, however, that the distribution skews toward long periods (Figure \ref{fig:pobs_dist}), suggesting the possible effects of red noise \citep{MacLeod2010, Vaughan2016}. Tests of these periodic candidates against red noise will thus be performed in Section \ref{sec:lkhd}.


\subsection{Extended Baseline Analysis and a Maximum Likelihood Approach}\label{sec:lkhd}

As has been pointed out by \cite{Vaughan2016}, red noise can easily mimic a periodic variation over a small number of cycles ($N_{\rm cycle} \sim 3$), especially when the sampling is sparse and uneven and the photometric uncertainty is large. Therefore, efforts to systematically search for periodically varying quasars (e.g., \citealt{Graham2015, Charisi2016}) are limited by the several-years-long baseline of the survey, and it is essential to test the persistence of periodicity with long-term monitoring. Our extended baseline analysis of the periodic candidates from MD09 presented in L16 and of PG 1302$-$102 in \cite{Liu2018a} further demonstrated the necessity. Thus, in this work, we put our full sample of candidates to the test over an extended baseline, using the new imaging data we have described in Section \ref{sec:extended}.

In Figure \ref{fig:hist_ncycles}, we demonstrate the improvement on the temporal coverage of the candidates: while most PS1-only light curves only have $\sim$ two cycles, the LMI and LCO monitoring data extended the baseline to about three to four cycles, and, in the cases where archival SDSS Stripe 82 light curves are also available, as long as $\approx 15$ cycles\footnote{We stress here again that the number of cycles quantifies the total length of the light curve and that a ``cycle'' over the extended baseline does not imply temporal coverage comparable to PS1 MDS.}. We show the PS1 and extended light curves of the full sample in Appendix \ref{app:lc}.

Additionally, we have assumed the null hypothesis of white noise when searching for a periodic signal with the LS periodogram (Section \ref{sec:sel_period}). However,  quasar variability is known to be stochastic and has the characteristic of ``red noise,'' where variability power increases on longer timescales. Therefore, we will reevaluate the significance of our periodic candidates using a maximum likelihood method and investigate whether a periodic component is justified if a red noise background is also present. A similar approach has been applied to the periodic quasar candidate PG 1302$-$102 by \cite{D'Orazio2015}, and here we leverage our newly obtained monitoring data to put a more rigorous test on the periodic candidate. We refer the reader to \cite{Liu2018a} for details on this procedure, which is also described below using the widely adopted damped random walk (DRW; \citealt{Kelly2009}) model of stochastic AGN variability for illustration. 

We first assume the null hypothesis that the light curve is characterized by the DRW process, which has a short-timescale variation parameter and a characteristic timescale. The power spectral density (PSD) of a DRW process is in the form of a bending power law parameterized by a normalization and a break frequency, and its low- and high-frequency slopes are fixed at $\alpha = 0$ and $2$, respectively ($P(f) \propto f^{-\alpha}$). The PSD is then used to calculate the likelihood function ($\ln \mathcal{L}$) given the data. A model in which a periodic signal is superimposed on DRW noise (``DRW+periodic'') includes two additional parameters: amplitude and period of the signal. Note that the simpler model is nested within the more complex model. We therefore down-select candidates that meet the following criteria:


\begin{figure}[h] 
\centering
\epsfig{file=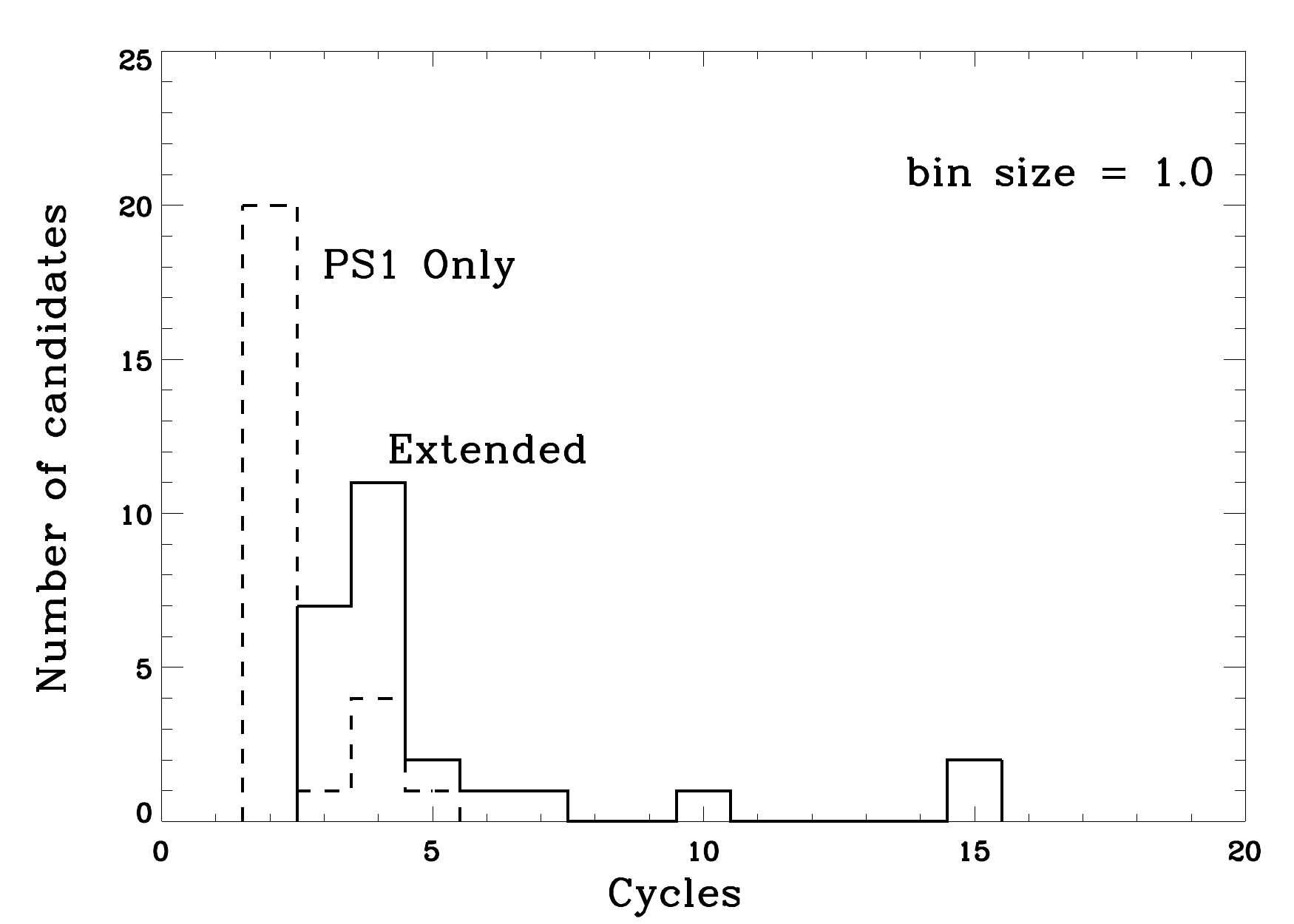,width=0.45\textwidth,clip=}
\caption{While most candidates have only $\sim$ two cycles in their PS1-only light curves (dashed histogram), we have extended the baseline to $>$ three cycles with new imaging data from DCT/LMI and LCO/Spectral and archival SDSS data (solid histogram).}
\label{fig:hist_ncycles}
\end{figure} 


\begin{enumerate}
\item $\ln\mathcal{L}_{\rm DRW+periodic}>\ln\mathcal{L}_{\rm DRW}$ for both PS1-only and extended light curves;
\item ($\ln\mathcal{L}_{\rm DRW+periodic}-\ln\mathcal{L}_{\rm DRW})_{\rm extended} >  \\ (\ln\mathcal{L}_{\rm DRW+periodic}-\ln\mathcal{L}_{\rm DRW})_{\rm PS1-only}$ or, equivalently, $p_{\rm extended}<p_{\rm PS1-only}$;
\item $P_{\rm extended}$ = $P_{\rm PS1-only}$ = $P_{\rm LS}$ within their uncertainties; and
\item $p<\frac{1}{N}$; where $N = 9314$ is the size of the initial sample of quasars,
\end{enumerate}

\noindent where the maximum likelihoods ($\ln\mathcal{L}_{\rm DRW}$ and \\ $\ln\mathcal{L}_{\rm DRW+periodic}$) were obtained by exploring the parameter space using a Markov chain Monte Carlo sampler. While the DRW+periodic model may be preferred by the data (criterion 1), the chance probability of mistaking pure DRW noise for a signal\footnote{Here the signal is superimposed on red noise.} can be quantified by a $p$-value, since $-2 \Delta\ln\mathcal{L}$ is $\chi^{2}$ distributed where the degree of freedom is the number of additional parameters in the more complex model. Based on our expectations for a true periodic signal, $p$ should decrease over a longer baseline (criterion 2). Additionally, we impose that the period should be consistent with the one determined by the LS periodogram (criteria 3), and that the candidate should be statistically significant, having been selected from a large sample of quasars (criterion 4).


\begin{figure}[h] 
\centering
\epsfig{file=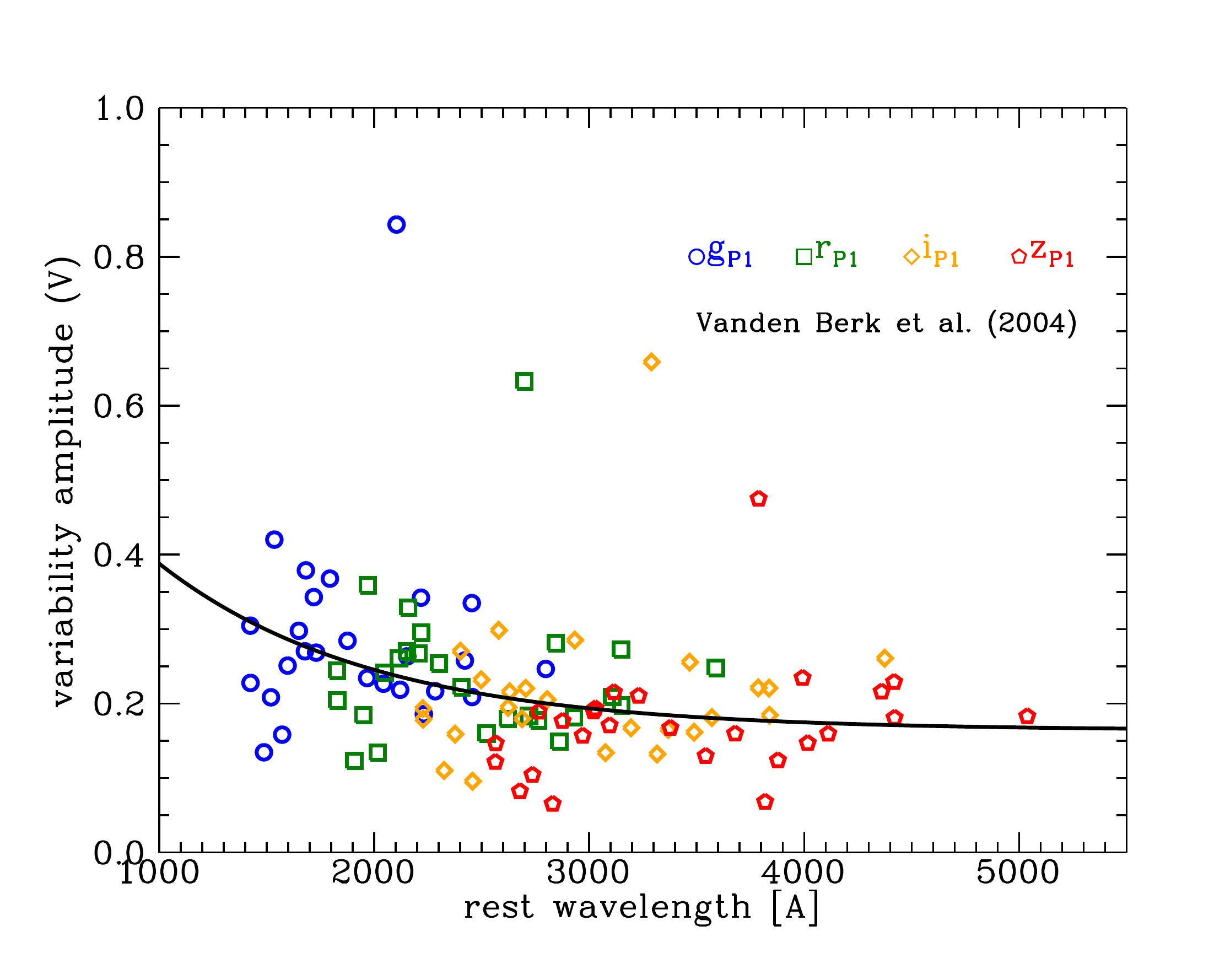,width=0.45\textwidth,clip=}
\caption{We measure the variability amplitude of each candidate in each filter after subtracting the measurement uncertainty in quadrature ($g_{\rm P1}$: blue circles; $r_{\rm P1}$: green squares; $i_{\rm P1}$: orange diamonds; $z_{\rm P1}$: red pentagons). The amplitude $V$ decreases with longer rest wavelength, consistent with the exponential relation from \cite{VandenBerk2004} (black curve).}
\label{fig:varamp}
\end{figure} 


We note that by applying our methods above, we have limited our periodicity search to simple sinusoids. There are ample possibilities where the periodic variation of an SMBHB can deviate from a simple sinusoid: when the orbital eccentricity is imprinted on the line-of-sight velocity and hence the Doppler modulation \citep{D'Orazio2015}, or the ``bursty'' variations predicted in hydrodynamical simulations of binaries of various binary mass ratios \citep{D'Orazio2013,Farris2015}. However, those deviations should only be a second-order effect in our analysis.

\section{Results}\label{sec:cand}

\subsection{Full Sample: Variability Amplitudes}\label{sec:var}

To compare with the relation of variability amplitude vs. rest-frame wavelength of the full sample of $26$ candidates with the previous study of normal AGNs by \cite{VandenBerk2004}, we calculate the rest wavelength of a PS1 filter at the redshift of each quasar ($\lambda_{\rm eff} (g) = 4810$  \AA, $\lambda_{\rm eff} (r) = 6170$  \AA, $\lambda_{\rm eff} (i) = 7520$  \AA, $\lambda_{\rm eff} (z) = 8660$  \AA) and define an intrinsic variability amplitude $V = \sqrt{\pi(\Delta m)^{2}/2-\sigma^2}$, where $\Delta m$ is the amplitude $A_{0}$ obtained from our sinusoidal fit, and the magnitude--dependent observed scatter from stars is used as a proxy for $\sigma$ (see L16). The intrinsic variability amplitude $V$ of our candidates decreases with longer rest wavelength, which is consistent with the empirical relation from \cite{VandenBerk2004} and has no apparent deviation from regular AGNs (Figure \ref{fig:varamp}). We note, however, the exception of PSO J334.0298+0.9687, which shows much larger variability amplitudes in all filters and an apparently steeper amplitude--wavelength trend; a visual inspection of its light curves also shows a large variation ($\sim 0.8$ mag in the $g$ band). The amplitudes of the best-fit sinusoids ($A_{0}$), as well as mean PS1 magnitudes, are listed in Table \ref{tab:var_amp}.

We note that the variability amplitude of a Doppler--boosted SMBHB should follow the relation $A_{\rm UV}/A_{\rm opt} = (3-\alpha_{\rm UV})/(3-\alpha_{\rm opt})$, where $\alpha_{\rm UV}$ and $\alpha_{\rm opt}$ are the spectral slopes in the UV and optical bands, respectively \citep{D'Orazio2015}. In fact, this relation has been applied in \cite{Charisi2018} to test the Doppler-boost hypothesis of reported periodic candidates whenever their UV data are available. However, we do not see evidence that the wavelength-dependent variability amplitudes of our candidates deviate from those of normal quasars, as shown in Figure \ref{fig:varamp}.


\begin{deluxetable*}{lll}
\tablecaption{PS1 Mean Magnitudes and Variability Amplitudes of Periodic Quasar Candidates \label{tab:var_amp}}
\tablehead{
\colhead{PS1 Designation} & \colhead{$m$ ($g,r,i,z$)} & \colhead{$A_{0}$ ($g,r,i,z$)} 
}
\startdata
PSO J35.7068--4.23144 & (19.69, 19.64, 19.69, 19.53) & (0.23, 0.18, 0.23, 0.14) \\
PSO J35.8704--4.0263 & (19.52, 19.46, 19.52, 19.23) & (0.24, 0.21, 0.24, 0.13) \\
PSO J52.6172--27.6268 & (20.37, 20.20, 20.14, 19.93) & (0.34, 0.29, 0.22, 0.16) \\
PSO J129.4288+43.8234 & (19.53, 19.37, 19.50, 19.48) & (0.17, 0.16, 0.15, 0.15) \\
PSO J130.9953+43.7685 & (19.88, 19.65, 19.81, 19.88) & (0.21, 0.17, 0.18, 0.18) \\
PSO J131.1273+44.8582 & (20.57, 20.42, 20.12, 19.87) & (0.21, 0.20, 0.19, 0.15) \\
PSO J131.7789+45.0939 & (20.62, 20.29, 20.29, 20.37) & (0.22, 0.15, 0.14, 0.12) \\
PSO J148.8485+1.8124 & (20.43, 20.17, 20.10, 19.88) & (0.25, 0.20, 0.16, 0.11) \\
PSO J149.4989+2.7827 & (20.34, 20.25, 20.24, 20.04) & (0.19, 0.17, 0.15, 0.13) \\
PSO J149.2447+3.1393 & (20.72, 20.72, 20.48, 20.45) & (0.31, 0.27, 0.18, 0.17) \\
PSO J149.9400+1.5090 & (20.17, 19.91, 20.00, 20.09) & (0.18, 0.15, 0.15, 0.14) \\
PSO J149.6873+1.7192 & (20.42, 20.12, 20.08, 20.08) & (0.19, 0.15, 0.14, 0.14) \\
PSO J150.9191+3.3880 & (19.63, 19.49, 19.39, 19.20) & (0.20, 0.20, 0.21, 0.15) \\
PSO J160.6037+56.9160 & (19.52, 19.33, 19.28, 19.33) & (0.19, 0.13, 0.11, 0.11) \\
PSO J161.2980+57.4038 & (20.45, 20.44, 20.18, 20.22) & (0.28, 0.22, 0.15, 0.15) \\
PSO J163.2331+58.8626 & (19.59, 19.48, 19.43, 19.19) & (0.17, 0.15, 0.13, 0.09) \\
PSO J185.8689+46.9752 & (20.54, 20.50, 20.23, 20.28) & (0.30, 0.21, 0.17, 0.18) \\
PSO J213.9985+52.7527 & (19.94, 20.13, 19.90, 19.89) & (0.22, 0.22, 0.16, 0.16) \\
PSO J214.9172+53.8166 & (20.53, 20.32, 20.39, 20.44) & (0.28, 0.23, 0.21, 0.20) \\
PSO J242.5040+55.4391 & (20.17, 20.17, 19.91, 19.95) & (0.22, 0.24, 0.18, 0.18) \\
PSO J242.8039+54.05853 & (19.72, 19.64, 19.87, 19.89) & (0.27, 0.22, 0.18, 0.19) \\
PSO J243.5676+54.9741 & (19.97, 19.64, 19.58, 19.61) & (0.18, 0.15, 0.11, 0.07) \\
PSO J333.0298+0.9687 & (21.42, 20.94, 20.96, 20.95) & (0.68, 0.51, 0.53, 0.39) \\
PSO J333.9832+1.0242 & (18.97, 18.85, 18.79, 18.57) & (0.11, 0.10, 0.09, 0.07) \\
PSO J334.2028+1.4075 & (19.38, 19.28, 19.14, 18.94) & (0.13, 0.11, 0.08, 0.06) \\
PSO J351.5679--1.6795 & (18.91, 18.56, 18.54, 18.67) & (0.15, 0.12, 0.13, 0.12) \\
\enddata
\end{deluxetable*}


\subsection{Full Sample: Spectroscopy and Black Hole Mass}\label{sec:bhmass}

We retrieved archival spectra of $16$ candidates from the SDSS Science Archive Server. The remaining candidates with no archival spectra were observed at the Gemini-South Telescope (PI: Liu) or the DCT. The Gemini spectra were obtained with the R400 slit with GMOS, while the DCT spectra were obtained with the DeVeny spectrograph with a 300 g mm$^{-1}$ grating. We summarize the details of the observations in Table \ref{tab:gspec_dspec}. The Gemini/GMOS spectra were reduced with the Gemini \texttt{IRAF} package, and the DCT/DeVeny data were reduced with standard \texttt{IRAF} procedures.


\begin{deluxetable*}{llcccc}
\tablecaption{Spectroscopic Follow-ups \label{tab:gspec_dspec}}
\tablehead{
\colhead{PS1 Designation} & \colhead{Telescope/Instrument} & \colhead{Semester or Quarter} & \colhead{Grating} & \colhead{Slit Width} & \colhead{Exposure Time} \\
\colhead{} & \colhead{} & \colhead{} & \colhead{} & \colhead{(arcsec)} & \colhead{(s)}
}
\startdata
PSO J52.6172--27.6268 & Gemini/GMOS & 16B (Gemini ID: GS-2016B-Q-50) & R400 & 0.75 & 2$\times$1000 \\
PSO J149.2447+3.1393 & Gemini/GMOS & 15B (Gemini ID: GS-2015B-Q-42) & R400 & 0.75 & 2$\times$1000 \\
PSO J149.6873+1.7192 & DCT/DeVeny & 17Q1 & 300 g mm$^{-1}$ & 1.5 & 2$\times$2000 \\
PSO J161.2980+57.4038 & DCT/DeVeny & 17Q1 & 300 g mm$^{-1}$ & 1.5 & 2$\times$1700 \\
PSO J163.2331+58.8626 & DCT/DeVeny & 17Q1 & 300 g mm$^{-1}$ & 1.5 & 2$\times$1800 \\
PSO J242.5040+55.4391  & DCT/DeVeny & 17Q1 & 300 g mm$^{-1}$ & 1.5 & 2100 \\
PSO J243.5676+54.9741 & DCT/DeVeny & 16Q3 & 300 g mm$^{-1}$ & 1.5 & 2$\times$900 \\
PSO J333.0298+0.9687 & DCT/DeVeny & 15Q3 & 300 g mm$^{-1}$ & 1.5 & 1400 \\
PSO J334.2028+1.4075 & Gemini/GMOS & 15A (Gemini ID: GS-2015A-Q-17) & R400 & 0.75 & 720 \\
PSO J351.5679--1.6795 & DCT/DeVeny & 17Q2 & 300 g mm$^{-1}$ & 1.5 & 1200 \\
\enddata
\end{deluxetable*}

\begin{deluxetable*}{lllccccccc}
\tablecaption{Spectroscopic Measurements and Inferred Binary Parameters \label{tab:spec}}
\tablehead{
\colhead{PS1 Designation} & \colhead{Spectroscopy} & \colhead{$M_{\rm BH}$} & \colhead{$f_{\lambda}$} & \colhead{$\rm FWHM$} & \colhead{$\log{(M_{\rm BH})}$} & \colhead{$z$} & \colhead{$P_{\rm rest}$} & \colhead{$a$}  &  \colhead{$a$} \\
\colhead{} & \colhead{} & \colhead{Estimator}&  \colhead{$(\rm erg\,s^{-1}cm^{-2}\AA^{-1})$} & \colhead{$\rm (km\,s^{-1})$} & \colhead{($M_{\odot}$)} & \colhead{} & \colhead{$(\rm day)$} & \colhead{$(\rm pc)$} & \colhead{$\rm (R_{\rm s})$}
}
\startdata
PSO J35.7068--4.23144 & SDSS & \ion{Mg}{2} & 1.4$\times10^{-17}$ & 5185 & 8.7 & 1.564 & 167 & 0.002 & 47 \\
PSO J35.8704--4.0263 & SDSS & \ion{Mg}{2} & 3.3$\times10^{-17}$ & 3810 & 8.8 & 1.916 & 284 & 0.004 & 55 \\
PSO J52.6172--27.6268 & GS16B & \ion{Mg}{2} & $1.3\times10^{-17}$ & 7384 & 9.2 & 2.134 & 317 & 0.005 & 32 \\
PSO J129.4288+43.8234 & SDSS & \ion{Mg}{2} & 4.5$\times10^{-17}$ & 3744 & 8.3 & 0.959 & 160 & 0.002 & 80 \\
PSO J130.9953+43.7685 & SDSS & \ion{Mg}{2} & 4.1$\times10^{-17}$ & 3850 & 8.4 & 0.986 & 361 & 0.003 & 133 \\
PSO J131.1273+44.8582 & SDSS & \ion{Mg}{2} & 1.6$\times10^{-17}$ & 2450 & 8.3 & 2.011 & 280 & 0.002 & 126 \\
PSO J131.7789+45.0939 & SDSS & \ion{Mg}{2} & 2.0$\times10^{-17}$ & 6773 & 8.8 & 1.233 & 312 & 0.004 & 58 \\ 
PSO J148.8485+1.8124 & SDSS & \ion{Mg}{2}  & 7$\times10^{-18}$ & 5402 & 8.9 & 2.378 & 242 & 0.003 & 45 \\
PSO J149.4989+2.7827 & SDSS & \ion{C}{4} & 3.4$\times10^{-17}$ & 5173 & 9.1 & 2.376 & 284 & 0.004 & 38 \\
PSO J149.2447+3.1393 & GS15B & \ion{Mg}{2} & $8.6\times10^{-17}$ & 1955 & 8.5 & 1.859 & 283 & 0.003 & 94 \\
PSO J149.9400+1.5090 & SDSS & \ion{Mg}{2} & 2.4$\times10^{-17}$ & 3715 & 8.3 & 1.106 & 198 & 0.002 & 102 \\
PSO J149.6873+1.7192 & DCT17Q1 & \ion{Mg}{2} & 1.3$\times10^{-17}$ (n) & 5755 & 8.6 & 1.354 & 348 & 0.004 & 85  \\ 
PSO J150.9191+3.3880 & SDSS & \ion{Mg}{2} & 6.9$\times10^{-17}$ & 1995 & 7.7 & 0.719 & 431 & 0.002 & 426 \\
PSO J160.6037+56.9160 & SDSS & \ion{Mg}{2} & 3.7$\times10^{-17}$ & 3251 & 8.5 & 1.445 & 404 & 0.004 & 119 \\
PSO J161.2980+57.4038 & DCT17Q1 & \ion{Mg}{2} & 2.0$\times10^{-17}$(n) & 3043 & 8.5 & 1.798 & 351 & 0.003 & 114 \\
PSO J163.2331+58.8626 & DCT17Q1 & \ion{C}{4} & 6.7$\times10^{-17}$(n) & 5611 & 9.2 & 2.165 & 316 & 0.005 & 33 \\
PSO J185.8689+46.9752 & SDSS & \ion{Mg}{2} & 1.3$\times10^{-17}$ & 6070 & 8.9 & 1.681 & 357 & 0.004 & 59 \\
PSO J213.9985+52.7527 & SDSS & \ion{Mg}{2} & 1.5$\times10^{-17}$ & 4123 & 8.7 & 1.867 & 253 & 0.003 & 67 \\
PSO J214.9172+53.8166 & SDSS & \ion{Mg}{2} & 1.5$\times10^{-17}$ & 4907 & 8.4 & 1.169 & 462 & 0.004 & 142 \\
PSO J242.5040+55.4391  & DCT17Q1 & \ion{Mg}{2} & $1.9\times10^{-17}$ (n) & 5547 & 8.9 & 1.780 & 310 & 0.004 & 53 \\ 
PSO J242.8039+54.0585 & SDSS & \ion{Mg}{2} & 3.6$\times10^{-17}$ & 6581 & 8.8 & 0.960 & 375 & 0.004 & 70 \\
PSO J243.5676+54.9741 & DCT16Q3  & \ion{Mg}{2} & $3.5\times10^{-17}$(n) & 2041 & 8.0 & 1.268 & 434 & 0.002 & 280 \\
PSO J333.0298+0.9687 & DCT15Q3 & \ion{Mg}{2} & $2.4\times10^{-17}$ & 8851 & 9.2 & 1.284 & 244  & 0.004 & 28 \\
PSO J333.9833+1.0242 & SDSS & \ion{Mg}{2} & 4.2$\times10^{-17}$ & 6157 & 9.5 & 2.234 & 144 & 0.003 & 13 \\
PSO J334.2028+1.4075 & GS15A & \ion{Mg}{2} & $1.9\times10^{-17}$ & 5492 & 9.1 & 2.070 & 182 & 0.003 & 28 \\
PSO J351.5679--1.6795 & DCT17Q2  & \ion{Mg}{2} & $10.7\times10^{-17}$ & 4702 & 8.9 & 1.156 & 373 & 0.005 & 59 \\
\enddata
\tablecomments{Those flux measurements that were made from the re-normalized DeVeny spectra are indicated by (n).}
\end{deluxetable*}


Due to the variable weather conditions under which the spectra were taken, a standard star may not accurately calibrate the science object's flux. Therefore, in addition to the standard procedures to reduce the spectroscopic data, we also calibrate the object's flux to its latest photometric measurement. We first convolve the DeVeny spectrum with the SDSS $r$-filter sensitivity curve to calculate a synthetic magnitude $r'_{\rm SDSS}$; if it differs from the latest photometric measurement $r_{\rm SDSS}$ by more than the variability amplitude of the object --- where $r_{\rm SDSS}$ is either observed with DCT/LMI (see Section \ref{sec:extended}) or, in the absence of new observations, obtained from the SDSS Science Archive Server --- we then renormalize the spectrum to match its synthetic magnitude to $r_{\rm SDSS}$. The procedure is repeated iteratively until $|r'_{\rm SDSS}-r_{\rm SDSS}| < 0.05$ mag. We note that this renormalization procedure is unlikely to significantly bias our black hole mass estimates: a $\Delta m \sim 0.8$ mag intrinsic variability (which is on the order of the maximum variability amplitude in our sample of candidates) translates to a factor of $\sim 2$ difference in the continuum luminosity (assuming $z=1$), which in turn corresponds to an $\sim 0.2$ dex error on the black hole mass -- much smaller than the systematic uncertainty of black hole mass estimates. The spectra of all candidates (including the renormalized DeVeny spectra) are presented in Appendix \ref{app:spec}.

To measure a virial black hole mass from the spectrum, we first use the following procedure to measure the broad-line width of \ion{Mg}{2}: we fit a power-law continuum in the range [2200, 2675] and [2925, 3090] \AA\ and subtract it from the spectrum. We then broaden and scale the iron emission template from \cite{Vestergaard2001} by fitting it to the range [2250, 2650] \AA\ where iron emission is strong, which is then subtracted from the spectrum. In those spectra where S/N is low, we do not fit the iron emission to avoid overfitting and subtracting. Next, we fit a single Gaussian to the emission line in the range [2700, 2900] \AA\ and measure an FWHM. Although \cite{McLure2004} fit two components (broad and narrow) to the \ion{Mg}{2} line and adopted the broad component in the black hole mass estimate, we do not find the clear presence of a narrow component in every spectrum and thus only fit a single Gaussian. Then, we measure the flux density $f_{\lambda}$ at 3000 \AA\ in the fitted continuum and convert to a continuum luminosity: $\lambda L_{\lambda} = \lambda 4 \pi D_{\rm L}^2 f_\lambda (1+z)$. We also correct for Galactic extinction using the dust map by \cite{SF2011} and the extinction curve of \cite{Cardelli1989}. Finally, we substitute the FWHM and $\lambda L_{\lambda}$ into the following equation from \cite{McLure2004} to calculate the black hole mass:

\begin{equation*}
\frac{M_{\rm BH}}{M_\odot} = 3.2\,\Big(\frac{\rm FWHM(Mg II)}{\rm km\,s^{-1}}\Big)^2\Big(\frac{\lambda L_{\lambda} (3000\rm \AA)}{10^{44}\,\rm ergs\,s^{-1}}\Big)^{0.62}\quad.
\label{eqn:mgii}
\end{equation*}

In a spectrum where \ion{C}{4} is the black hole mass estimator, we fit the continuum in the range [1445, 1465] and [1700, 1705] \AA, and after subtracting the continuum, we adopt the procedure in \cite{Shen2008} and use a three-component fit to fully characterize the \ion{C}{4} line profile: a narrow component with FWHM $<$ 1200 km s$^{-1}$, a broad component with FWHM $>$ 1200 km s$^{-1}$, and a broader hump component. We then measure the FWHM from the fitted profile. The corresponding continuum luminosity is calculated from the mean flux density in the range [1340, 1360] \AA, and the black hole mass estimate is adopted from \cite{Vestergaard2006}:

\begin{align*}
\log\Big(\frac{M_{\rm BH}}{M_\odot}\Big) &= \log\Big[\Big(\frac{\rm FWHM(CIV)}{1000\,\rm km\,s^{-1}}\Big)^{2}\Big(\frac{\lambda L_{\lambda} (1350\rm \AA)}{10^{44}\,\rm ergs\,s^{-1}}\Big)^{0.53}\Big] \\
&\quad+6.66 \quad.
\label{eqn:civ}
\end{align*}

Typical examples from the above fitting procedures are demonstrated in the last two panels in Appendix \ref{app:spec} (\ion{Mg}{2} and \ion{C}{4}, respectively), and the measurements of $z$, $f_{\lambda}$, FWHM, and $M_{\rm BH}$ are listed in Table \ref{tab:spec}. 

We note that there are two caveats of our black hole mass estimate: first, the virial black hole masses obtained from \ion{Mg}{2} or \ion{C}{4} have a large systematic uncertainty of $\sim 0.3$ dex, and there are systematic biases between the two mass estimators (e.g., \citealt{Shen2008}). In addition, while \ion{Mg}{2} is considered a more reliable mass estimator than \ion{C}{4}, a fraction of objects have atypically broad \ion{Mg}{2} lines, i.e., FWHM(\ion{Mg}{2}) $>$ FWHM(H$\alpha$, H$\beta$), which cannot be used to reliably measure the black hole mass (e.g., \citealt{Mejia2016}). Second, by working under the SMBHB hypothesis, we are only able to obtain an estimate of the total black hole mass. In an unequal-mass binary, the secondary black hole is expected to be more actively accreting due to its easier access to gas \citep{Cuadra2009,Farris2015}. In this picture, the broad lines are assumed to be associated with the secondary, \footnote{In fact, this is the assumption in the spectroscopic search for SMBHBs by measuring the offsets and shifts of broad H$\beta$ lines (e.g., \citealt{Eracleous2012, Runnoe2017}).} and therefore the black hole mass estimated from \ion{Mg}{2} or \ion{C}{4} does not represent the total mass of the hypothesized binary system.

Nevertheless, we use the obtained black hole estimates to calculate inferred binary separations, noting that they are systematically underestimated under the above assumption. We calculate the separation $a$ via Kepler's law by assuming the variation is exactly on the rest-frame orbital period timescale $a^{3}/t_{\rm orb}^{2} = GM/4\pi^{2}$, where $t_{\rm orb} = P_{\rm obs}/(1+z)$ is the rest-frame orbital period. Those separations (in units of pc and $R_{s}$) are also included in Table \ref{tab:spec}, and they confirm that our time-domain search for SMBHBs is sensitive to milliparsec separations, which would correspond to the gravitational wave--emitting regime. However, we are unable to measure any period derivative due to gravitational radiation, likely due to the photometric error and short baseline of the available data and that the binaries have not evolved into the final inspiral stage.

As we also show in Table \ref{tab:spec}, the inferred separations of the candidates are more compact than the binary separations that current spectroscopic searches are sensitive to: distinct broad-line regions associated with the two members of the binary may be identified via the broad-line profile in a binary at an $\sim 0.01-0.1$ pc separation \citep{Shen2010}, while offset broad lines with shifts measured over $\sim$ years-long temporal baselines may indicate binaries at separations $>10^{2} r_{\rm g}$ \citep{Pflueger2018}. Thus, those inferred separations of our candidates are also consistent with the lack of unusual spectroscopic features in their spectra (Appendix \ref{app:spec}).


\begin{figure}[h]
\centering
\epsfig{file=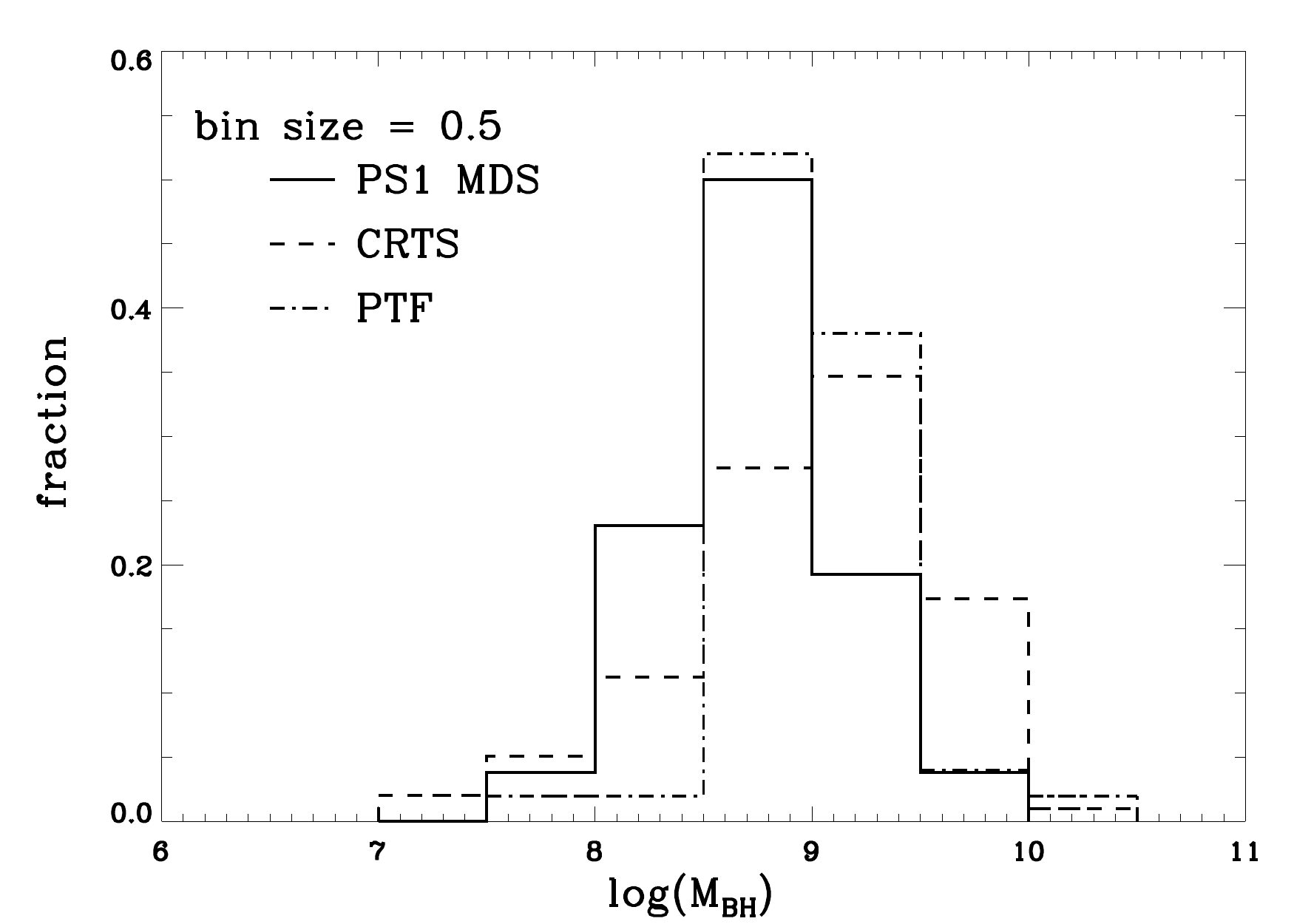,width=0.45\textwidth,clip=}
\epsfig{file=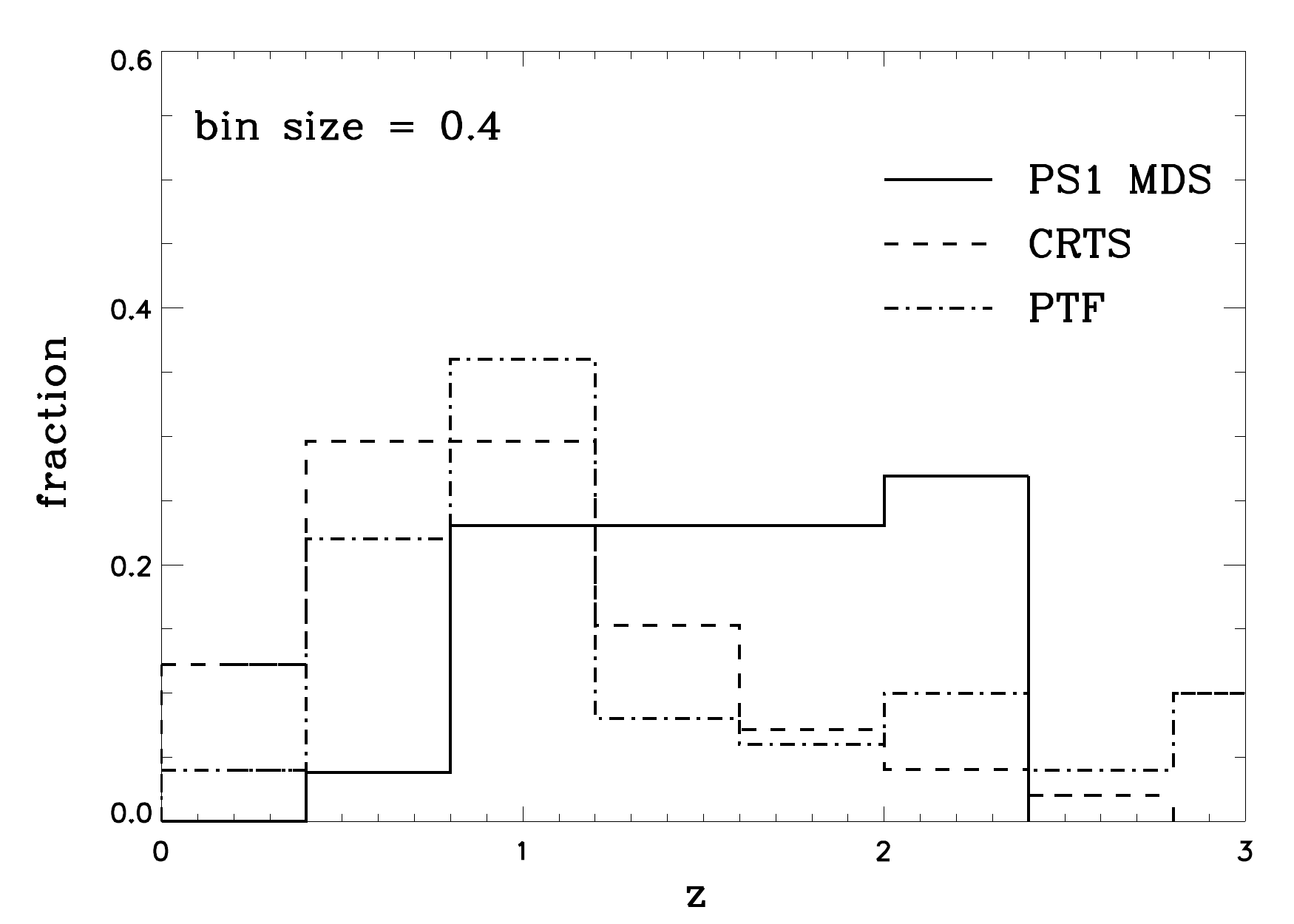,width=0.45\textwidth,clip=}
\caption{Upper panel: The black hole mass distribution of the candidates from PS1 MDS (solid histogram). It is similar to that of the candidates from \cite{Graham2015} and \cite{Charisi2016} (dashed and dashed-dotted histograms, respectively).
Lower panel: The redshift distribution of the candidates from PS1 MDS (solid histogram). Our selection is sensitive out to $z\sim2$, while the redshift distributions of the periodic candidates from CRTS and PTF peak at $z\sim1$ (dashed and dashed-dotted histograms, respectively).}
\label{fig:mbh_z_dist}
\end{figure} 

\subsection{Full Sample: Comparing with Previous Work}\label{sec:compare}

We now compare the physical properties of our candidates from PS1 MDS with those previously identified in CRTS \citep{Graham2015} and PTF \citep{Charisi2016}. The black hole masses in all three samples are in the range $\log (M_{BH}/M_{\odot}) \approx 8 - 10$, although our sample appears to include more objects with lower black hole masses (Figure \ref{fig:mbh_z_dist}, upper panel). As we also show (Figure \ref{fig:mbh_z_dist}, lower panel), our search with PS1 MDS is more sensitive to candidates at higher redshifts ($\langle\bar{z}\rangle \sim 2$) than CRTS or PTF ($\langle\bar{z}\rangle \sim 1$). In fact, the redshifts of MDS candidates follow an opposite trend to those of the variable quasars that our selection pipeline can detect (see L16), suggesting a selection bias toward high redshifts.

\begin{figure}[h] 
\centering
\epsfig{file=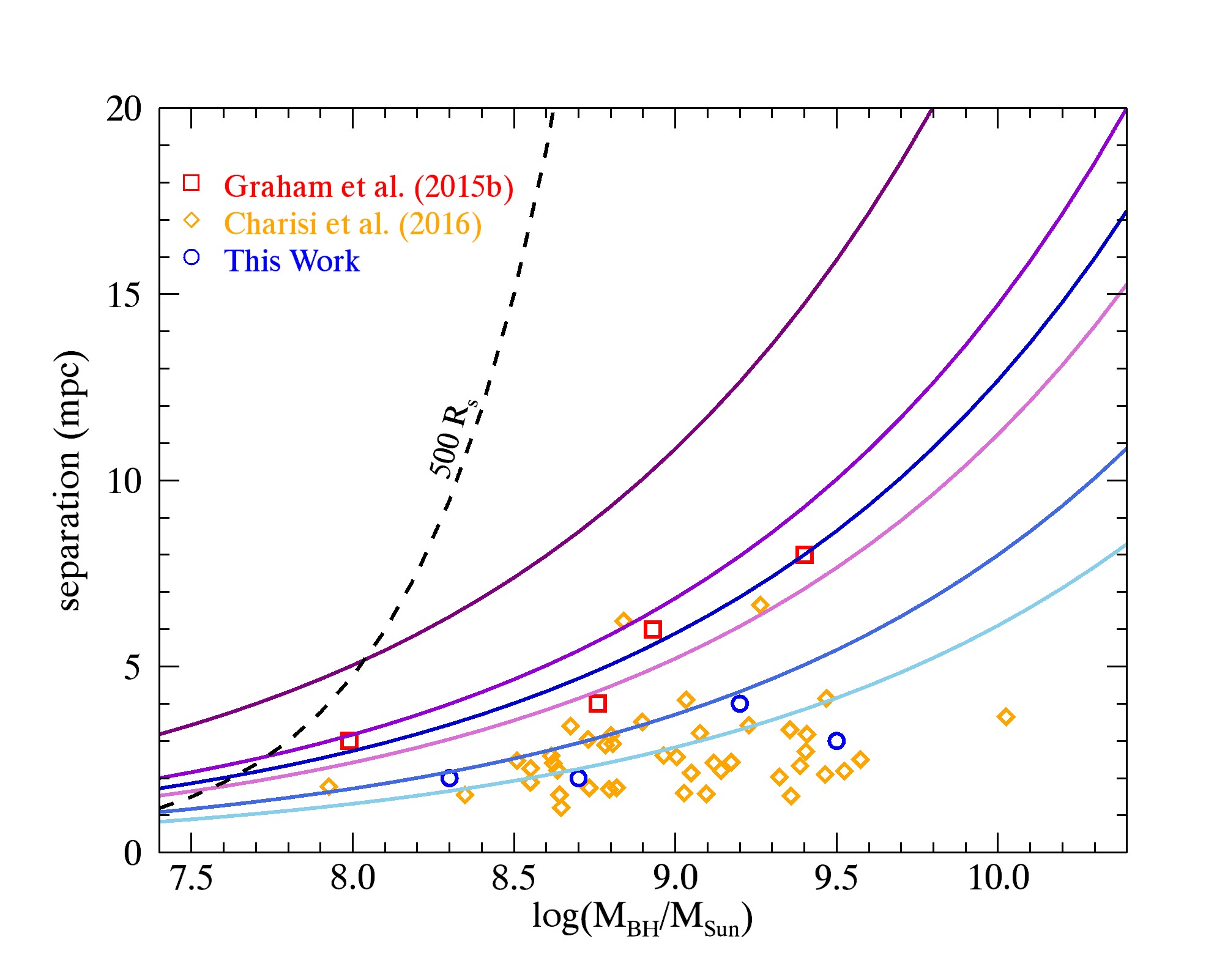,width=0.45\textwidth,clip=}
\caption{Black hole masses ($M_{\rm BH}$) and separations ($a$) of the periodic candidates from CRTS, PTF, and PS1 MDS with at least three cycles over their respective baselines (red squares, orange diamonds, and blue circles, respectively). The blue solid curves represent the parameter space occupied by periodic sources with three cycles over an $\sim 4$ yr baseline (e.g., PS1 MDS and PTF); from dark to light shades, $z=0$, $1$, and $2$. The purple solid curves correspond to three cycles over $\sim 10$ yr (e.g., CRTS and the future LSST); from dark to light shades: $z=0,1,2$. The black dashed curve represents a binary separation of 500 $R_{S}$, our fiducial value within which the binary is in the gravitational wave-driven regime (i.e., candidates that lie below the black dashed curve).}
\label{fig:mbh_sep}
\end{figure} 

In Figure \ref{fig:mbh_sep}, the $M_{\rm BH}$--$a$ parameter space occupied by the SMBHB candidates with more than three cycles from \cite{Graham2015}, \cite{Charisi2016}, and this work show that those short-period candidates could already be in the gravitational wave-dominated regime of orbital decay. While the temporal baseline of the upcoming LSST is comparable to that of CRTS, it will probe a much larger sky volume and therefore explore a much larger parameter space than any of the three surveys. We will further explore the capabilities of the LSST in detecting SMBHBs in Section \ref{sec:lsst}.


\subsection{Down-selected Sample: Statistical Significance}\label{sec:reanalysis}

Applying the method in Section \ref{sec:lkhd} to the full sample and assuming an underlying DRW red noise model, we find that $11$ candidates satisfy criteria (1)-(3) (Table \ref{tab:zoghbi_drw}), and one of them meets all criteria (PSO J185.8689+46.9752, hereafter PSO J185), having a highly statistical significant $p$-value of $<\frac{1}{9000}$.


\begin{rotatetable*}
\begin{deluxetable*}{lccccccccc}
\tablecaption{Reanalyses Using the Maximum Likelihood Method under the DRW Model \label{tab:zoghbi_drw}}
\tablehead{
\colhead{PS1 Designation} & \colhead{$\ln \mathcal{L}_{\rm DRW}$} & \colhead{$\ln \mathcal{L}_{\rm DRW+p}$}  & \colhead{P (day)}  & \colhead{$p$-value} & \colhead{$\ln \mathcal{L}_{\rm DRW}$} & \colhead{$\ln \mathcal{L}_{\rm DRW+p}$}  & \colhead{P (day)} & \colhead{$p$-value} \\ 
\colhead{} & \colhead{(PS1 Only)} & \colhead{(PS1 Only)} & \colhead{(PS1 Only)} & \colhead{(PS1 Only)} & \colhead{(Extended)} & \colhead{(Extended)} & \colhead{(Extended)}  & \colhead{(Extended)} }
\startdata
\textbf{PSO J35.7068--4.2314} & 123.199 & 125.139 & $436.10^{+350.50}_{-169.49}$ & 0.143 & 146.020 & 152.284 & $428.34^{+222.52}_{-7.48}$ & 0.00190 \\
PSO J35.8704--4.0263 & 131.743 & 137.596 &  $953.43^{+9.75}_{-240.24}$ & 0.00287 &  151.181 & 151.941 & $334.72^{+550.73}_{-159.26}$ & 0.467 \\ 
PSO J52.6172--27.6268 & 88.442 & 95.024 & $953.20^{+15.16}_{-34.83}$  &  0.00138 & \nodata  & \nodata & \nodata  & \nodata \\ 
\textbf{PSO J129.4288+43.8234} & 146.559 & 146.866 &  $412.99^{+192.45}_{-277.54}$ & 0.735  &  145.918  & 147.322 & $229.49^{+237.12}_{-222.88}$  & 0.245  \\ 
\textbf{PSO J130.9953+43.7685} & 139.766 & 142.989 & $658.15^{+310.40}_{-349.59}$& 0.0398 &  153.903 & 159.197 & $734.91^{+150.92}_{-249.08}$  & 0.00502  \\ 
\textbf{PSO J131.1273+44.8582} & 121.928 & 125.289 & $845.20^{+113.61}_{-566.38}$ & 0.0347 &  136.755 & 142.457 & $963.49^{+2.08}_{-37.91}$  & 0.00333  \\ 
PSO J131.7789+45.0939 & 113.227 & 122.755 & $694.46^{+259.30}_{-40.69}$  & 7.27$\times$10$^{-5}$ & 140.334 & 145.664 & $801.40^{+49.08}_{-140.911}$  & 0.00484 \\ 
PSO J148.8485+1.8124 & 115.019 & 120.755 &  $829.88^{+50.10}_{-59.89}$  & 0.00322 & 134.768 & 139.345 & $904.74^{+61.39}_{-128.61}$ & 0.0103 \\ 
PSO J149.4989+2.7827 & 113.043 & 121.036 & $873.75^{+35.77}_{-44.22}$   &  3.37$\times$10$^{-4}$ & 127.629 & 131.386 & $966.54^{+0}_{-80.76}$ & 0.0233 \\ 
PSO J149.2447+3.1393 & 101.544 & 108.689 & $929.64^{+41.27}_{-28.72}$  &  7.88$\times$10$^{-4}$ & 99.631  & 103.688 & $936.39^{+29.77}_{-40.23}$  & 0.0173 \\
\textbf{PSO J149.9400+1.5090} & 131.136 & 131.962 & $332.23^{+340.78}_{-329.21}$  & 0.437 & 149.546 & 151.097 & $408.28^{+238.52}_{-381.47}$ & 0.212 \\ 
\textbf{PSO J149.6873+1.7192} & 105.086 & 111.844 & $812.11^{+93.47}_{-56.52}$  & 0.00116 & 137.971 & 145.578 & $891.25^{+44.45}_{-15.54}$  & 4.96$\times$10$^{-4}$ \\ 
\textbf{PSO J150.9191+3.3880} & 94.701 & 95.448 & $768.34^{+134.88}_{-145.11}$  & 0.4737 & 110.450 & 111.276 & $673.37^{+135.73}_{-104.27}$ & 0.438 \\
\textbf{PSO J160.6037+56.9160} & 160.342 & 162.407 & $840.95^{+124.35}_{-135.64}$  & 0.126 & 159.801 & 161.953 & $957.86^{+1.96}_{-78.03}$ & 0.116 \\  
PSO J161.2980+57.4038 & 120.002 & 127.036 & $850.52^{+79.49}_{-180.50}$  & 8.8$\times$10$^{-4}$ & 104.344 & 105.932 & $967.41^{+0}_{-231.59}$ & 0.204 \\ 
PSO J163.2331+58.8626 & 157.293 & 158.086 & $948.16^{+22.23}_{-177.76}$  & 0.452 & 159.232 & 160.174 & $294.44^{+350.86}_{-109.13}$ & 0.389 \\  
\textbf{PSO J185.8689+46.9752$\star$} & 112.525 & 114.019 & $279.05^{+603.95}_{-166.04}$  & 0.224 & 112.853 & 122.156 & $962.30^{+1.32}_{-8.67}$ & 9.11$\times$10$^{-5}$ \\ 
PSO J213.9985+52.7527 & 146.229 & 154.131 & $768.64^{+159.95}_{-410.04}$  & 3.7$\times$10$^{-4}$ & 142.168 & 148.269 & $686.37^{+20.72}_{-29.27}$ & 0.00224 \\ 
PSO J214.9172+53.8166 & 118.971 & 120.498 & $89.91^{+315.98}_{-44.01}$  & 0.217 & 120.820 & 122.185 & $892.77^{+70.37}_{-129.62}$ & 0.255 \\  
PSO J242.5040+55.4391 & 150.588 & 155.394 &  $950.25^{+20.73}_{-39.26}$  & 0.00818 & 173.236 & 175.409 & $830.45^{+139.56}_{-310.43}$ & 0.113 \\
PSO J242.8039+54.0585 & 140.902 & 146.990 & $720.49^{+40.56}_{-39.43}$  & 0.00226 & 169.062 & 173.653 & $746.24^{+100.70}_{-359.29}$ & 0.0101 \\
PSO J243.5676+54.9741 & 152.054 & 152.959 & $113.69^{+295.66}_{-84.33}$  & 0.404 & 184.189 & 186.841 & $189.08^{+155.22}_{-164.77}$ & 0.0705 \\ 
PSO J333.0298+0.9687 & 62.335 & 63.660 & $408.09^{+365.78}_{-254.21}$  & 0.265 & 106.265 & 107.822 & $971.06^{+0}_{-230.31}$ & 0.210 \\ 
\textbf{PSO J333.9832+1.0242} & 181.015 & 184.466 & $466.40^{+298.54}_{-441.45}$  & 0.0317 & 244.015 & 248.686 & $442.76^{+402.56}_{-17.43}$ & 0.00936 \\ 
PSO J334.2028+1.4075 & 164.979 & 167.303 & $578.32^{+108.16}_{-441.83}$  & 0.0978 & 214.130 & 215.850 & $695.27^{+262.37}_{-127.62}$ & 0.179 \\ 
\textbf{PSO J351.5679--1.6795} & 117.924 & 124.945 & $814.33^{+140.21}_{-89.78}$  & 8.92$\times$10$^{-4}$ & 81.783 & 88.879 & $826.42^{+67.75}_{-22.24}$ & 8.28$\times$10$^{-4}$ \\
\enddata
\tablecomments{Candidates that satisfy criteria (1)-(2) are in bold. The candidate that met criteria (1)-(3) (PSO J185) is marked with a star.}
\end{deluxetable*}
\end{rotatetable*}

\begin{rotatetable*}
\begin{deluxetable*}{lcccccccccc}
\tablecaption{Reanalyses Using the Maximum Likelihood Method under the BPL Model \label{tab:zoghbi_bpl}}
\tablehead{
\colhead{PS1 Designation} & \colhead{$\ln \mathcal{L}_{\rm BPL}$} & \colhead{$\ln \mathcal{L}_{\rm BPL+p}$}  & \colhead{P (day)} & \colhead{$p$-value} & \colhead{$\ln \mathcal{L}_{\rm BPL}$ } & \colhead{$\ln \mathcal{L}_{\rm BPL+p}$ }  & \colhead{P (day) } & \colhead{$p$-value} \\ 
\colhead{} & \colhead{(PS1 Only)} & \colhead{(PS1 Only)} & \colhead{(PS1 Only)} & \colhead{(PS1 Only)} & \colhead{(Extended)} & \colhead{(Extended)} & \colhead{(Extended)}  & \colhead{(Extended)} }
\startdata
PSO J35.7068--4.2314 & 122.084 & 123.290 & $440.65^{+272.31}_{-197.69}$  & 0.299 & 147.69 & 148.57 & $427.01^{+536.13}_{-153.86}$  & 0.412  \\
PSO J35.8704--4.0263 & 135.932 & 139.463 & $933.95^{+37.15}_{-42.84}$  & 0.0292 & 159.216 & 160.058 & $108.835^{+355.02}_{-104.97}$  &  0.431 \\ 
PSO J52.6172--27.6268 & 92.067 & 94.015 & $920.65^{+42.86}_{-327.13}$  & 0.142 & \nodata & \nodata & \nodata & \nodata \\ 
PSO J129.4288+43.8234 & 149.500 & 151.079 & $39.67^{+286.67}_{-23.32}$  & 0.206 & 145.452 & 145.884 & $226.73^{+182.15}_{-217.84}$  & 0.649 \\ 
PSO J130.9953+43.7685 & 144.857 & 146.891 & $676.61^{+125.09}_{-94.90}$  & 0.130 & 164.045 & 164.739 & $745.551^{+103.57}_{-126.42}$  & 0.499 \\ 
PSO J131.1273+44.8582 & 122.882 & 126.346 & $854.51^{+63.68}_{-76.31}$  & 0.0313 & 139.925 & 142.186 & $961.517^{+6.13}_{-213}$  & 0.104 \\ 
PSO J131.7789+45.0939 & 116.029 & 122.908 & $683.18^{+35.24}_{-564.75}$  & 0.00102 & 144.822 & 147.025 & $791.39^{+161.90}_{-138.09}$  & 0.110 \\ 
PSO J148.8485+1.8124 & 118.235 & 121.033 & $843.44^{+124.29}_{-115.70}$  & 0.0609 & 137.189 & 139.656 & $895.10^{+70.17}_{-59.82}$  & 0.0848 \\ 
\textbf{PSO J149.4989+2.7827} & 117.984 & 115.298 & $837.46^{+36.91}_{-33.08}$  & \nodata & 133.597 & 134.509 & $970.23^{+0}_{-187.79}$  & 0.401  \\ 
PSO J149.2447+3.1393 & 103.452 & 107.834 & $937.46^{+25.36}_{-54.63}$  & 0.0125 & 100.726 & 103.061 & $944.05^{+22.39}_{-27.60}$  & 0.0968 \\
PSO J149.9400+1.5090 & 132.670 & 135.961 & $139.77^{+221.62}_{-58.38}$  & 0.0372 & 150.163 & 150.793 & $410.55^{+260.84}_{-399.15}$  & 0.532  \\ 
\textbf{PSO J149.6873+1.7192} & 106.677 & 111.590 & $809.69^{+124.54}_{-225.45}$  & 0.00735 & 140.048 & 144.992 & $896.740^{+37.90}_{-72.09}$  & 0.00712  \\ 
PSO J150.9191+3.3880 & 91.667 & 96.013 & $750.01^{+215.48}_{-184.51}$  & 0.0129 & 108.706 & 110.942 & $16.93^{+88.65}_{-11.34}$  & 0.106 \\
PSO J160.6037+56.9160 & 163.097 & 165.788 & $940.31^{+27.71}_{-362.28}$  & 0.0678 & 160.066 & 162.094 & $961.193^{+6.47}_{-103.52}$  & 0.131  \\  
PSO J161.2980+57.4038 & 121.914 & 127.821 & $889.98^{+80.16}_{-59.83}$  & 0.00272 & 102.754 & 103.929 & $799.60^{+162.911}_{-227.08}$  & 0.308  \\ 
PSO J163.2331+58.8626 & 157.344 & 157.899 & $114.96^{+329.92}_{-110.07}$  & 0.574 & 159.140 & 159.884 & $459.93^{+412.34}_{-317.65}$  & 0.475 \\  
\textbf{PSO J185.8689+46.9752} & 115.700 & 117.084 & $276.42^{+282.49}_{-17.50}$  & 0.250 & 115.750 & 120.851 & $945.85^{+22.95}_{-17.04}$  & 0.00609 \\ 
PSO J213.9985+52.7527 & 149.063 & 151.434 & $252.65^{+591.97}_{-88.02}$  & 0.0933 & 151.416 & 153.182 & $691.74^{+178.37}_{-261.62}$  & 0.171 \\ 
\textbf{PSO J214.9172+53.8166} & 120.828 & 121.999 & $89.51^{+283.66}_{-46.33}$  & 0.310 & 119.487 & 122.671 & $859.68^{+40.56}_{-289.43}$  & 0.0414 \\  
PSO J242.5040+55.4391 & 156.506 & 157.818 & $923.49^{+44.85}_{-125.14}$  & 0.269 & 178.838 & 179.623 & $96.47^{+548.49}_{-71.50}$  & 0.456 \\
PSO J242.8039+54.0585 & 141.217 & 145.791 & $11.11^{+289.78}_{-0.21}$  & 0.0103 & 175.006 & 177.065 & $741.22^{+225.30}_{-264.69}$  & 0.127 \\
PSO J243.5676+54.9741 & 152.160 & 152.454 & $957.06^{+7.42}_{-262.57}$  & 0.745 & 184.377 & 186.607 & $127.27^{+288.24}_{-111.75}$  & 0.107 \\
PSO J333.0298+0.9687 & 63.575 & 63.509 & $417.03^{+408.59}_{-131.40}$  & \nodata & 109.700 & 110.376 & $62.03^{+297.64}_{-52.35}$  & 0.508 \\ 
PSO J333.9832+1.0242 & 179.692 & 185.789 & $472.80^{+314.55}_{-15.44}$  & 0.00224 & 247.255 & 241.241 & $790.410^{+179.756}_{-260.244}$  & \nodata \\ 
PSO J334.2028+1.4075 & 165.162 & 166.541 & $573.64^{+106.54}_{-153.45}$  & 0.251& 215.223 & 216.272 & $131.73^{+132.76}_{-127.23}$  & 0.350 \\ 
PSO J351.5679--1.6795 & 124.080 & 134.519 & $832.49^{+81.66}_{-28.33}$  & 2.92$\times$10$^{-5}$  & 113.713 & 120.584 & $827.56^{+70.45}_{-19.54}$  & 0.00103 \\
\enddata
\tablecomments{Candidates that satisfy criteria (1)-(2) are in bold.}
\end{deluxetable*}
\end{rotatetable*}


However, this analysis is dependent on the assumption of the red noise model, or the PSD. If we instead adopt a PSD whose power-law slopes are steeper than DRW (hereafter the broken power-law, or BPL, model), then only four candidates satisfy criteria (1)-(3) and none of them have $p<\frac{1}{9000}$ (Table \ref{tab:zoghbi_bpl}). We note that PSO J185 met criteria (1)-(3) independent of the assumed underlying red noise model, and a total of $12$ candidates are consistent with criteria (1)-(3), since we have chosen the BPL parameters so that they do not overlap with those of DRW. 

Among the candidates that met criteria (1)-(3), PSO J185 also has the largest decrease in its $p$-value, despite the fact that other candidates have a similar number of new observations. It is further evidence that the behavior of PSO J185 is consistent with the expectation that the false-alarm probability sharply decreases with a longer baseline for a periodic signal \citep{Liu2018a}. However, we note the caveat that the cadences of our follow-up observations are inhomogeneous among candidates due to scheduling, and therefore claiming the level of evidence for periodicity in the individual candidates is beyond the scope of this work.

Although PSO J185 is the most statistically significant candidate under the DRW model, it does not satisfy criteria (4) under the BPL model ($p = 0.006$), and the statistical significance of all candidates has decreased overall, which again indicates that the assumption of the underlying red noise model is important when determining the significance of the periodic signal. 


\subsection{Down-selected Sample: Alternative Interpretations}\label{sec:alternative}

While periodic variability is a predicted signature of an SMBHB, we must consider the possibility that it can also be produced in an AGN powered by a single black hole. This is analogous to the phenomenon of quasi-periodic oscillation (QPO) found in Galactic X-ray binaries (XRBs) and, in rare cases, AGNs. A highly significant X-ray QPO signature is detected in the \emph{XMM-Newton} light curve of the active galaxy RE J1034+396 \citep{Gierlinski2008}, but a candidate optical QPO was only recently identified in the high-precision Kepler light curve of an AGN, and its frequency is consistent with an inverse scaling relation with black hole mass extrapolated from low-frequency X-ray QPOs \citep{Smith2018QPO}. Therefore, here we explore the possibility that the $12$ down-selected candidates are optical analogs of X-ray QPOs, which could originate from the accretion disk and are not due to the presence of a putative binary.

\begin{figure}[h] 
\centering
\epsfig{file=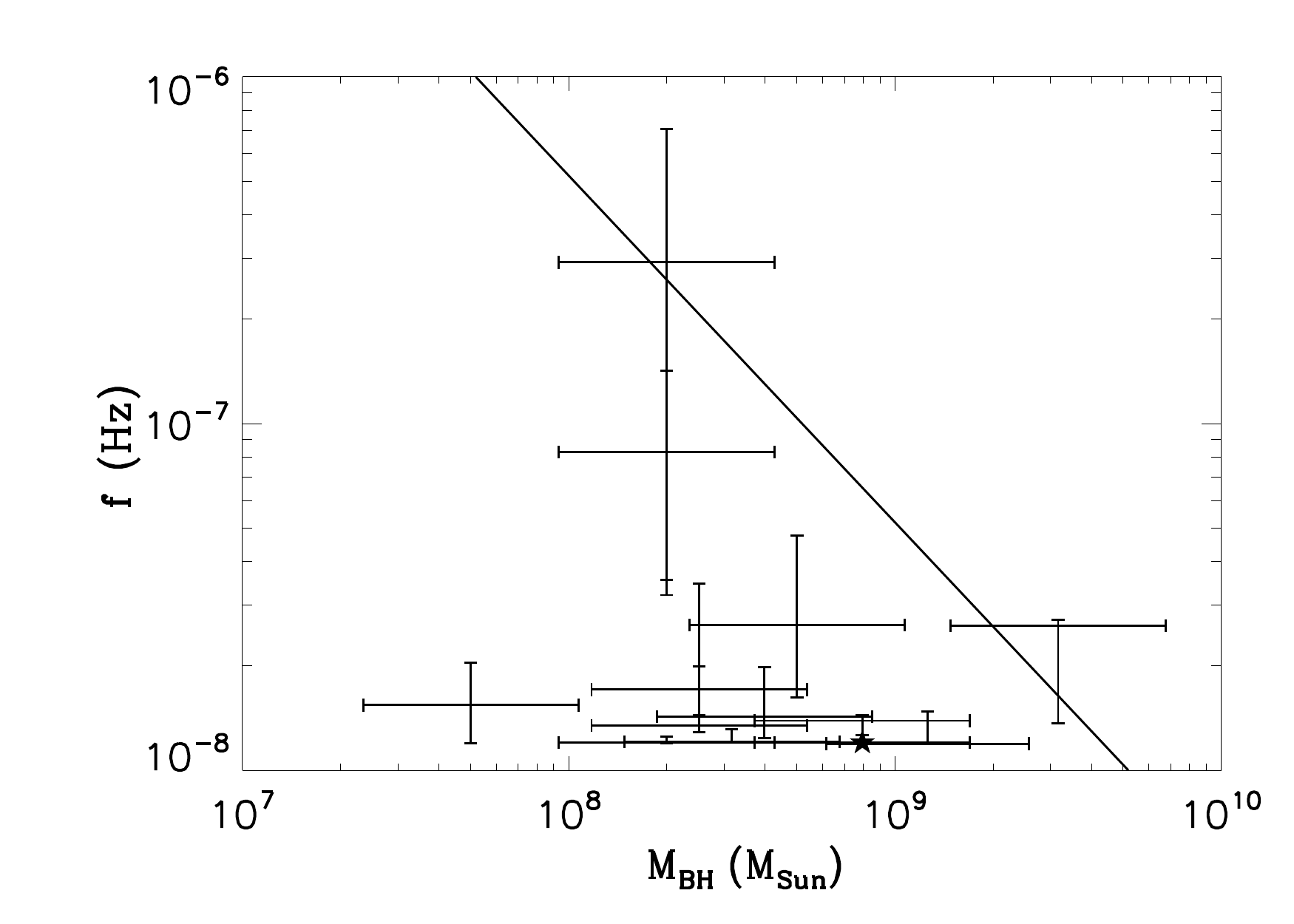,width=0.45\textwidth,clip=}
\caption{We show the frequencies of the $12$ periodic candidates vs. their black hole masses. The best candidate (PSO J185) in our sample is indicated by a star. The majority of candidates are inconsistent with the $f - M_{\rm BH}$ relation expected for an optical QPO (solid line).}
\label{fig:mbh_qpo}
\end{figure} 

In Figure \ref{fig:mbh_qpo}, we show their frequencies versus virial black hole mass. The uncertainty in frequency is determined from the middle 68\% of the posterior distribution of $P$, and the error on the black hole mass estimate is the systematic uncertainty of the \ion{Mg}{2} (0.33 dex) or \ion{C}{4}  (0.31 dex) estimator \citep{Shen2008}. We then adopt the best-fit $f-M_{\rm BH}$ relation from \cite{Smith2018QPO}, $f$ (Hz) $= 51.9 (M_{\rm BH}/M_{\odot})^{-1}$, and extrapolate to higher masses. Only two candidates are consistent with this relation, while the others do not show a correlation between  frequency and black hole mass. While this lack of correlation does not confirm the binary origin of the periodicity, it disfavors a disk origin for our sample of candidates. We also note that a sample of true SMBHBs should have a weak (if any) correlation between their orbital frequencies and black hole masses, as the frequency is also dependent on the orbital separation.


\section{Discussion}\label{sec:discuss}

In this section, we discuss the astrophysical implications of our most statistically significant candidates: how does our detection rate of SMBHB candidates compare with previous work? Given the capabilities of the LSST, how many periodic quasars can it detect? Can we look for complementary evidence for an SMBHB?

\subsection{The Detection Rate of SMBHBs}\label{sec:rate}

\citet[hereafter BL09]{Boroson2009} searched for SDSS quasars that have multiple redshift systems, which could indicate the presence of a binary, and there are two candidates that show such features from $\sim 17,500$ SDSS quasars at $z<0.7$. This rate ($\sim 0.01\%$) is consistent with the results from \citet[hereafter VMD09]{Volonteri2009}, who predicted an upper limit of $\sim 0.1\%$ per quasar for $z<0.7$ or $\sim 1\%$ for $z<1$. 

\begin{figure}[h] 
\centering
\epsfig{file=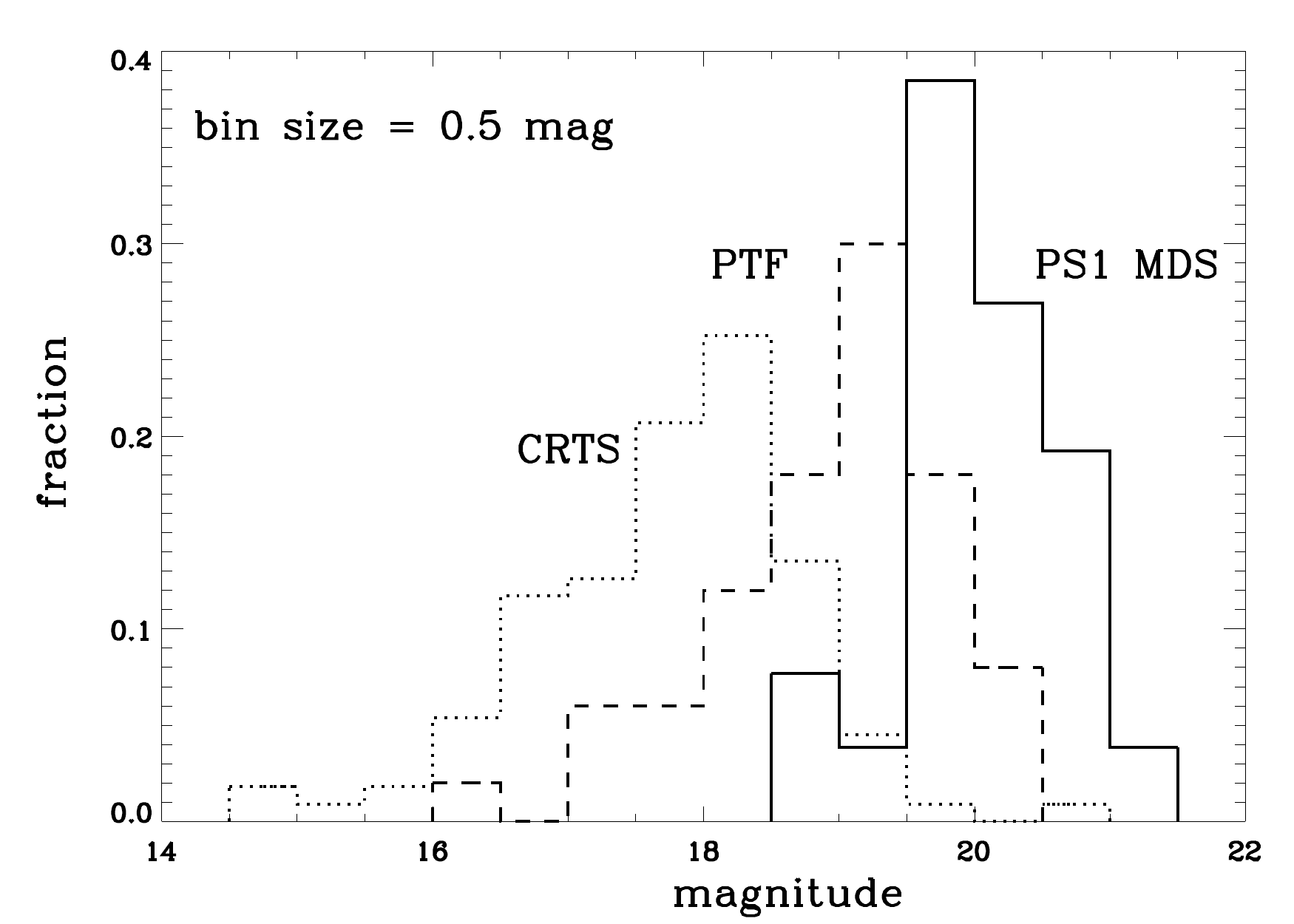,width=0.45\textwidth,clip=}
\caption{Dotted histogram: $V$-band magnitude distribution of the candidates from CRTS (G15). Dashed histogram: the $R$ magnitude distribution of the candidates from PTF (C16). Solid histogram: the g$_{\rm P1}$ magnitude distribution of candidates from this work.}
\label{fig:mag_dist_crts_ptf}
\end{figure}

To compare with the results from BL09, we calculate the cumulative number of SMBHB candidates ($N(<z)$) per $1000$ quasars from this work. We also compare with previous work by \citet[hereafter G15]{Graham2015} and \citet[hereafter C16]{Charisi2016}: G15 searched among $\approx 243,000$ spectroscopically confirmed quasars and claimed $111$ candidates, and $50$ candidates from C16 were selected among $\approx 35,000$ spectroscopic quasars ($33$ after reanalysis with extended data).

\begin{figure}[h] 
\centering
\epsfig{file=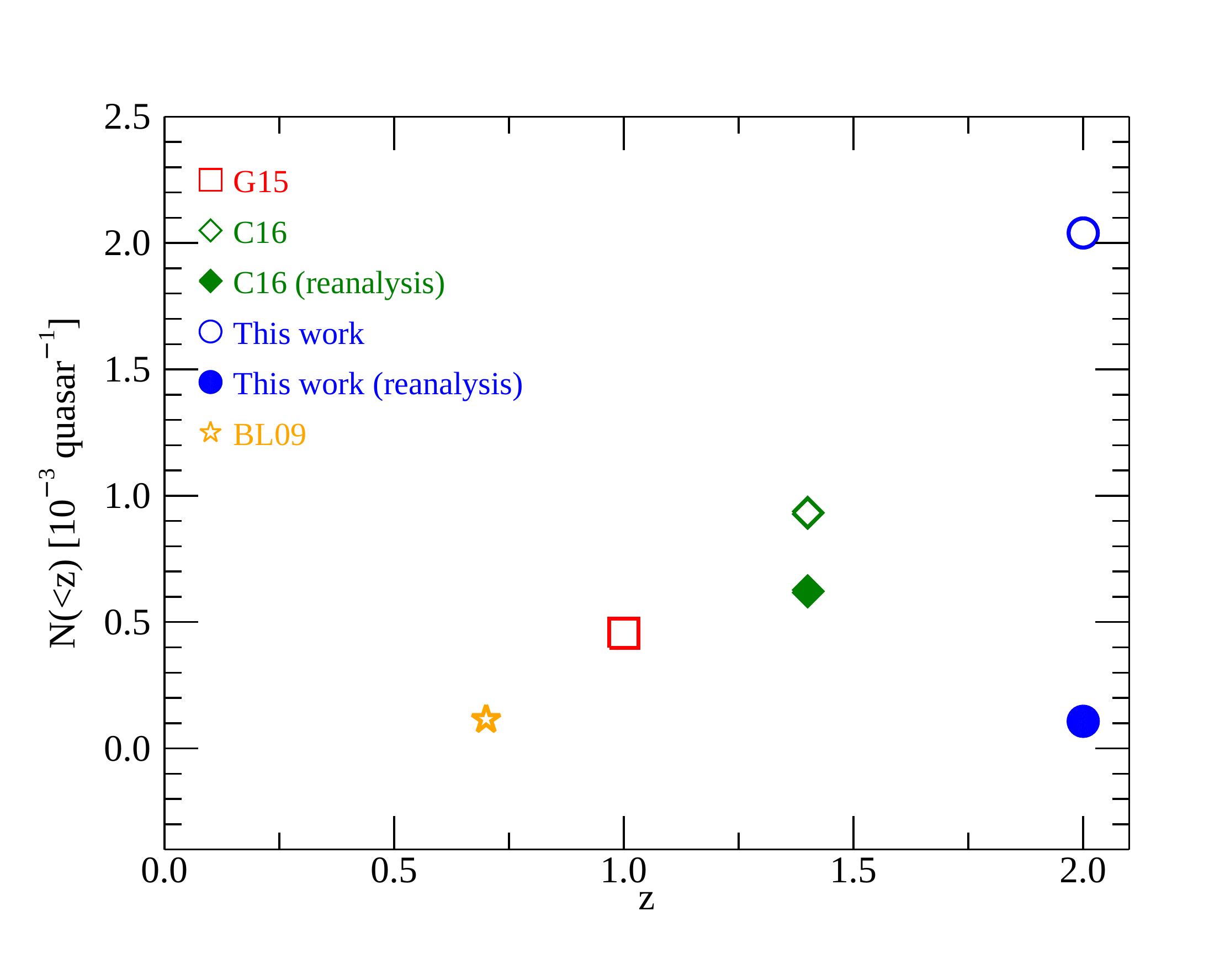,width=0.5\textwidth,clip=}
\epsfig{file=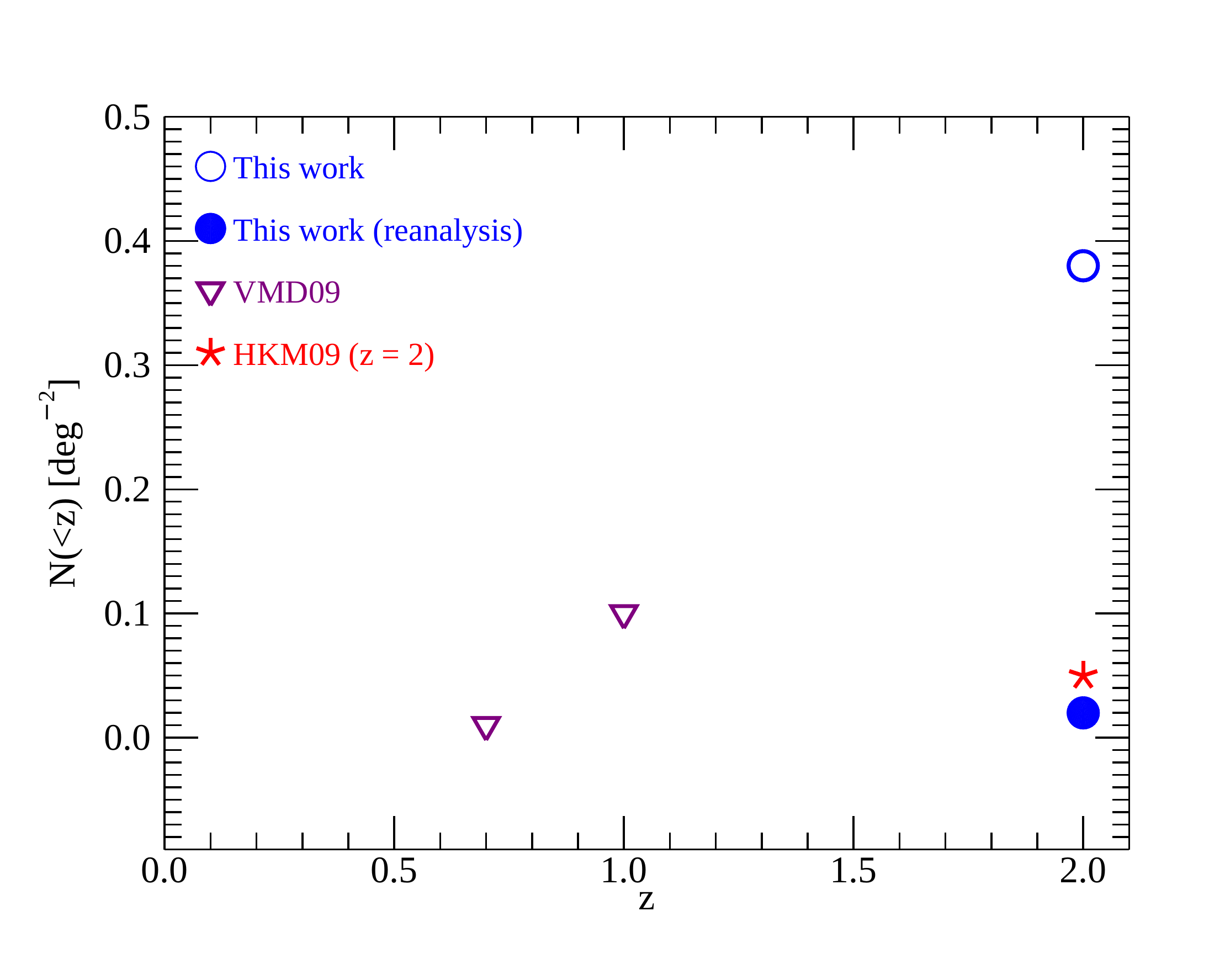,width=0.5\textwidth,clip=}
\caption{Upper panel: the cumulative number of SMBHB candidates per $1000$ quasars from this work (blue circles), C16 (green diamonds), G15 (red square), and BL09 (orange star). The rates inferred from C16 and this work after extended baseline analysis are indicated with filled symbols.
Lower panel: We compare the cumulative number of SMBHB candidates per square degree of sky area from this work (open and filled blue circles) with the predicted rates by VMD09 (purple triangles) and HKM09 (red star).}
\label{fig:smbhb_rate}
\end{figure}

We first calibrate the completeness of G15 and C16 in detecting periodic quasars relative to this work: our candidates have a magnitude cutoff at $m \sim 20$ mag (Figure \ref{fig:mag_dist_crts_ptf}), which results in our sensitivity out to $z\sim2$. Assuming that this work is complete out to $z \sim 2$ and the candidates from G15 and C16 are relatively complete down to $m \sim 18$ and $19$ mag, respectively, that translates to a redshift limit at $z \sim 1.0$ for G15 and $z \sim 1.4$ for C16. We then count the total number of $<z$ candidates that are in the respective sample by assuming that the full quasar sample follows the same redshift distribution and drawing from the distribution. Since we tentatively identify one statistically significant candidate in our sample, this corresponds to an SMBHB rate of $0.1$ per $1000$ quasars. However, the cumulative rates inferred from G15 and C16 have higher values out to lower redshifts and are therefore in potential tension with our rate (Figure \ref{fig:smbhb_rate}, upper panel).


We also compare the number of SMBHB candidates per square degree of sky area searched. We performed our search in the cross-matched area between the PS1 MDS and PS1$\times$CFHT catalogs, which covers an area of $\sim 50$ $\degr^{2}$. This corresponds to a rate of $0.02$ SMBHBs deg$^{-2}$ (out to $z\sim2$). To compare with the predicted observability of periodic sources, we adopt the fiducial values in \citet[hereafter HKM09]{Haiman2009}, for which we expect $20$ sources varying at $245$ days in a $10^{4}$ deg$^{2}$ sky area for a survey magnitude depth of $22$ mag (which is the magnitude limit of our candidate selection). Since most of our candidates vary on a timescale of $\sim 800$ days, we then apply the scaling relation for a population of purely gravitational wave-driven SMBHBs to calculate the expected number of periodic quasars, i.e., $(800/245)^{8/3}\times0.002 = 0.05$, which is largely consistent with our detection rate (Figure \ref{fig:smbhb_rate}, lower panel). We note the caveat, however, that the redshift of the sources from HKM09 is fixed at $z=2$, while we have measured a cumulative rate out to $z=2$. We have also compared with the predicted upper limit from VMD09 of $\sim 0.1$ SMBHBs deg$^{-2}$ out to $z=1$, and our measured rate is still consistent with this upper limit\footnote{However, we note that the VMD09 prediction is motivated by SMBHBs with broad emission line features and not optical periodicity.}.

In a recent study, \cite{Kelley2019} incorporated the predictions for periodic variability due to Doppler boosting and modulated accretion into synthetic AGN spectra and, from a population of SMBHBs from the Illustris cosmological simulation, predict the number of binaries observable as periodic AGNs in time-domain surveys. In particular, for a magnitude depth of $\sim 22$ mag, it is expected that $\sim 50$ binaries could be detected out to $z\sim2$ on the full sky, or $\sim 0.06$ in an $\sim 50$ $\degr^{2}$ sky area. Our upper limit is therefore also consistent with this prediction. 

Given the high efficiency of color selection at $z<2.7$ ($98$\% of known quasars are correctly classified as quasars; \citealt{Peters2015}), the fraction of our parent sample of $\sim9000$ color-selected quasars that is contaminated by stars is negligible in our upper limit rate estimate. However, color selection is only $93$\% complete in this redshift range, causing the observed number rate of SMBHB candidates to be higher than the actual rate. As such, our upper limit still holds.


\subsection{Periodic Quasar Detections in the LSST Era}\label{sec:lsst}

Expected to start its operation in about 2022, the LSST \citep{Ivezic2008} will be thousands of times more powerful than PS1 MDS, thanks to its magnitude depth, photometric precision, and large survey area (Table \ref{tab:lsst}). Here we explore its capabilities to detect periodic quasars by using our results from PS1 MDS as a benchmark. The notation $\tilde{N}$ represents the number of quasars from a simulated population, while $N$ is the observed or expected number from a survey.

\begin{figure}[h] 
\centering
\epsfig{file=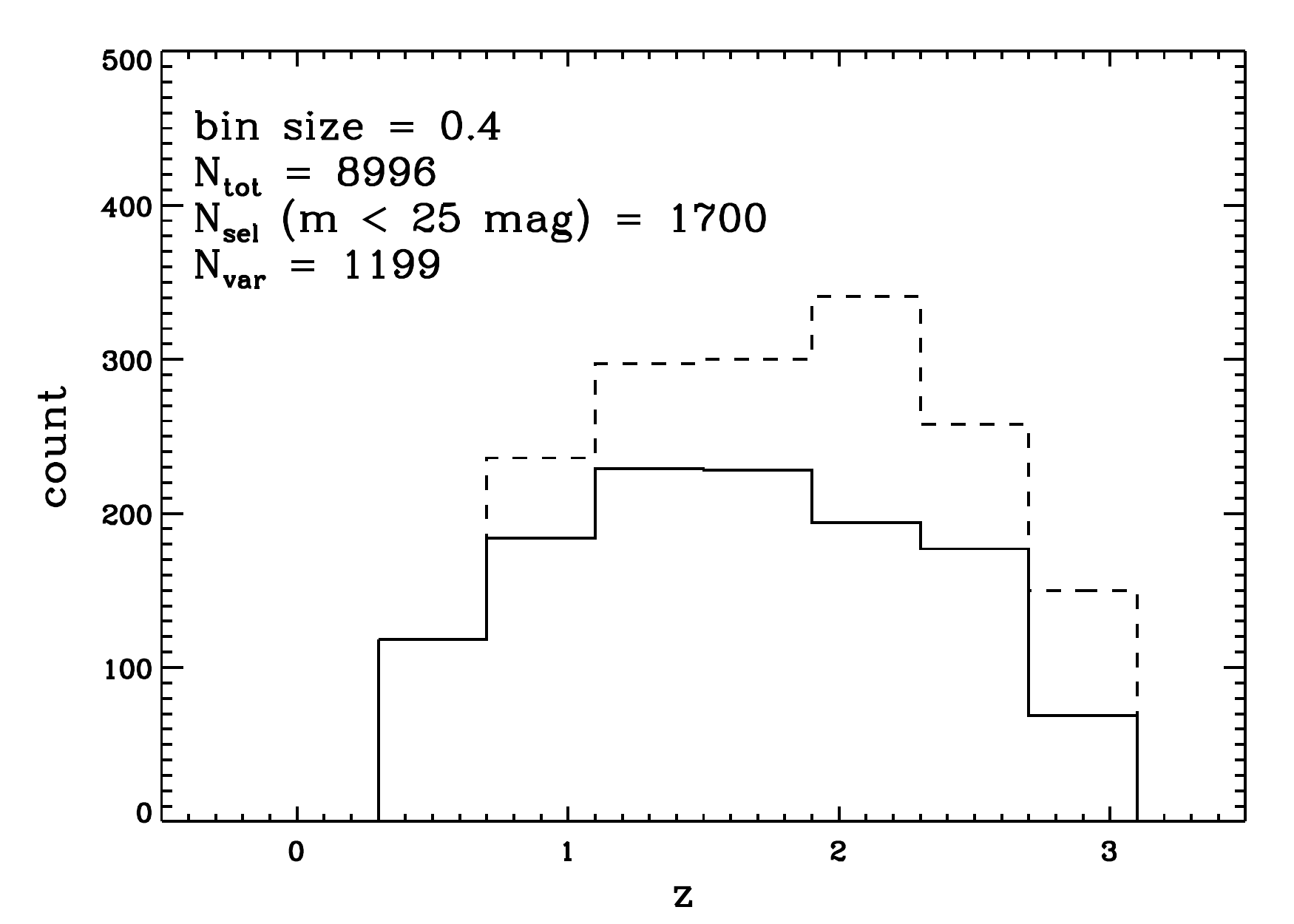,width=0.45\textwidth,clip=}
\epsfig{file=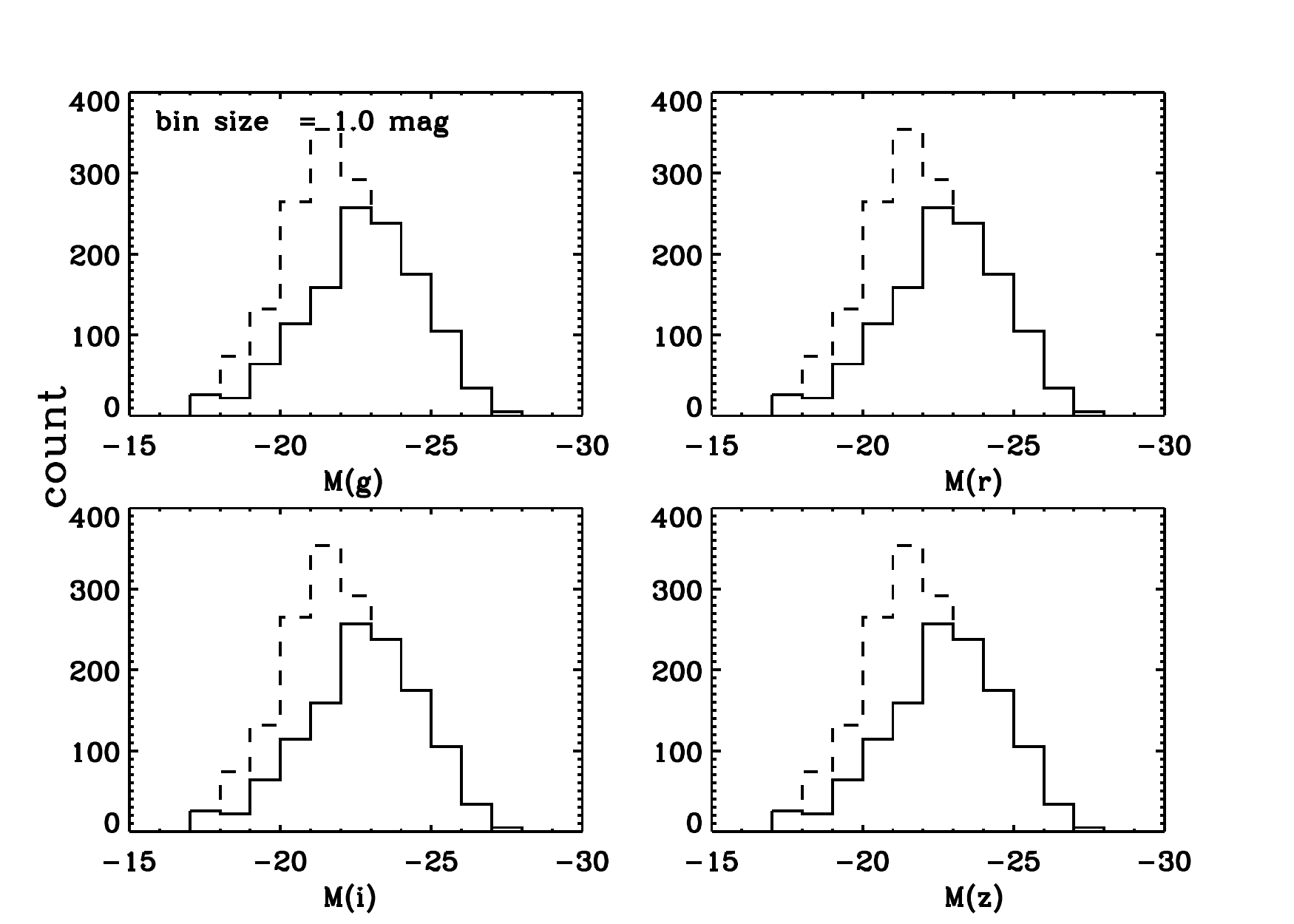,width=0.5\textwidth,clip=}
\caption{From an initial sample of $8996$ quasars drawn from the quasar luminosity function, $1700$ can be detected by LSST (dashed histograms). Assuming they follow the variability amplitude--absolute magnitude relation in \cite{Heinis2016AGN}, $1199$ can be detected as variable quasars (solid histograms).}
\label{fig:lsst_sim}
\end{figure}

Following the method in L16, we first simulate a population of quasars from $0.3<z<3.1$ given the quasar luminosity function. We then apply the magnitude cut at $m<25$ mag; from $\tilde{N}_{\rm tot, LSST} = 8996$ simulated quasars, $\tilde{N}_{\rm sel, LSST} = 1700$ quasars can be ``visible'' in the survey (Figure \ref{fig:lsst_sim}). Next, we assign a variability amplitude to each quasar based on the same amplitude--absolute magnitude relation from \cite{Heinis2016AGN}. To determine the variability detection threshold, we adopt the expected photometric error as a function of magnitude from \cite{Ivezic2008}:

\begin{equation*}
\sigma^{2} = \sigma^{2}_{\rm sys} + (0.04-\gamma)10^{0.4(m-m5)} + \gamma 10^{0.8(m-m5)}\quad.
\end{equation*}

From the same simulation performed for MDS in L16, $\tilde{N}_{\rm sel, MDS} = 924$ quasars are selected from an initial sample of $\tilde{N}_{\rm tot, MDS} = 8996$. To estimate the total number of quasars in the LSST footprint, we simply scale up the number of quasars selected in MDS ($N_{\rm sel, MDS} = 9314$) by the survey area $A$:

\begin{equation*}
N_{\rm tot, LSST} =  N_{\rm sel, MDS}\times\frac{\tilde{N}_{\rm tot, MDS}}{\tilde{N}_{\rm sel, MDS}}\times\frac{\rm A(LSST)}{\rm A(MDS)} = 3.63\times10^{7}\,.
\end{equation*}


\begin{deluxetable}{lcc}
\tablecaption{Comparing PS1 MDS and LSST Capabilities \label{tab:lsst}}
\tablehead{\colhead{} & \colhead{PS1 MDS} & \colhead{LSST} }
\startdata
Single-visit $5\sigma$ magnitude depth &  &  \\
\quad in $g$ band (mag) & 22.5 & 25.0 \\
Expected photometric error &  &  \\
\quad at $g = 17$ mag (mag) & 0.02 & 0.005 \\
Sky coverage (deg$^2$) & 50 & 20,000 \\
\enddata
\tablecomments{The total sky area in PS1 MDS is $\sim 80$ $\degr^2$, however, we have crossed-matched with the PS1$\times$CFHT, and therefore the effective sky area in our study is $\sim 50$ $\degr^2$.}
\end{deluxetable}

Since $\tilde{N}_{\rm var, LSST} = 1199$ quasars are selected as variables from $\tilde{N}_{\rm tot, LSST} = 8996$ quasars from our simulation, the number of variable quasars that can be detected by LSST is

\begin{equation*}
N_{\rm var, LSST} = N_{\rm tot, LSST}\times\frac{\tilde{N}_{\rm var, LSST}}{\tilde{N}_{\rm tot, LSST}} = 4.84\times10^{6}\quad.
\end{equation*}

Assuming that the same periodic candidate selection method (which selected $26$ candidates from $N_{\rm var, MDS} = 1369$ variable quasars, out of which $N_{\rm cand, MDS} = 1$ is statistically significant) is applied to LSST variable quasars, the number of periodic candidates it could yield is

\begin{equation*}
N_{\rm cand, LSST} = N_{\rm var, LSST}\times\frac{N_{\rm cand, MDS}}{N_{\rm var, MDS}} \approx 3500\quad,
\end{equation*}

\noindent a factor of $\sim 20$ more than the number of SMBHB candidates from G15, C16, and this work combined. We note that our prediction is much more optimistic than that of \cite{Kelley2019}, as we have identified one statistically significant candidate in PS1 MDS. Interestingly, if we adopt the expectation value of $N_{\rm cand, MDS} \approx 0.06$ instead, we would obtain $N_{\rm cand, LSST} \approx 200$, which is consistent with their prediction.


\subsection{Probing the SED and Spectral Properties of SMBHBs}\label{sec:multi}

While a long baseline is essential to break false signals due to red noise and help to verify the variability behavior of SMBHB candidates (Sections \ref{sec:extended} and \ref{sec:lkhd}), analyses of these systems based on optical variability alone may not suffice to identify robust SMBHB candidates, and follow-up multiwavelength studies are needed to independently verify an SMBHB candidate.


\begin{deluxetable}{lccc}
\tablecaption{The SED of PSO J185 \label{tab:sed}}
\tablehead{ \colhead{Catalog} & \colhead{Filter/Band} &  \colhead{$\nu$ (Hz)} & \colhead{$\nu L_{\nu}$ (erg s$^{-1}$)}
}
\startdata
FIRST & 1.4 GHz & 3.75$\times$10$^{9}$ & (2.50$\times$10$^{41}$) \\
\emph{AllWISE} & W1 & 2.40$\times$10$^{14}$ & 9.38$\times$10$^{44}$ \\
\emph{AllWISE} & W2 & 1.75$\times$10$^{14}$ & 7.96$\times$10$^{44}$ \\
\emph{AllWISE} & W3 & 6.93$\times$10$^{13}$ & (1.68$\times$10$^{45}$) \\
\emph{AllWISE} & W4 & 3.64$\times$10$^{13}$ & (4.29$\times$10$^{45}$) \\
SDSS  & $u$  & 2.27$\times$10$^{15}$  & 3.07$\times$10$^{45}$ \\
SDSS  & $g$  & 1.69$\times$10$^{15}$  & 2.58$\times$10$^{45}$ \\
SDSS  & $r$  & 1.29$\times$10$^{15}$ & 1.80$\times$10$^{45}$ \\
SDSS  & $i$  & 1.05$\times$10$^{15}$ & 2.05$\times$10$^{45}$ \\
SDSS  & $z$  & 8.81$\times$10$^{14}$ & 1.67$\times$10$^{45}$ \\
\emph{GALEX} & NUV  & 3.55$\times$10$^{15}$ & 2.02$\times$10$^{45}$ \\
\emph{XMM-Newton}  & 1.5 keV  & 9.72$\times$10$^{17}$  & (4.89$\times$10$^{46}$) \\
\enddata
\tablecomments{Values in parentheses represent upper limits.}
\end{deluxetable}

\begin{figure}[h] 
\centering
\epsfig{file=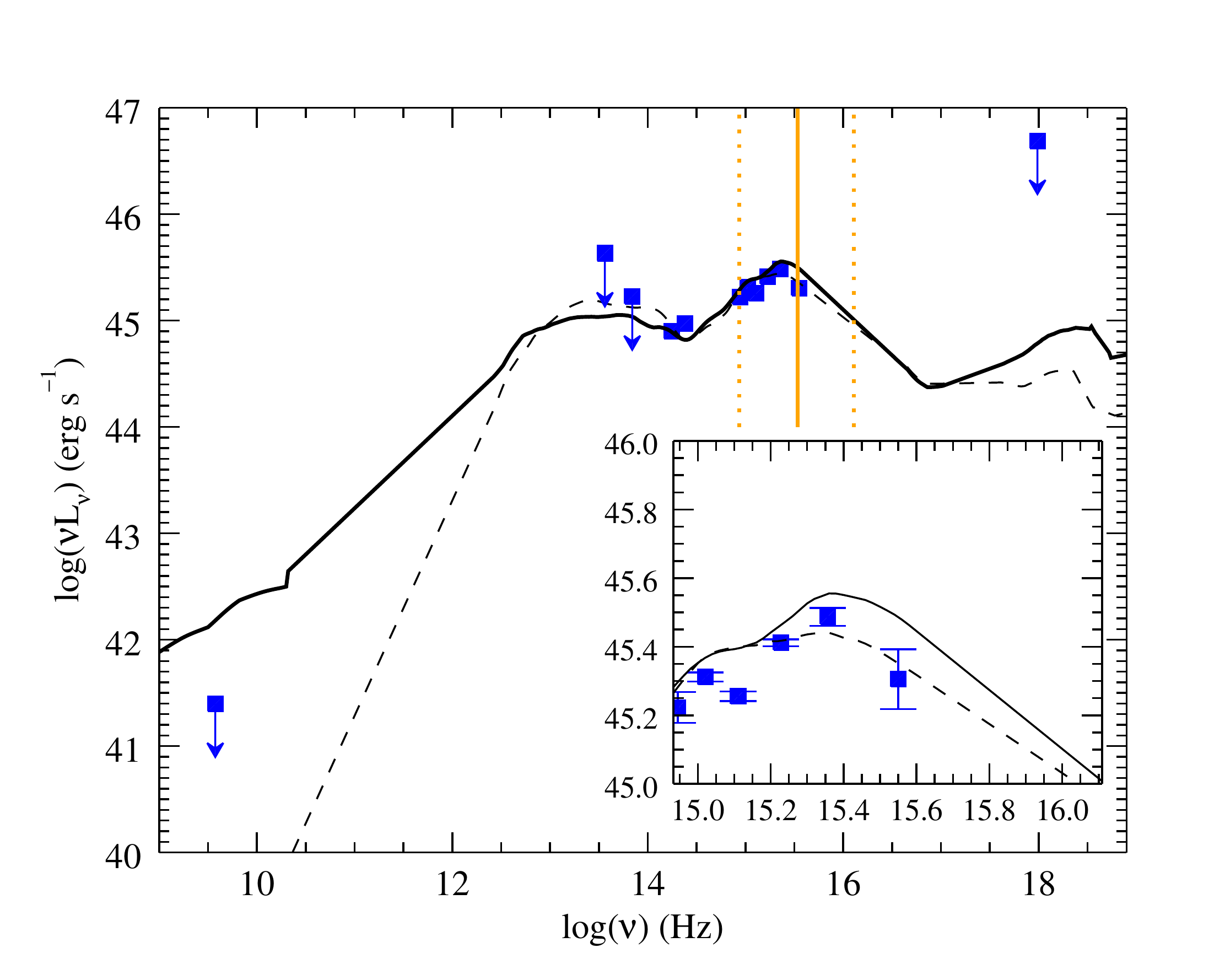,width=0.45\textwidth,clip=}
\caption{We construct the SED of PSO J185 using multiband archival data (or upper limits; blue squares) and compare with the mean SEDs of radio-loud and radio-quiet quasars (solid and dashed curves, respectively) from \cite{Elvis1994}. The mean SED has been normalized to PSO J185 at $\lambda_{\rm rest} = 2000$\AA. The SED of PSO J185 is consistent with that of a radio-quiet quasar and shows no evidence for a spectral notch in the $kT_{0}$--$15 kT_{0}$ range (marked by dashed orange lines; also shown in the inset). The expected energy of the largest deficit ($4kT_{0}$) is marked with a solid orange line.}
\label{fig:sed}
\end{figure}


\begin{figure}[h] 
\centering
\epsfig{file=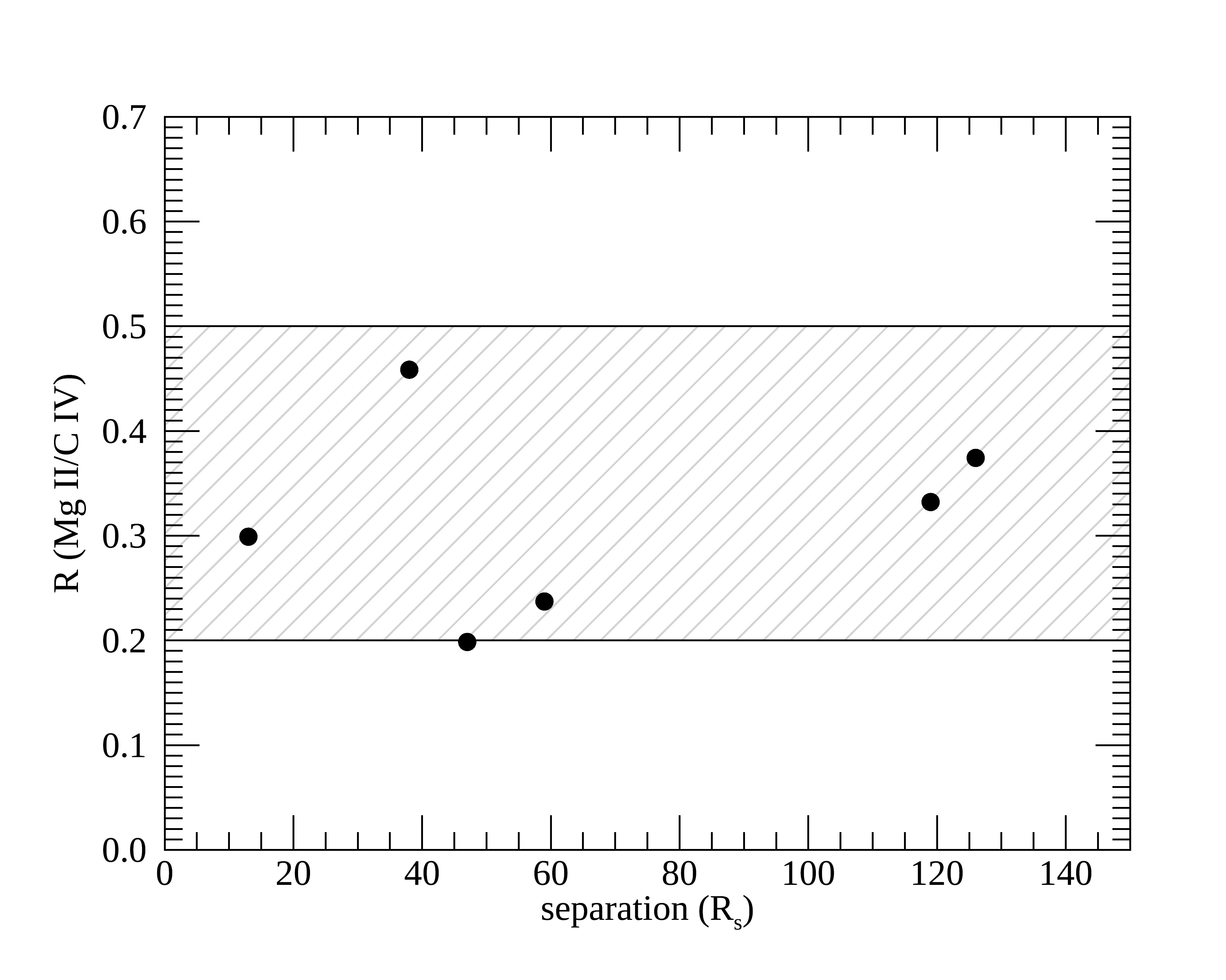,width=0.45\textwidth,clip=}
\caption{We show the flux ratio between \ion{Mg}{2} and \ion{C}{4} for each candidate that has both broad lines captured in its spectrum (black filled circles). The hatched area represents the FWHM of the flux ratio distribution of a large sample of SDSS quasars from \cite{Montuori2011}.}
\label{fig:flux_ratio}
\end{figure}


For example, \cite{Roedig2014} and \cite{Shi2016} have predicted a deficit in the spectrum (``notch'') due to missing radiation from the cavity in the circumbinary disk\footnote{However, see \cite{Farris2015}, who predicted that the notch is likely unnoticeable.}. The wavelength range of the notch is expected in the optical-to-UV band, depending on the binary parameters. A multiwavelength study of the SMBHB candidate PSO J334.2028+1.4075 (hereafter PSO J334; L15) by \cite{Foord2017} explored the possibility of such a notch. They showed that its spectral energy distribution (SED) constructed using multiband data is consistent with that of a radio-quiet quasar\footnote{With $R\sim17$ \citep{Foord2017}, PSO J334 is technically classified as radio-loud.} and does not show evidence for any deviations from a conventional AGN.

We here explore any possible notch for the best candidate from our PS1 MDS sample, PSO J185. We query the archival photometry data from the \emph{AllWISE} \citep{Cutri2013}, SDSS, and \emph{Galaxy Evolution Explorer} (\emph{GALEX}; \citealt{Bianchi2011}) catalogs; in the radio and X-ray bands, where no detections are reported, we instead use their respective upper limits. We summarize the calculated rest-frame $\nu$ and $\nu L_{\nu}$ in Table \ref{tab:sed}. We then calculate the temperature range $kT_{0} $--$15 kT_{0}$, where the spectral notch is expected, where $T_{0}$ is the characteristic temperature of the notch: $T_{0}=3.3\times10^{4}[\dot{m}(\eta/0.1)^{-1}M_{8}^{-1}(a/100R_{g})^{-3}]^{1/4}$K (we have assumed a radiative efficiency $\eta=0.1$) and the largest deficit is expected at $\sim 4 kT_{0}$ \citep{Roedig2014}. As we show in Figure \ref{fig:sed}, the SED of PSO J185 is consistent with that of a radio-quiet quasar and does not show evidence for a spectral deficit. We note, however, that at binary separations as close as those of our candidates ($\sim$ a few tens of $R_{g}$), the temperature contrast between the minidisks and the circumbinary disk is small, and the notch is consequentially likely to be unnoticeable \citep{dAscoli2018}.

Another possible signature that could arise from the binary picture and accompanies any periodic variation is a lower flux ratio between low- and high-ionization lines due to the tidal truncation of the broad-line region of the secondary \citep{Montuori2011}. Furthermore, the truncation radius is even smaller in a low-mass ratio binary and should decrease toward closer binary separations. Among the $12$ candidates we identified in Section \ref{sec:lkhd}, six have both \ion{Mg}{2} and \ion{C}{4} lines in their SDSS spectra, which allows us to measure a flux ratio $R$(\ion{Mg}{2}/\ion{C}{4}). In Figure \ref{fig:flux_ratio}, we show the flux ratios of the six candidates as a function of inferred binary separation. The ratios are consistent with those of single AGNs and do not show any correlation with the separation. However, \cite{Montuori2012} emphasized that the above prediction is only applicable to separations of $\sim 0.01$--$0.2$ pc, below which the flux ratio would be indistinguishable from that of a single AGN, due to contributions from the circumbinary disk.


\section{Summary and Concluding Remarks}\label{sec:conclude}

We have conducted a systematic search for periodically varying quasars in PS1 MDS, following our previous work in L16. Periodic variability has been predicted as a signature of an SMBHB system, as the mass accretion is modulated by the binary's orbital motion; in an SMBHB viewed at a high inclination angle, periodic variation can also be produced by relativistic Doppler boosting. The SMBHBs at subparsec separations should be products of galaxy mergers; however, compelling observational evidence for their existence has been elusive. A systematic search for periodic quasars in the time domain is therefore a novel approach to identify SMBHB candidates that are not resolvable via direct imaging.

One challenge to the SMBHB candidates identified in systematic searches (e.g., G15; C16; L16) is a robust detection of periodicity, since stochastic, normal quasar variability can easily mimic periodic variation over a small number of cycles. To monitor the variability of our periodic candidates, we have initiated an imaging campaign to monitor their variability using the DCT and the LCO network telescopes and are able to extend the total baseline of observations to 3--15 cycles. We then adopt a more rigorous, maximum likelihood approach and search for a periodic signal in the presence of red noise, which is modeled by the DRW process, or a BPL model with a steeper power spectrum. Only one candidate is statistically significant when DRW red noise is assumed, but none are significant when BPL is assumed instead. This translates to an SMBHB rate of $0.1$ per $1000$ quasars, or $0.02$ deg$^{-2}$, which is largely consistent with theoretical predictions but is lower than the rates inferred by previous searches.

We have also looked for corroborating evidence for an SMBHB by examining the SED of the most statistically significant periodic candidate from our sample. However, the apparent lack of evidence thus far signals that further multiwavelength follow-up of variability-selected SMBHB candidates is still needed in order to confirm these elusive objects.

We have developed a progressively computationally intensive pipeline for our periodicity search: from identifying quasars by their colors and variability, to computing the LS periodogram, to the more computationally expensive maximum likelihood analysis. While there exist alternative period-searching techniques, we argue that our approach is easily scalable to a much larger dataset (such as the ongoing Zwicky Transient Facility \citep{Bellm2019} and the upcoming LSST) without requiring intensive Monte Carlo simulations of light curves (which rely heavily on an assumed PSD and its parameters) and only applies the most costly analysis to the most promising candidates. As we have estimated from our down-selected rate from PS1 MDS, and as \cite{Kelley2019} recently predicted, the orders-of-magnitude more powerful LSST promises to transform the search for periodic quasars as SMBHB candidates.


\acknowledgements

T.L. thanks Cole Miller for helpful discussions and the referee(s) for their comments. S.G. is supported in part by NSF AAG grant 1616566. Partial support for T.L. was provided by the NANOGrav NSF Physics Frontiers Center award No. 1430284.

This research has made use of the VizieR catalog access tool, CDS, Strasbourg, France.

This work makes use of observations from the LCO network.

These results made use of the Discovery Channel Telescope at Lowell Observatory. Lowell is a private, nonprofit institution dedicated to astrophysical research and public appreciation of astronomy and operates the DCT in partnership with Boston University, the University of Maryland, the University of Toledo, Northern Arizona University, and Yale University.

The LMI construction was supported by a grant AST-1005313 from the National Science Foundation. The upgrade of the DeVeny optical spectrograph has been funded by a generous grant from John and Ginger Giovale.

Based on observations obtained at the Gemini Observatory (acquired through the Gemini Science Archive and processed using the Gemini IRAF package), which is operated by the Association of Universities for Research in Astronomy, Inc., under a cooperative agreement with the NSF on behalf of the Gemini partnership: the National Science Foundation (United States), the National Research Council (Canada), CONICYT (Chile), Ministerio de Ciencia, Tecnolog\'{i}a e Innovaci\'{o}n Productiva (Argentina), and Minist\'{e}rio da Ci\^{e}ncia, Tecnologia e Inova\c{c}\~{a}o (Brazil).

The Pan-STARRS1 surveys (PS1) have been made possible through contributions of the IfA, the University of Hawaii, the Pan-STARRS Project Office, the Max Planck Society and its participating institutes, MPIA, Heidelberg and MPE, Garching, JHU, Durham University, the University of Edinburgh, QUB, the Harvard-Smithsonian CfA, LCOGT Inc., the National Central University of Taiwan, STScI, NASA under grant No. NNX08AR22G issued through the Planetary Science Division of the NASA Science Mission Directorate, the NSF under grant No. AST-1238877, the University of Maryland, and Eotvos Lorand University.

Funding for SDSS-III has been provided by the Alfred P. Sloan Foundation, the Participating Institutions, the National Science Foundation, and the U.S. Department of Energy Office of Science. The SDSS-III website is http://www.sdss3.org/.

The SDSS-III is managed by the Astrophysical Research Consortium for the Participating Institutions of the SDSS-III Collaboration, including the University of Arizona, the Brazilian Participation Group, Brookhaven National Laboratory, Carnegie Mellon University, the University of Florida, the French Participation Group, the German Participation Group, Harvard University, the Instituto de Astrofisica de Canarias, the Michigan State/Notre Dame/JINA Participation Group, Johns Hopkins University, Lawrence Berkeley National Laboratory, the Max Planck Institute for Astrophysics, the Max Planck Institute for Extraterrestrial Physics, New Mexico State University, New York University, Ohio State University, Pennsylvania State University, the University of Portsmouth, Princeton University, the Spanish Participation Group, the University of Tokyo, the University of Utah, Vanderbilt University, the University of Virginia, the University of Washington, and Yale University.


\software{Astropy \citep{Astropy}, create\_fringes \citep{Snodgrass2013}, IRAF \citep{Tody1986,Tody1993}, remove\_fringes \citep{Snodgrass2013}, SCAMP \citep{BertinSCAMP}, SExtractor \citep{BertinSE}, SWARP \citep{BertinSWARP}}



\begin{thebibliography}{}
\expandafter\ifx\csname natexlab\endcsname\relax\def\natexlab#1{#1}\fi
\providecommand{\url}[1]{\href{#1}{#1}}
\providecommand{\dodoi}[1]{doi:~\href{http://doi.org/#1}{\nolinkurl{#1}}}
\providecommand{\doeprint}[1]{\href{http://ascl.net/#1}{\nolinkurl{http://ascl.net/#1}}}
\providecommand{\doarXiv}[1]{\href{https://arxiv.org/abs/#1}{\nolinkurl{https://arxiv.org/abs/#1}}}

\bibitem[{{Alam} {et~al.}(2015){Alam}, {Albareti}, {Allende Prieto}, {Anders},
  {Anderson}, {Anderton}, {Andrews}, {Armengaud}, {Aubourg}, {Bailey}, \&
  et~al.}]{SDSSDR12}
{Alam}, S., {Albareti}, F.~D., {Allende Prieto}, C., {et~al.} 2015, \apjs, 219,
  12, \dodoi{10.1088/0067-0049/219/1/12}

\bibitem[{{Astropy Collaboration} {et~al.}(2013){Astropy Collaboration},
  {Robitaille}, {Tollerud}, {Greenfield}, {Droettboom}, {Bray}, {Aldcroft},
  {Davis}, {Ginsburg}, {Price-Whelan}, {Kerzendorf}, {Conley}, {Crighton},
  {Barbary}, {Muna}, {Ferguson}, {Grollier}, {Parikh}, {Nair}, {Unther},
  {Deil}, {Woillez}, {Conseil}, {Kramer}, {Turner}, {Singer}, {Fox}, {Weaver},
  {Zabalza}, {Edwards}, {Azalee Bostroem}, {Burke}, {Casey}, {Crawford},
  {Dencheva}, {Ely}, {Jenness}, {Labrie}, {Lim}, {Pierfederici}, {Pontzen},
  {Ptak}, {Refsdal}, {Servillat}, \& {Streicher}}]{Astropy}
{Astropy Collaboration}, {Robitaille}, T.~P., {Tollerud}, E.~J., {et~al.} 2013,
  \aap, 558, A33, \dodoi{10.1051/0004-6361/201322068}

\bibitem[{{Begelman} {et~al.}(1980){Begelman}, {Blandford}, \&
  {Rees}}]{Begelman1980}
{Begelman}, M.~C., {Blandford}, R.~D., \& {Rees}, M.~J. 1980, \nat, 287, 307,
  \dodoi{10.1038/287307a0}

\bibitem[{{Bellm} {et~al.}(2019){Bellm}, {Kulkarni}, {Graham}, {Dekany},
  {Smith}, {Riddle}, {Masci}, {Helou}, {Prince}, {Adams}, {Barbarino},
  {Barlow}, {Bauer}, {Beck}, {Belicki}, {Biswas}, {Blagorodnova}, {Bodewits},
  {Bolin}, {Brinnel}, {Brooke}, {Bue}, {Bulla}, {Burruss}, {Cenko}, {Chang},
  {Connolly}, {Coughlin}, {Cromer}, {Cunningham}, {De}, {Delacroix}, {Desai},
  {Duev}, {Eadie}, {Farnham}, {Feeney}, {Feindt}, {Flynn}, {Franckowiak},
  {Frederick}, {Fremling}, {Gal-Yam}, {Gezari}, {Giomi}, {Goldstein},
  {Golkhou}, {Goobar}, {Groom}, {Hacopians}, {Hale}, {Henning}, {Ho}, {Hover},
  {Howell}, {Hung}, {Huppenkothen}, {Imel}, {Ip}, {Ivezi{\'c}}, {Jackson},
  {Jones}, {Juric}, {Kasliwal}, {Kaspi}, {Kaye}, {Kelley}, {Kowalski},
  {Kramer}, {Kupfer}, {Landry}, {Laher}, {Lee}, {Lin}, {Lin}, {Lunnan},
  {Giomi}, {Mahabal}, {Mao}, {Miller}, {Monkewitz}, {Murphy}, {Ngeow},
  {Nordin}, {Nugent}, {Ofek}, {Patterson}, {Penprase}, {Porter}, {Rauch},
  {Rebbapragada}, {Reiley}, {Rigault}, {Rodriguez}, {van Roestel}, {Rusholme},
  {van Santen}, {Schulze}, {Shupe}, {Singer}, {Soumagnac}, {Stein}, {Surace},
  {Sollerman}, {Szkody}, {Taddia}, {Terek}, {Van Sistine}, {van Velzen},
  {Vestrand}, {Walters}, {Ward}, {Ye}, {Yu}, {Yan}, \& {Zolkower}}]{Bellm2019}
{Bellm}, E.~C., {Kulkarni}, S.~R., {Graham}, M.~J., {et~al.} 2019, \pasp, 131,
  018002, \dodoi{10.1088/1538-3873/aaecbe}

\bibitem[{{Bertin}(2006)}]{BertinSCAMP}
{Bertin}, E. 2006, in Astronomical Society of the Pacific Conference Series,
  Vol. 351, Astronomical Data Analysis Software and Systems XV, ed.
  C.~{Gabriel}, C.~{Arviset}, D.~{Ponz}, \& S.~{Enrique}, 112

\bibitem[{{Bertin} \& {Arnouts}(1996)}]{BertinSE}
{Bertin}, E., \& {Arnouts}, S. 1996, \aaps, 117, 393,
  \dodoi{10.1051/aas:1996164}

\bibitem[{{Bertin} {et~al.}(2002){Bertin}, {Mellier}, {Radovich}, {Missonnier},
  {Didelon}, \& {Morin}}]{BertinSWARP}
{Bertin}, E., {Mellier}, Y., {Radovich}, M., {et~al.} 2002, in Astronomical
  Society of the Pacific Conference Series, Vol. 281, Astronomical Data
  Analysis Software and Systems XI, ed. D.~A. {Bohlender}, D.~{Durand}, \&
  T.~H. {Handley}, 228

\bibitem[{{Bianchi} {et~al.}(2011){Bianchi}, {Herald}, {Efremova}, {Girardi},
  {Zabot}, {Marigo}, {Conti}, \& {Shiao}}]{Bianchi2011}
{Bianchi}, L., {Herald}, J., {Efremova}, B., {et~al.} 2011, \apss, 335, 161,
  \dodoi{10.1007/s10509-010-0581-x}

\bibitem[{{Boroson} \& {Lauer}(2009)}]{Boroson2009}
{Boroson}, T.~A., \& {Lauer}, T.~R. 2009, \nat, 458, 53,
  \dodoi{10.1038/nature07779}

\bibitem[{{Cardelli} {et~al.}(1989){Cardelli}, {Clayton}, \&
  {Mathis}}]{Cardelli1989}
{Cardelli}, J.~A., {Clayton}, G.~C., \& {Mathis}, J.~S. 1989, \apj, 345, 245,
  \dodoi{10.1086/167900}

\bibitem[{{Chambers} {et~al.}(2016){Chambers}, {Magnier}, {Metcalfe},
  {Flewelling}, {Huber}, {Waters}, {Denneau}, {Draper}, {Farrow}, {Finkbeiner},
  {Holmberg}, {Koppenhoefer}, {Price}, {Saglia}, {Schlafly}, {Smartt},
  {Sweeney}, {Wainscoat}, {Burgett}, {Grav}, {Heasley}, {Hodapp}, {Jedicke},
  {Kaiser}, {Kudritzki}, {Luppino}, {Lupton}, {Monet}, {Morgan}, {Onaka},
  {Stubbs}, {Tonry}, {Banados}, {Bell}, {Bender}, {Bernard}, {Botticella},
  {Casertano}, {Chastel}, {Chen}, {Chen}, {Cole}, {Deacon}, {Frenk},
  {Fitzsimmons}, {Gezari}, {Goessl}, {Goggia}, {Goldman}, {Grebel}, {Hambly},
  {Hasinger}, {Heavens}, {Heckman}, {Henderson}, {Henning}, {Holman}, {Hopp},
  {Ip}, {Isani}, {Keyes}, {Koekemoer}, {Kotak}, {Long}, {Lucey}, {Liu},
  {Martin}, {McLean}, {Morganson}, {Murphy}, {Nieto-Santisteban}, {Norberg},
  {Peacock}, {Pier}, {Postman}, {Primak}, {Rae}, {Rest}, {Riess}, {Riffeser},
  {Rix}, {Roser}, {Schilbach}, {Schultz}, {Scolnic}, {Szalay}, {Seitz},
  {Shiao}, {Small}, {Smith}, {Soderblom}, {Taylor}, {Thakar}, {Thiel},
  {Thilker}, {Urata}, {Valenti}, {Walter}, {Watters}, {Werner}, {White},
  {Wood-Vasey}, \& {Wyse}}]{Chambers2016}
{Chambers}, K.~C., {Magnier}, E.~A., {Metcalfe}, N., {et~al.} 2016, ArXiv
  e-prints.
\newblock \doarXiv{1612.05560}

\bibitem[{{Charisi} {et~al.}(2016){Charisi}, {Bartos}, {Haiman},
  {Price-Whelan}, {Graham}, {Bellm}, {Laher}, \& {M{\'a}rka}}]{Charisi2016}
{Charisi}, M., {Bartos}, I., {Haiman}, Z., {et~al.} 2016, \mnras, 463, 2145,
  \dodoi{10.1093/mnras/stw1838}

\bibitem[{{Charisi} {et~al.}(2018){Charisi}, {Haiman}, {Schiminovich}, \&
  {D'Orazio}}]{Charisi2018}
{Charisi}, M., {Haiman}, Z., {Schiminovich}, D., \& {D'Orazio}, D.~J. 2018,
  \mnras, 476, 4617, \dodoi{10.1093/mnras/sty516}

\bibitem[{{Comerford} {et~al.}(2015){Comerford}, {Pooley}, {Barrows}, {Greene},
  {Zakamska}, {Madejski}, \& {Cooper}}]{Comerford2015}
{Comerford}, J.~M., {Pooley}, D., {Barrows}, R.~S., {et~al.} 2015, \apj, 806,
  219, \dodoi{10.1088/0004-637X/806/2/219}

\bibitem[{{Cuadra} {et~al.}(2009){Cuadra}, {Armitage}, {Alexander}, \&
  {Begelman}}]{Cuadra2009}
{Cuadra}, J., {Armitage}, P.~J., {Alexander}, R.~D., \& {Begelman}, M.~C. 2009,
  \mnras, 393, 1423, \dodoi{10.1111/j.1365-2966.2008.14147.x}

\bibitem[{{Cutri} \& {et al.}(2013)}]{Cutri2013}
{Cutri}, R.~M., \& {et al.} 2013, VizieR Online Data Catalog, 2328

\bibitem[{{d'Ascoli} {et~al.}(2018){d'Ascoli}, {Noble}, {Bowen}, {Campanelli},
  {Krolik}, \& {Mewes}}]{dAscoli2018}
{d'Ascoli}, S., {Noble}, S.~C., {Bowen}, D.~B., {et~al.} 2018, \apj, 865, 140,
  \dodoi{10.3847/1538-4357/aad8b4}

\bibitem[{{D'Orazio} {et~al.}(2013){D'Orazio}, {Haiman}, \&
  {MacFadyen}}]{D'Orazio2013}
{D'Orazio}, D.~J., {Haiman}, Z., \& {MacFadyen}, A. 2013, \mnras, 436, 2997,
  \dodoi{10.1093/mnras/stt1787}

\bibitem[{{D'Orazio} {et~al.}(2015){D'Orazio}, {Haiman}, \&
  {Schiminovich}}]{D'Orazio2015}
{D'Orazio}, D.~J., {Haiman}, Z., \& {Schiminovich}, D. 2015, \nat, 525, 351,
  \dodoi{10.1038/nature15262}

\bibitem[{{Elvis} {et~al.}(1994){Elvis}, {Wilkes}, {McDowell}, {Green},
  {Bechtold}, {Willner}, {Oey}, {Polomski}, \& {Cutri}}]{Elvis1994}
{Elvis}, M., {Wilkes}, B.~J., {McDowell}, J.~C., {et~al.} 1994, \apjs, 95, 1,
  \dodoi{10.1086/192093}

\bibitem[{{Eracleous} {et~al.}(2012){Eracleous}, {Boroson}, {Halpern}, \&
  {Liu}}]{Eracleous2012}
{Eracleous}, M., {Boroson}, T.~A., {Halpern}, J.~P., \& {Liu}, J. 2012, \apjs,
  201, 23, \dodoi{10.1088/0067-0049/201/2/23}

\bibitem[{{Farris} {et~al.}(2014){Farris}, {Duffell}, {MacFadyen}, \&
  {Haiman}}]{Farris2014}
{Farris}, B.~D., {Duffell}, P., {MacFadyen}, A.~I., \& {Haiman}, Z. 2014, ApJ,
  783, 134, \dodoi{10.1088/0004-637X/783/2/134}

\bibitem[{{Farris} {et~al.}(2015){Farris}, {Duffell}, {MacFadyen}, \&
  {Haiman}}]{Farris2015}
---. 2015, \mnras, 446, L36, \dodoi{10.1093/mnrasl/slu160}

\bibitem[{{Foord} {et~al.}(2017){Foord}, {G{\"u}ltekin}, {Reynolds}, {Ayers},
  {Liu}, {Gezari}, \& {Runnoe}}]{Foord2017}
{Foord}, A., {G{\"u}ltekin}, K., {Reynolds}, M., {et~al.} 2017, \apj, 851, 106,
  \dodoi{10.3847/1538-4357/aa9a39}

\bibitem[{{Gierli{\'n}ski} {et~al.}(2008){Gierli{\'n}ski}, {Middleton}, {Ward},
  \& {Done}}]{Gierlinski2008}
{Gierli{\'n}ski}, M., {Middleton}, M., {Ward}, M., \& {Done}, C. 2008, \nat,
  455, 369, \dodoi{10.1038/nature07277}

\bibitem[{{Gold} {et~al.}(2014){Gold}, {Paschalidis}, {Etienne}, {Shapiro}, \&
  {Pfeiffer}}]{Gold2014}
{Gold}, R., {Paschalidis}, V., {Etienne}, Z.~B., {Shapiro}, S.~L., \&
  {Pfeiffer}, H.~P. 2014, \prd, 89, 064060, \dodoi{10.1103/PhysRevD.89.064060}

\bibitem[{{Graham} {et~al.}(2015{\natexlab{a}}){Graham}, {Djorgovski}, {Stern},
  {Drake}, {Mahabal}, {Donalek}, {Glikman}, {Larson}, \&
  {Christensen}}]{Graham2015}
{Graham}, M.~J., {Djorgovski}, S.~G., {Stern}, D., {et~al.} 2015{\natexlab{a}},
  \mnras, 453, 1562, \dodoi{10.1093/mnras/stv1726}

\bibitem[{{Graham} {et~al.}(2015{\natexlab{b}}){Graham}, {Djorgovski}, {Stern},
  {Glikman}, {Drake}, {Mahabal}, {Donalek}, {Larson}, \&
  {Christensen}}]{Graham2015Nat}
---. 2015{\natexlab{b}}, \nat, 518, 74, \dodoi{10.1038/nature14143}

\bibitem[{{Haiman} {et~al.}(2009){Haiman}, {Kocsis}, \& {Menou}}]{Haiman2009}
{Haiman}, Z., {Kocsis}, B., \& {Menou}, K. 2009, \apj, 700, 1952,
  \dodoi{10.1088/0004-637X/700/2/1952}

\bibitem[{{Heinis} {et~al.}(2016{\natexlab{a}}){Heinis}, {Gezari}, {Kumar},
  {Burgett}, {Flewelling}, {Huber}, {Kaiser}, {Wainscoat}, \&
  {Waters}}]{Heinis2016AGN}
{Heinis}, S., {Gezari}, S., {Kumar}, S., {et~al.} 2016{\natexlab{a}}, \apj,
  826, 62, \dodoi{10.3847/0004-637X/826/1/62}

\bibitem[{{Heinis} {et~al.}(2016{\natexlab{b}}){Heinis}, {Kumar}, {Gezari},
  {Burgett}, {Chambers}, {Draper}, {Flewelling}, {Kaiser}, {Magnier},
  {Metcalfe}, \& {Waters}}]{Heinis2016SVM}
{Heinis}, S., {Kumar}, S., {Gezari}, S., {et~al.} 2016{\natexlab{b}}, \apj,
  821, 86, \dodoi{10.3847/0004-637X/821/2/86}

\bibitem[{{Horne} \& {Baliunas}(1986)}]{Horne1986}
{Horne}, J.~H., \& {Baliunas}, S.~L. 1986, \apj, 302, 757,
  \dodoi{10.1086/164037}

\bibitem[{{Ivezic} {et~al.}(2008){Ivezic}, {Tyson}, {Abel}, {Acosta},
  {Allsman}, {AlSayyad}, {Anderson}, {Andrew}, {Angel}, {Angeli}, {Ansari},
  {Antilogus}, {Arndt}, {Astier}, {Aubourg}, {Axelrod}, {Bard}, {Barr},
  {Barrau}, {Bartlett}, {Bauman}, {Beaumont}, {Becker}, {Becla}, {Beldica},
  {Bellavia}, {Blanc}, {Blandford}, {Bloom}, {Bogart}, {Borne}, {Bosch},
  {Boutigny}, {Brandt}, {Brown}, {Bullock}, {Burchat}, {Burke}, {Cagnoli},
  {Calabrese}, {Chandrasekharan}, {Chesley}, {Cheu}, {Chiang}, {Claver},
  {Connolly}, {Cook}, {Cooray}, {Covey}, {Cribbs}, {Cui}, {Cutri}, {Daubard},
  {Daues}, {Delgado}, {Digel}, {Doherty}, {Dubois}, {Dubois-Felsmann},
  {Durech}, {Eracleous}, {Ferguson}, {Frank}, {Freemon}, {Gangler}, {Gawiser},
  {Geary}, {Gee}, {Geha}, {Gibson}, {Gilmore}, {Glanzman}, {Goodenow},
  {Gressler}, {Gris}, {Guyonnet}, {Hascall}, {Haupt}, {Hernandez}, {Hogan},
  {Huang}, {Huffer}, {Innes}, {Jacoby}, {Jain}, {Jee}, {Jernigan},
  {Jevremovic}, {Johns}, {Jones}, {Juramy-Gilles}, {Juric}, {Kahn}, {Kalirai},
  {Kallivayalil}, {Kalmbach}, {Kantor}, {Kasliwal}, {Kessler}, {Kirkby},
  {Knox}, {Kotov}, {Krabbendam}, {Krughoff}, {Kubanek}, {Kuczewski},
  {Kulkarni}, {Lambert}, {Le Guillou}, {Levine}, {Liang}, {Lim}, {Lintott},
  {Lupton}, {Mahabal}, {Marshall}, {Marshall}, {May}, {McKercher}, {Migliore},
  {Miller}, {Mills}, {Monet}, {Moniez}, {Neill}, {Nief}, {Nomerotski},
  {Nordby}, {O'Connor}, {Oliver}, {Olivier}, {Olsen}, {Ortiz}, {Owen}, {Pain},
  {Peterson}, {Petry}, {Pierfederici}, {Pietrowicz}, {Pike}, {Pinto}, {Plante},
  {Plate}, {Price}, {Prouza}, {Radeka}, {Rajagopal}, {Rasmussen}, {Regnault},
  {Ridgway}, {Ritz}, {Rosing}, {Roucelle}, {Rumore}, {Russo}, {Saha},
  {Sassolas}, {Schalk}, {Schindler}, {Schneider}, {Schumacher}, {Sebag},
  {Sembroski}, {Seppala}, {Shipsey}, {Silvestri}, {Smith}, {Smith}, {Strauss},
  {Stubbs}, {Sweeney}, {Szalay}, {Takacs}, {Thaler}, {Van Berg}, {Vanden Berk},
  {Vetter}, {Virieux}, {Xin}, {Walkowicz}, {Walter}, {Wang}, {Warner},
  {Willman}, {Wittman}, {Wolff}, {Wood-Vasey}, {Yoachim}, {Zhan}, \& {for the
  LSST Collaboration}}]{Ivezic2008}
{Ivezic}, Z., {Tyson}, J.~A., {Abel}, B., {et~al.} 2008, ArXiv e-prints.
\newblock \doarXiv{0805.2366}

\bibitem[{Kaiser {et~al.}(2010)Kaiser, Burgett, Chambers, Denneau, Heasley,
  Jedicke, Magnier, Morgan, Onaka, \& Tonry}]{Kaiser2010}
Kaiser, N., Burgett, W., Chambers, K., {et~al.} 2010, in Ground-based and
  Airborne Telescopes III, ed. L.~M. Stepp, R.~Gilmozzi, \& H.~J. Hall, Vol.
  7733, International Society for Optics and Photonics (SPIE), 159 -- 172.
\newblock \url{https://doi.org/10.1117/12.859188}

\bibitem[{{Kelley} {et~al.}(2019){Kelley}, {Haiman}, {Sesana}, \&
  {Hernquist}}]{Kelley2019}
{Kelley}, L.~Z., {Haiman}, Z., {Sesana}, A., \& {Hernquist}, L. 2019, \mnras,
  485, 1579, \dodoi{10.1093/mnras/stz150}

\bibitem[{{Kelly} {et~al.}(2009){Kelly}, {Bechtold}, \&
  {Siemiginowska}}]{Kelly2009}
{Kelly}, B.~C., {Bechtold}, J., \& {Siemiginowska}, A. 2009, \apj, 698, 895,
  \dodoi{10.1088/0004-637X/698/1/895}

\bibitem[{{Kochanek} {et~al.}(2017){Kochanek}, {Shappee}, {Stanek}, {Holoien},
  {Thompson}, {Prieto}, {Dong}, {Shields}, {Will}, {Britt}, {Perzanowski}, \&
  {Pojma{\'n}ski}}]{Kochanek2017}
{Kochanek}, C.~S., {Shappee}, B.~J., {Stanek}, K.~Z., {et~al.} 2017, \pasp,
  129, 104502, \dodoi{10.1088/1538-3873/aa80d9}

\bibitem[{{Liu} {et~al.}(2018){Liu}, {Gezari}, \& {Miller}}]{Liu2018a}
{Liu}, T., {Gezari}, S., \& {Miller}, M.~C. 2018, \apjl, 859, L12,
  \dodoi{10.3847/2041-8213/aac2ed}

\bibitem[{{Liu} {et~al.}(2015){Liu}, {Gezari}, {Heinis}, {Magnier}, {Burgett},
  {Chambers}, {Flewelling}, {Huber}, {Hodapp}, {Kaiser}, {Kudritzki}, {Tonry},
  {Wainscoat}, \& {Waters}}]{Liu2015}
{Liu}, T., {Gezari}, S., {Heinis}, S., {et~al.} 2015, \apjl, 803, L16,
  \dodoi{10.1088/2041-8205/803/2/L16}

\bibitem[{{Liu} {et~al.}(2016){Liu}, {Gezari}, {Burgett}, {Chambers}, {Draper},
  {Hodapp}, {Huber}, {Kudritzki}, {Magnier}, {Metcalfe}, {Tonry}, {Wainscoat},
  \& {Waters}}]{Liu2016}
{Liu}, T., {Gezari}, S., {Burgett}, W., {et~al.} 2016, \apj, 833, 6,
  \dodoi{10.3847/0004-637X/833/1/6}

\bibitem[{{Lomb}(1976)}]{Lomb1976}
{Lomb}, N.~R. 1976, \apss, 39, 447, \dodoi{10.1007/BF00648343}

\bibitem[{{MacFadyen} \& {Milosavljevi{\'c}}(2008)}]{MacFadyen2008}
{MacFadyen}, A.~I., \& {Milosavljevi{\'c}}, M. 2008, \apj, 672, 83,
  \dodoi{10.1086/523869}

\bibitem[{{MacLeod} {et~al.}(2010){MacLeod}, {Ivezi{\'c}}, {Kochanek},
  {Koz{\l}owski}, {Kelly}, {Bullock}, {Kimball}, {Sesar}, {Westman}, {Brooks},
  {Gibson}, {Becker}, \& {de Vries}}]{MacLeod2010}
{MacLeod}, C.~L., {Ivezi{\'c}}, {\v Z}., {Kochanek}, C.~S., {et~al.} 2010,
  \apj, 721, 1014, \dodoi{10.1088/0004-637X/721/2/1014}

\bibitem[{{Magnier}(2006)}]{Magnier2006IPP}
{Magnier}, E. 2006, in The Advanced Maui Optical and Space Surveillance
  Technologies Conference, E50

\bibitem[{{McLure} \& {Dunlop}(2004)}]{McLure2004}
{McLure}, R.~J., \& {Dunlop}, J.~S. 2004, \mnras, 352, 1390,
  \dodoi{10.1111/j.1365-2966.2004.08034.x}

\bibitem[{{Mej{\'{\i}}a-Restrepo} {et~al.}(2016){Mej{\'{\i}}a-Restrepo},
  {Trakhtenbrot}, {Lira}, {Netzer}, \& {Capellupo}}]{Mejia2016}
{Mej{\'{\i}}a-Restrepo}, J.~E., {Trakhtenbrot}, B., {Lira}, P., {Netzer}, H.,
  \& {Capellupo}, D.~M. 2016, \mnras, 460, 187, \dodoi{10.1093/mnras/stw568}

\bibitem[{{Montuori} {et~al.}(2011){Montuori}, {Dotti}, {Colpi}, {Decarli}, \&
  {Haardt}}]{Montuori2011}
{Montuori}, C., {Dotti}, M., {Colpi}, M., {Decarli}, R., \& {Haardt}, F. 2011,
  \mnras, 412, 26, \dodoi{10.1111/j.1365-2966.2010.17888.x}

\bibitem[{{Montuori} {et~al.}(2012){Montuori}, {Dotti}, {Haardt}, {Colpi}, \&
  {Decarli}}]{Montuori2012}
{Montuori}, C., {Dotti}, M., {Haardt}, F., {Colpi}, M., \& {Decarli}, R. 2012,
  \mnras, 425, 1633, \dodoi{10.1111/j.1365-2966.2012.21530.x}

\bibitem[{{Noble} {et~al.}(2012){Noble}, {Mundim}, {Nakano}, {Krolik},
  {Campanelli}, {Zlochower}, \& {Yunes}}]{Noble2012}
{Noble}, S.~C., {Mundim}, B.~C., {Nakano}, H., {et~al.} 2012, \apj, 755, 51,
  \dodoi{10.1088/0004-637X/755/1/51}

\bibitem[{{Peters} {et~al.}(2015){Peters}, {Richards}, {Myers}, {Strauss},
  {Schmidt}, {Ivezi{\'c}}, {Ross}, {MacLeod}, \& {Riegel}}]{Peters2015}
{Peters}, C.~M., {Richards}, G.~T., {Myers}, A.~D., {et~al.} 2015, \apj, 811,
  95, \dodoi{10.1088/0004-637X/811/2/95}

\bibitem[{{Pflueger} {et~al.}(2018){Pflueger}, {Nguyen}, {Bogdanovi{\'c}},
  {Eracleous}, {Runnoe}, {Sigurdsson}, \& {Boroson}}]{Pflueger2018}
{Pflueger}, B.~J., {Nguyen}, K., {Bogdanovi{\'c}}, T., {et~al.} 2018, \apj,
  861, 59, \dodoi{10.3847/1538-4357/aaca2c}

\bibitem[{{Rodriguez} {et~al.}(2006){Rodriguez}, {Taylor}, {Zavala}, {Peck},
  {Pollack}, \& {Romani}}]{Rodriguez2006}
{Rodriguez}, C., {Taylor}, G.~B., {Zavala}, R.~T., {et~al.} 2006, ApJ, 646, 49,
  \dodoi{10.1086/504825}

\bibitem[{{Roedig} {et~al.}(2014){Roedig}, {Krolik}, \& {Miller}}]{Roedig2014}
{Roedig}, C., {Krolik}, J.~H., \& {Miller}, M.~C. 2014, \apj, 785, 115,
  \dodoi{10.1088/0004-637X/785/2/115}

\bibitem[{{Runnoe} {et~al.}(2017){Runnoe}, {Eracleous}, {Pennell}, {Mathes},
  {Boroson}, {Sigurdsson}, {Bogdanovic}, {Halpern}, {Liu}, \&
  {Brown}}]{Runnoe2017}
{Runnoe}, J.~C., {Eracleous}, M., {Pennell}, A., {et~al.} 2017, MNRAS, 468,
  1683, \dodoi{10.1093/mnras/stx452}

\bibitem[{{Scargle}(1982)}]{Scargle1982}
{Scargle}, J.~D. 1982, \apj, 263, 835, \dodoi{10.1086/160554}

\bibitem[{{Schlafly} \& {Finkbeiner}(2011)}]{SF2011}
{Schlafly}, E.~F., \& {Finkbeiner}, D.~P. 2011, \apj, 737, 103,
  \dodoi{10.1088/0004-637X/737/2/103}

\bibitem[{{Sesana} {et~al.}(2018){Sesana}, {Haiman}, {Kocsis}, \&
  {Kelley}}]{Sesana2018}
{Sesana}, A., {Haiman}, Z., {Kocsis}, B., \& {Kelley}, L.~Z. 2018, \apj, 856,
  42, \dodoi{10.3847/1538-4357/aaad0f}

\bibitem[{{Sesar} {et~al.}(2007){Sesar}, {Ivezi{\'c}}, {Lupton}, {Juri{\'c}},
  {Gunn}, {Knapp}, {DeLee}, {Smith}, {Miknaitis}, {Lin}, {Tucker}, {Doi},
  {Tanaka}, {Fukugita}, {Holtzman}, {Kent}, {Yanny}, {Schlegel}, {Finkbeiner},
  {Padmanabhan}, {Rockosi}, {Bond}, {Lee}, {Stoughton}, {Jester}, {Harris},
  {Harding}, {Brinkmann}, {Schneider}, {York}, {Richmond}, \& {Vanden
  Berk}}]{Sesar2007}
{Sesar}, B., {Ivezi{\'c}}, {\v Z}., {Lupton}, R.~H., {et~al.} 2007, \aj, 134,
  2236, \dodoi{10.1086/521819}

\bibitem[{{Shappee} {et~al.}(2014){Shappee}, {Prieto}, {Grupe}, {Kochanek},
  {Stanek}, {De Rosa}, {Mathur}, {Zu}, {Peterson}, {Pogge}, {Komossa}, {Im},
  {Jencson}, {Holoien}, {Basu}, {Beacom}, {Szczygie{\l}}, {Brimacombe},
  {Adams}, {Campillay}, {Choi}, {Contreras}, {Dietrich}, {Dubberley},
  {Elphick}, {Foale}, {Giustini}, {Gonzalez}, {Hawkins}, {Howell}, {Hsiao},
  {Koss}, {Leighly}, {Morrell}, {Mudd}, {Mullins}, {Nugent}, {Parrent},
  {Phillips}, {Pojmanski}, {Rosing}, {Ross}, {Sand}, {Terndrup}, {Valenti},
  {Walker}, \& {Yoon}}]{Shappee2014}
{Shappee}, B.~J., {Prieto}, J.~L., {Grupe}, D., {et~al.} 2014, \apj, 788, 48,
  \dodoi{10.1088/0004-637X/788/1/48}

\bibitem[{{Shen} {et~al.}(2008){Shen}, {Greene}, {Strauss}, {Richards}, \&
  {Schneider}}]{Shen2008}
{Shen}, Y., {Greene}, J.~E., {Strauss}, M.~A., {Richards}, G.~T., \&
  {Schneider}, D.~P. 2008, \apj, 680, 169, \dodoi{10.1086/587475}

\bibitem[{{Shen} \& {Loeb}(2010)}]{Shen2010}
{Shen}, Y., \& {Loeb}, A. 2010, \apj, 725, 249,
  \dodoi{10.1088/0004-637X/725/1/249}

\bibitem[{{Shi} \& {Krolik}(2016)}]{Shi2016}
{Shi}, J.-M., \& {Krolik}, J.~H. 2016, \apj, 832, 22,
  \dodoi{10.3847/0004-637X/832/1/22}

\bibitem[{{Shi} {et~al.}(2012){Shi}, {Krolik}, {Lubow}, \& {Hawley}}]{Shi2012}
{Shi}, J.-M., {Krolik}, J.~H., {Lubow}, S.~H., \& {Hawley}, J.~F. 2012, \apj,
  749, 118, \dodoi{10.1088/0004-637X/749/2/118}

\bibitem[{{Smith} {et~al.}(2018){Smith}, {Mushotzky}, {Boyd}, \&
  {Wagoner}}]{Smith2018QPO}
{Smith}, K.~L., {Mushotzky}, R.~F., {Boyd}, P.~T., \& {Wagoner}, R.~V. 2018,
  \apjl, 860, L10, \dodoi{10.3847/2041-8213/aac88c}

\bibitem[{{Snodgrass} \& {Carry}(2013)}]{Snodgrass2013}
{Snodgrass}, C., \& {Carry}, B. 2013, The Messenger, 152, 14

\bibitem[{{Tang} {et~al.}(2018){Tang}, {Haiman}, \& {MacFadyen}}]{Tang2018}
{Tang}, Y., {Haiman}, Z., \& {MacFadyen}, A. 2018, \mnras, 476, 2249,
  \dodoi{10.1093/mnras/sty423}

\bibitem[{{Tody}(1986)}]{Tody1986}
{Tody}, D. 1986, in \procspie, Vol. 627, Instrumentation in astronomy VI, ed.
  D.~L. {Crawford}, 733

\bibitem[{{Tody}(1993)}]{Tody1993}
{Tody}, D. 1993, in Astronomical Society of the Pacific Conference Series,
  Vol.~52, Astronomical Data Analysis Software and Systems II, ed. R.~J.
  {Hanisch}, R.~J.~V. {Brissenden}, \& J.~{Barnes}, 173

\bibitem[{{Tonry} {et~al.}(2012){Tonry}, {Stubbs}, {Lykke}, {Doherty},
  {Shivvers}, {Burgett}, {Chambers}, {Hodapp}, {Kaiser}, {Kudritzki},
  {Magnier}, {Morgan}, {Price}, \& {Wainscoat}}]{Tonry2012photometry}
{Tonry}, J.~L., {Stubbs}, C.~W., {Lykke}, K.~R., {et~al.} 2012, \apj, 750, 99,
  \dodoi{10.1088/0004-637X/750/2/99}

\bibitem[{{Vanden Berk} {et~al.}(2001){Vanden Berk}, {Richards}, {Bauer},
  {Strauss}, {Schneider}, {Heckman}, {York}, {Hall}, {Fan}, {Knapp},
  {Anderson}, {Annis}, {Bahcall}, {Bernardi}, {Briggs}, {Brinkmann}, {Brunner},
  {Burles}, {Carey}, {Castander}, {Connolly}, {Crocker}, {Csabai}, {Doi},
  {Finkbeiner}, {Friedman}, {Frieman}, {Fukugita}, {Gunn}, {Hennessy},
  {Ivezi{\'c}}, {Kent}, {Kunszt}, {Lamb}, {Leger}, {Long}, {Loveday}, {Lupton},
  {Meiksin}, {Merelli}, {Munn}, {Newberg}, {Newcomb}, {Nichol}, {Owen}, {Pier},
  {Pope}, {Rockosi}, {Schlegel}, {Siegmund}, {Smee}, {Snir}, {Stoughton},
  {Stubbs}, {SubbaRao}, {Szalay}, {Szokoly}, {Tremonti}, {Uomoto}, {Waddell},
  {Yanny}, \& {Zheng}}]{VandenBerk2001}
{Vanden Berk}, D.~E., {Richards}, G.~T., {Bauer}, A., {et~al.} 2001, \aj, 122,
  549, \dodoi{10.1086/321167}

\bibitem[{{Vanden Berk} {et~al.}(2004){Vanden Berk}, {Wilhite}, {Kron},
  {Anderson}, {Brunner}, {Hall}, {Ivezi{\'c}}, {Richards}, {Schneider}, {York},
  {Brinkmann}, {Lamb}, {Nichol}, \& {Schlegel}}]{VandenBerk2004}
{Vanden Berk}, D.~E., {Wilhite}, B.~C., {Kron}, R.~G., {et~al.} 2004, \apj,
  601, 692, \dodoi{10.1086/380563}

\bibitem[{{Vaughan} {et~al.}(2016){Vaughan}, {Uttley}, {Markowitz},
  {Huppenkothen}, {Middleton}, {Alston}, {Scargle}, \& {Farr}}]{Vaughan2016}
{Vaughan}, S., {Uttley}, P., {Markowitz}, A.~G., {et~al.} 2016, \mnras, 461,
  3145, \dodoi{10.1093/mnras/stw1412}

\bibitem[{{Vestergaard} \& {Peterson}(2006)}]{Vestergaard2006}
{Vestergaard}, M., \& {Peterson}, B.~M. 2006, \apj, 641, 689,
  \dodoi{10.1086/500572}

\bibitem[{{Vestergaard} \& {Wilkes}(2001)}]{Vestergaard2001}
{Vestergaard}, M., \& {Wilkes}, B.~J. 2001, \apjs, 134, 1,
  \dodoi{10.1086/320357}

\bibitem[{{Volonteri} {et~al.}(2009){Volonteri}, {Miller}, \&
  {Dotti}}]{Volonteri2009}
{Volonteri}, M., {Miller}, J.~M., \& {Dotti}, M. 2009, \apjl, 703, L86,
  \dodoi{10.1088/0004-637X/703/1/L86}

\end{thebibliography}


\appendix
\renewcommand\thefigure{\thesection.\arabic{figure}}  
\section{PS1-only and Extended Light Curves of PS1 MDS Candidates}\label{app:lc}

Figure \ref{fig:lc} shows the PS1 and extended light curves of the candidates from PS1 MDS (Section \ref{sec:lkhd}). Sinusoids of periods determined from the periodogram are imposed to guide the eye (dashed lines). Different sources of archival or new monitoring data are represented by different symbols: \emph{GALEX} -- dots, SDSS/S82 -- stars, PS1/MDS -- circles, DCT/LMI -- squares, LCO/Spectral -- diamonds.
\setcounter{figure}{0}   
\begin{figure*}[h] 
\centering
\epsfig{file=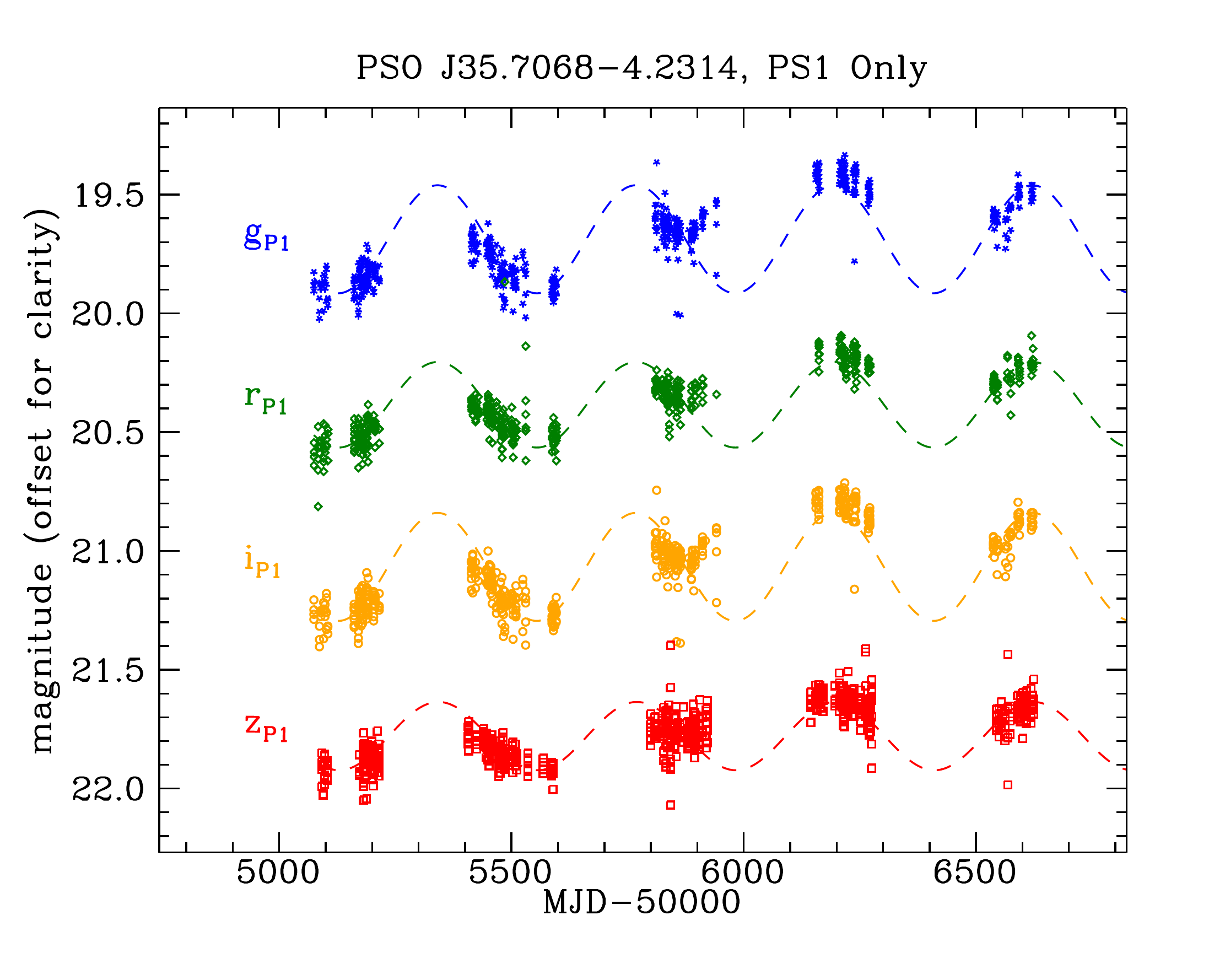,width=0.2\textwidth,clip=}
\epsfig{file=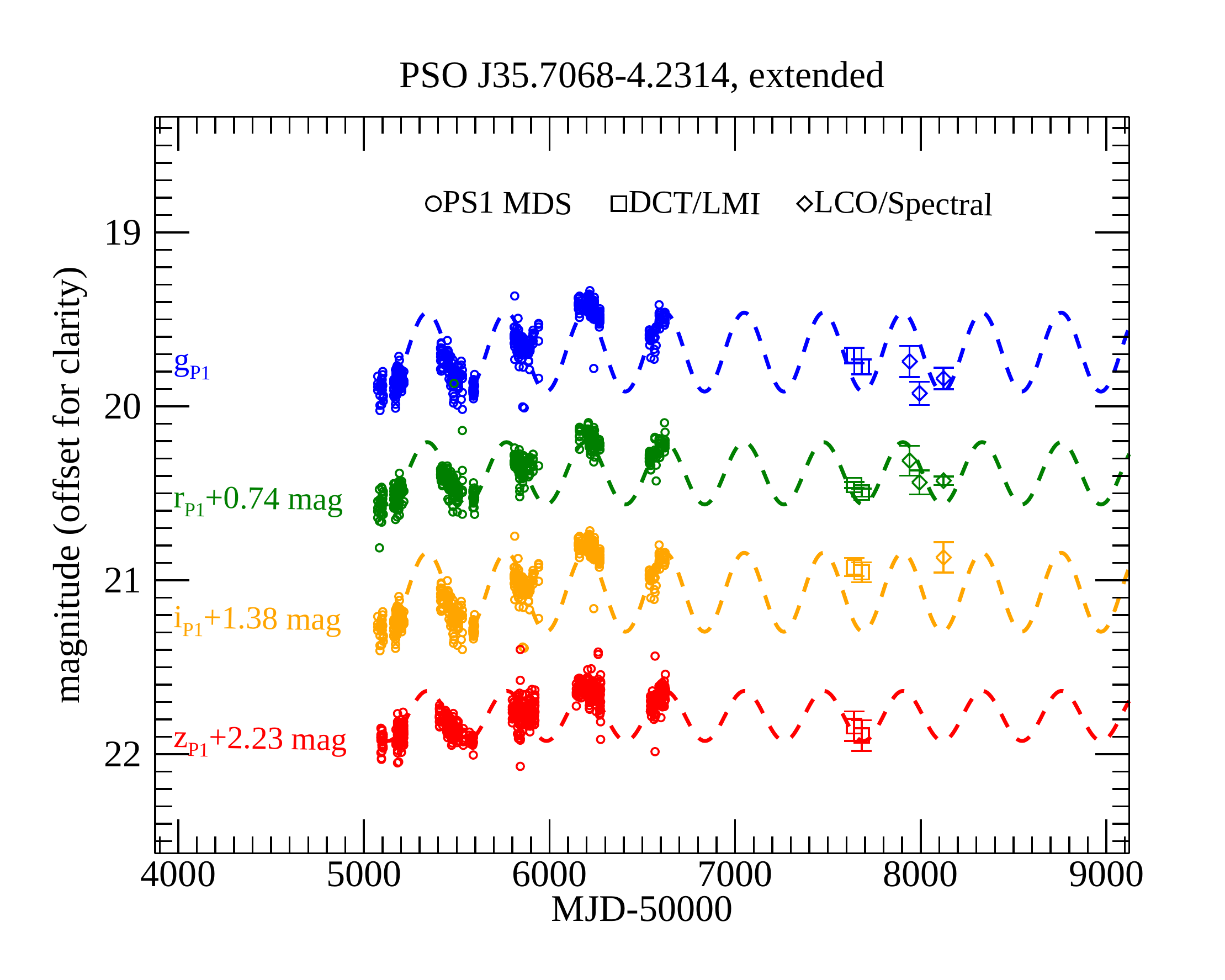,width=0.2\textwidth,clip=}
\epsfig{file=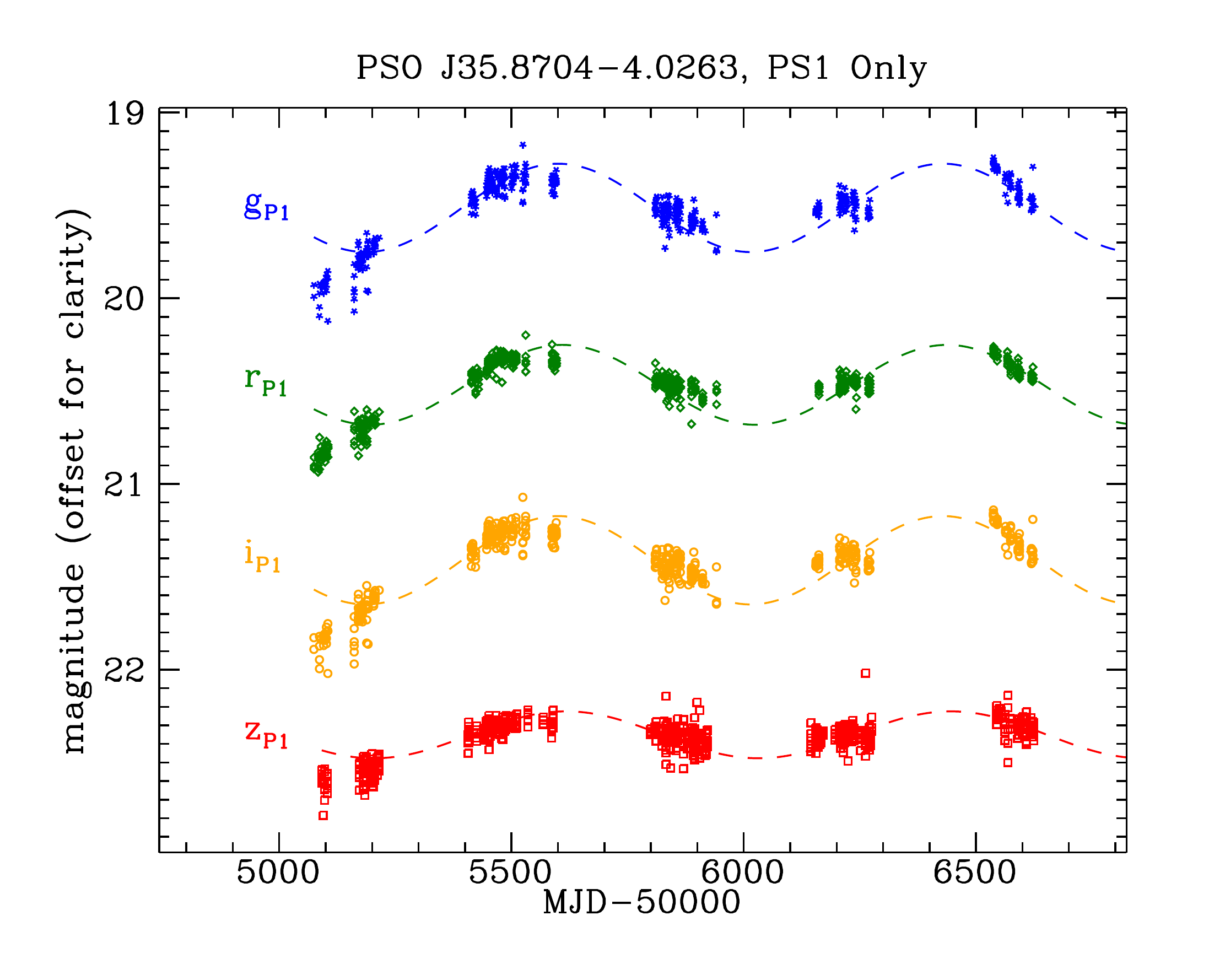,width=0.2\textwidth,clip=}
\epsfig{file=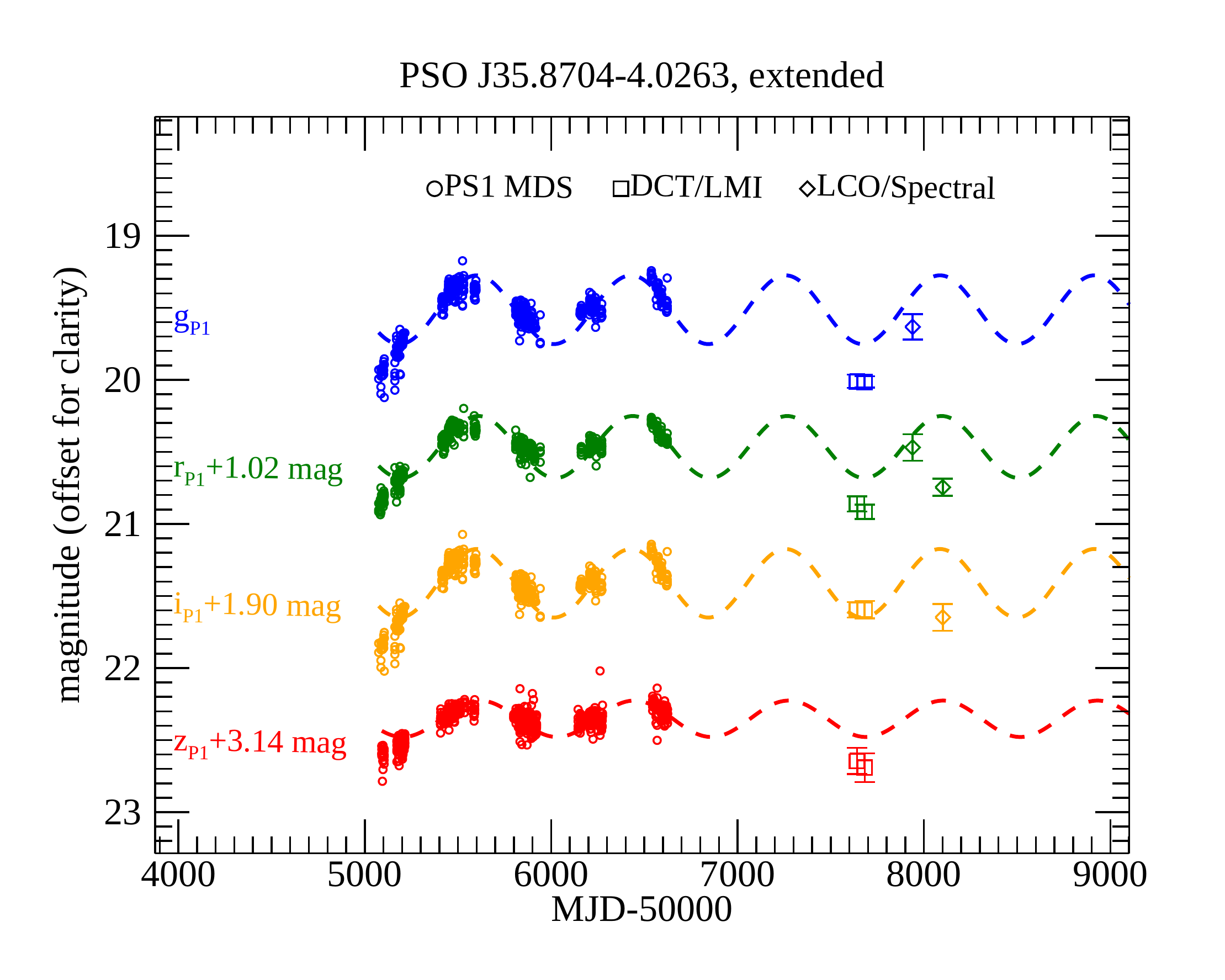,width=0.2\textwidth,clip=}
\epsfig{file=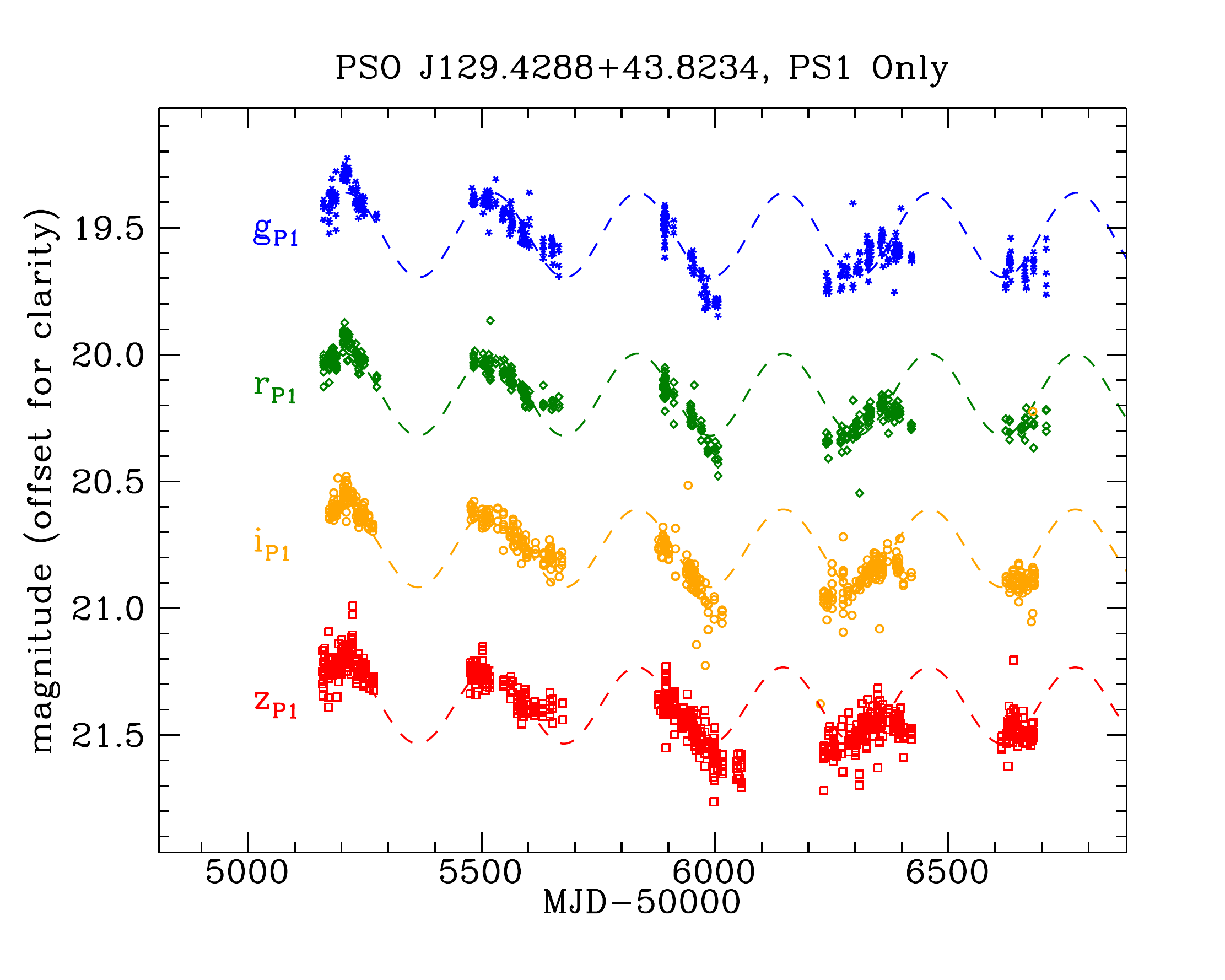,width=0.2\textwidth,clip=}
\epsfig{file=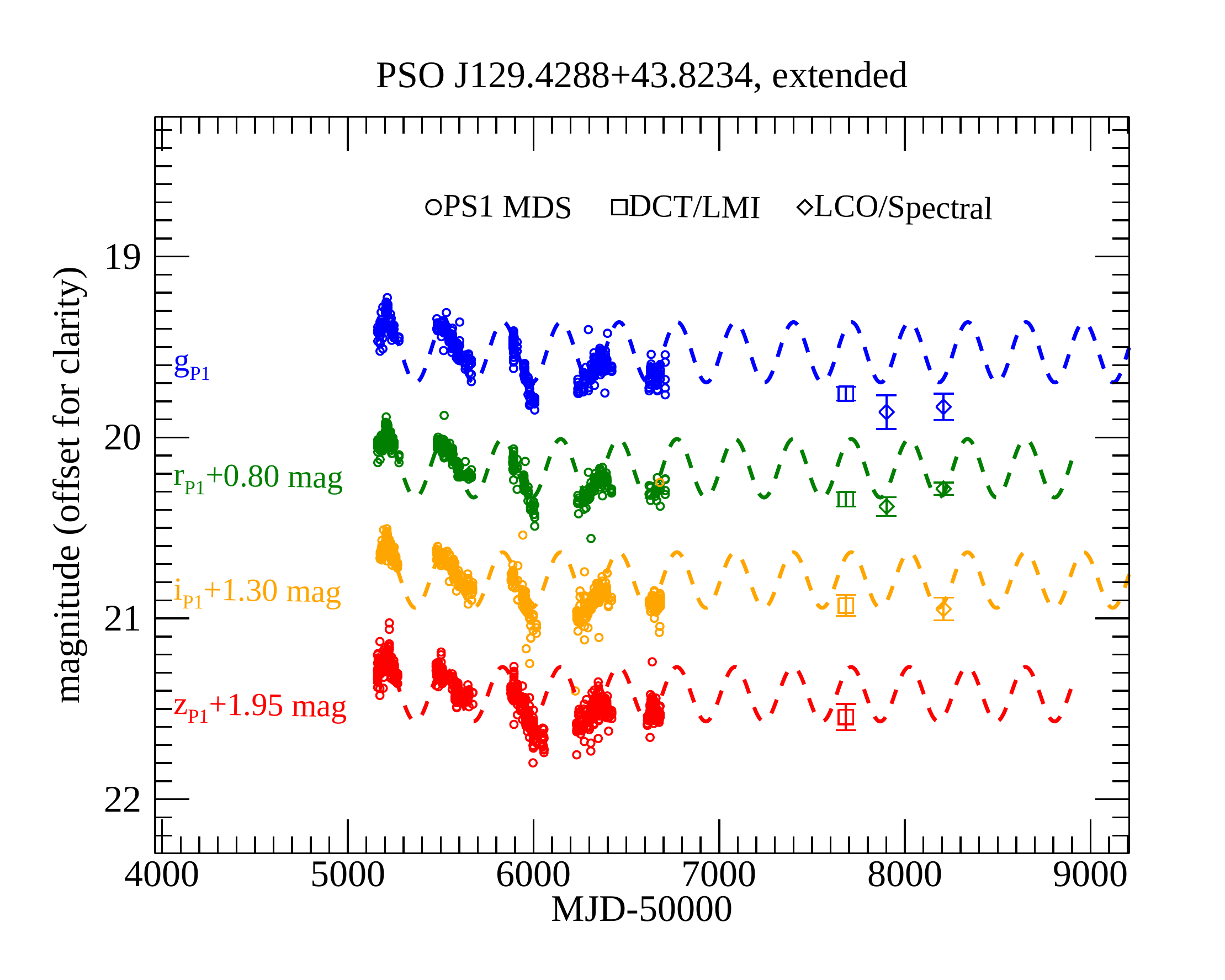,width=0.2\textwidth,clip=}
\epsfig{file=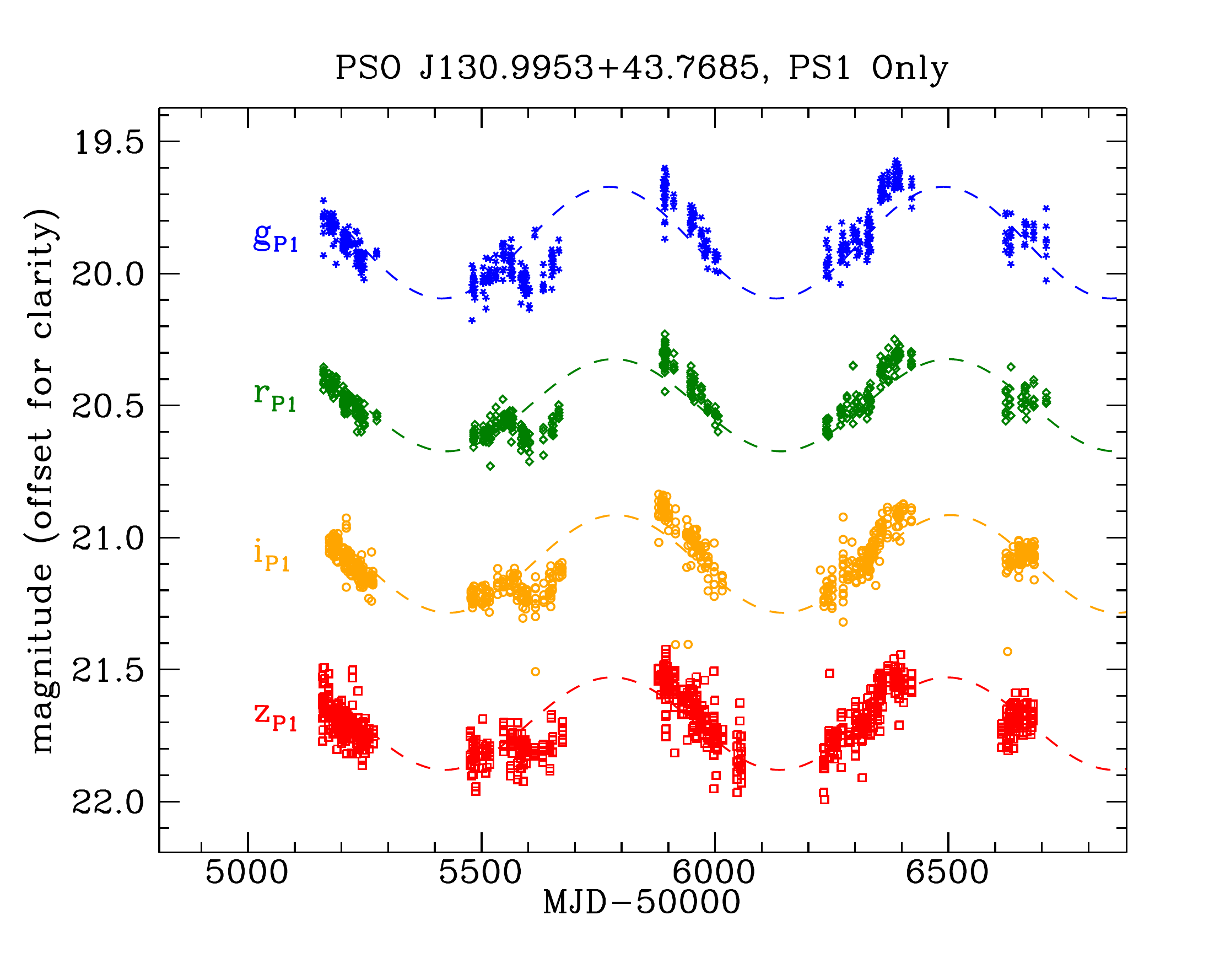,width=0.2\textwidth,clip=}
\epsfig{file=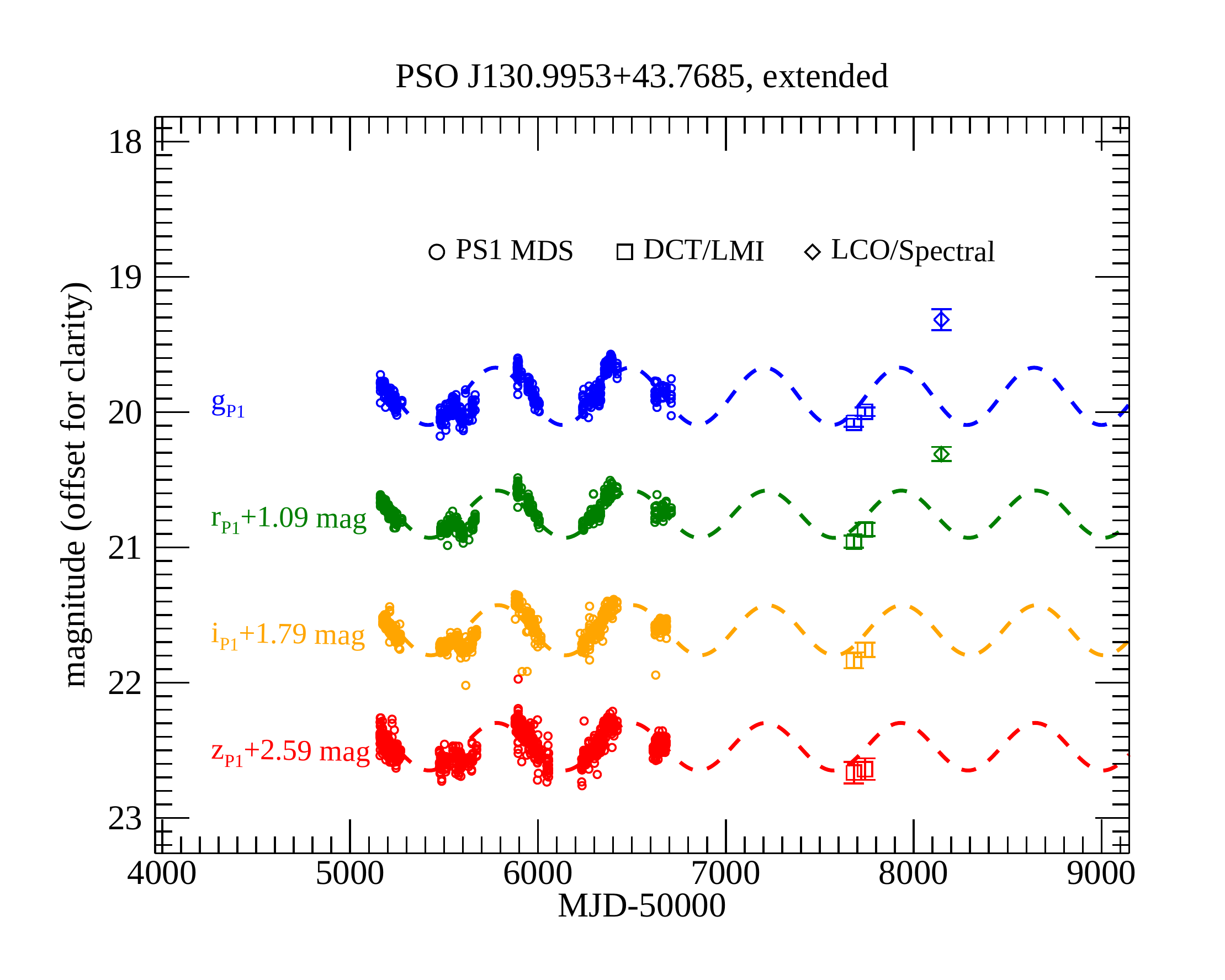,width=0.2\textwidth,clip=}
\epsfig{file=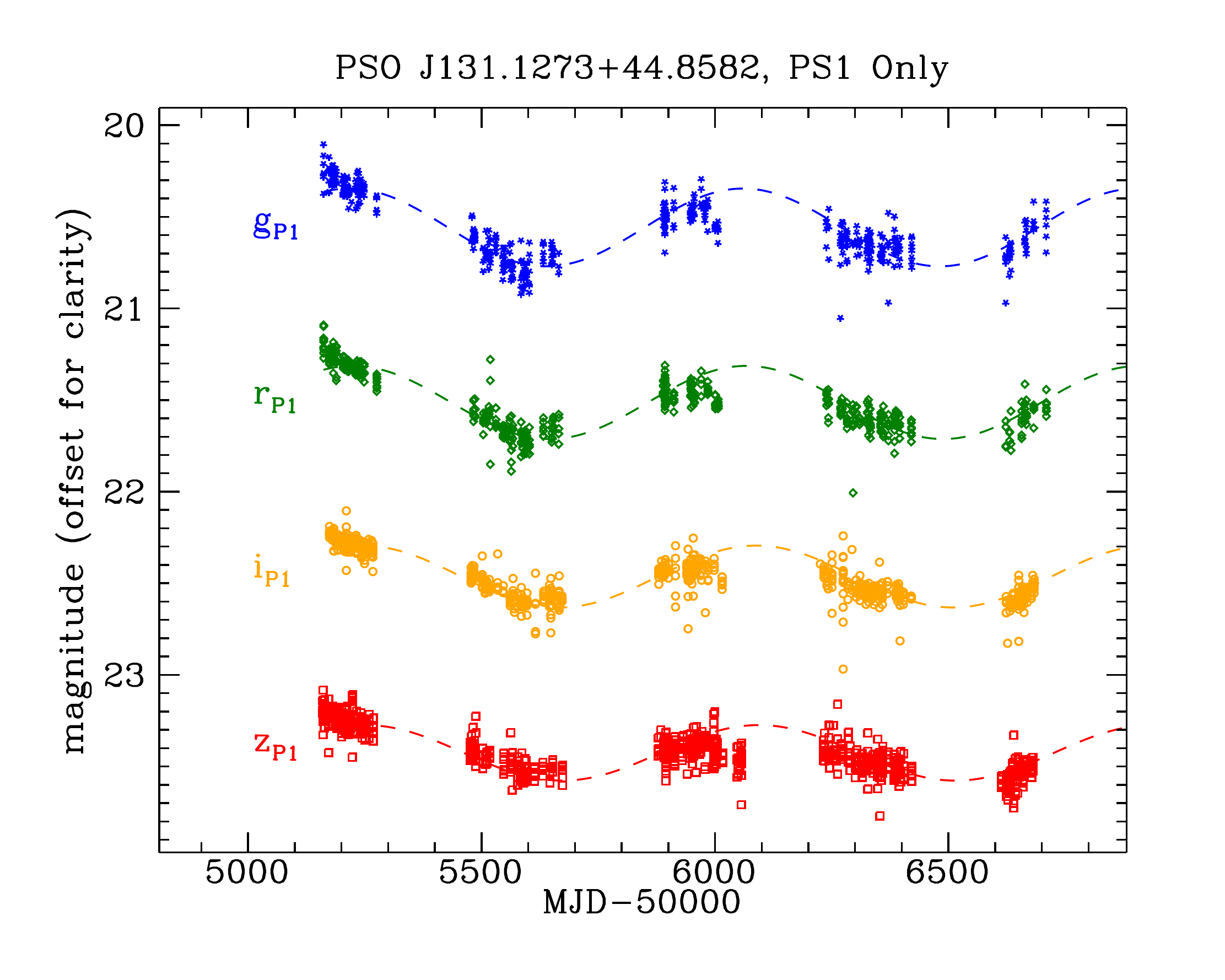,width=0.2\textwidth,clip=}
\epsfig{file=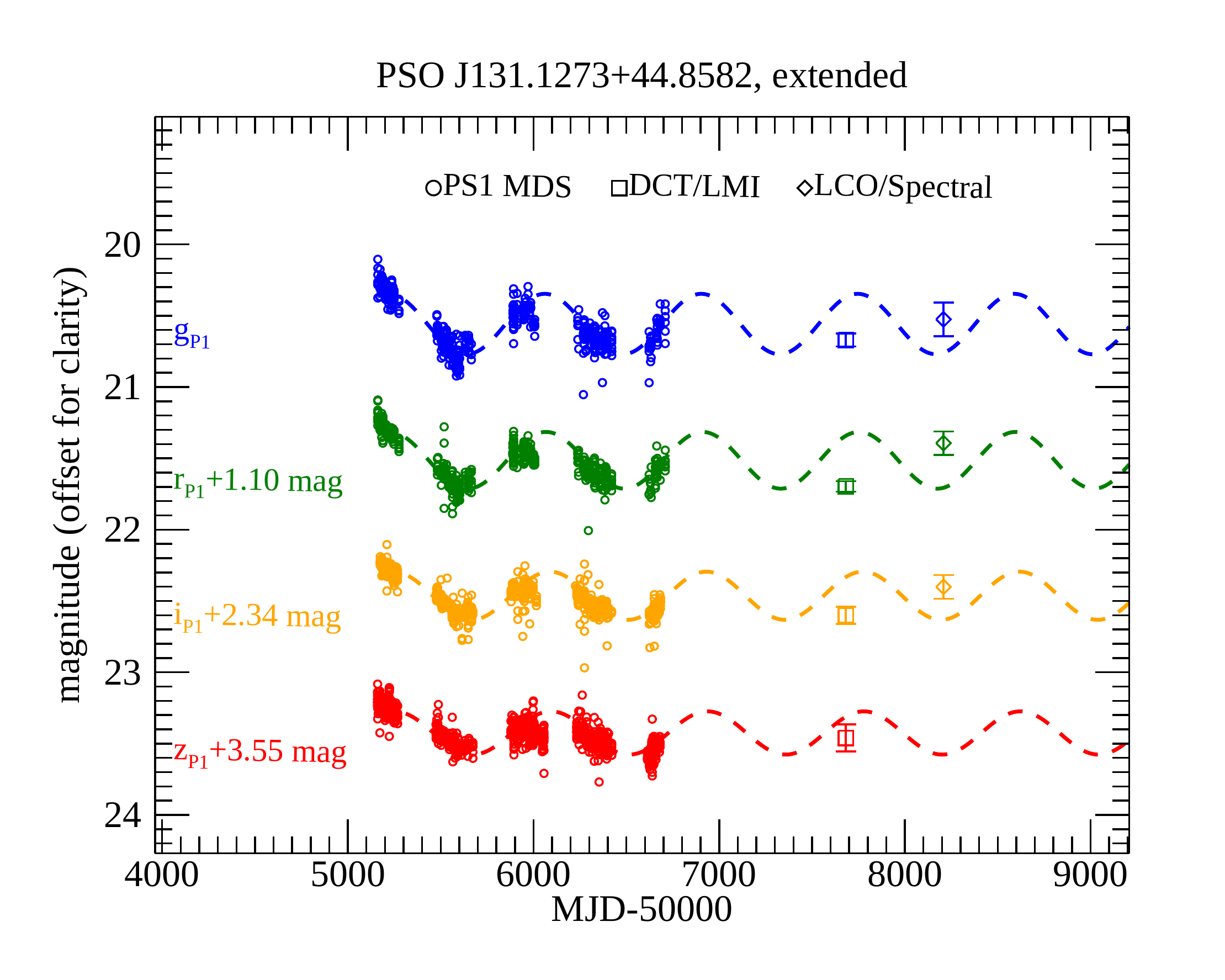,width=0.2\textwidth,clip=}
\epsfig{file=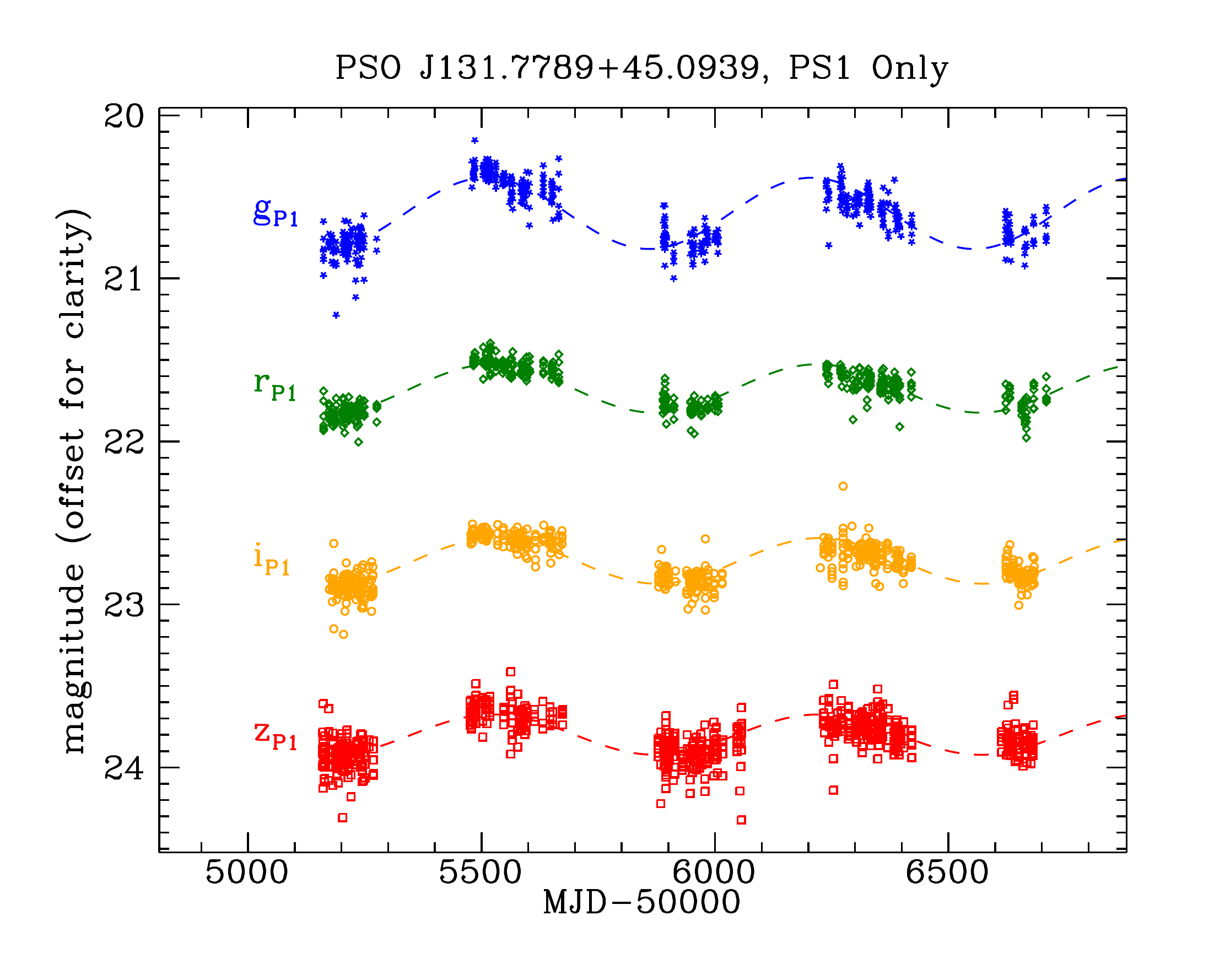,width=0.2\textwidth,clip=}
\epsfig{file=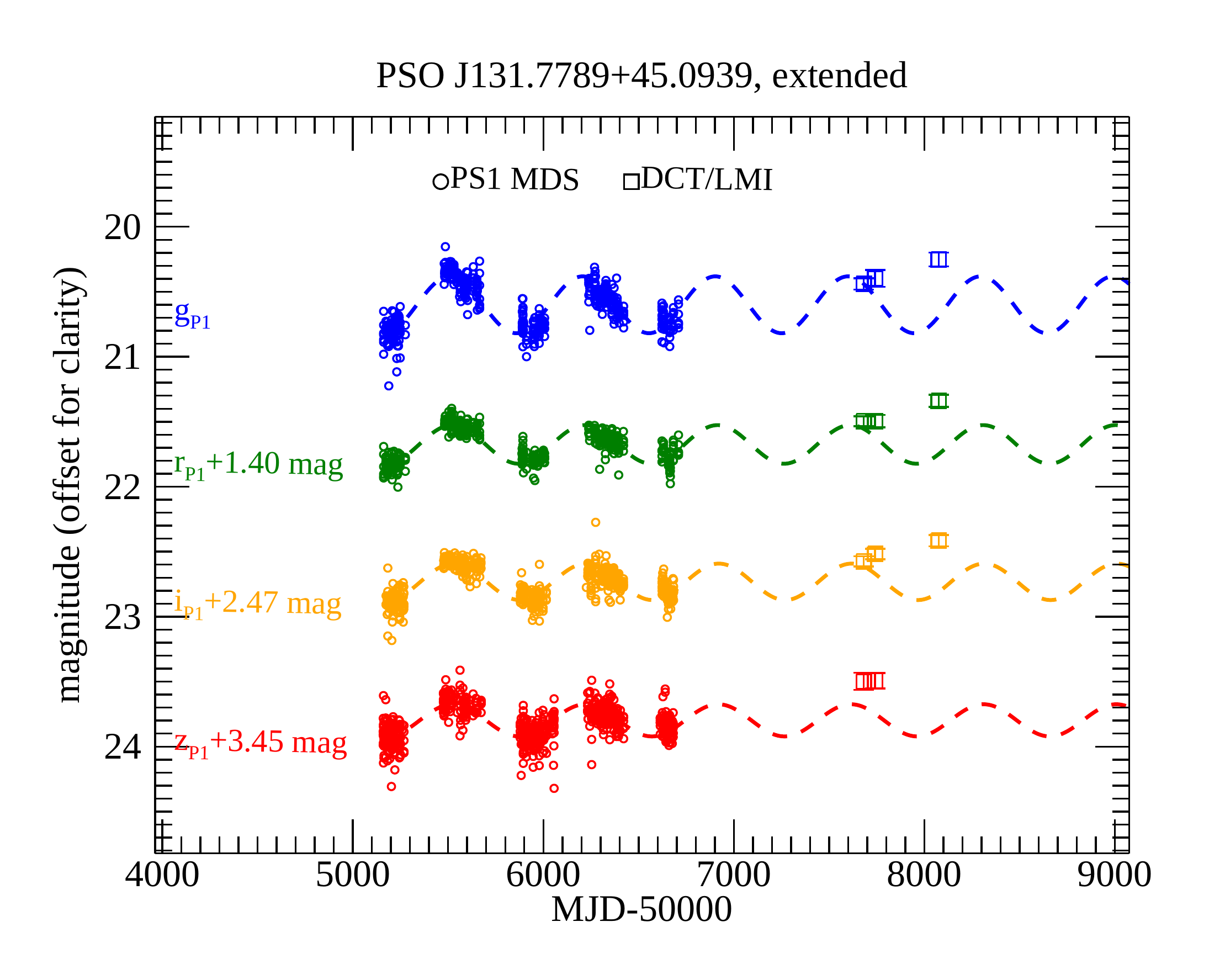,width=0.2\textwidth,clip=}
\epsfig{file=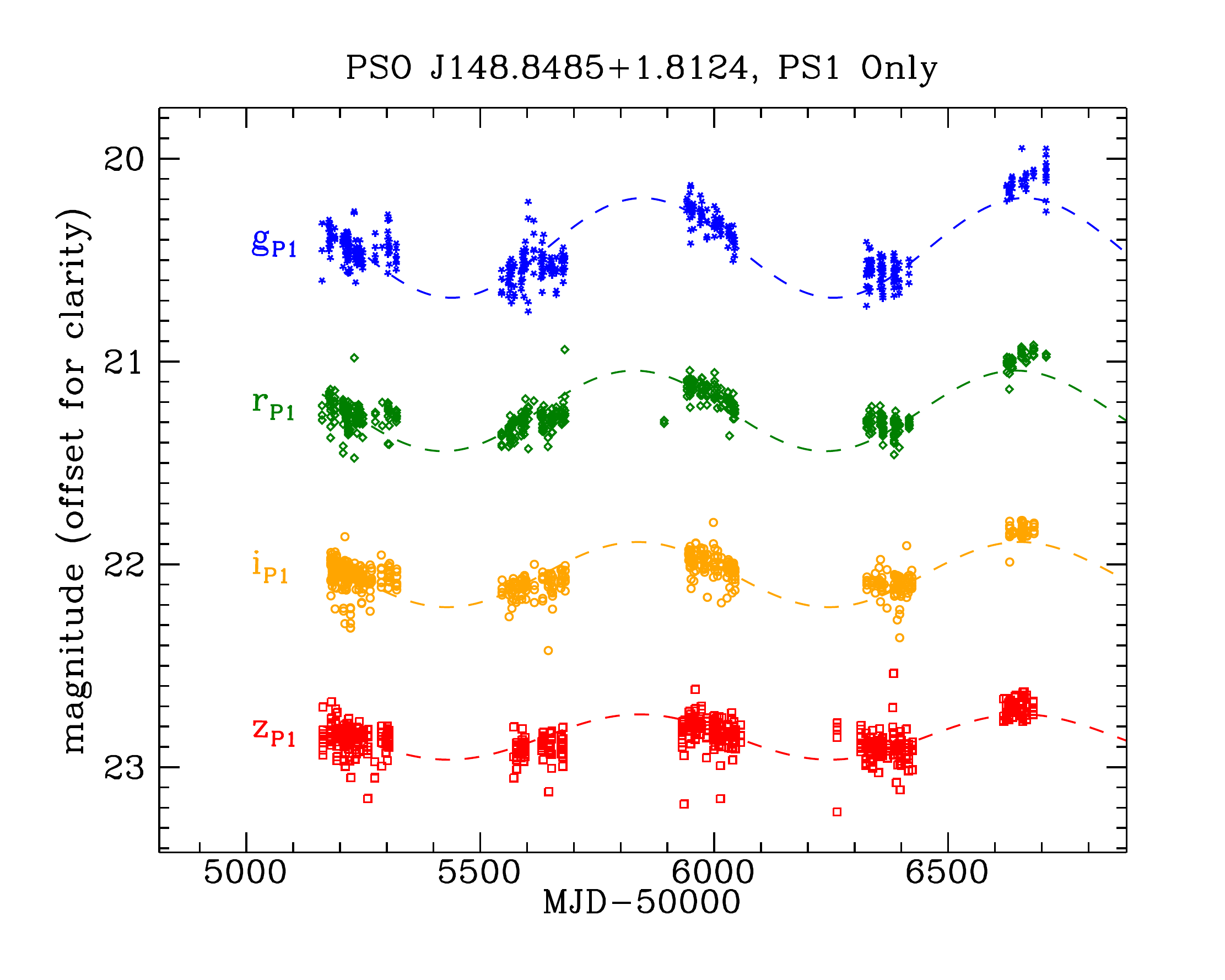,width=0.2\textwidth,clip=}
\epsfig{file=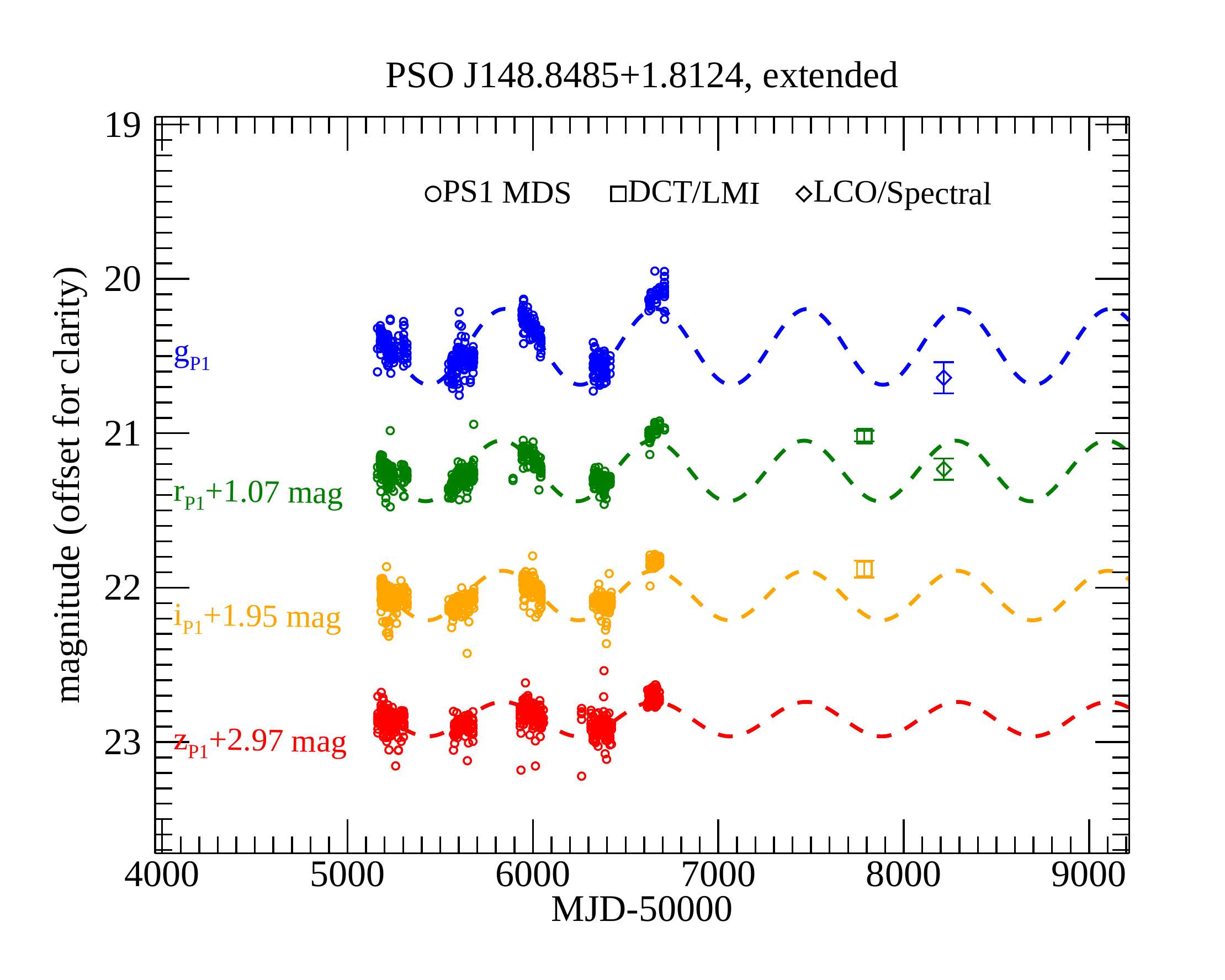,width=0.2\textwidth,clip=}
\epsfig{file=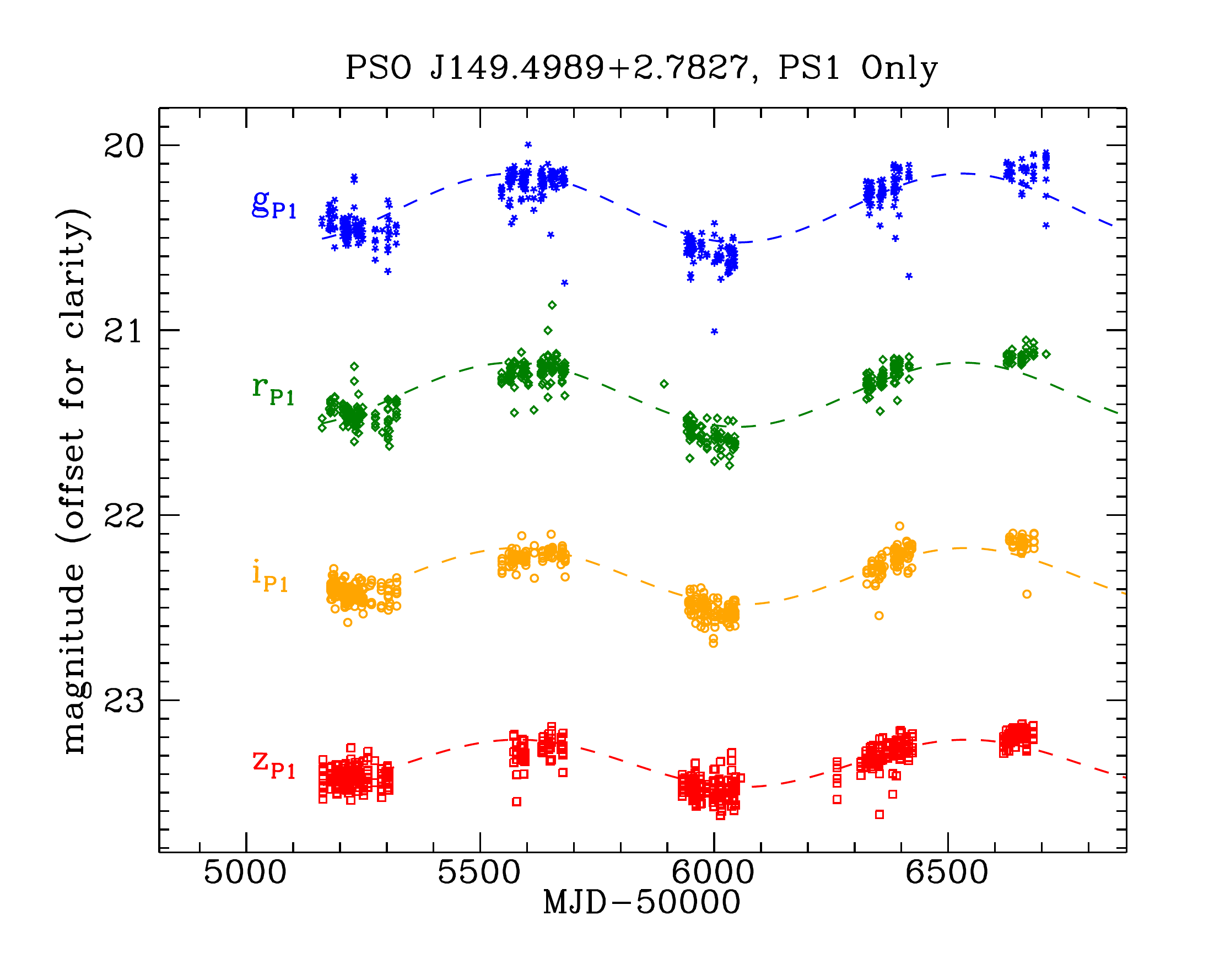,width=0.2\textwidth,clip=}
\epsfig{file=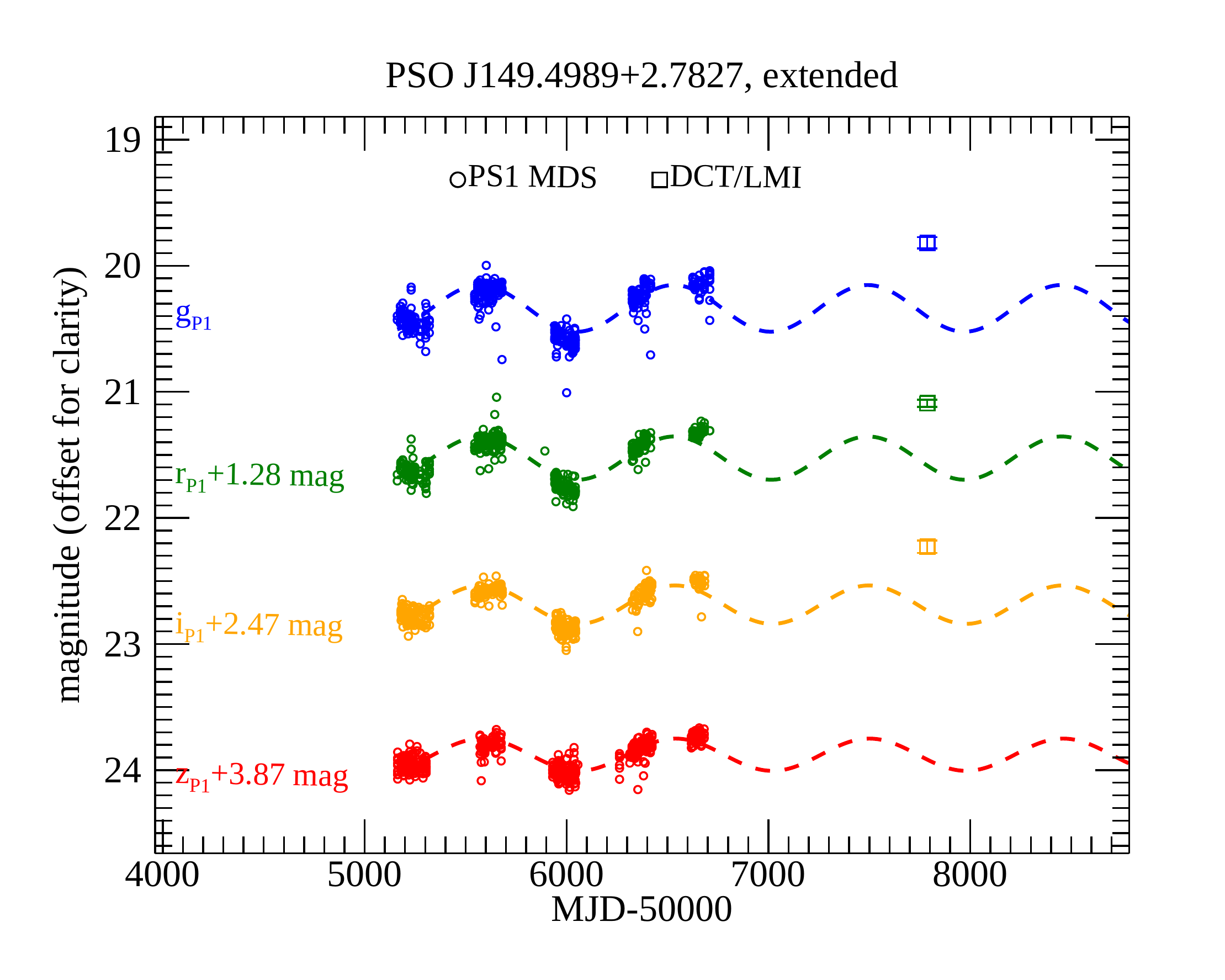,width=0.2\textwidth,clip=}
\epsfig{file=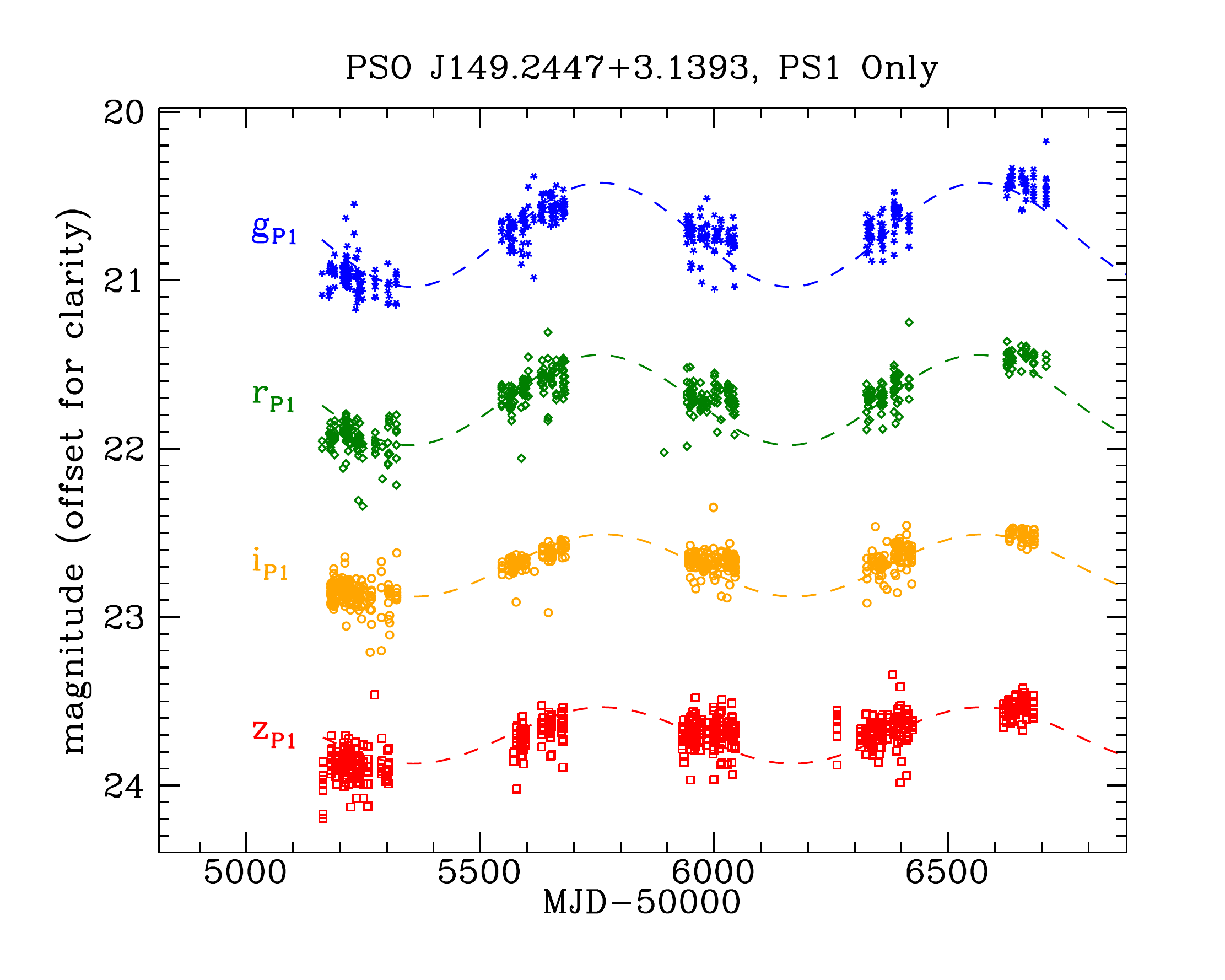,width=0.2\textwidth,clip=}
\epsfig{file=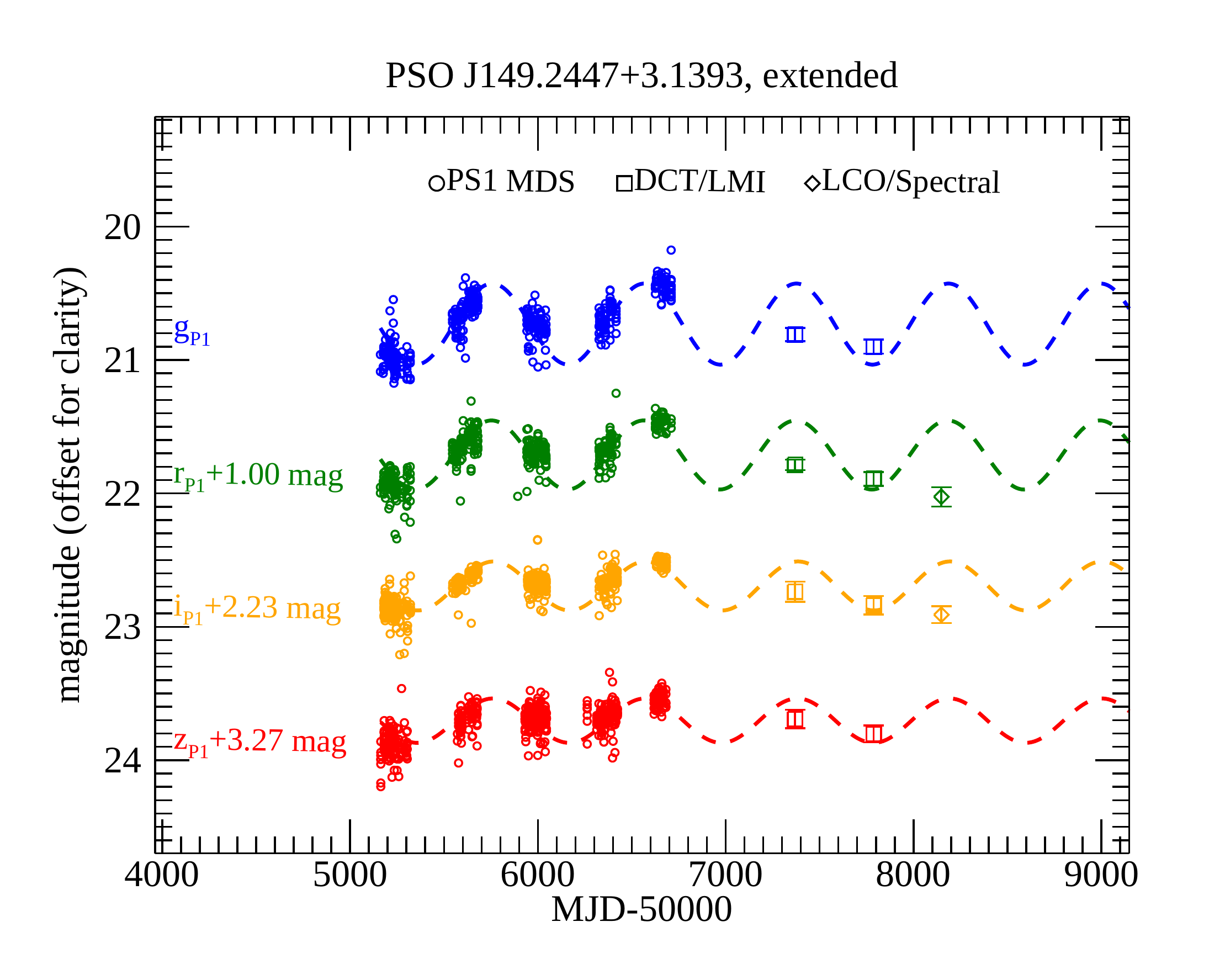,width=0.2\textwidth,clip=}
\epsfig{file=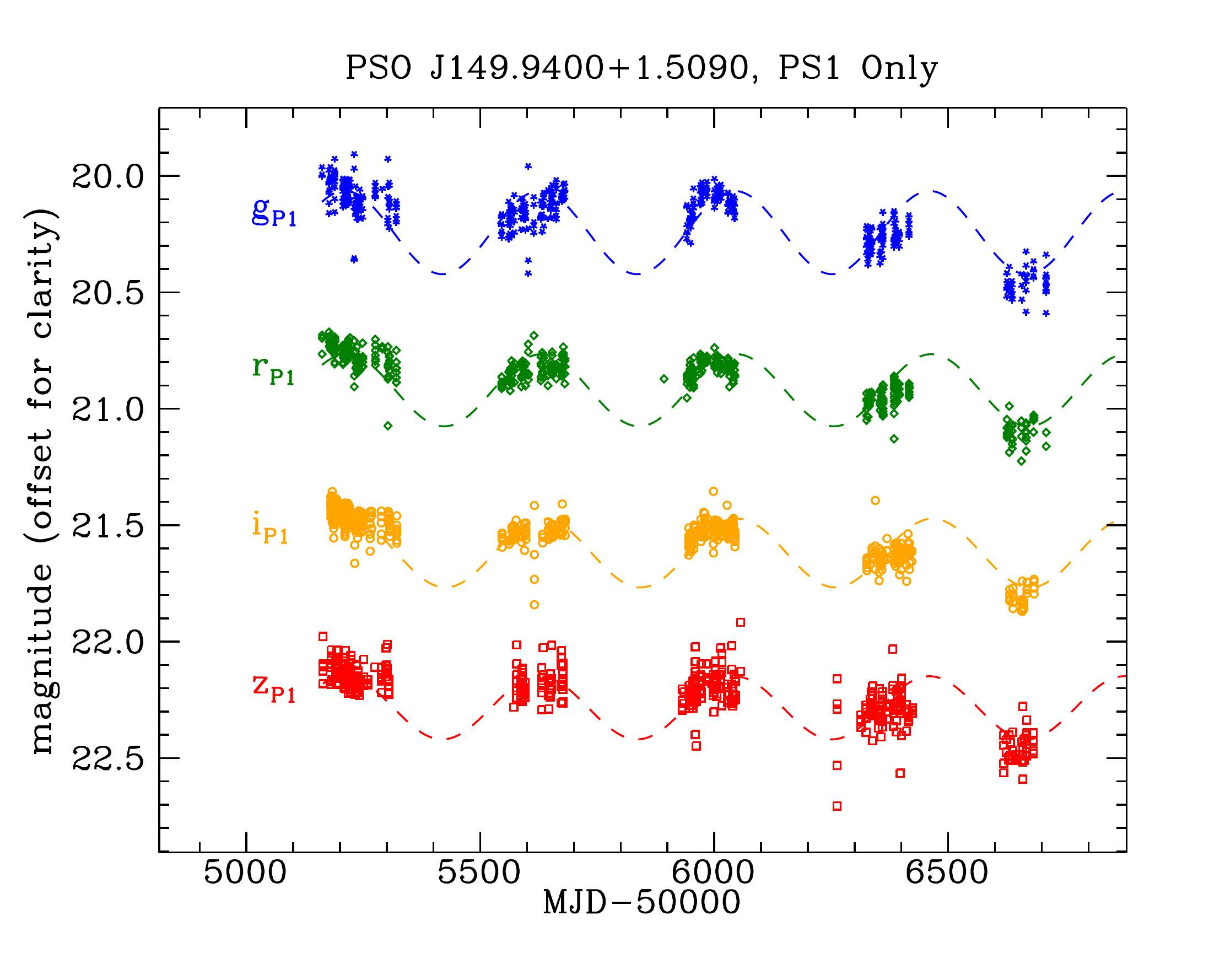,width=0.2\textwidth,clip=}
\epsfig{file=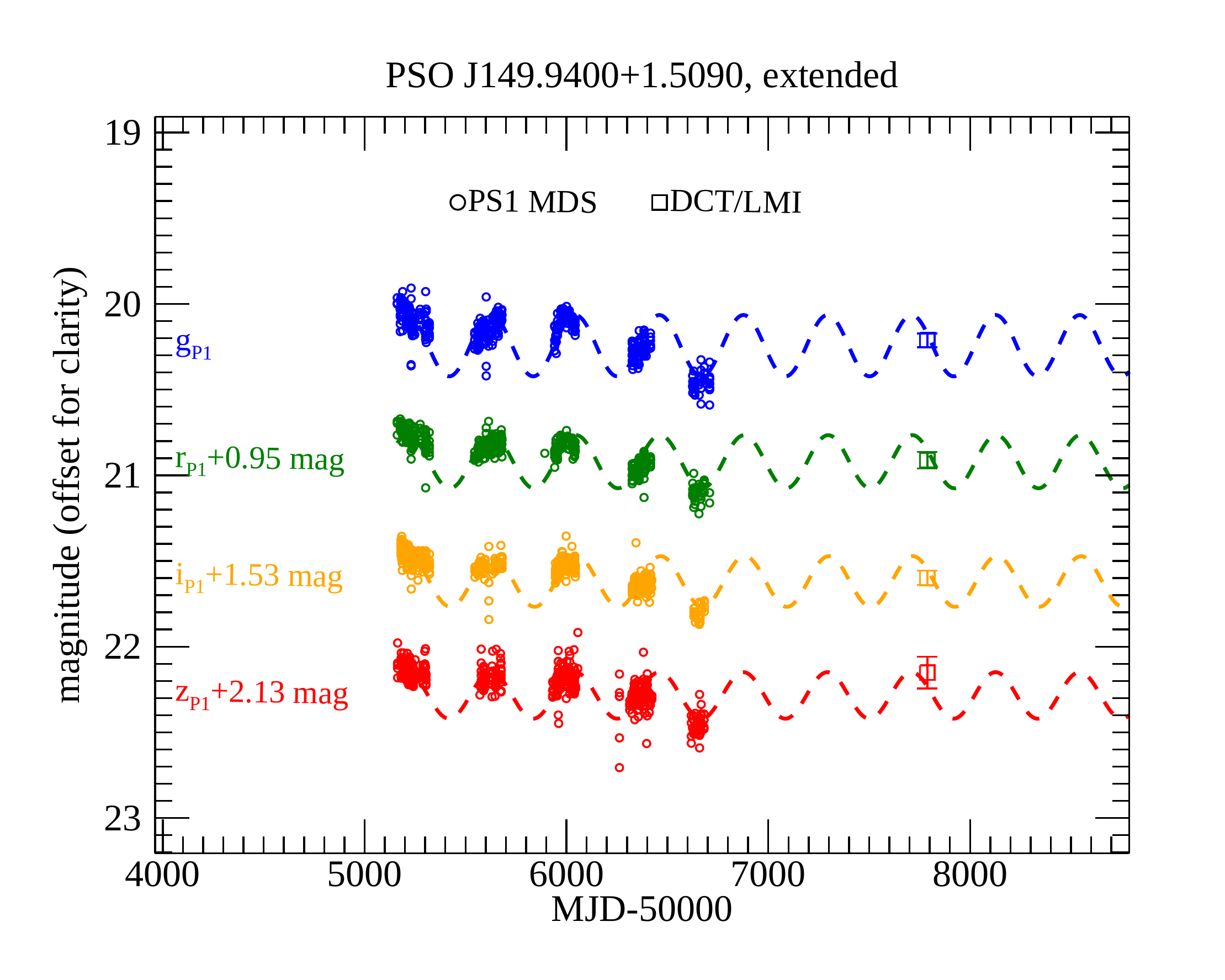,width=0.2\textwidth,clip=}
\epsfig{file=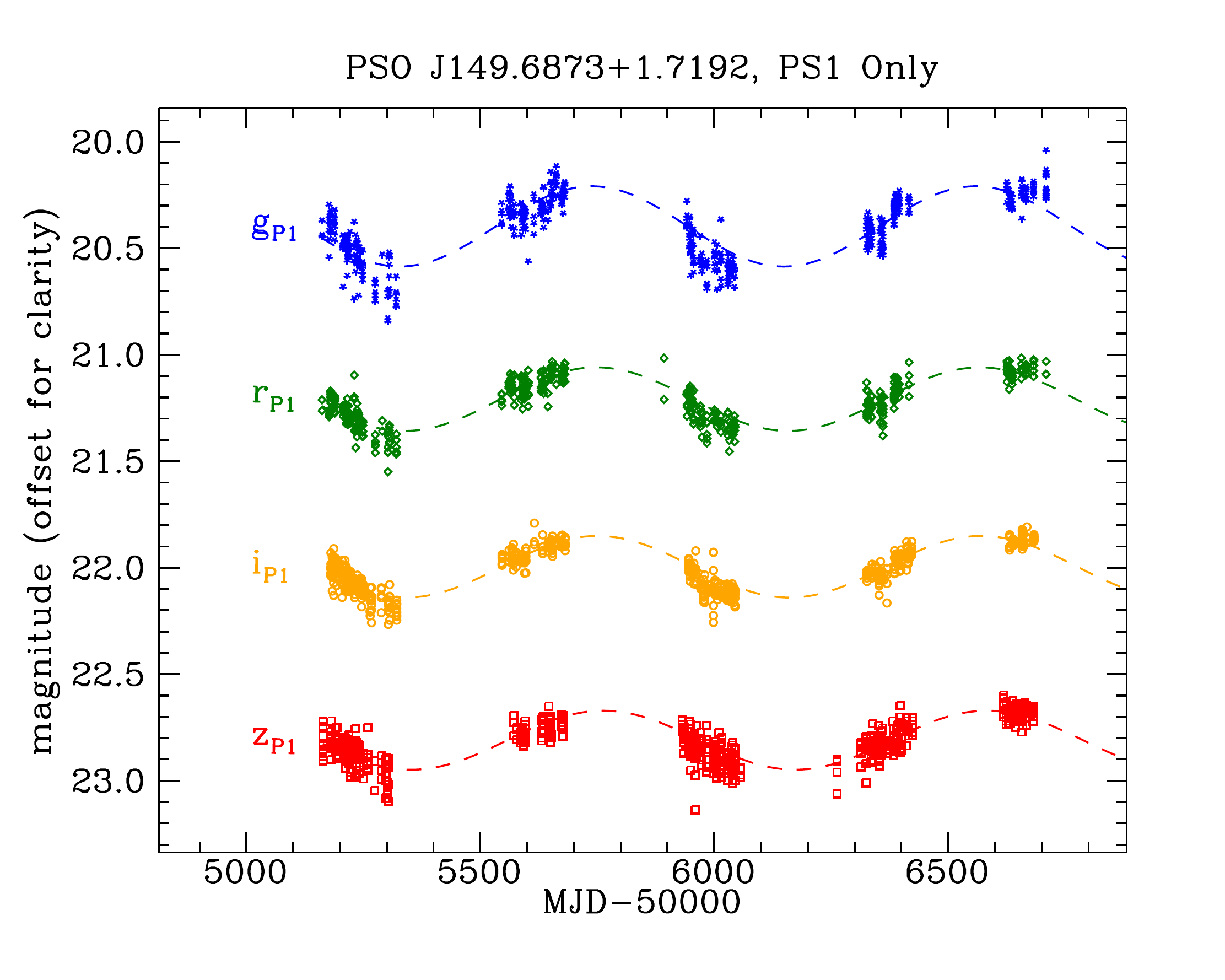,width=0.2\textwidth,clip=}
\epsfig{file=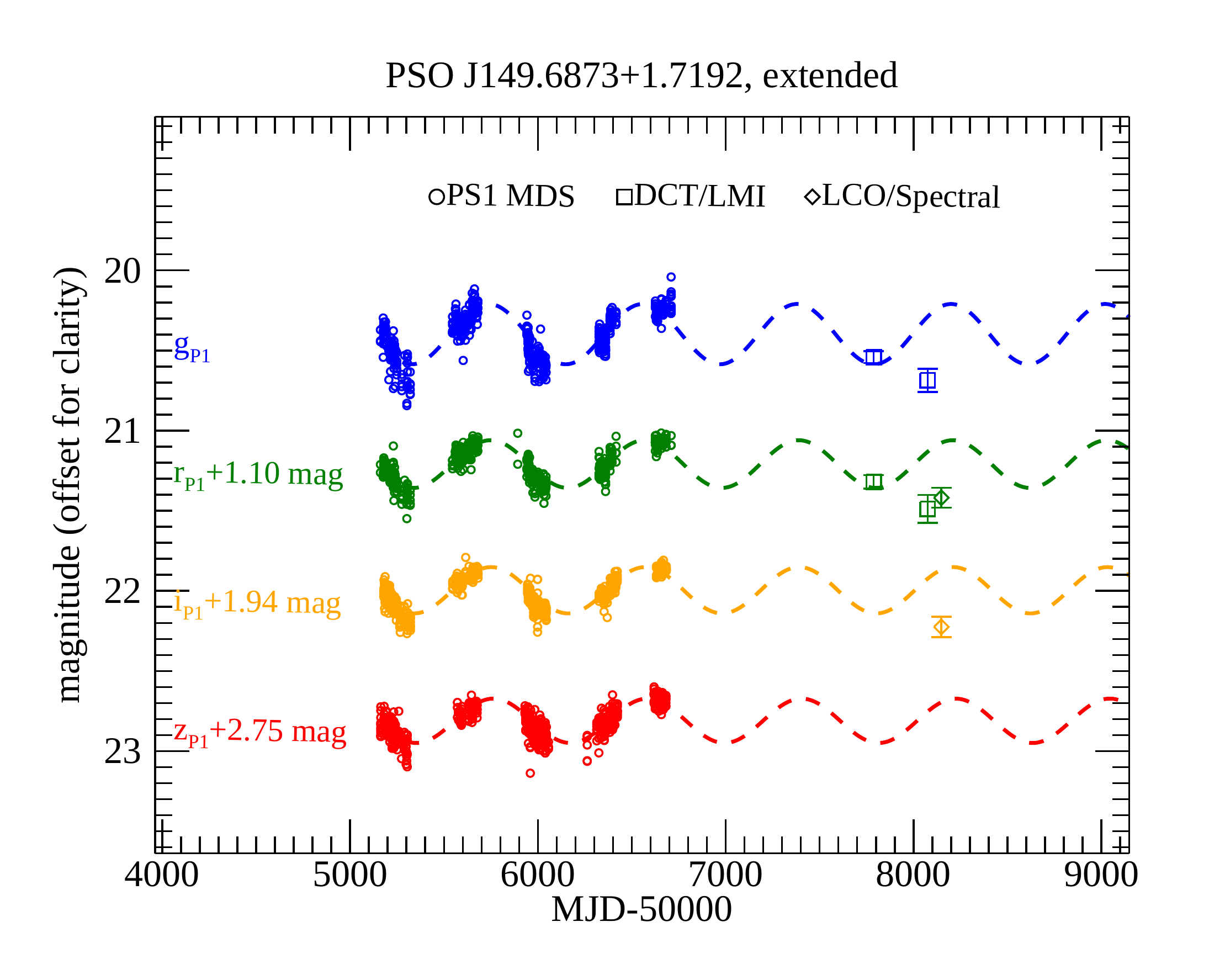,width=0.2\textwidth,clip=}
\epsfig{file=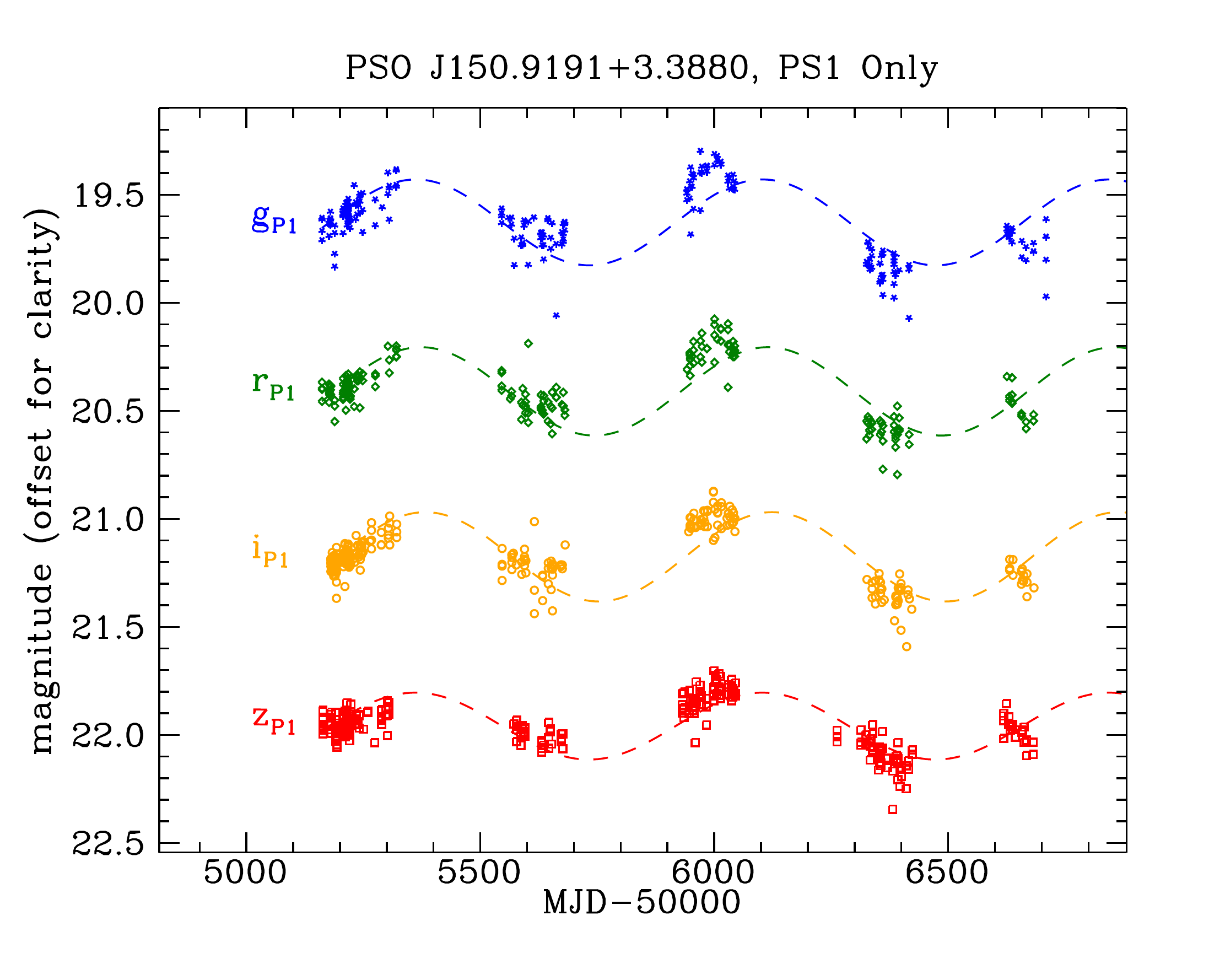,width=0.2\textwidth,clip=}
\epsfig{file=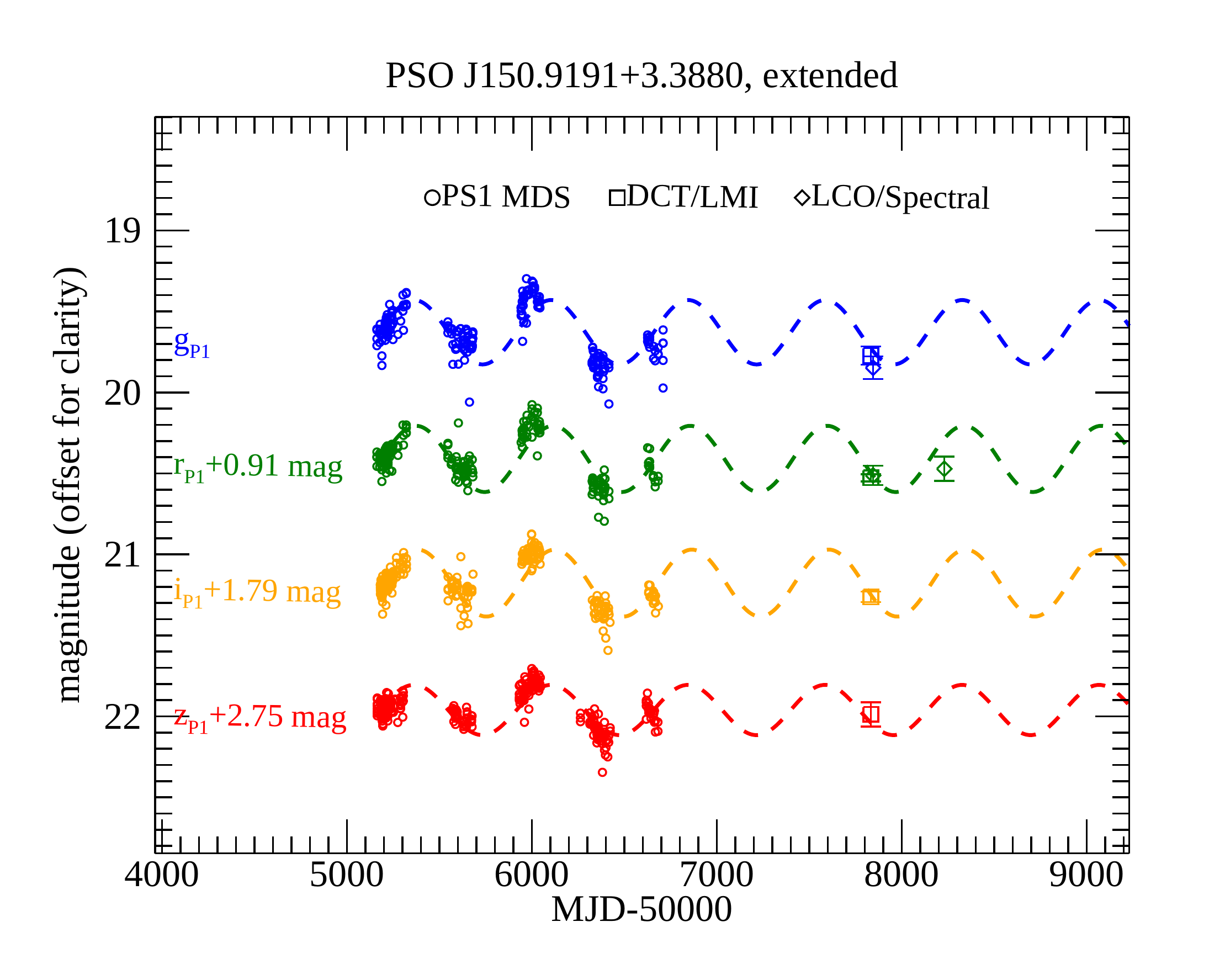,width=0.2\textwidth,clip=}
\epsfig{file=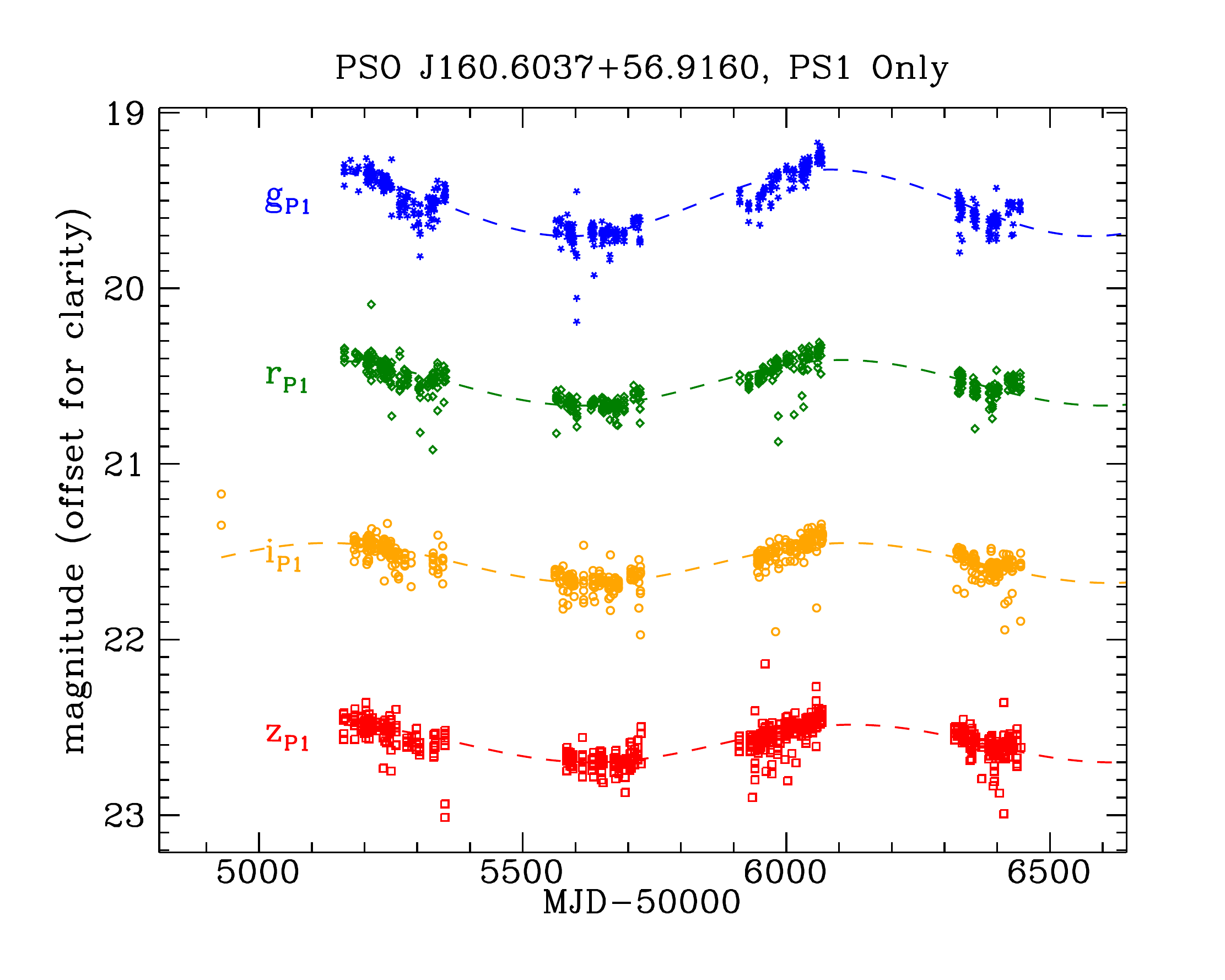,width=0.2\textwidth,clip=}
\epsfig{file=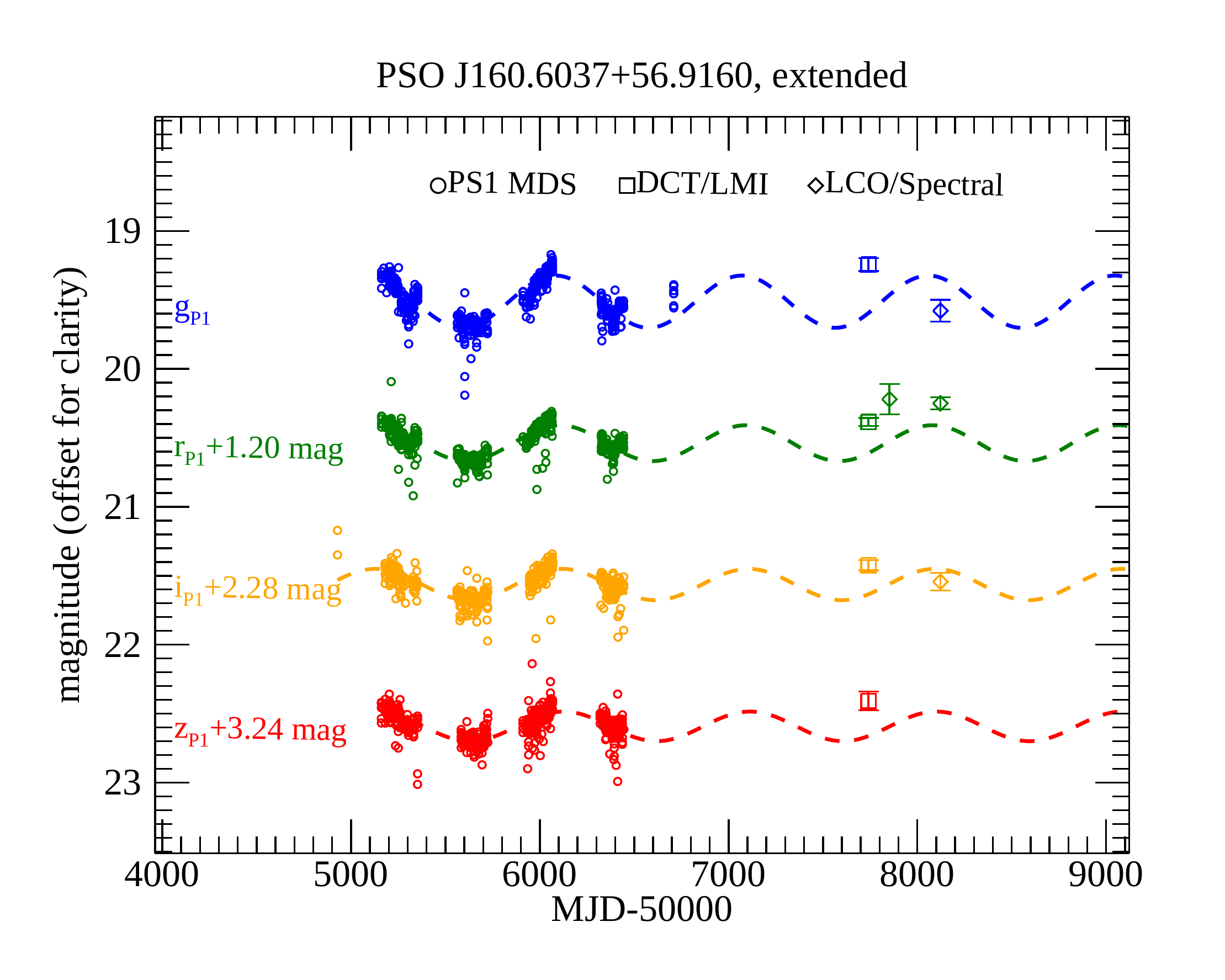,width=0.2\textwidth,clip=}
\epsfig{file=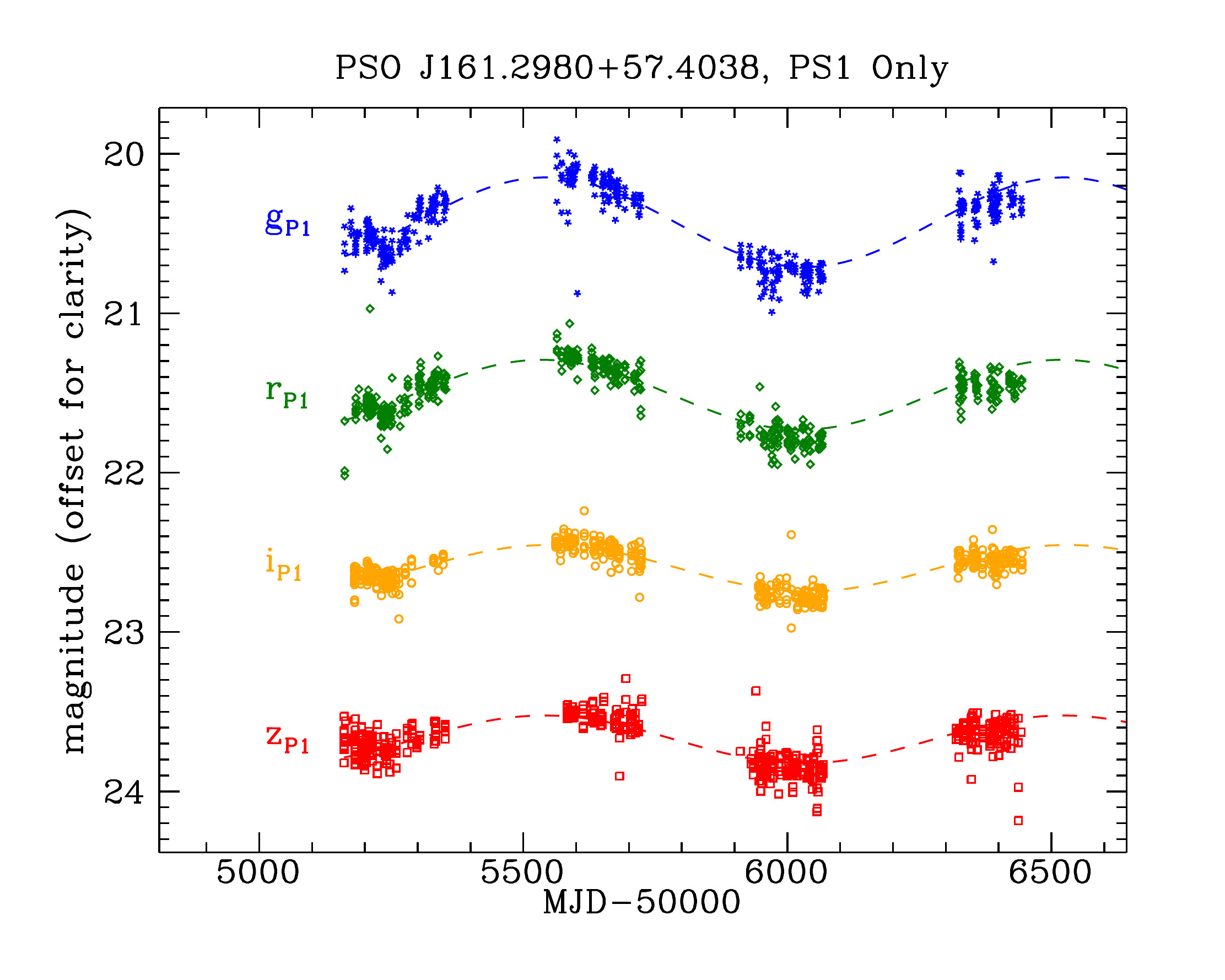,width=0.2\textwidth,clip=}
\epsfig{file=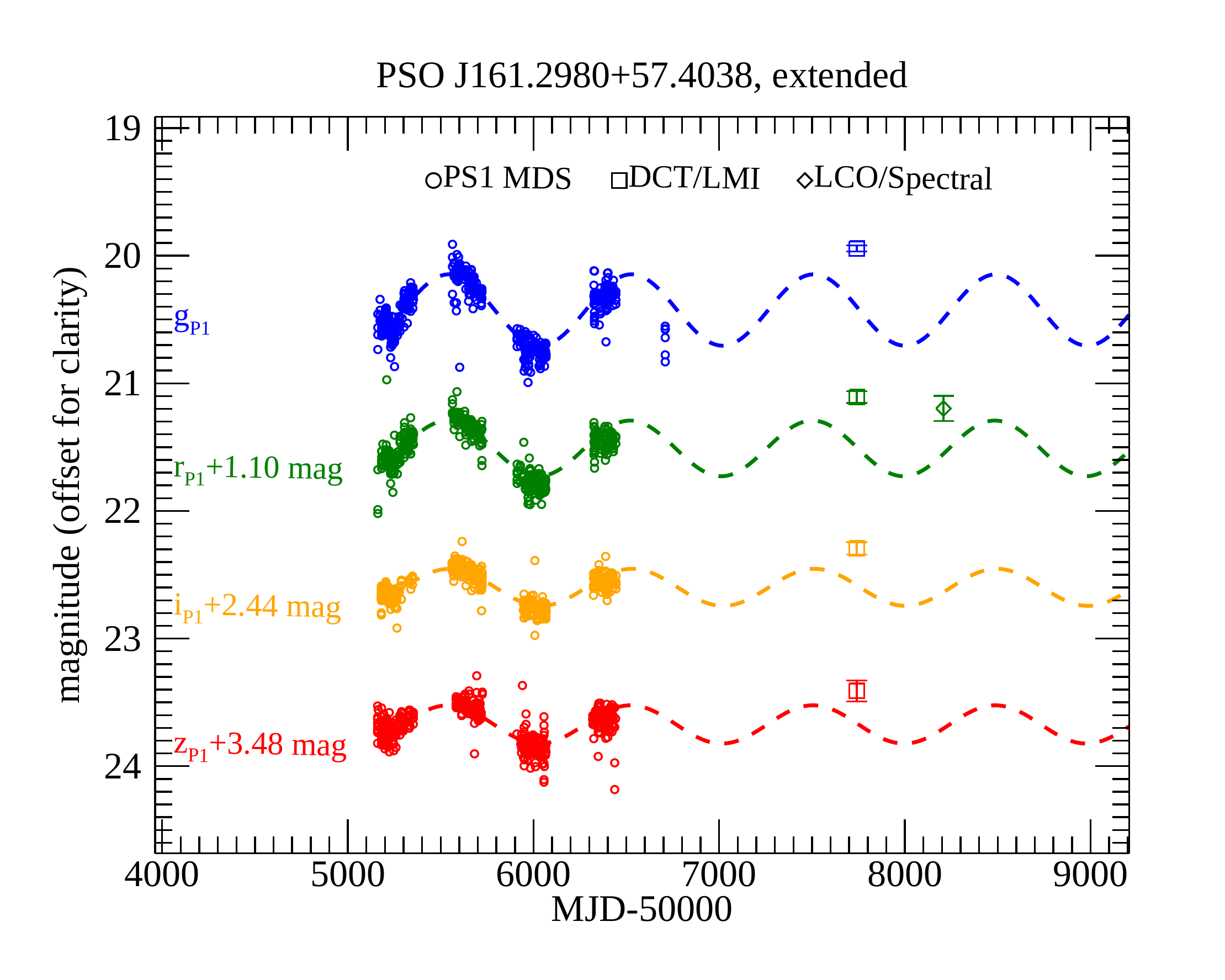,width=0.2\textwidth,clip=}
\label{fig:lc}
\caption{PS1-only and extended light curves of PS1 MDS candidates.}
\end{figure*} 

\begin{figure*}[h] 
\centering
\epsfig{file=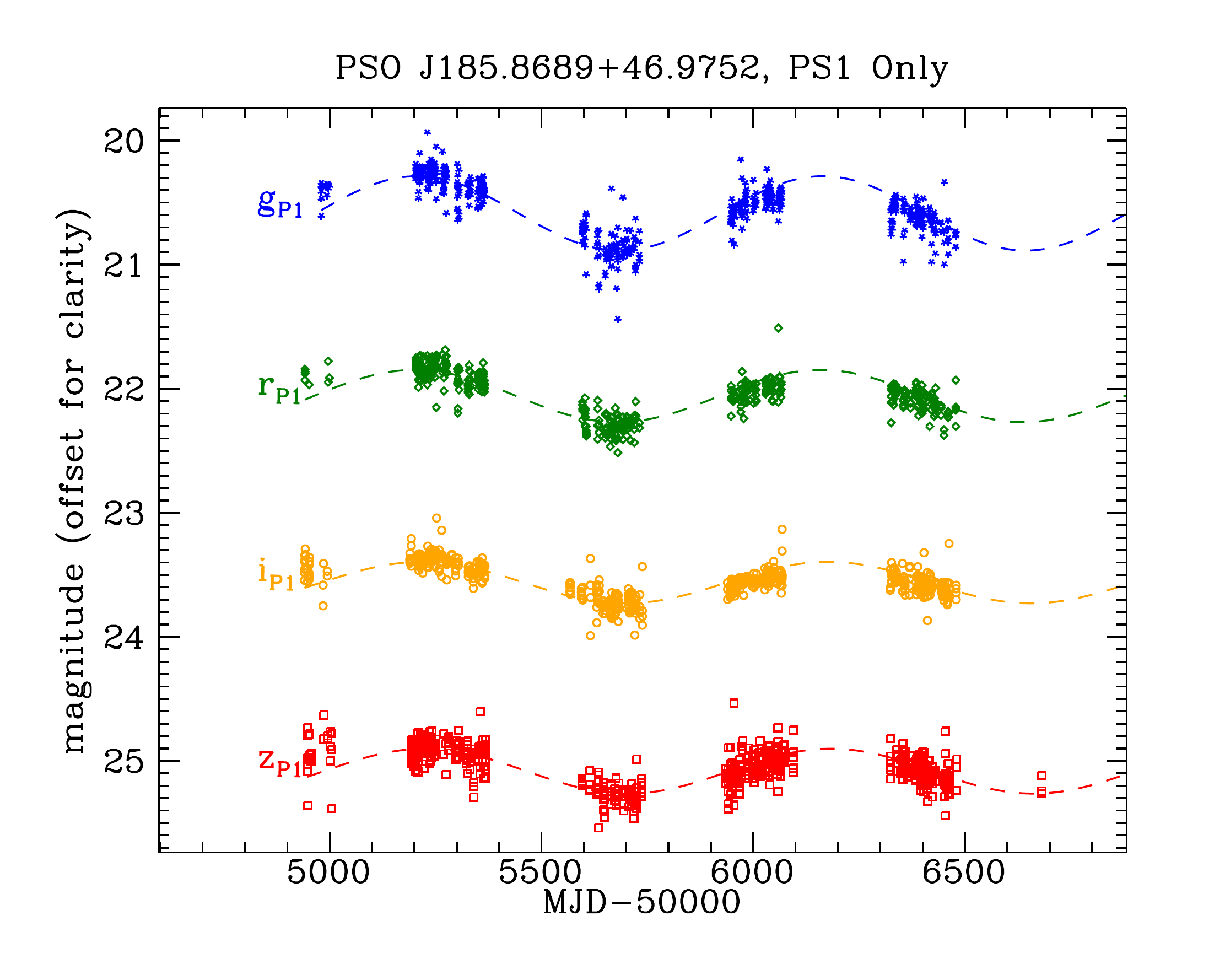,width=0.2\textwidth,clip=}
\epsfig{file=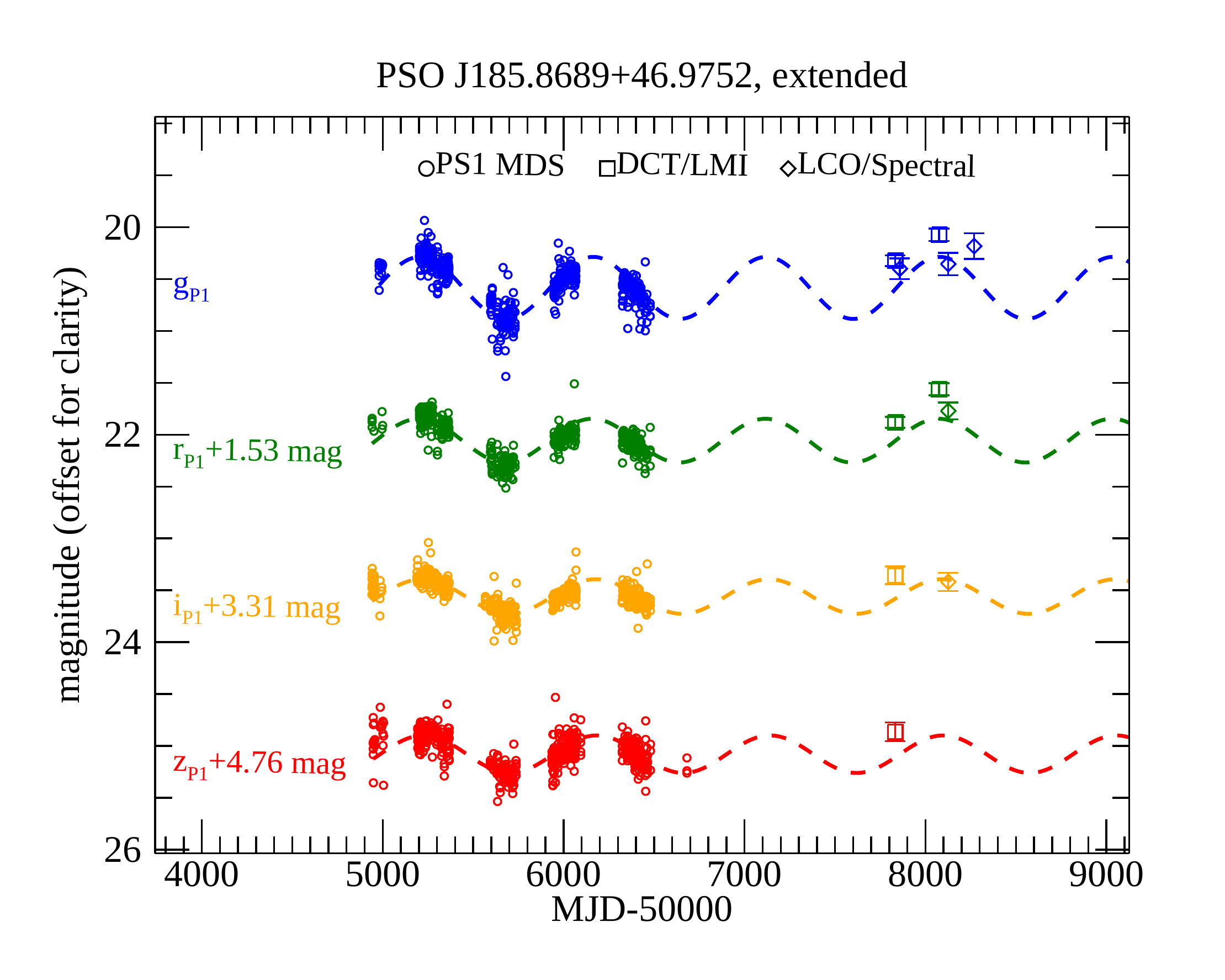,width=0.2\textwidth,clip=}
\epsfig{file=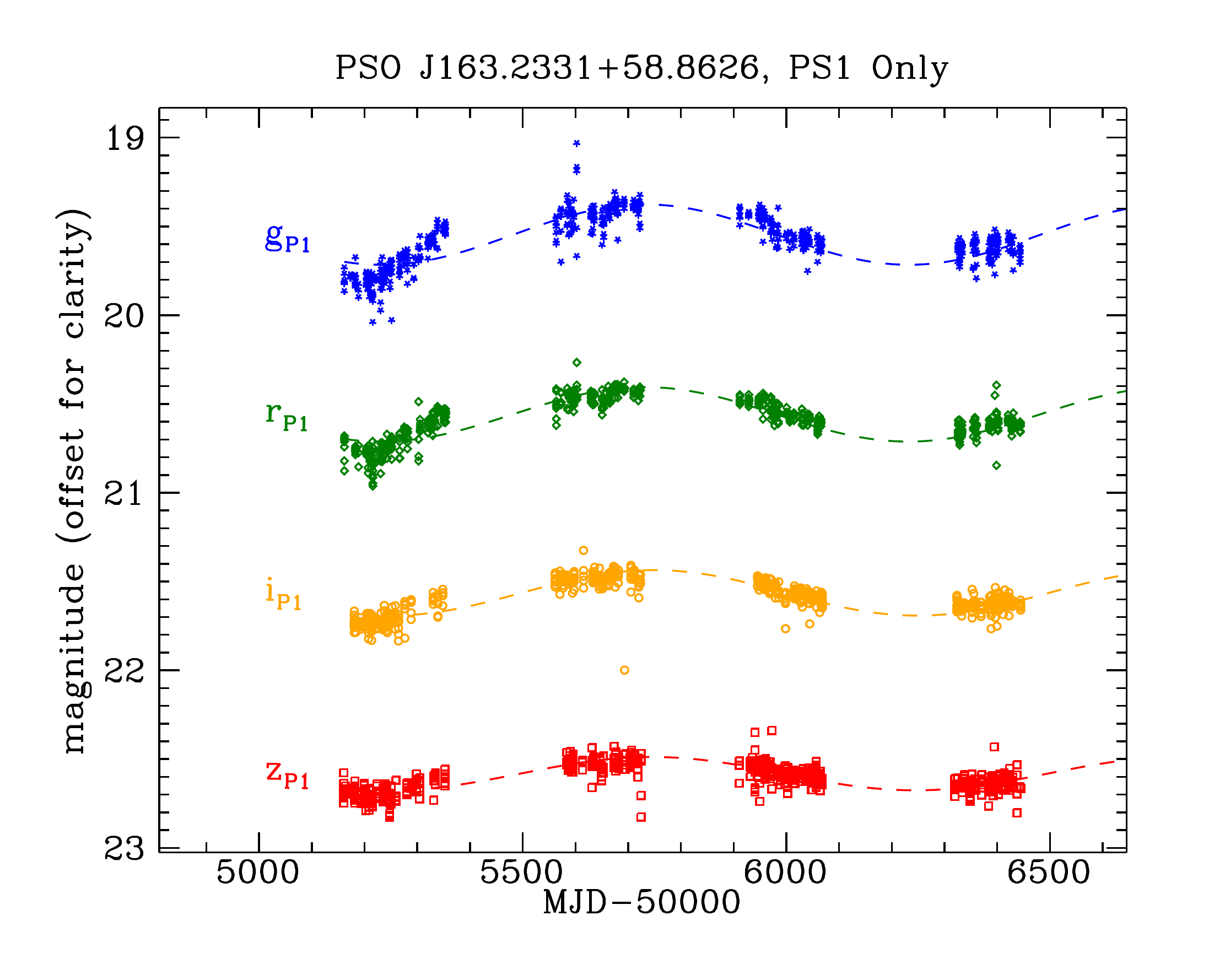,width=0.2\textwidth,clip=}
\epsfig{file=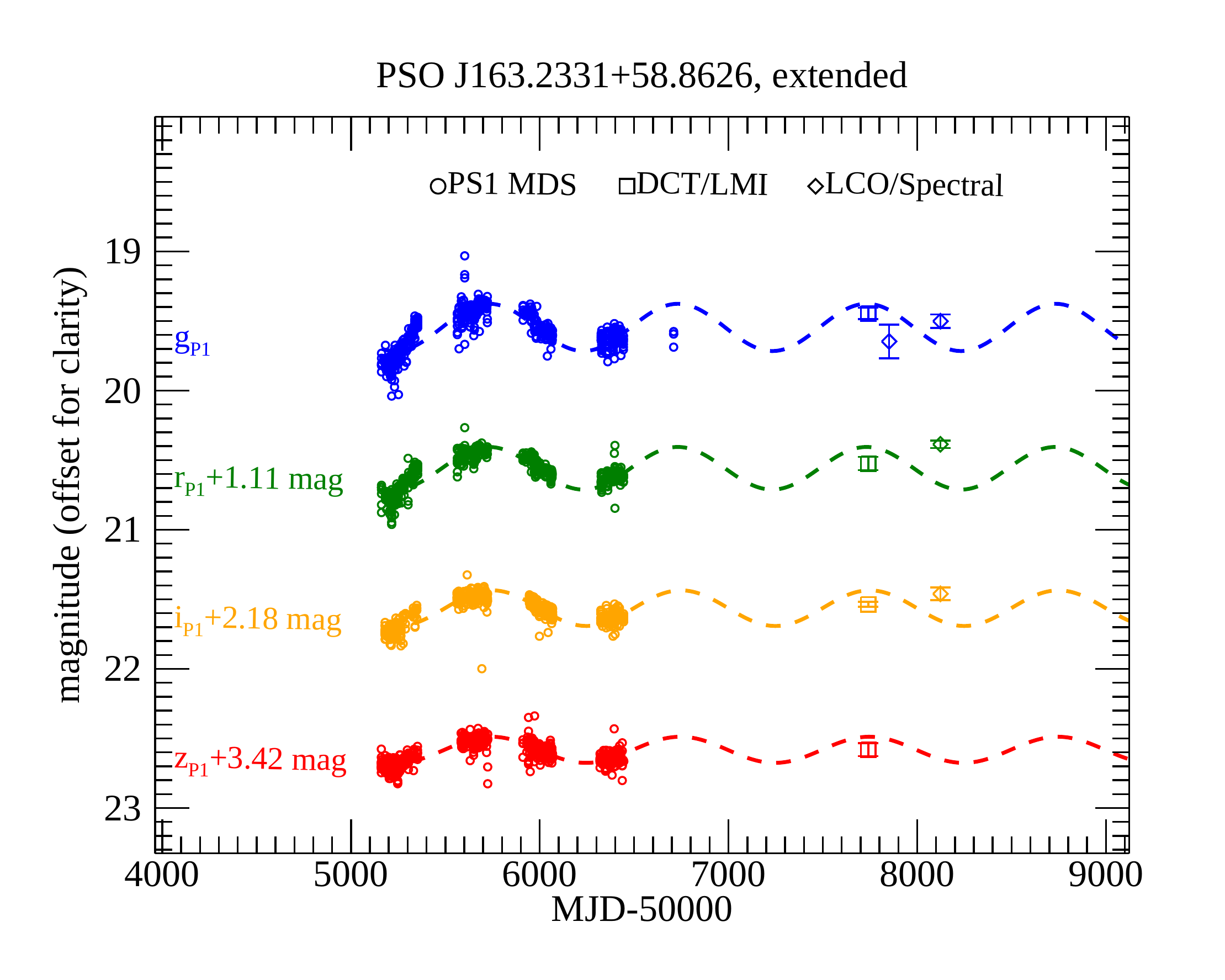,width=0.2\textwidth,clip=}
\epsfig{file=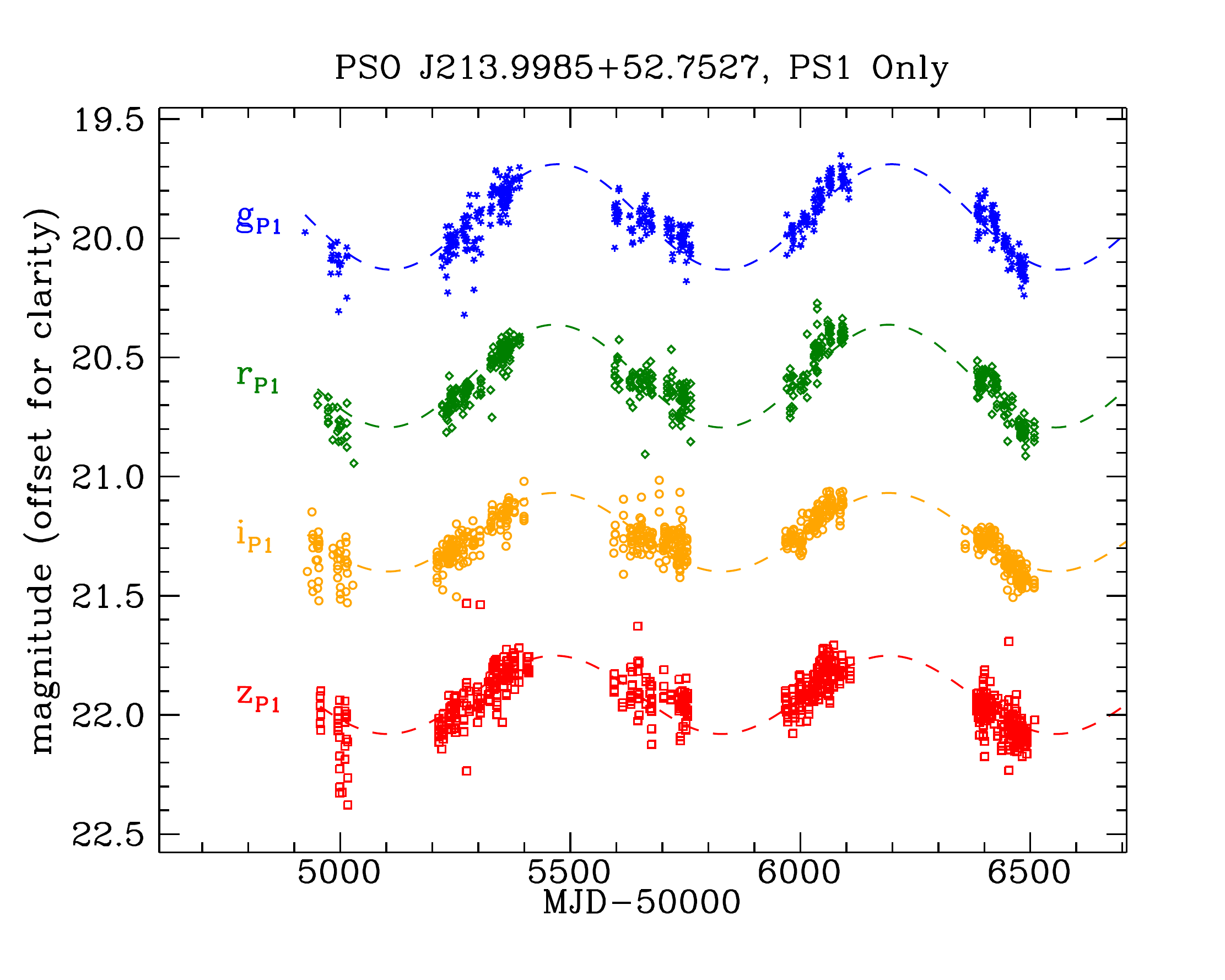,width=0.2\textwidth,clip=}
\epsfig{file=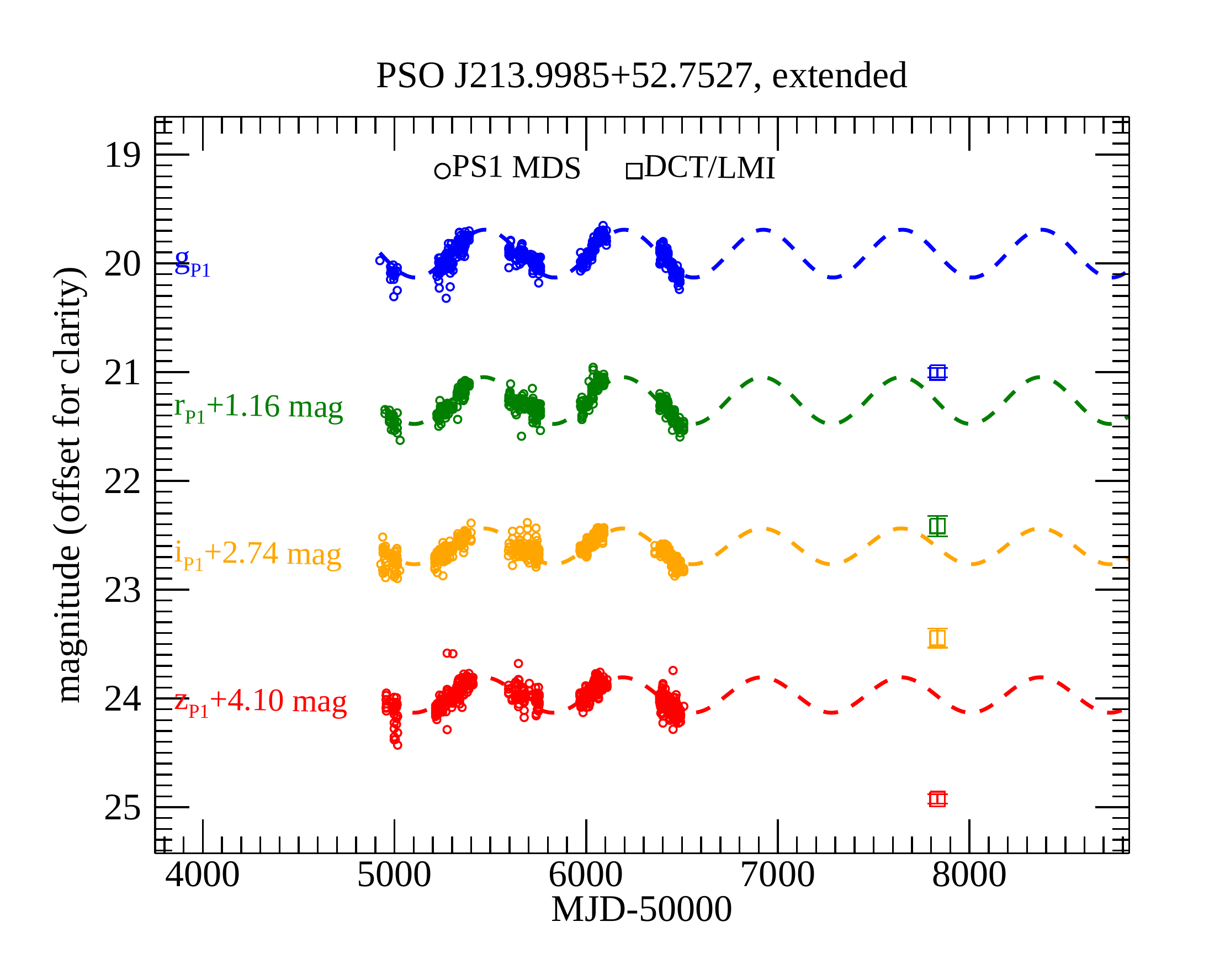,width=0.2\textwidth,clip=}
\epsfig{file=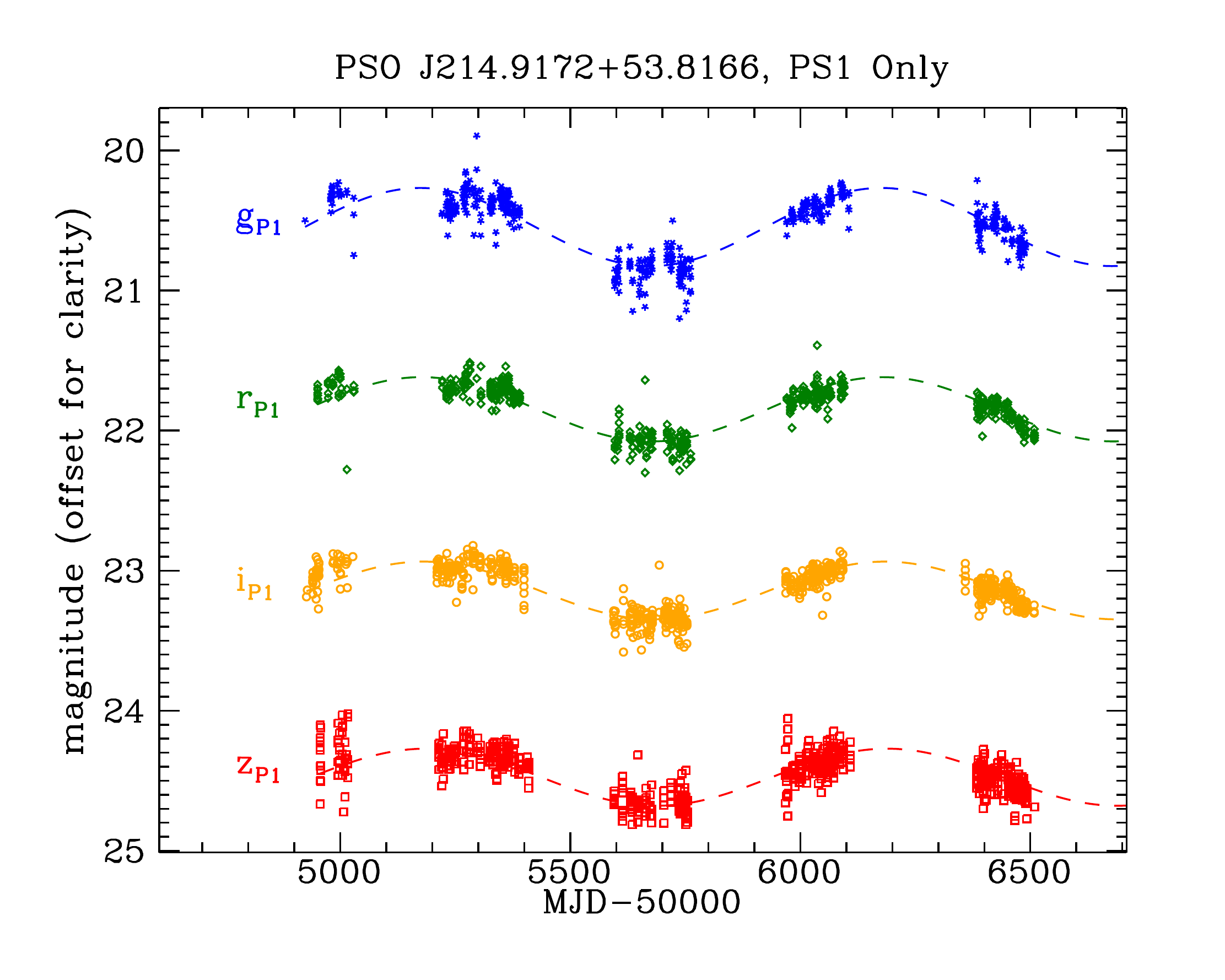,width=0.2\textwidth,clip=}
\epsfig{file=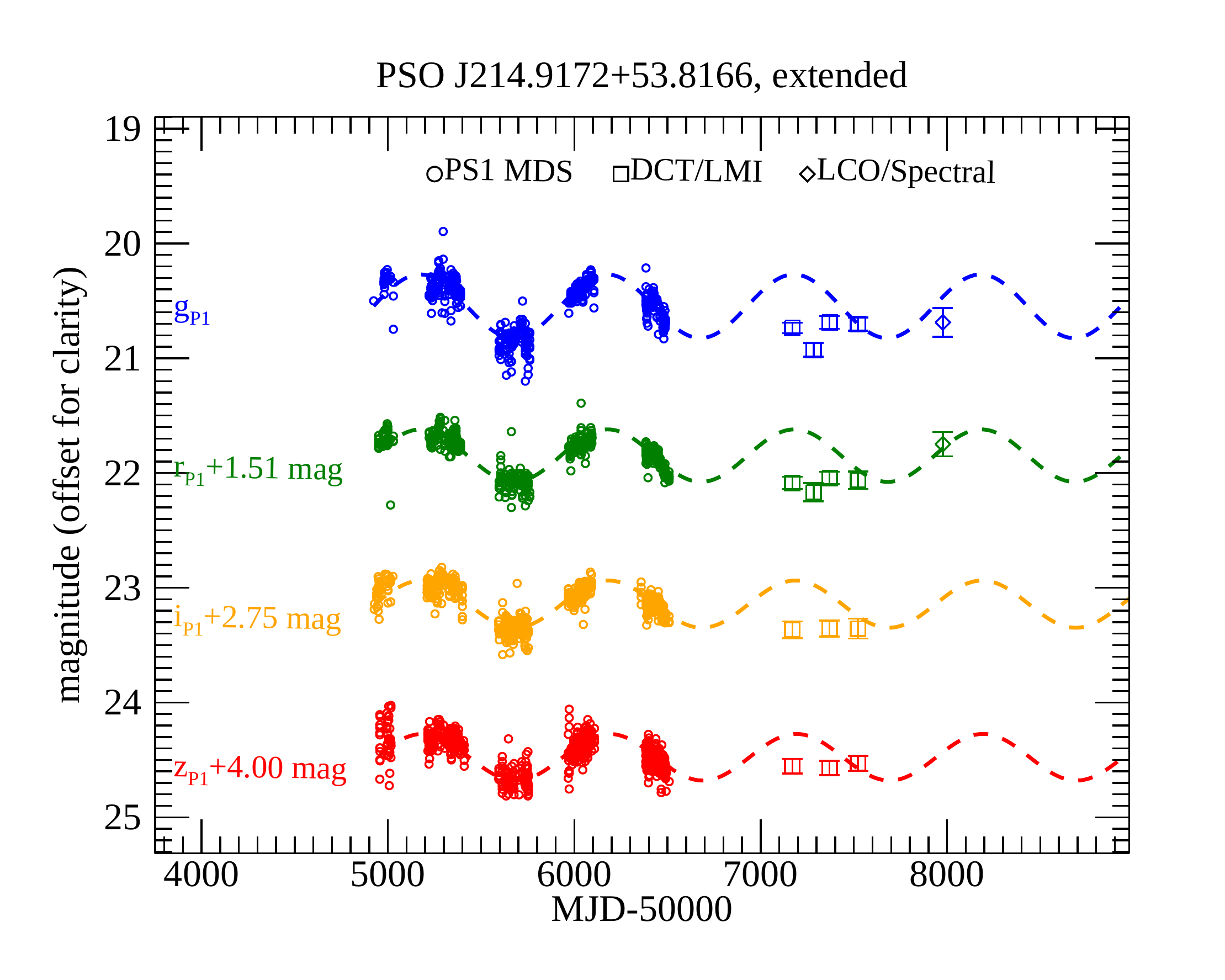,width=0.2\textwidth,clip=}
\epsfig{file=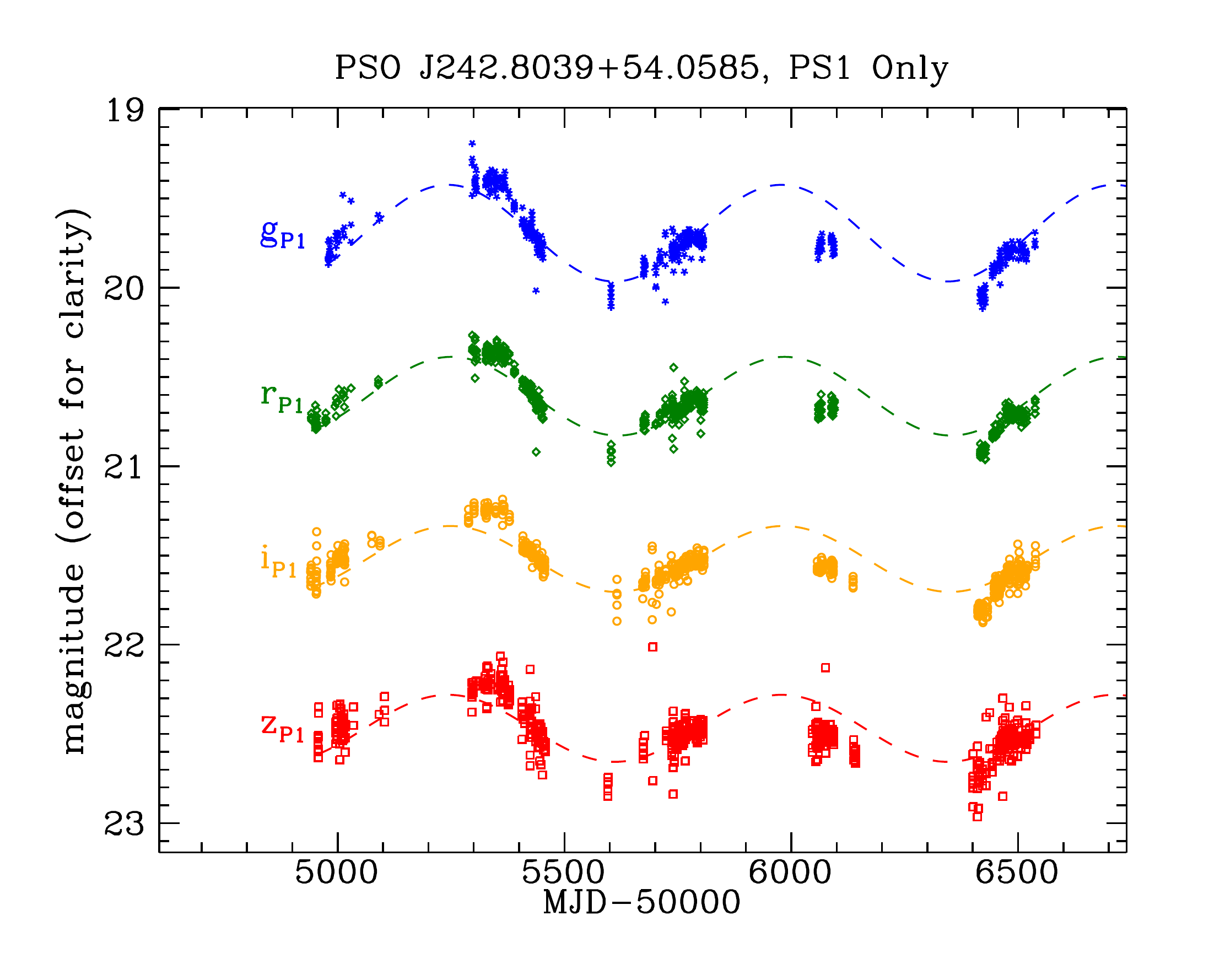,width=0.2\textwidth,clip=}
\epsfig{file=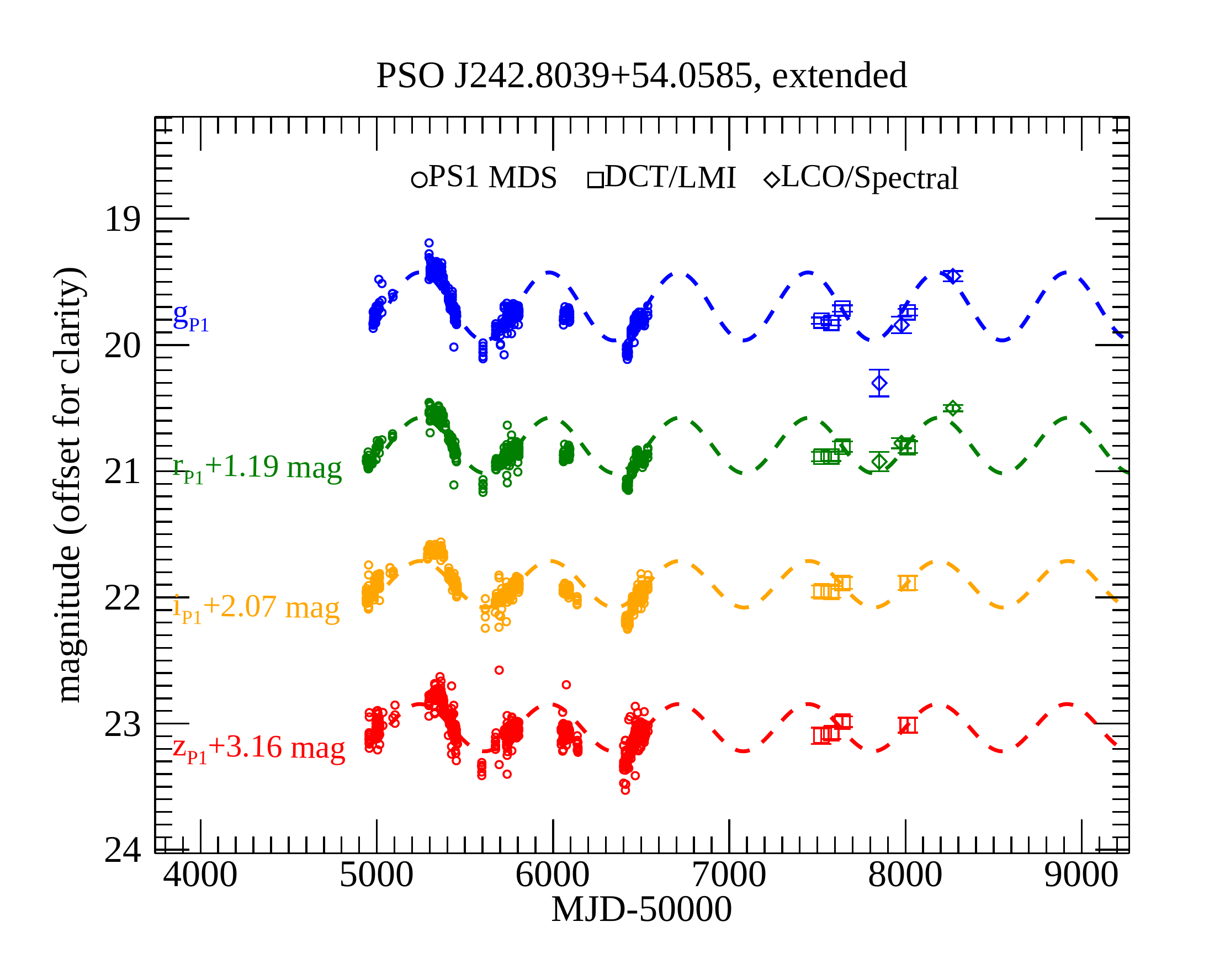,width=0.2\textwidth,clip=}
\epsfig{file=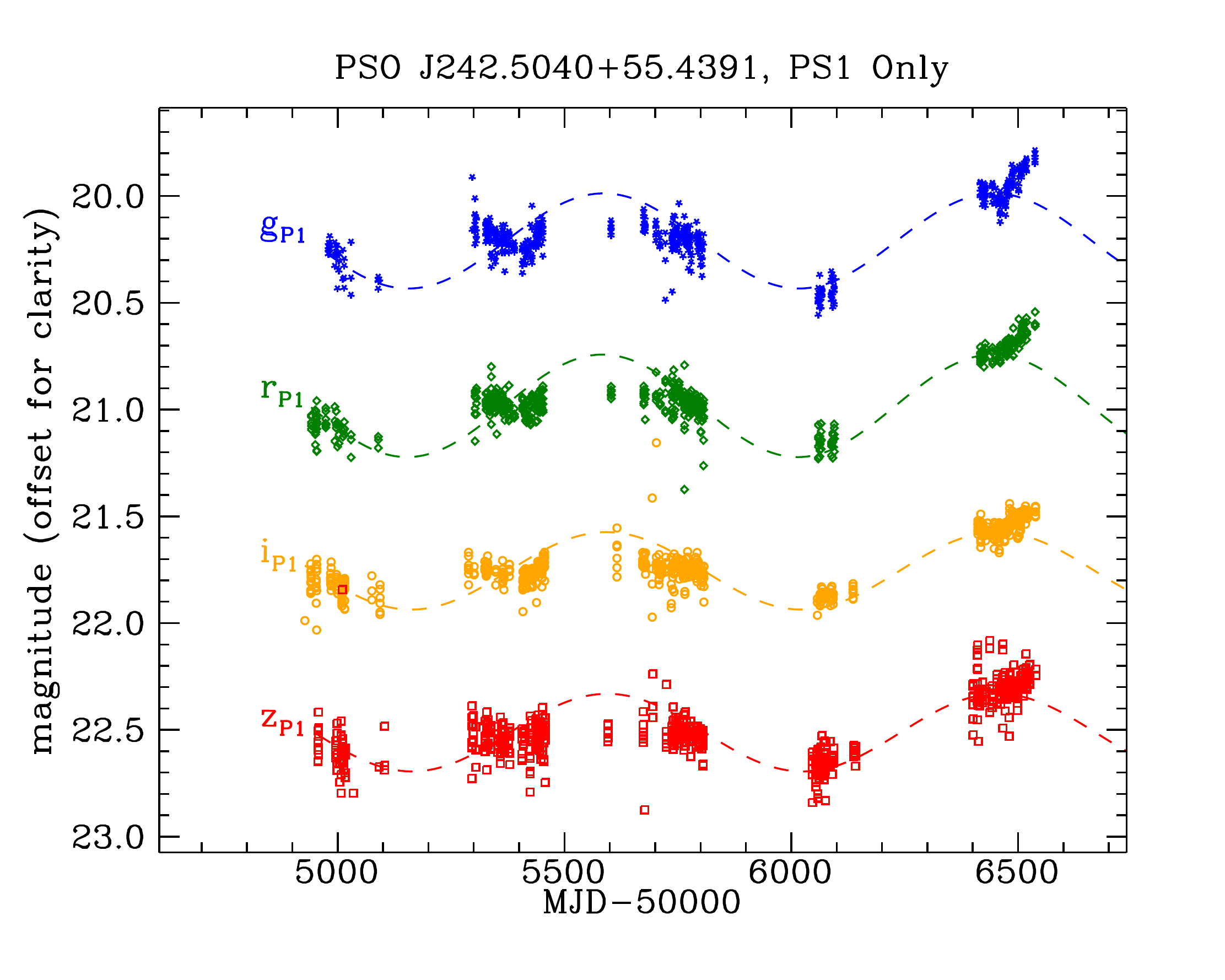,width=0.2\textwidth,clip=}
\epsfig{file=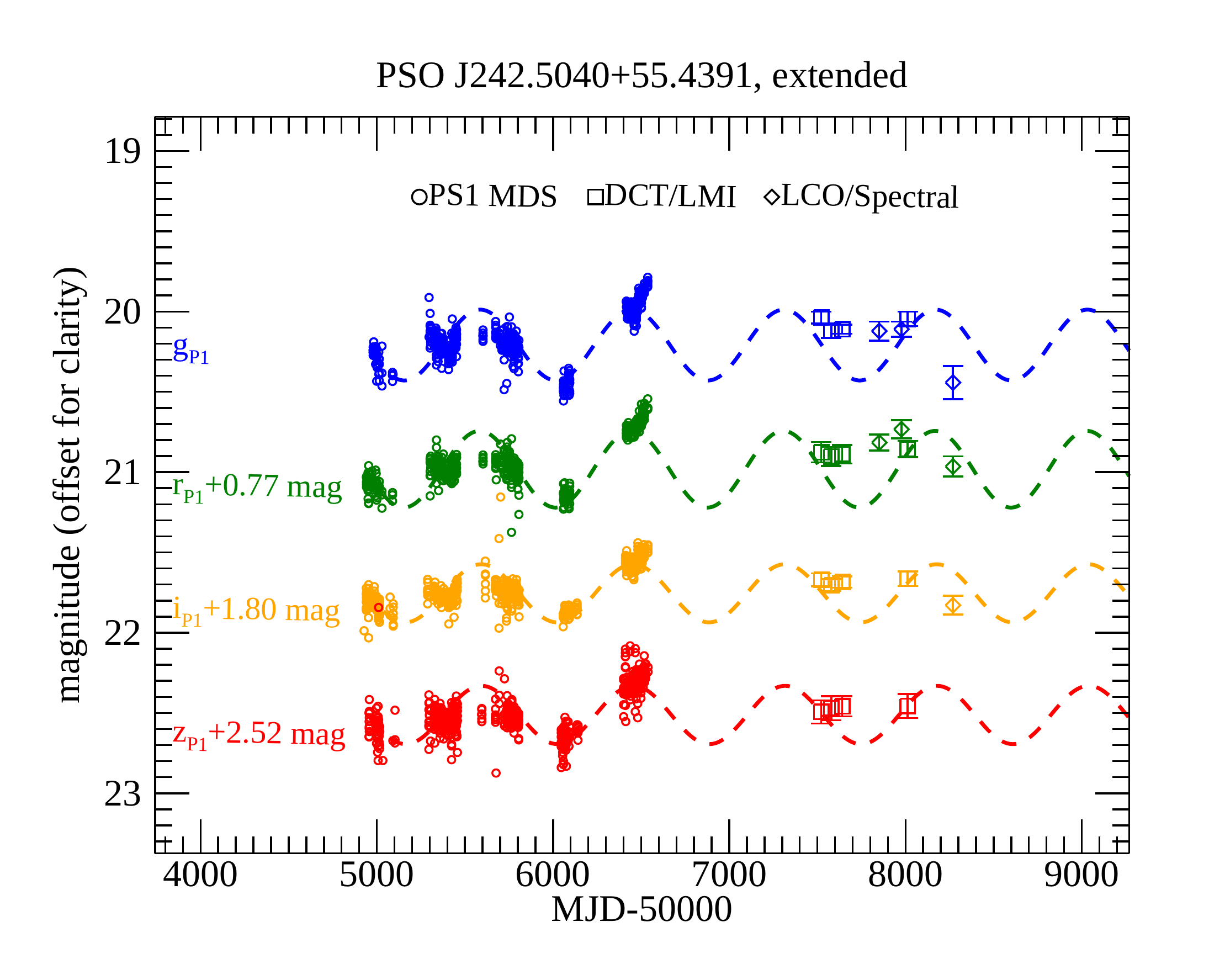,width=0.2\textwidth,clip=}
\epsfig{file=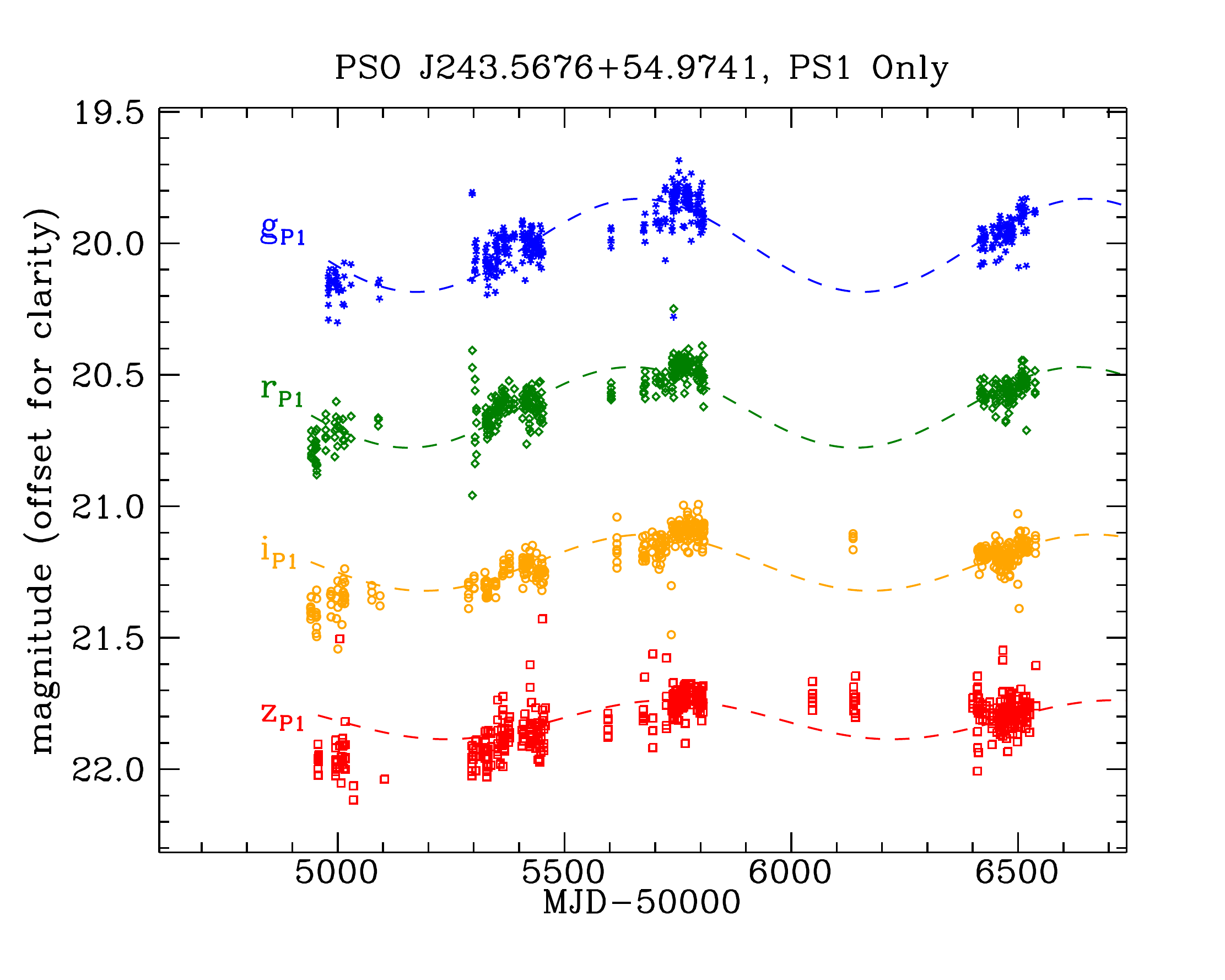,width=0.2\textwidth,clip=}
\epsfig{file=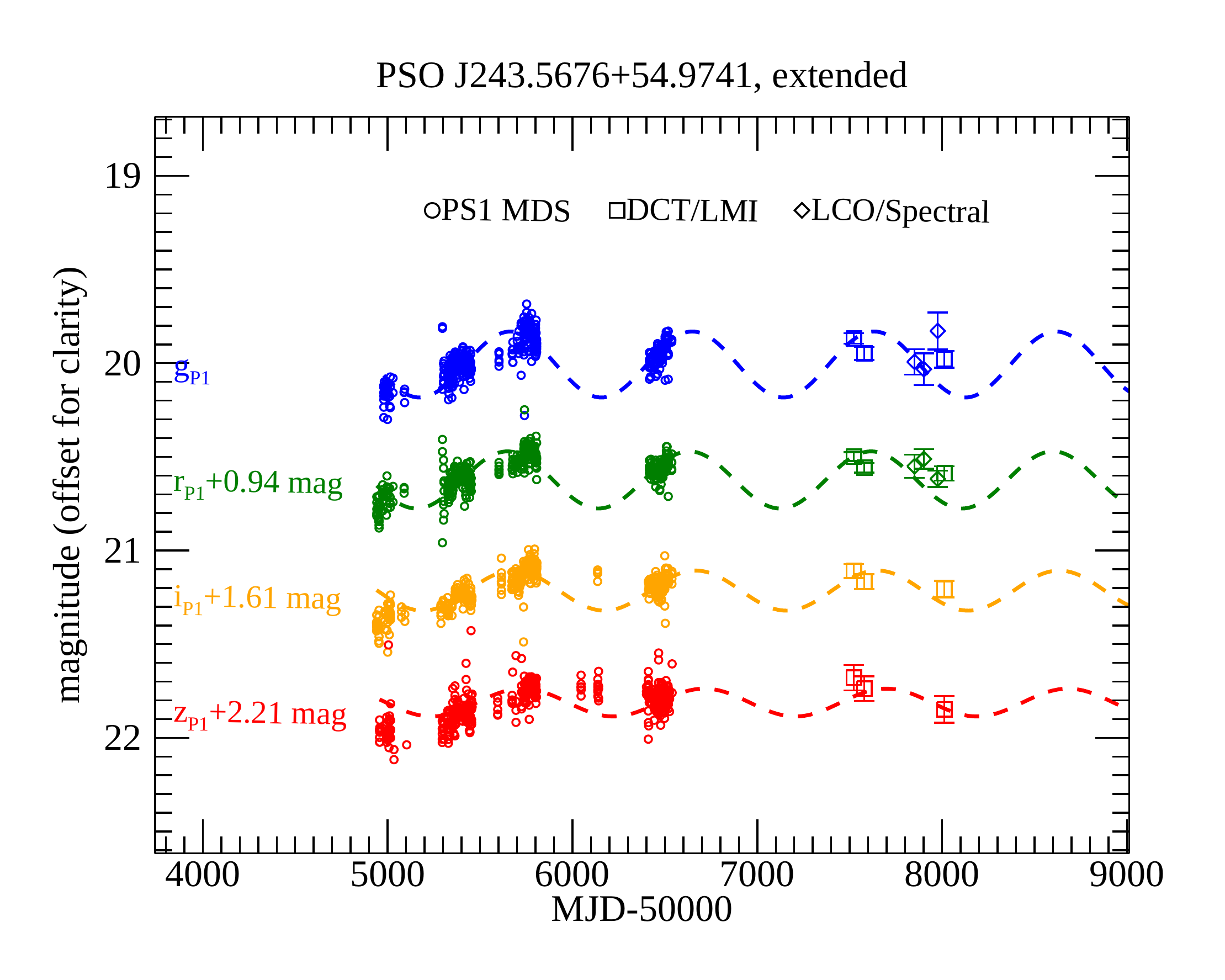,width=0.2\textwidth,clip=}
\epsfig{file=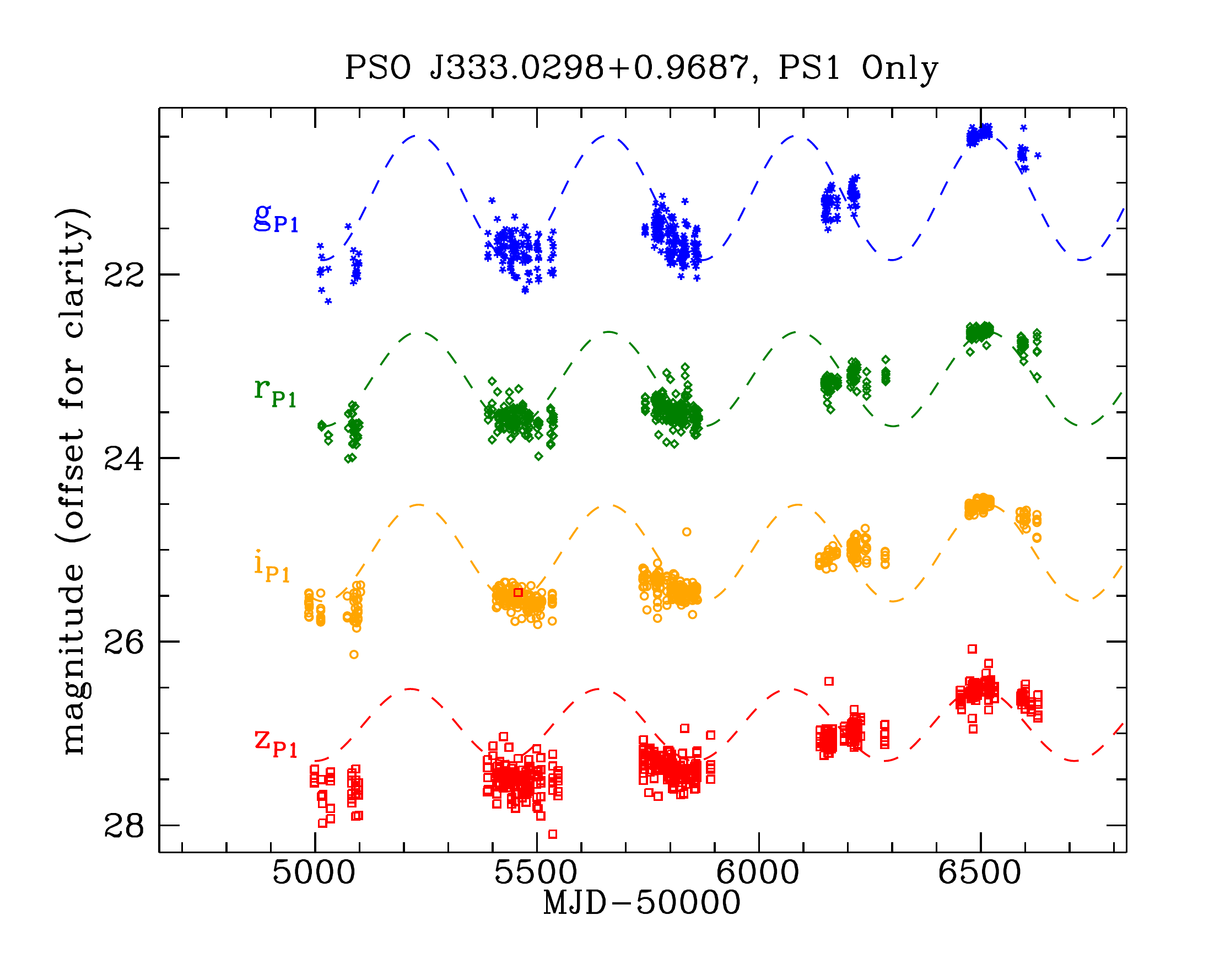,width=0.2\textwidth,clip=}
\epsfig{file=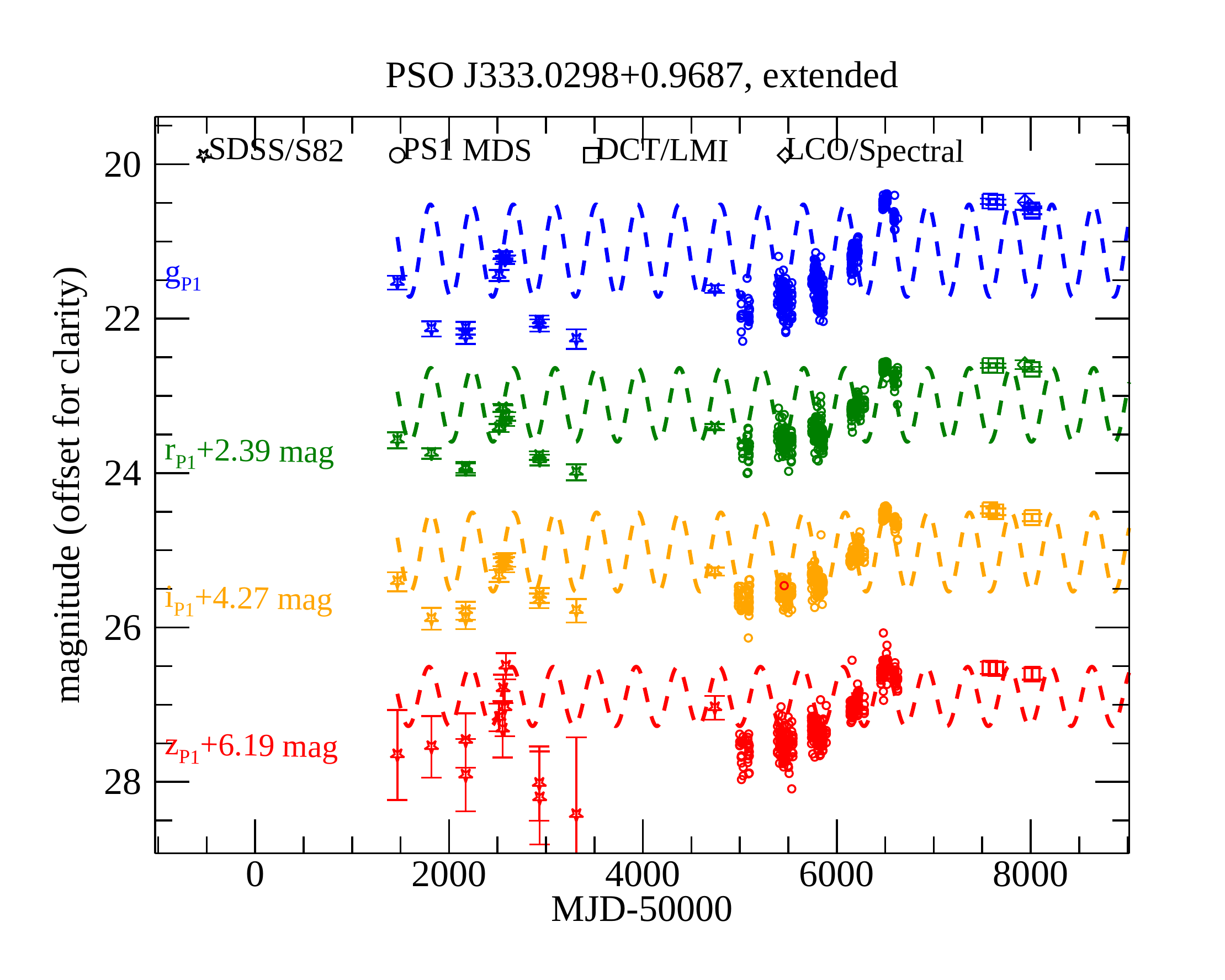,width=0.2\textwidth,clip=}
\epsfig{file=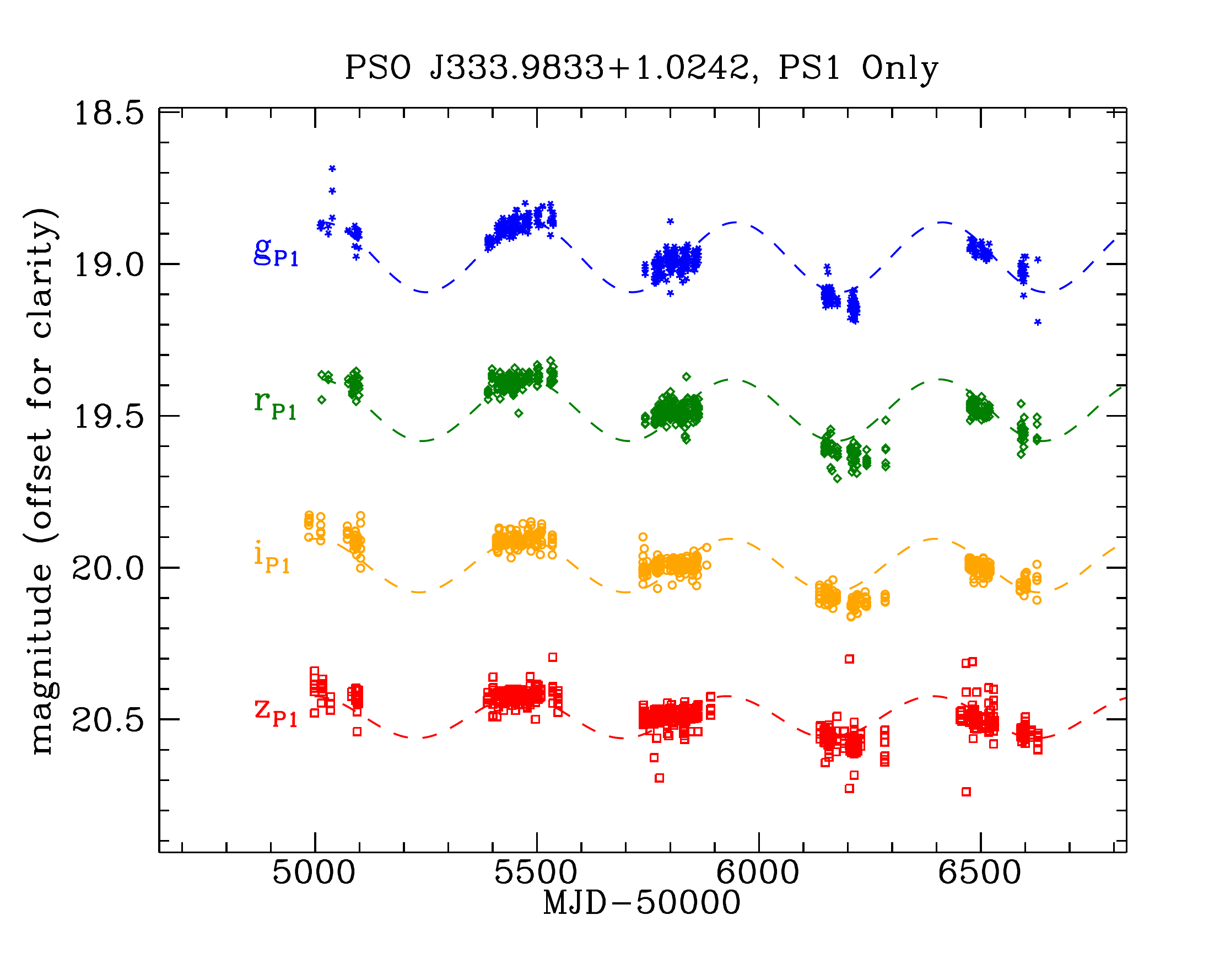,width=0.2\textwidth,clip=}
\epsfig{file=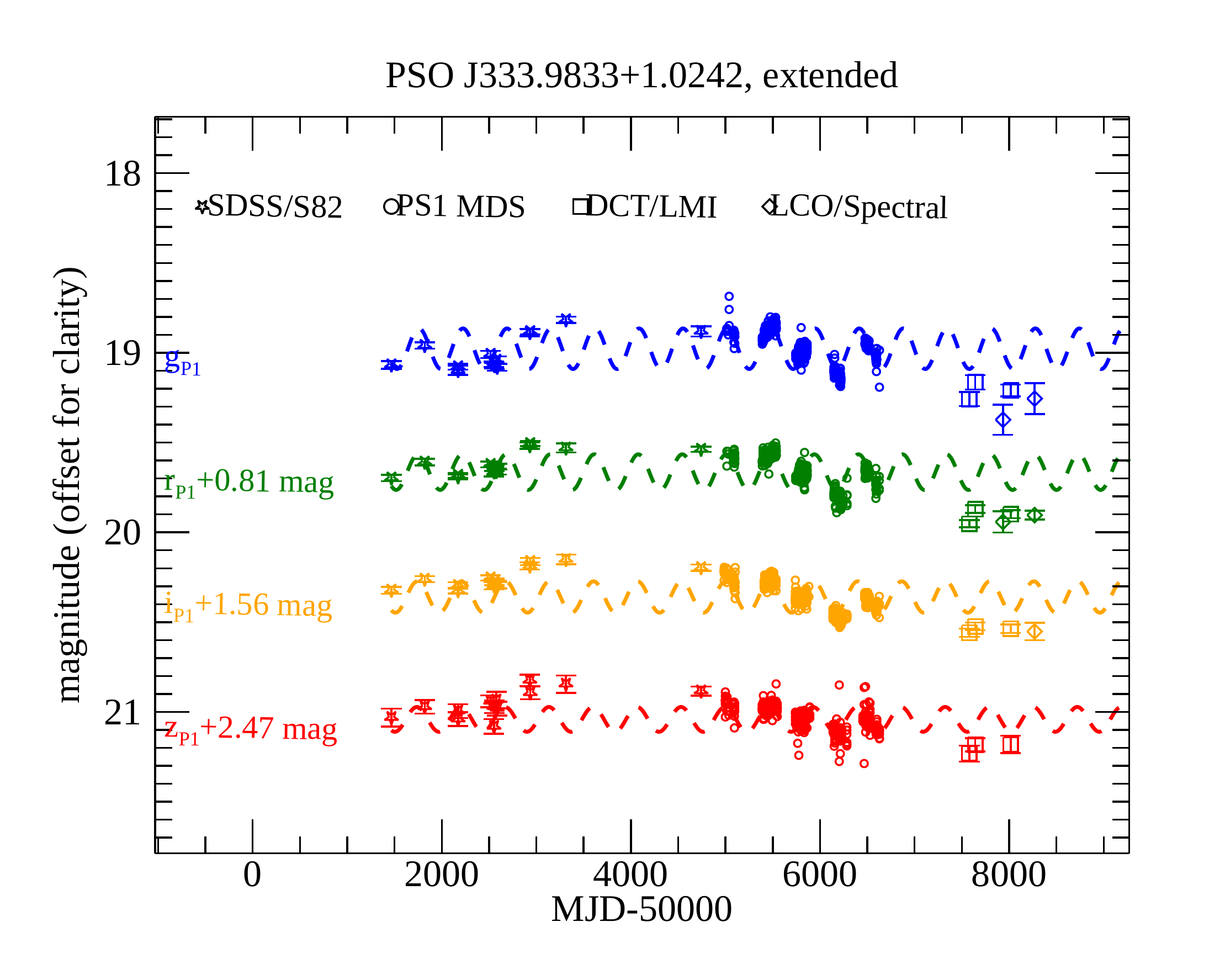,width=0.2\textwidth,clip=}
\epsfig{file=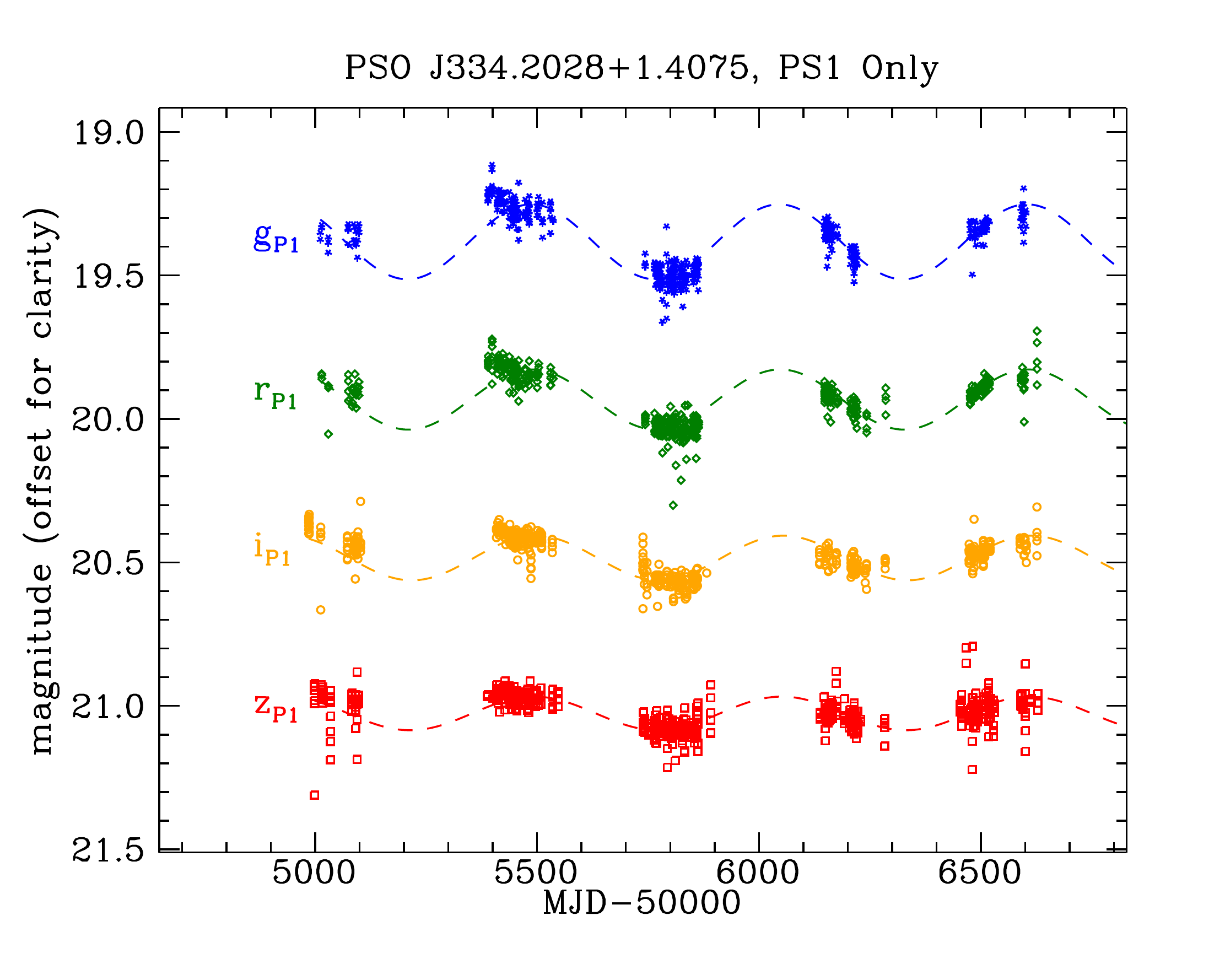,width=0.2\textwidth,clip=}
\epsfig{file=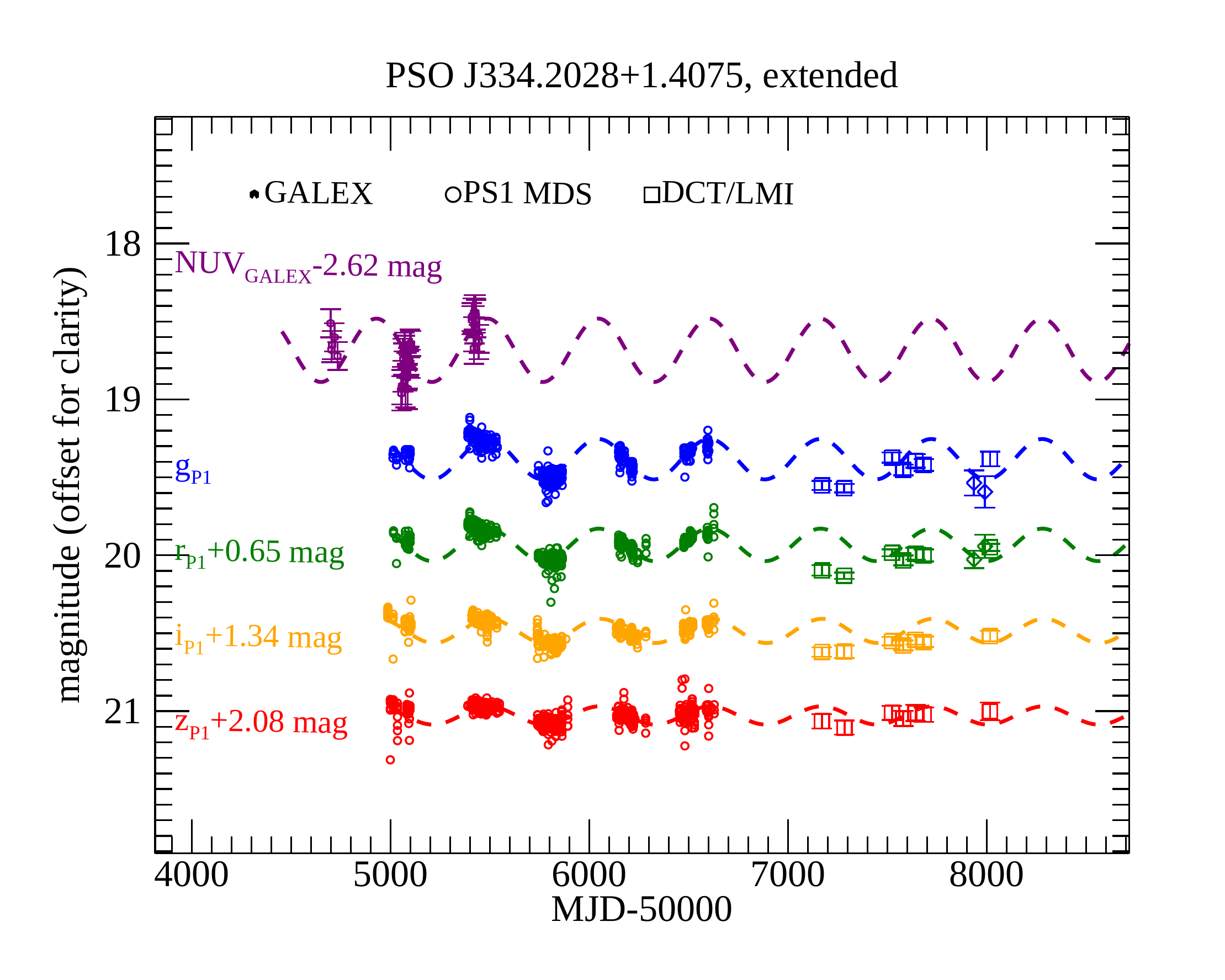,width=0.2\textwidth,clip=}
\epsfig{file=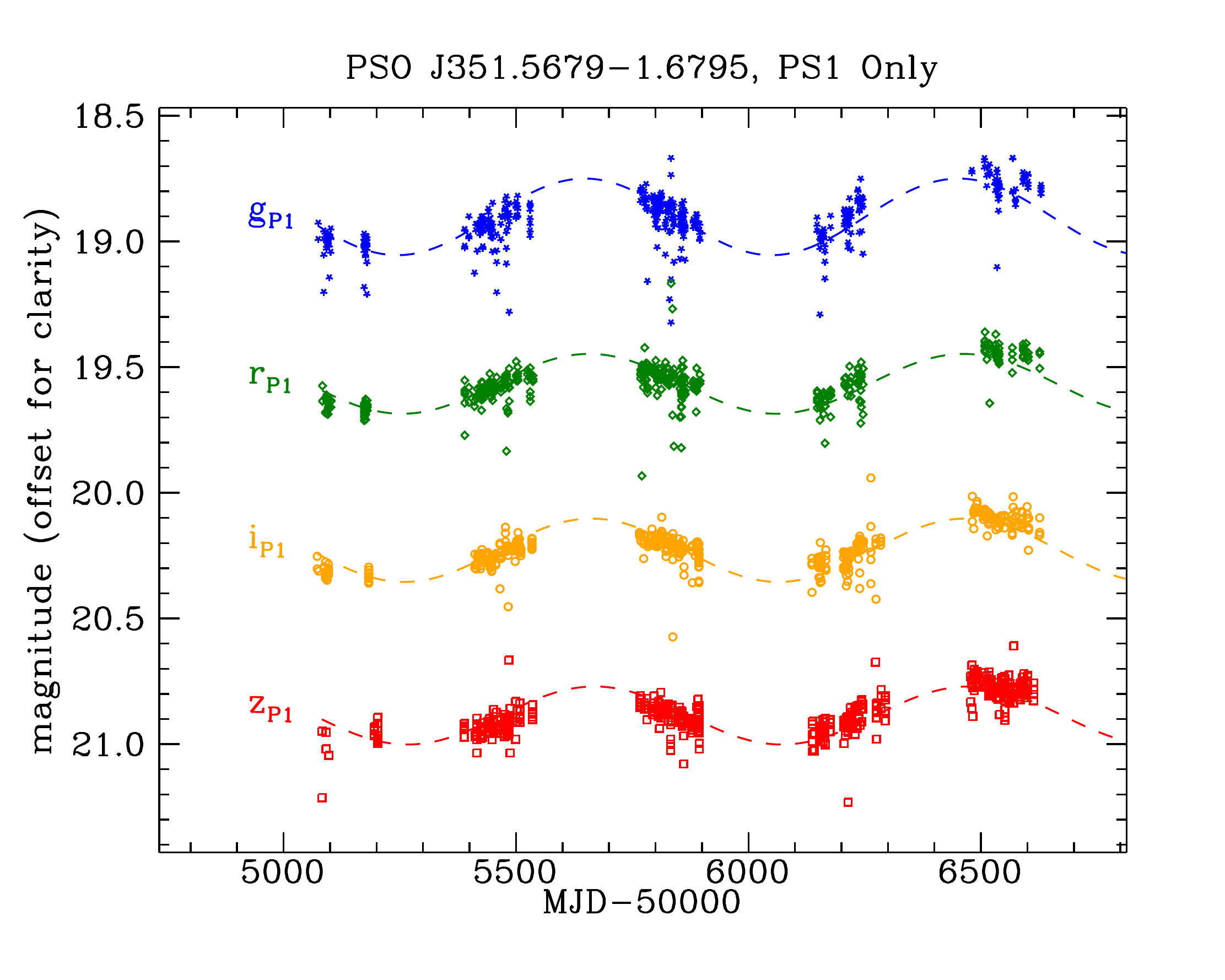,width=0.2\textwidth,clip=}
\epsfig{file=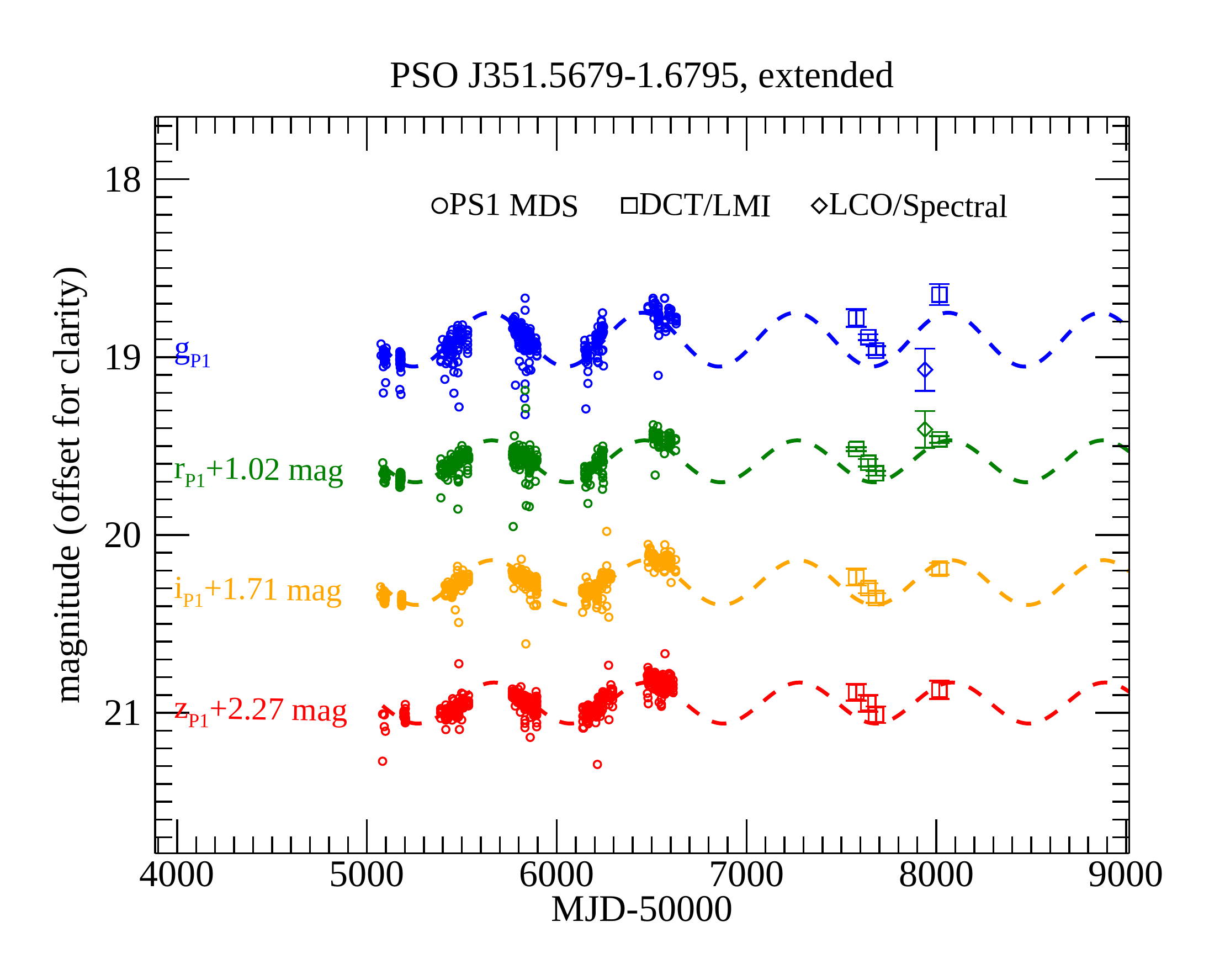,width=0.2\textwidth,clip=}
\epsfig{file=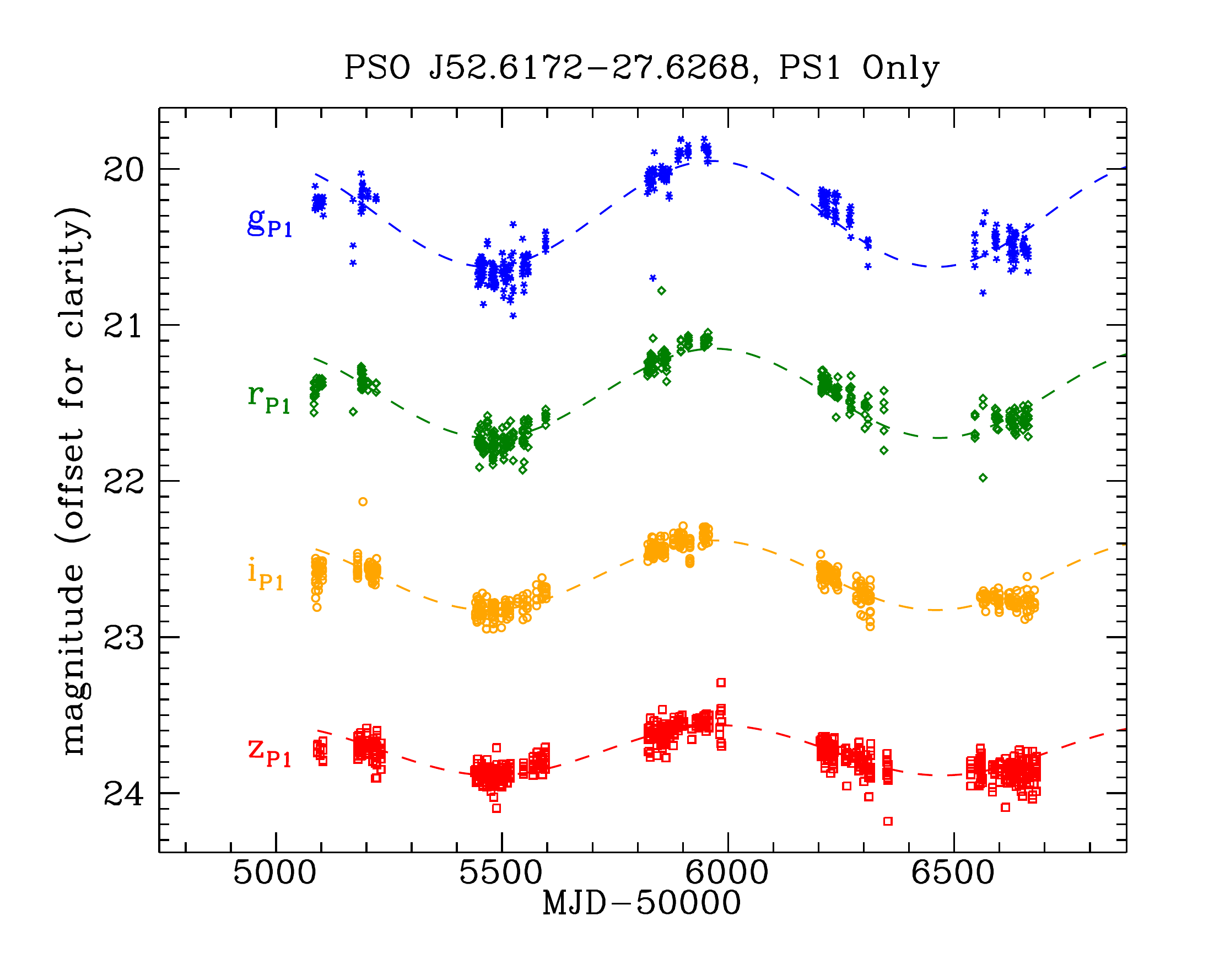,width=0.2\textwidth,clip=}
\epsfig{file=empty,width=0.2\textwidth,clip=} 
\epsfig{file=empty,width=0.2\textwidth,clip=}
\epsfig{file=empty,width=0.2\textwidth,clip=}
\label{fig:lc_cont}
\caption{PS1-only and extended light curves of PS1 MDS candidates (cont.).}
\end{figure*} 


\section{Archival and Follow-up Spectra of PS1 MDS Candidates}\label{app:spec}

We retrieved archival SDSS spectra from the SDSS Science Archive and obtained spectroscopic observations with Gemini/GMOS or DCT/DeVeny (Section \ref{sec:bhmass}). The spectra are presented in Figure \ref{fig:spec}. Prominent emission lines, including black hole mass estimators \ion{C}{4} and \ion{Mg}{2}, are indicated with red tick marks. The last two panels show typical example procedures of our spectral fitting of the continuum and the broad emission line (Section \ref{sec:bhmass}): fitting the \ion{Mg}{2} line of PSO J185, and \ion{C}{4} of PSO J149.4989+2.7827. We note that while both objects are considered statistically significant in our extended baseline analysis (Section \ref{sec:reanalysis}) and in particular, PSO J185 is our most significant candidate, no peculiar features are seen in their spectra (such as asymmetry in the broad emission line).
\setcounter{figure}{0}   
\begin{figure*}[h]
\centering
\epsfig{file=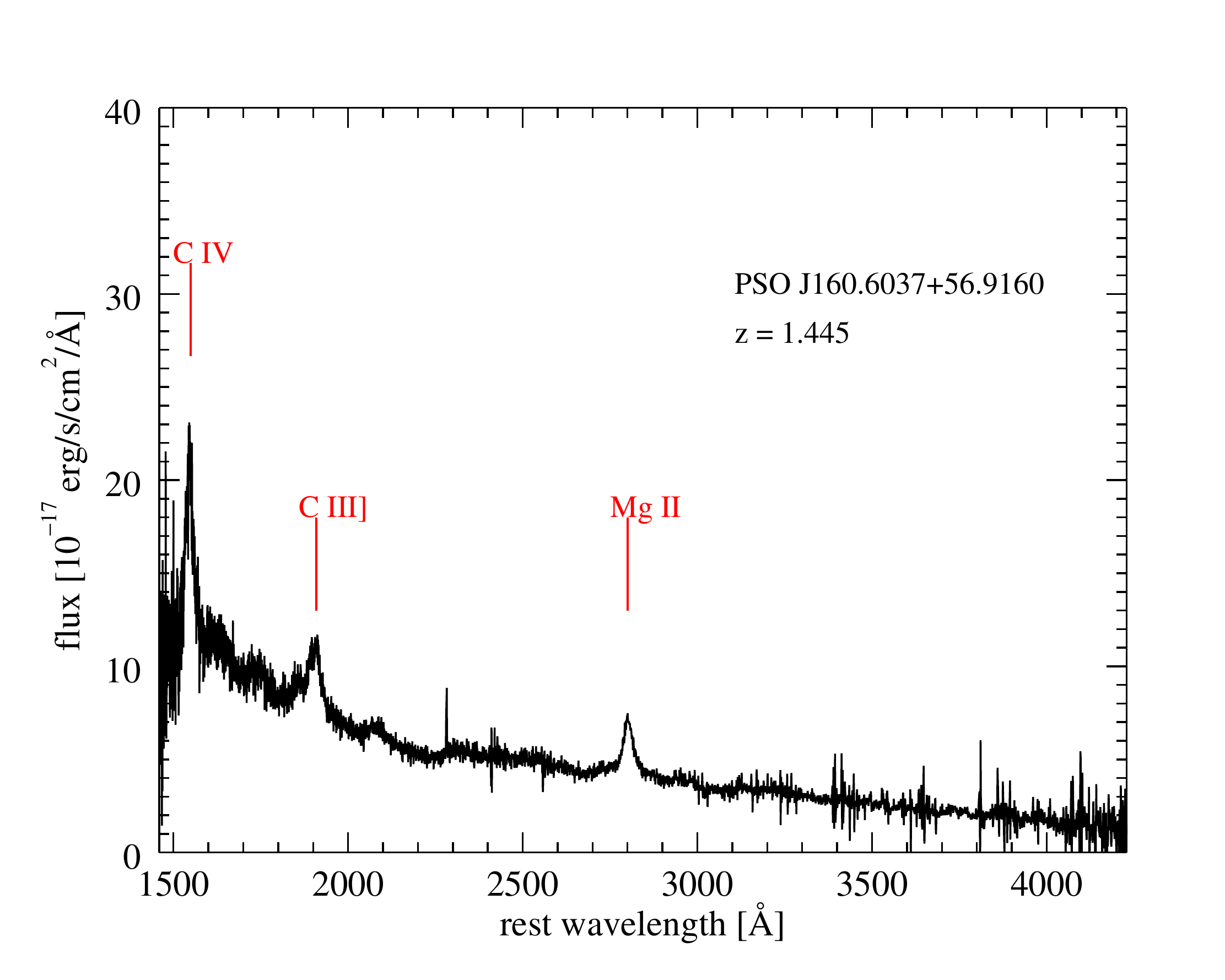,width=0.2\textwidth,clip=}
\epsfig{file=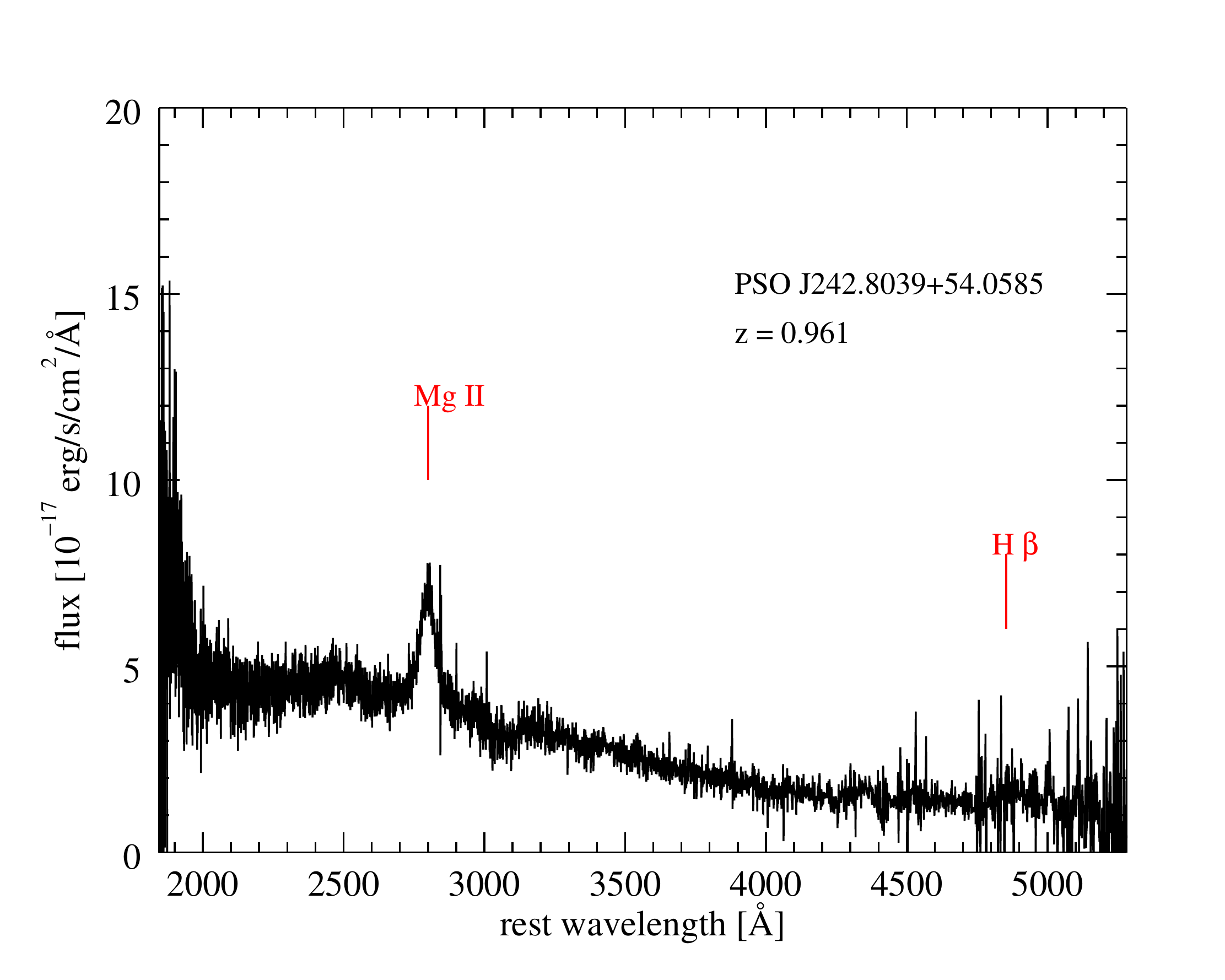,width=0.2\textwidth,clip=}
\epsfig{file=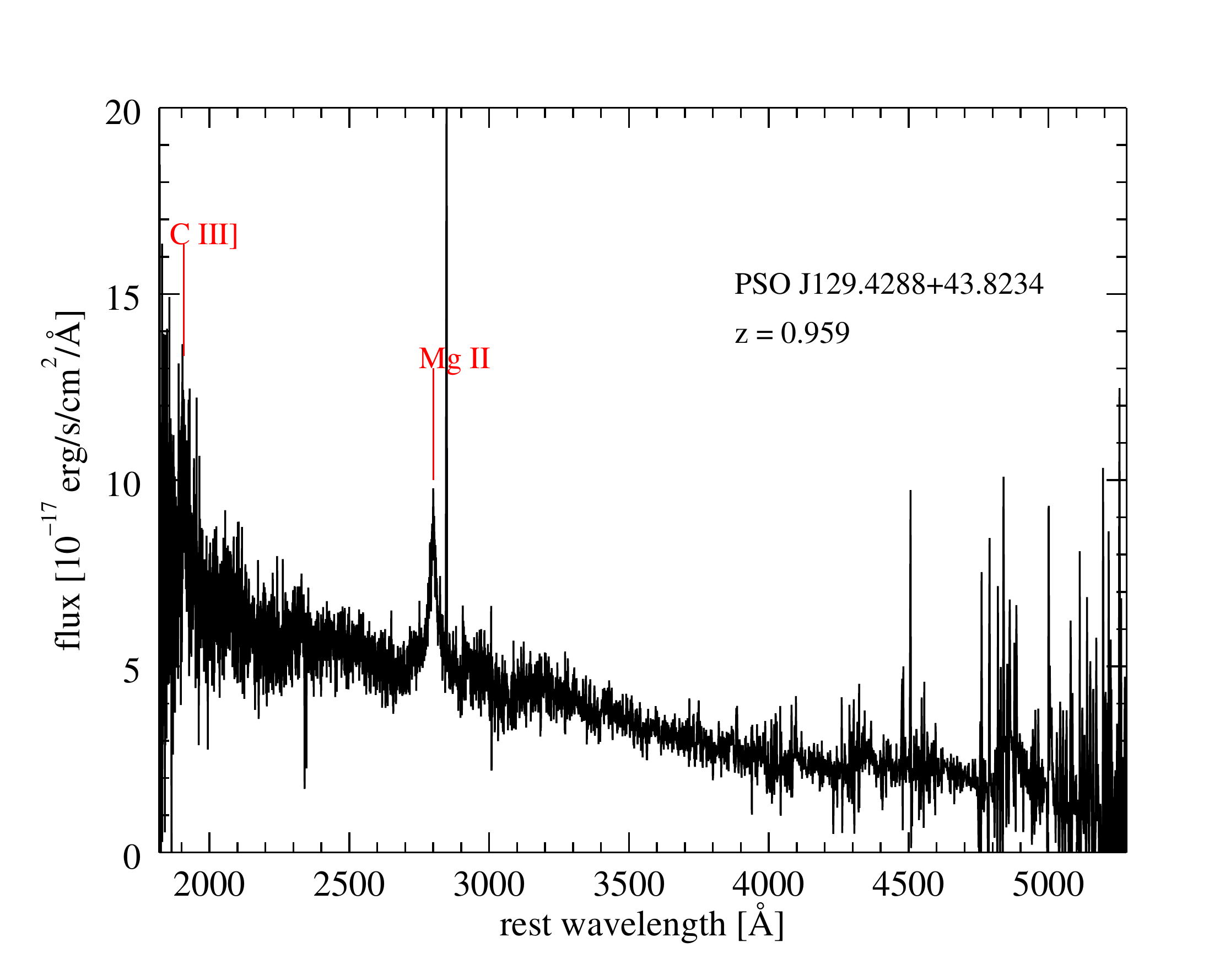,width=0.2\textwidth,clip=}
\epsfig{file=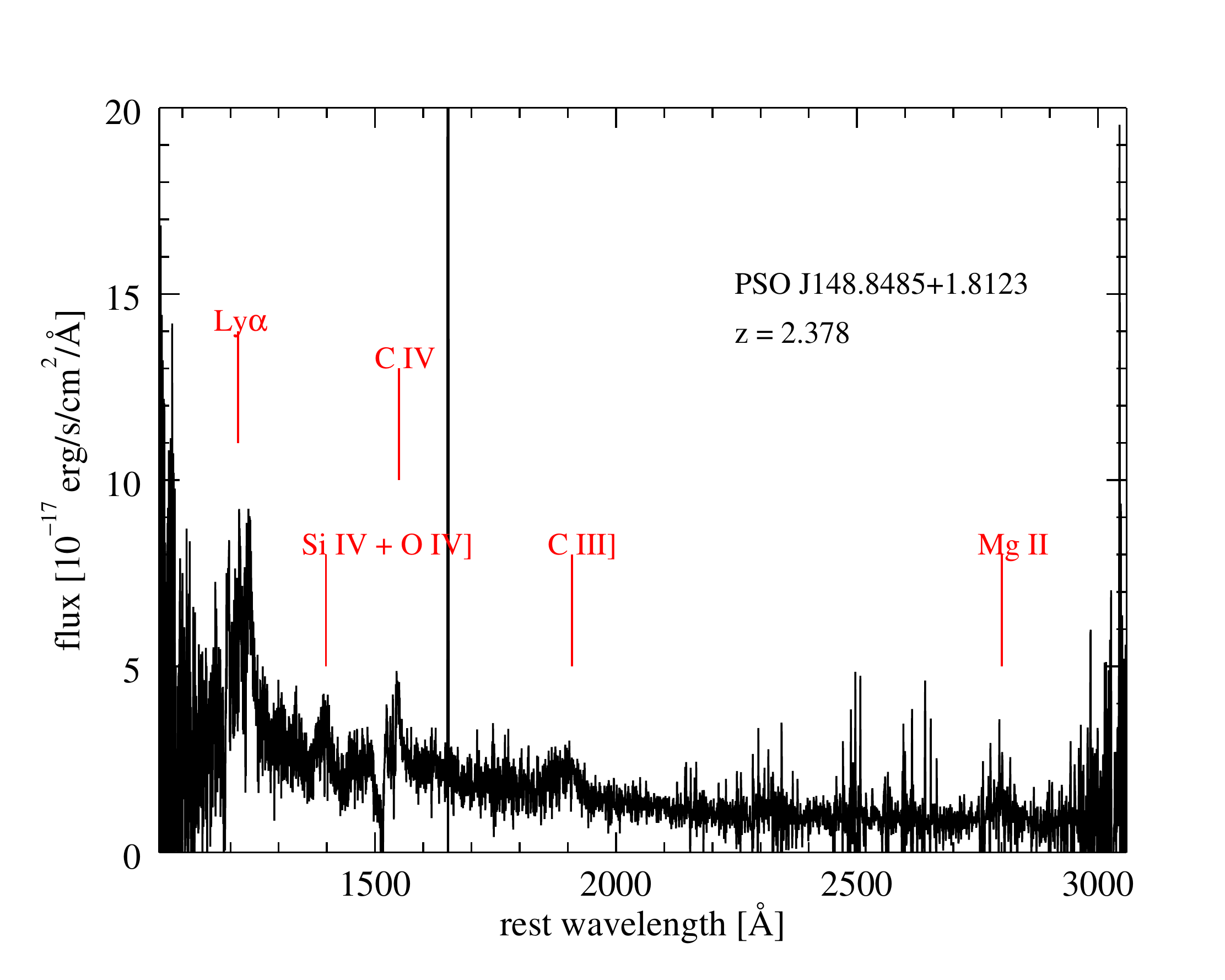,width=0.2\textwidth,clip=}
\epsfig{file=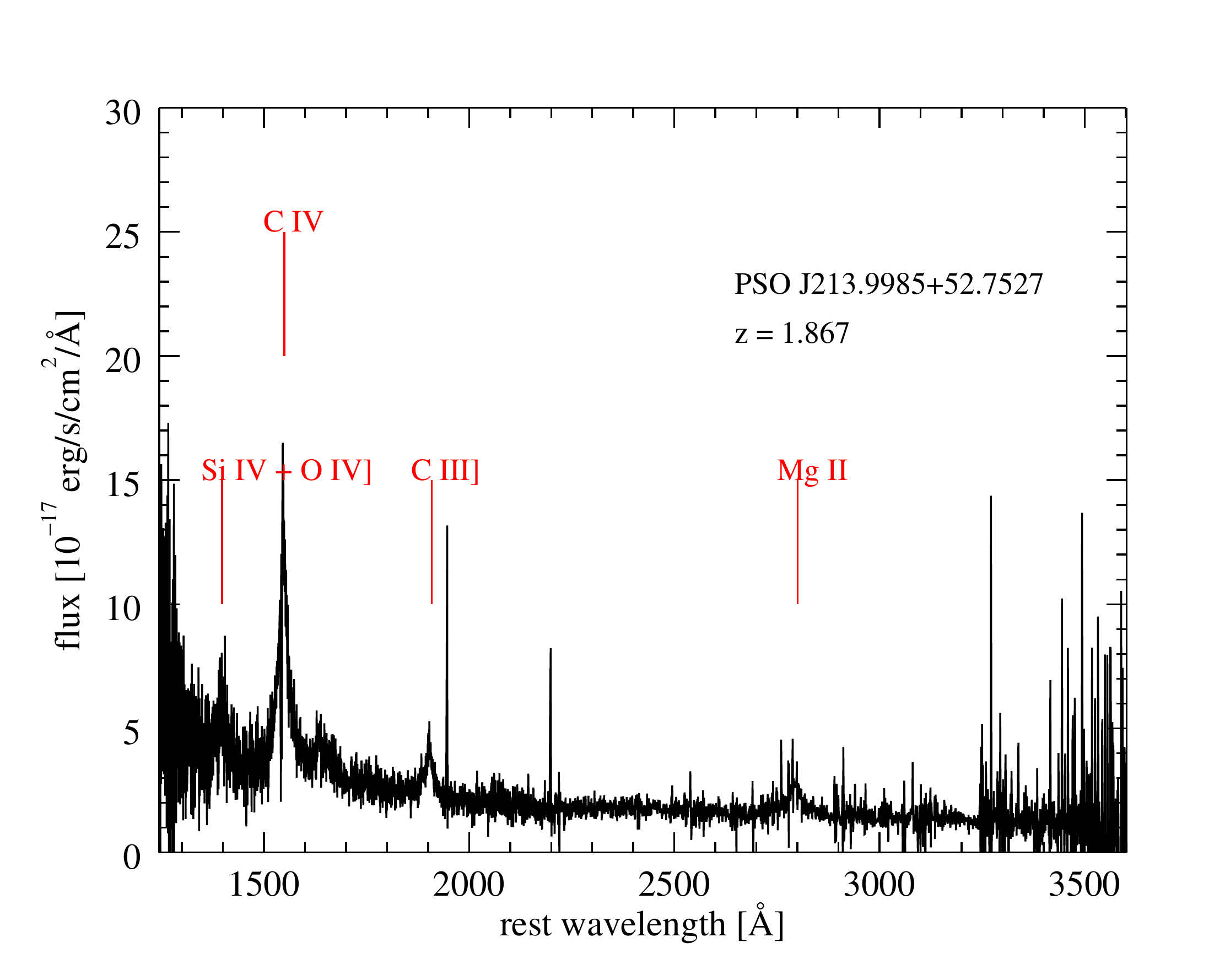,width=0.2\textwidth,clip=}
\epsfig{file=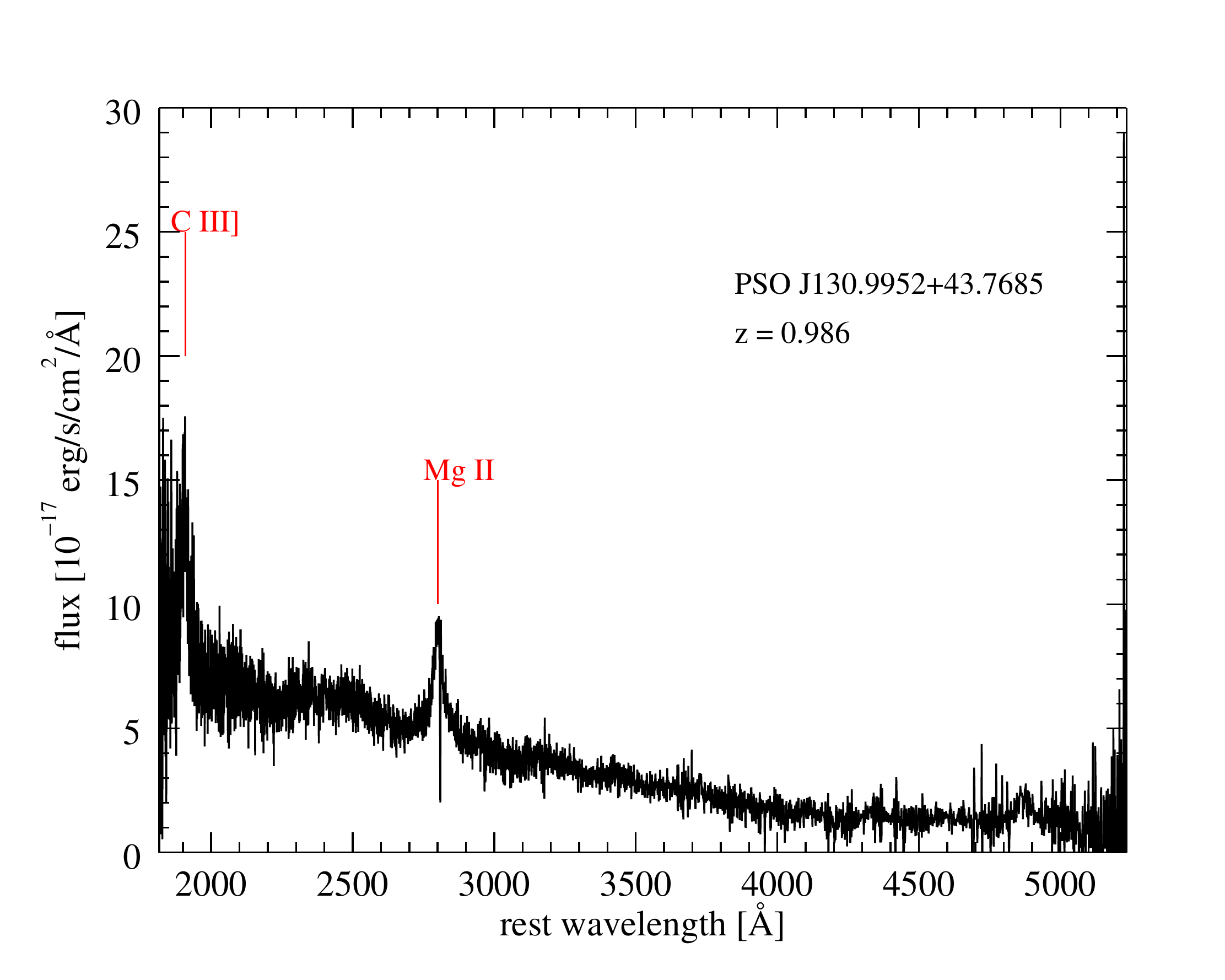,width=0.2\textwidth,clip=}
\epsfig{file=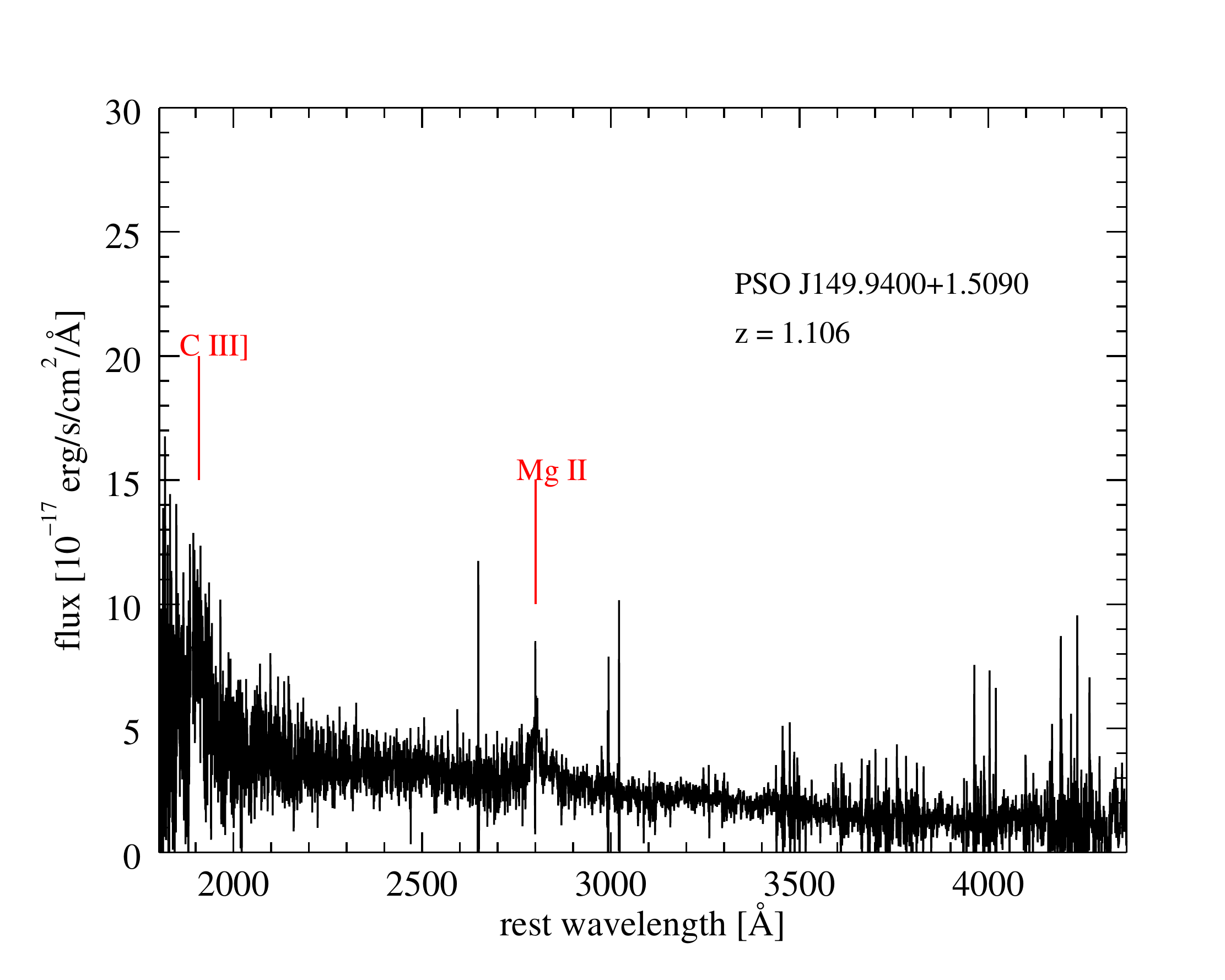,width=0.2\textwidth,clip=}
\epsfig{file=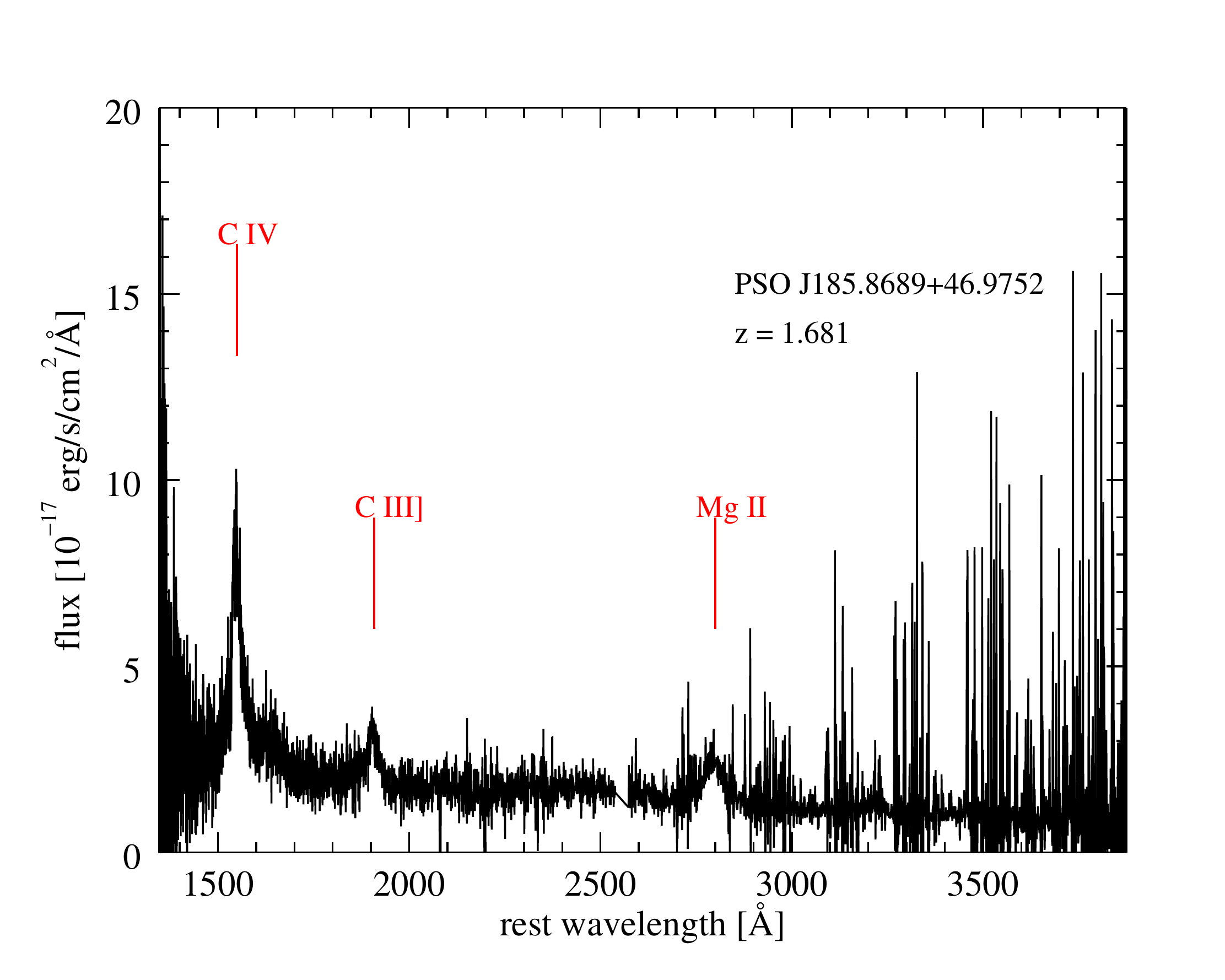,width=0.2\textwidth,clip=}
\epsfig{file=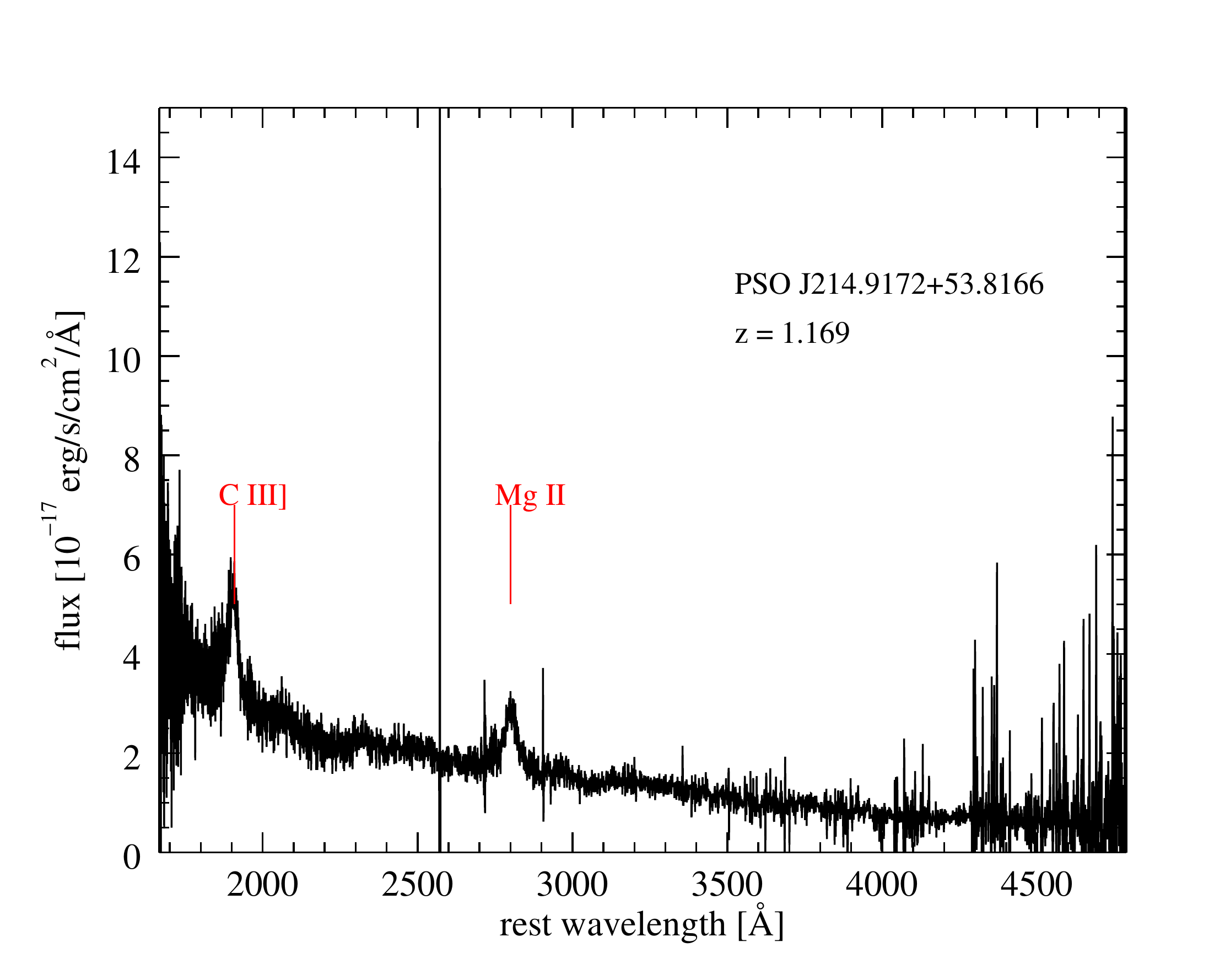,width=0.2\textwidth,clip=}
\epsfig{file=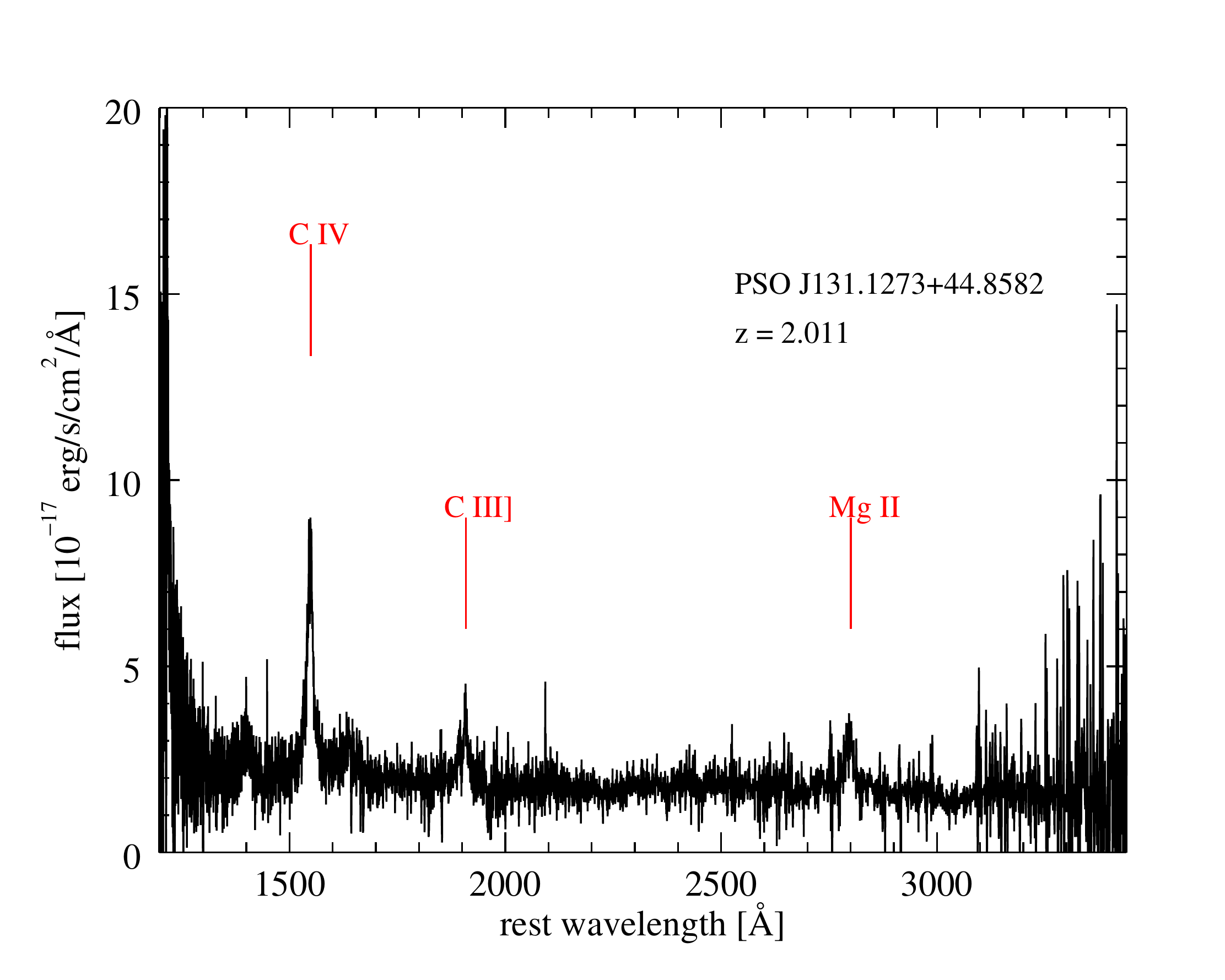,width=0.2\textwidth,clip=}
\epsfig{file=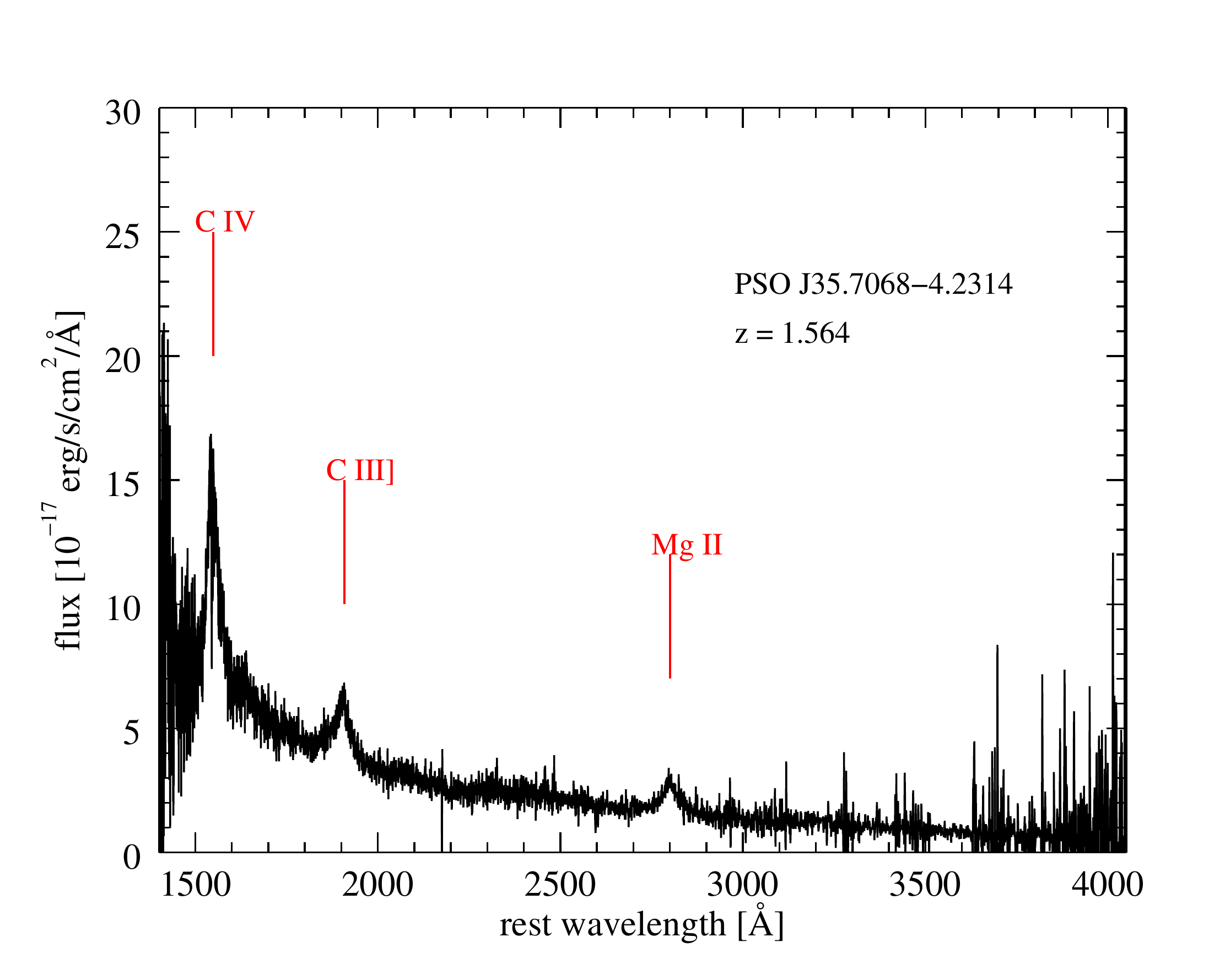,width=0.2\textwidth,clip=}
\epsfig{file=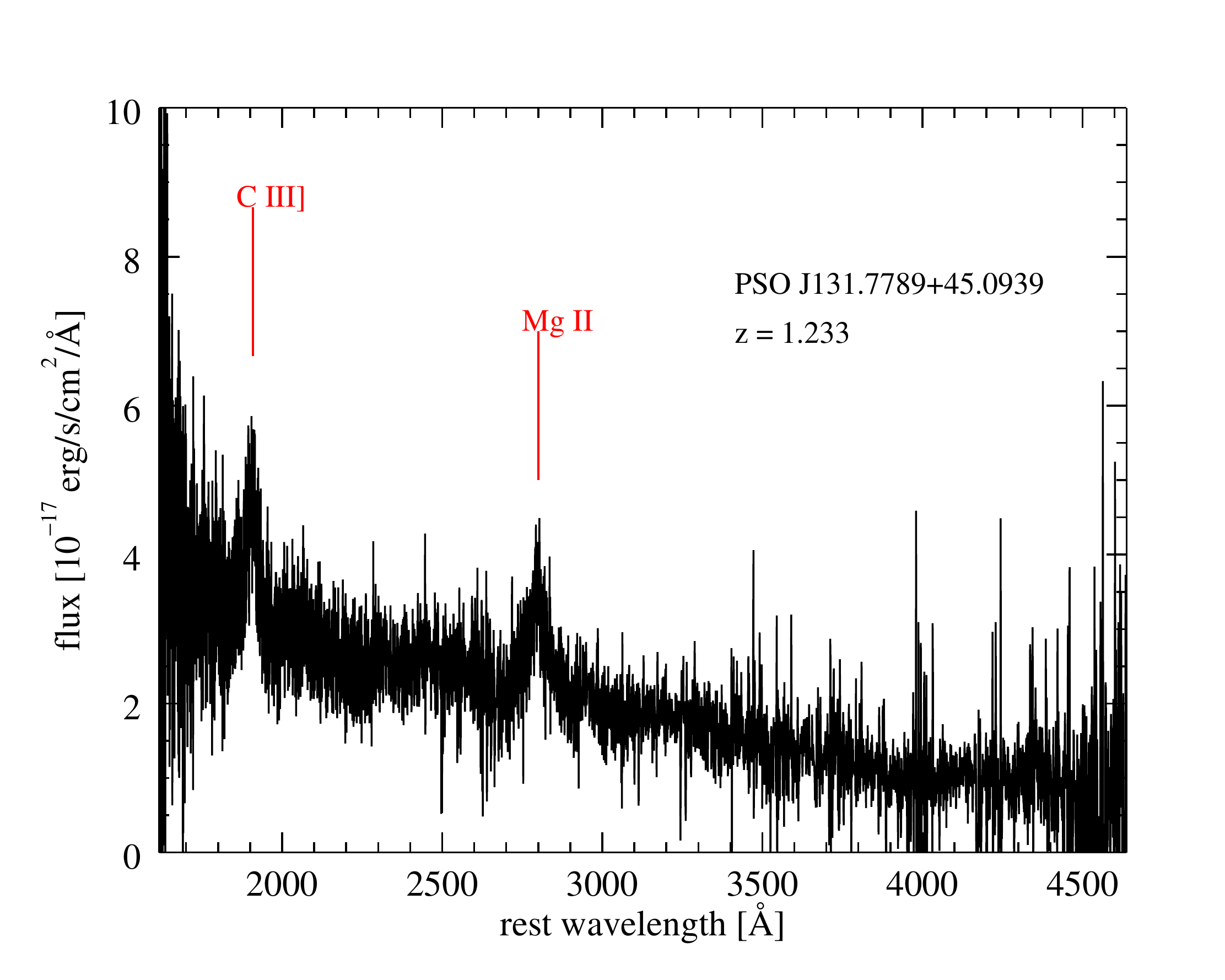,width=0.2\textwidth,clip=}
\epsfig{file=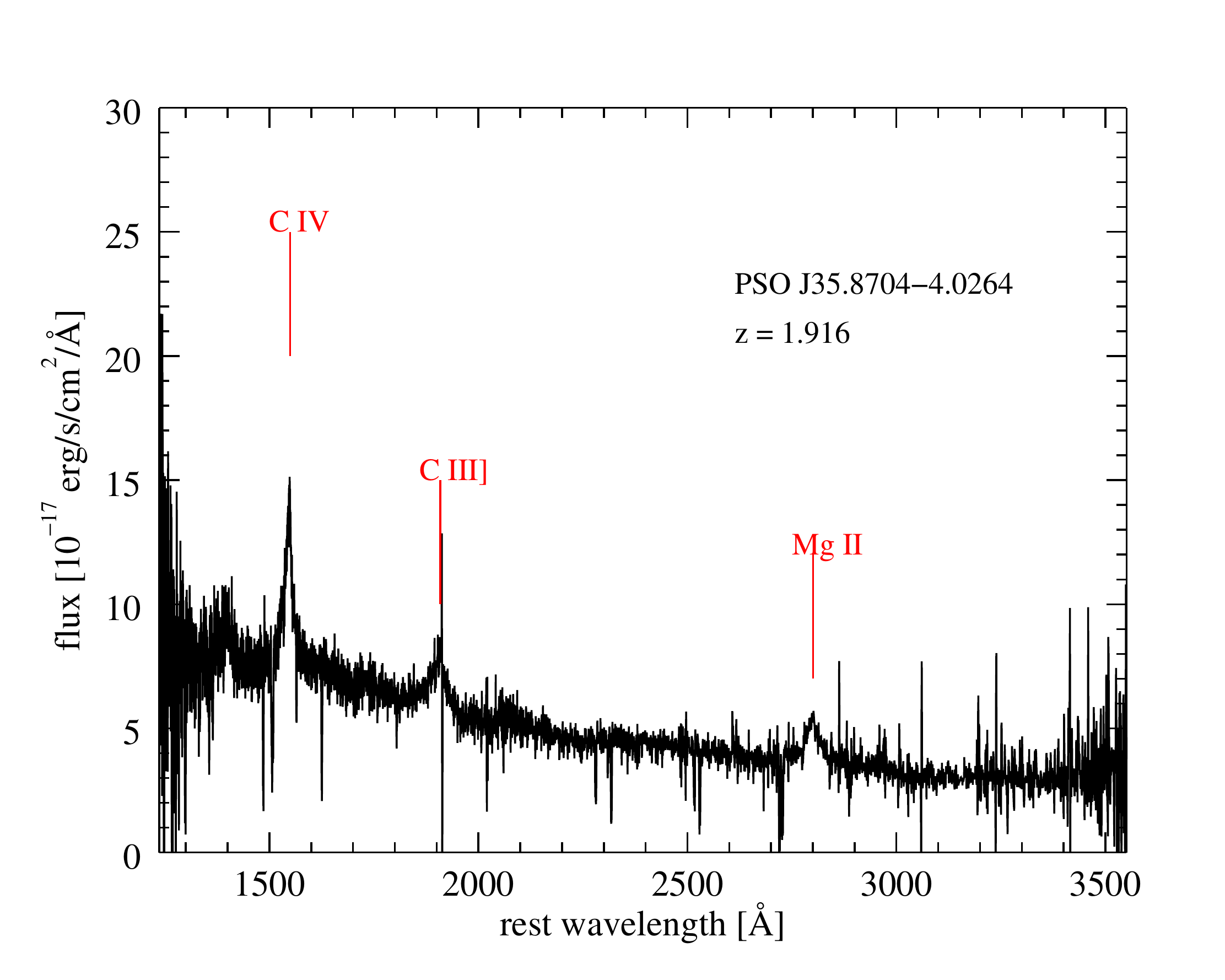,width=0.2\textwidth,clip=}
\epsfig{file=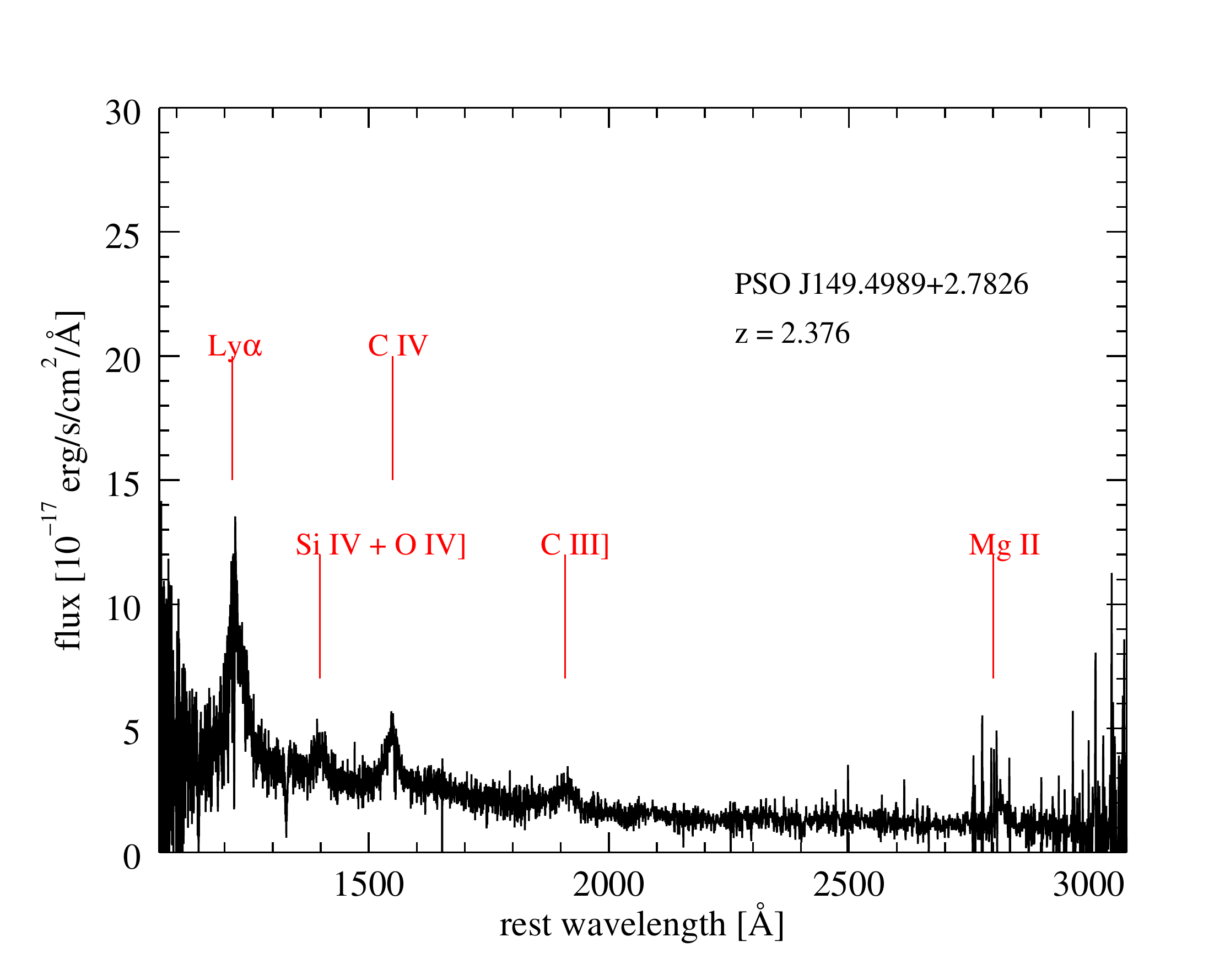,width=0.2\textwidth,clip=}
\epsfig{file=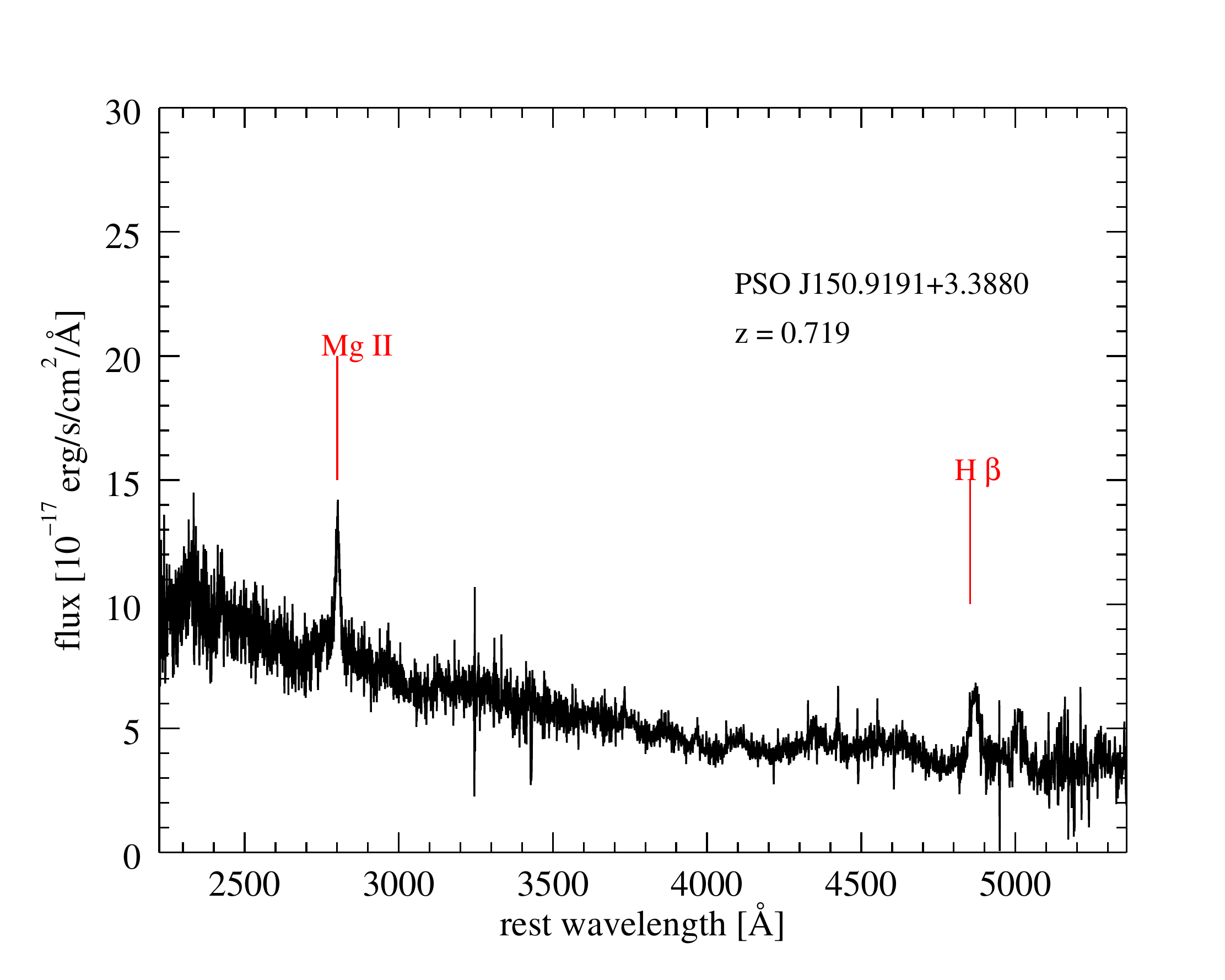,width=0.2\textwidth,clip=}
\epsfig{file=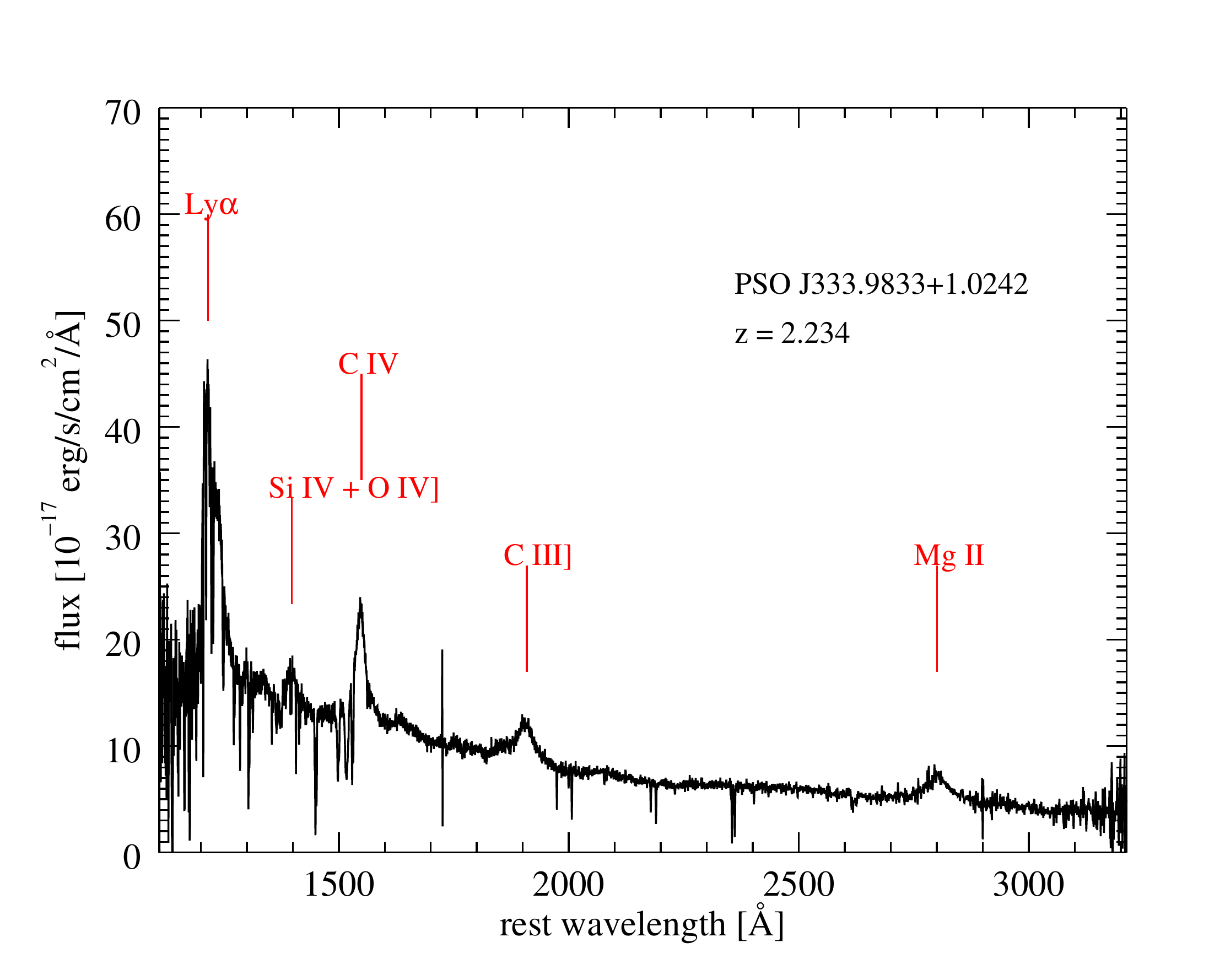,width=0.2\textwidth,clip=}
\epsfig{file=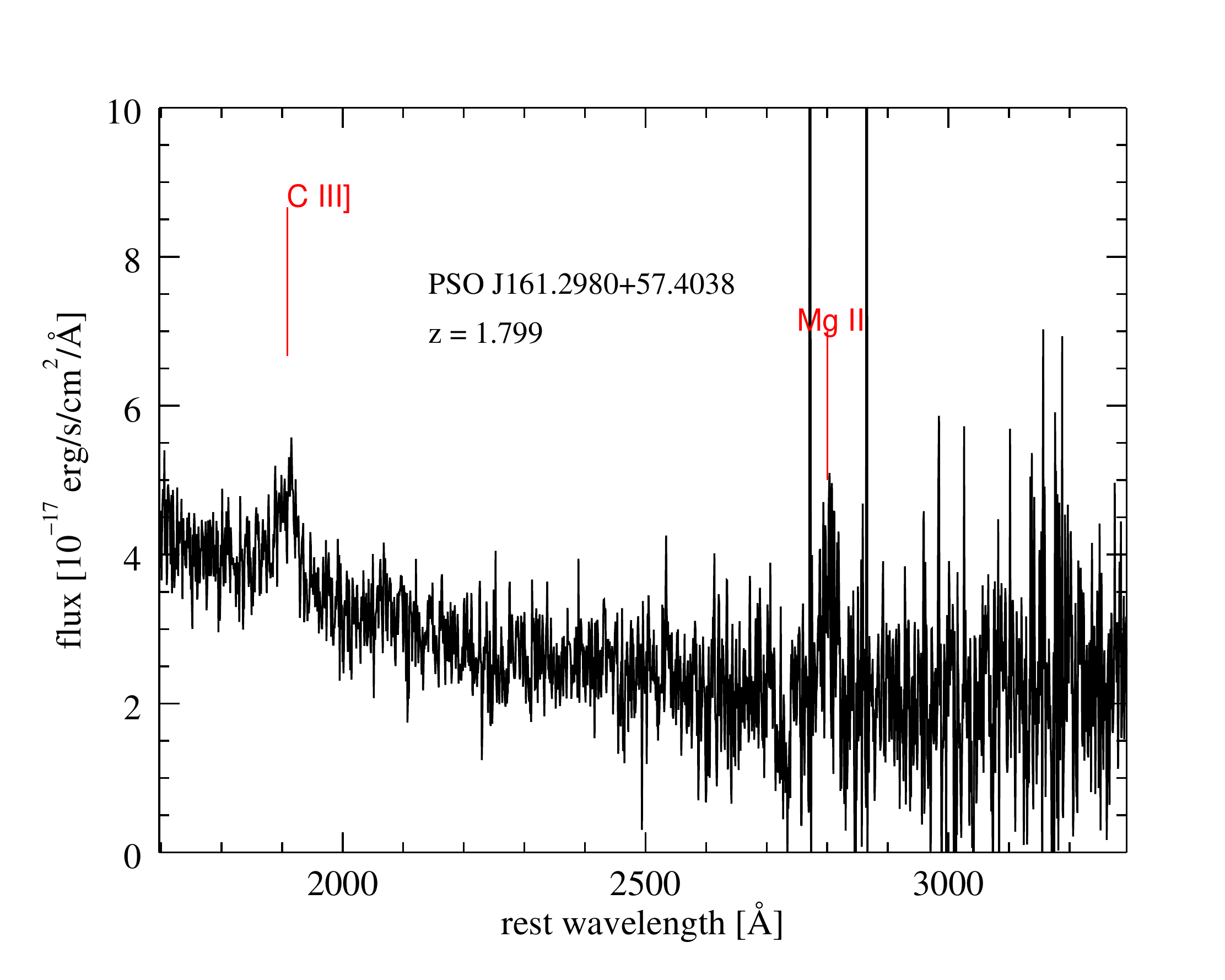,width=0.2\textwidth,clip=}
\epsfig{file=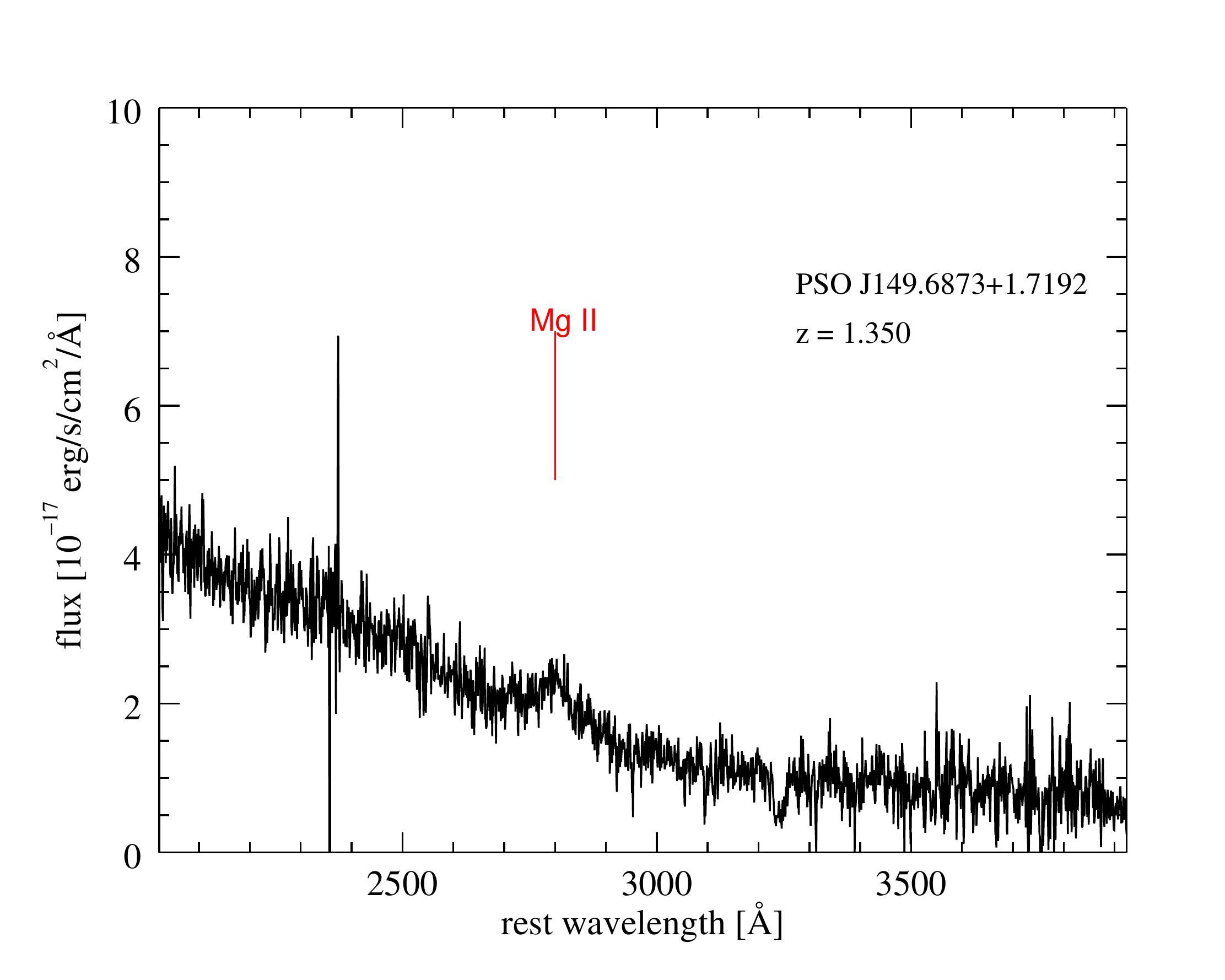,width=0.2\textwidth,clip=}
\epsfig{file=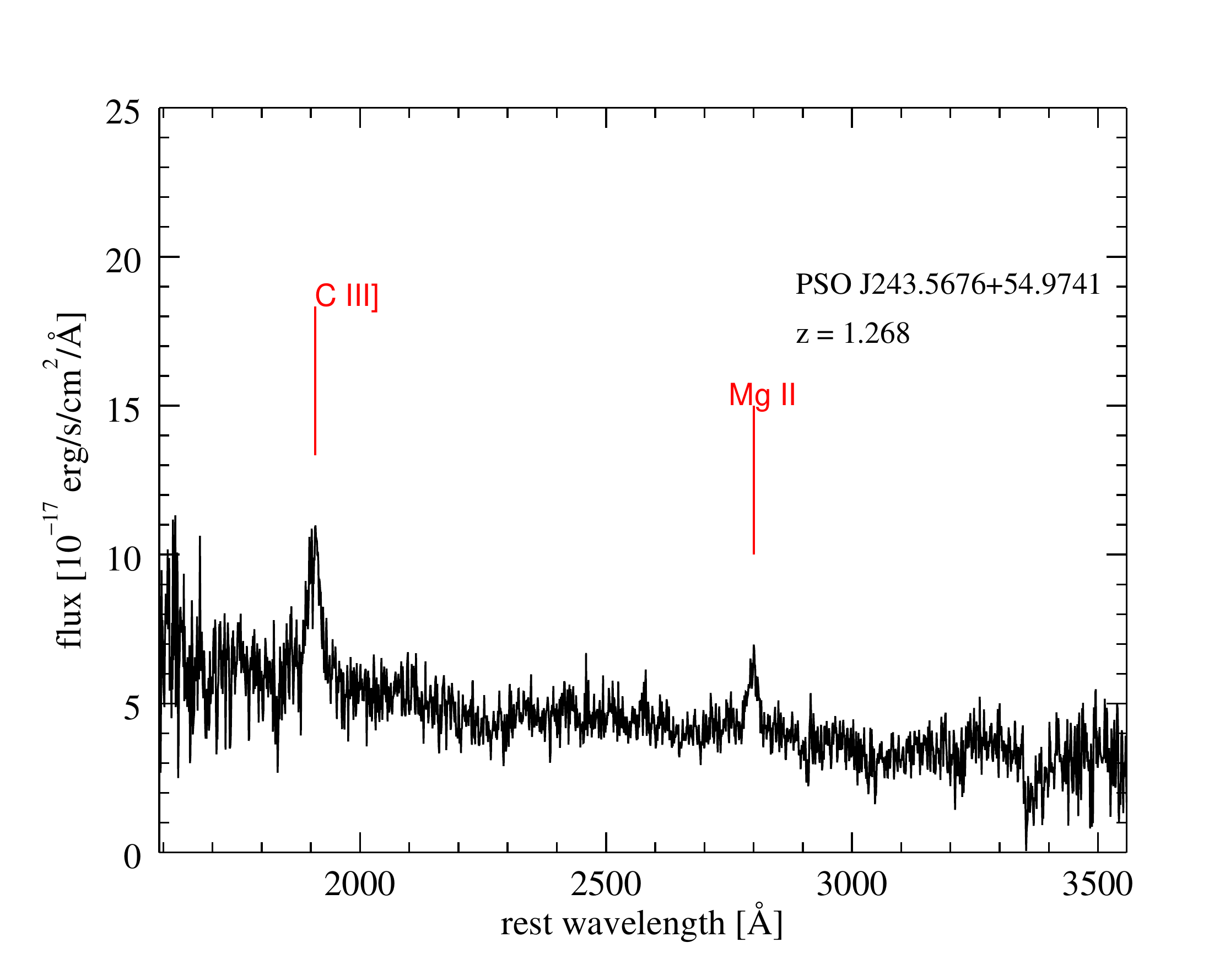,width=0.2\textwidth,clip=}
\epsfig{file=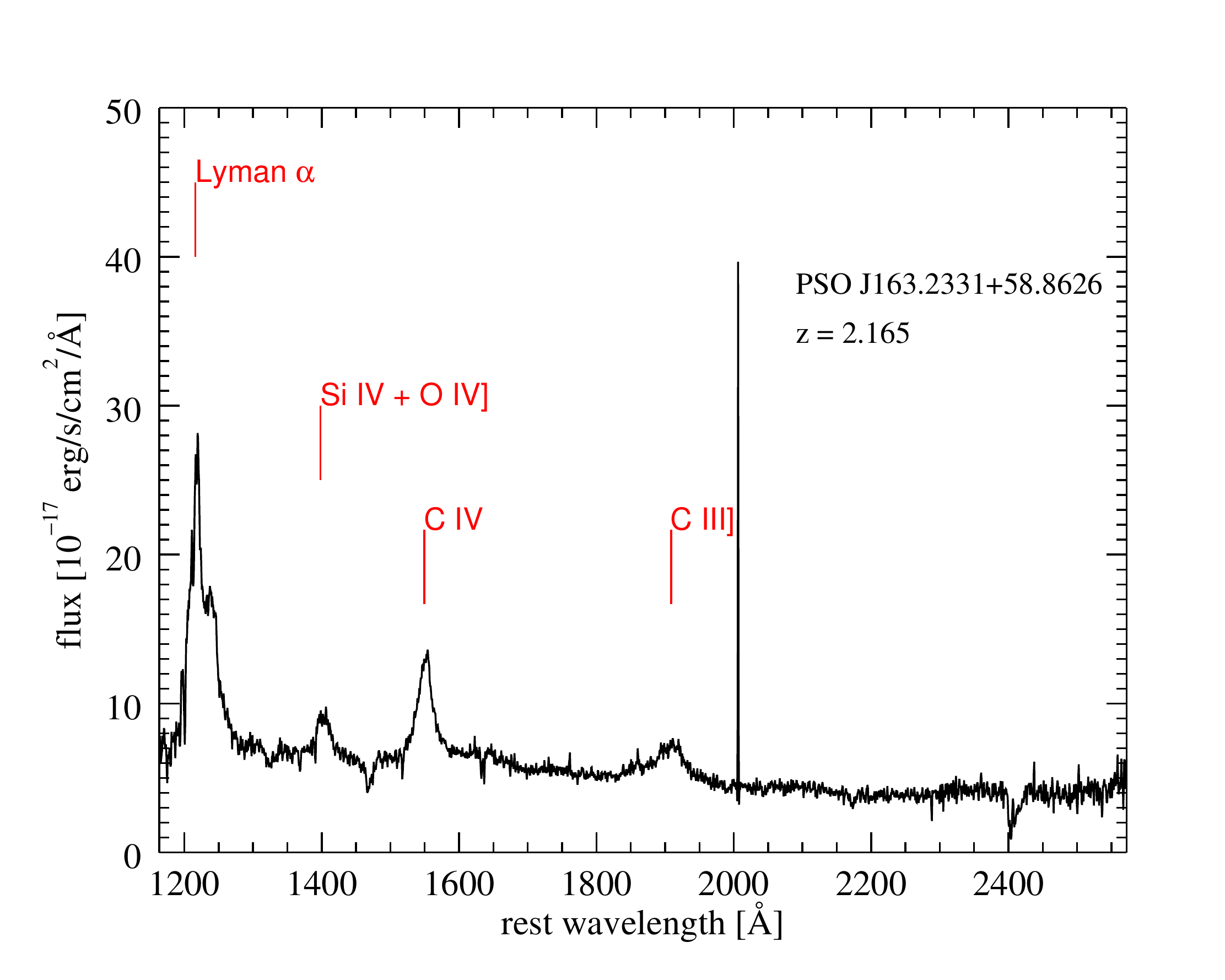,width=0.2\textwidth,clip=}
\epsfig{file=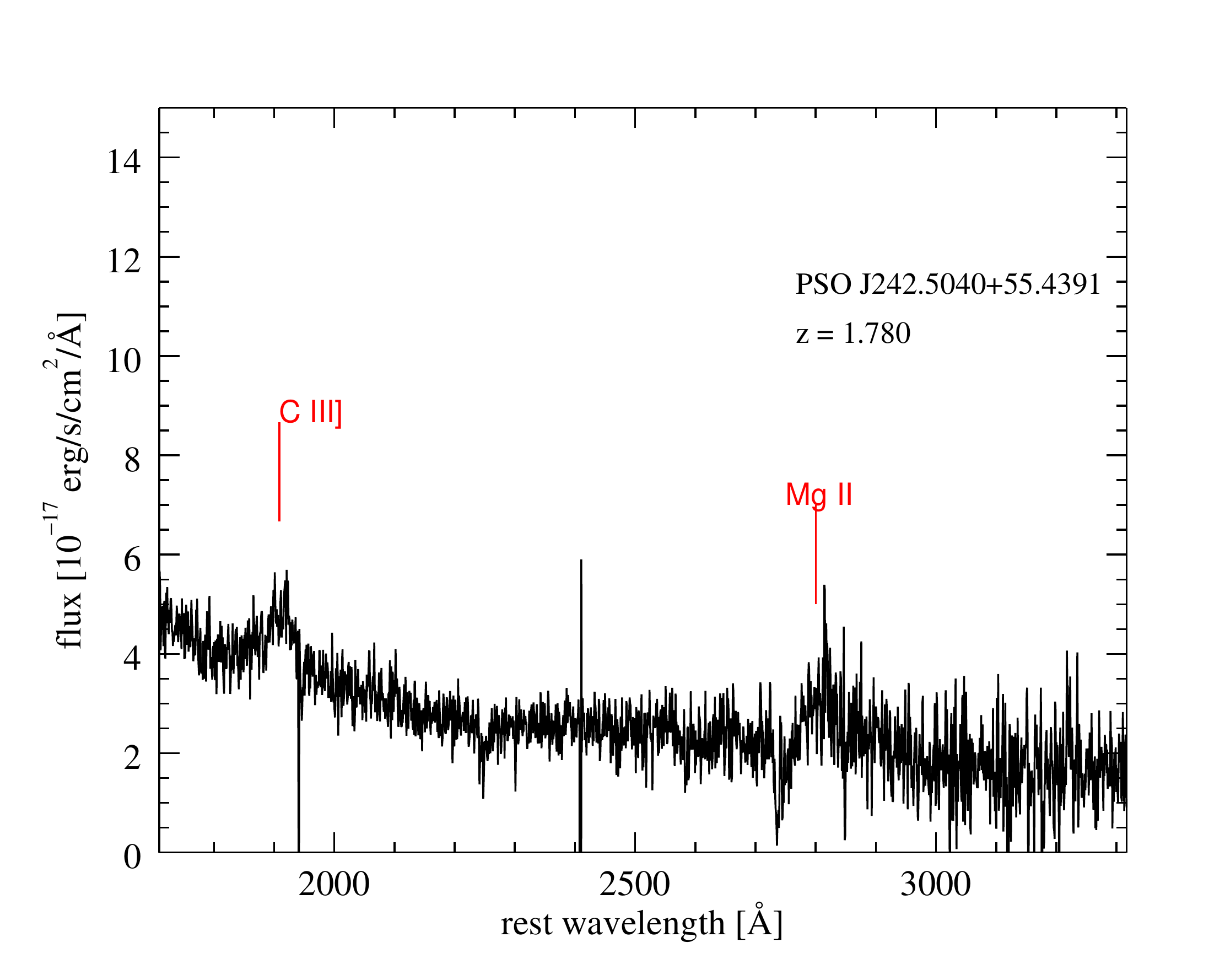,width=0.2\textwidth,clip=}
\epsfig{file=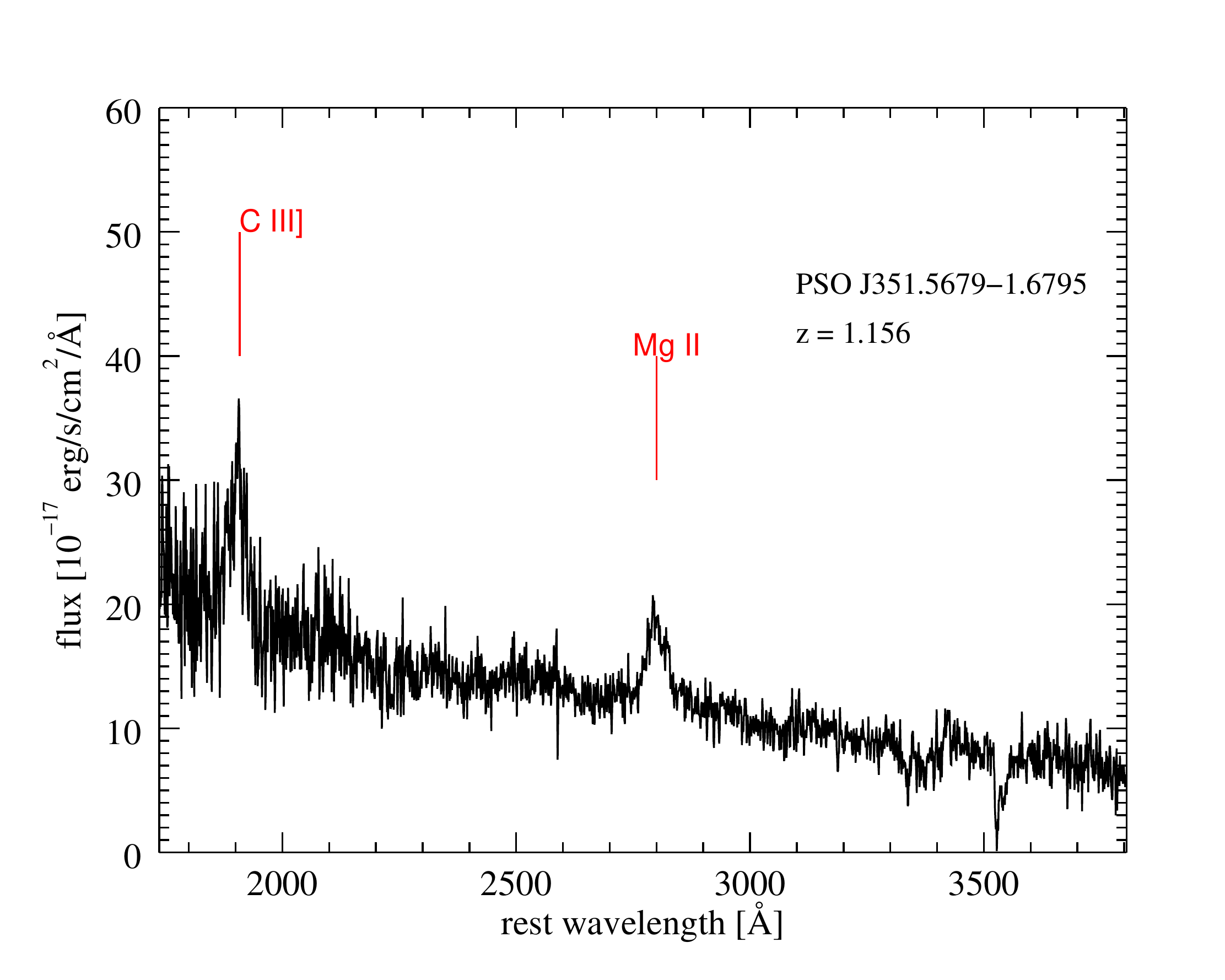,width=0.2\textwidth,clip=}
\epsfig{file=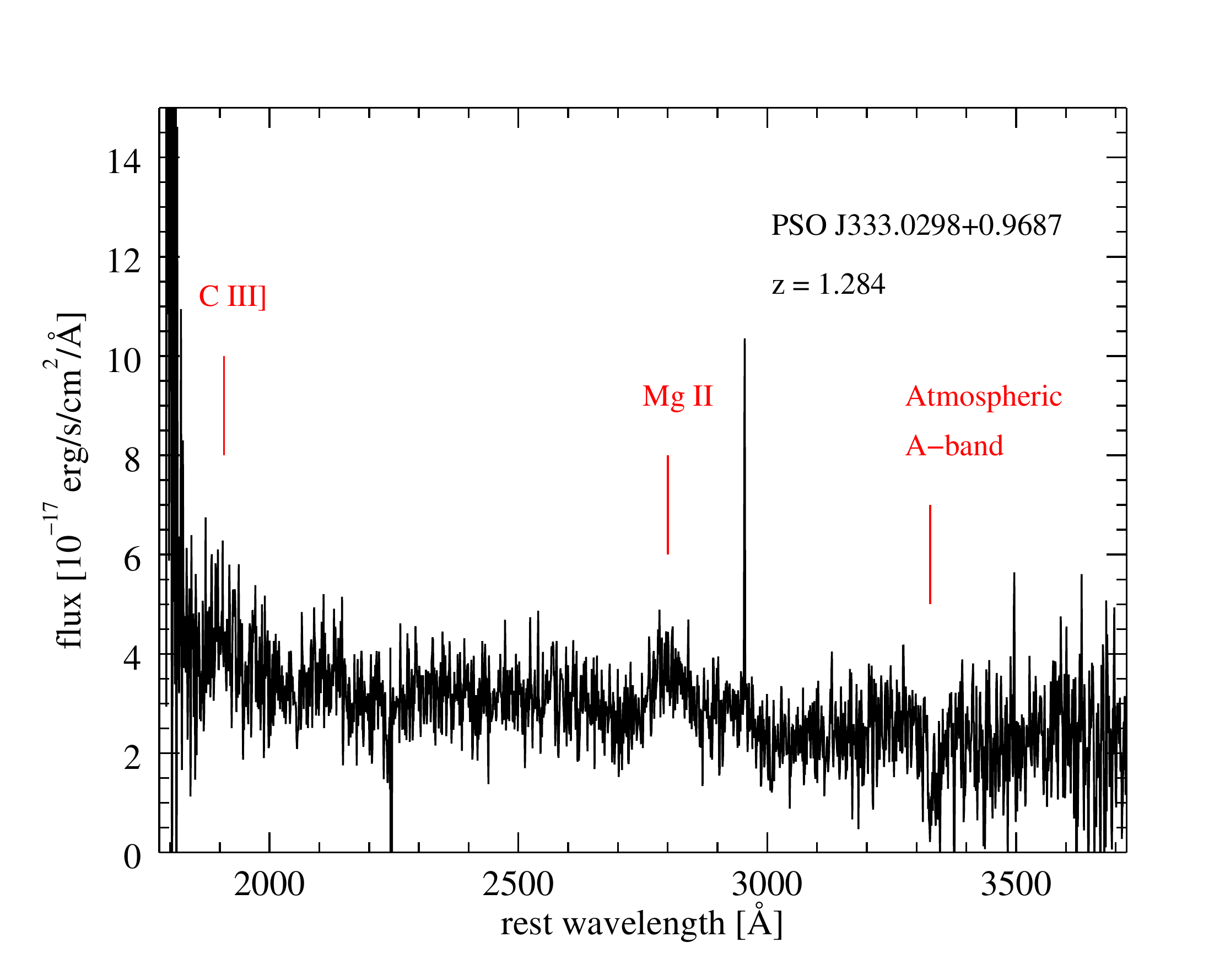,width=0.2\textwidth,clip=}
\epsfig{file=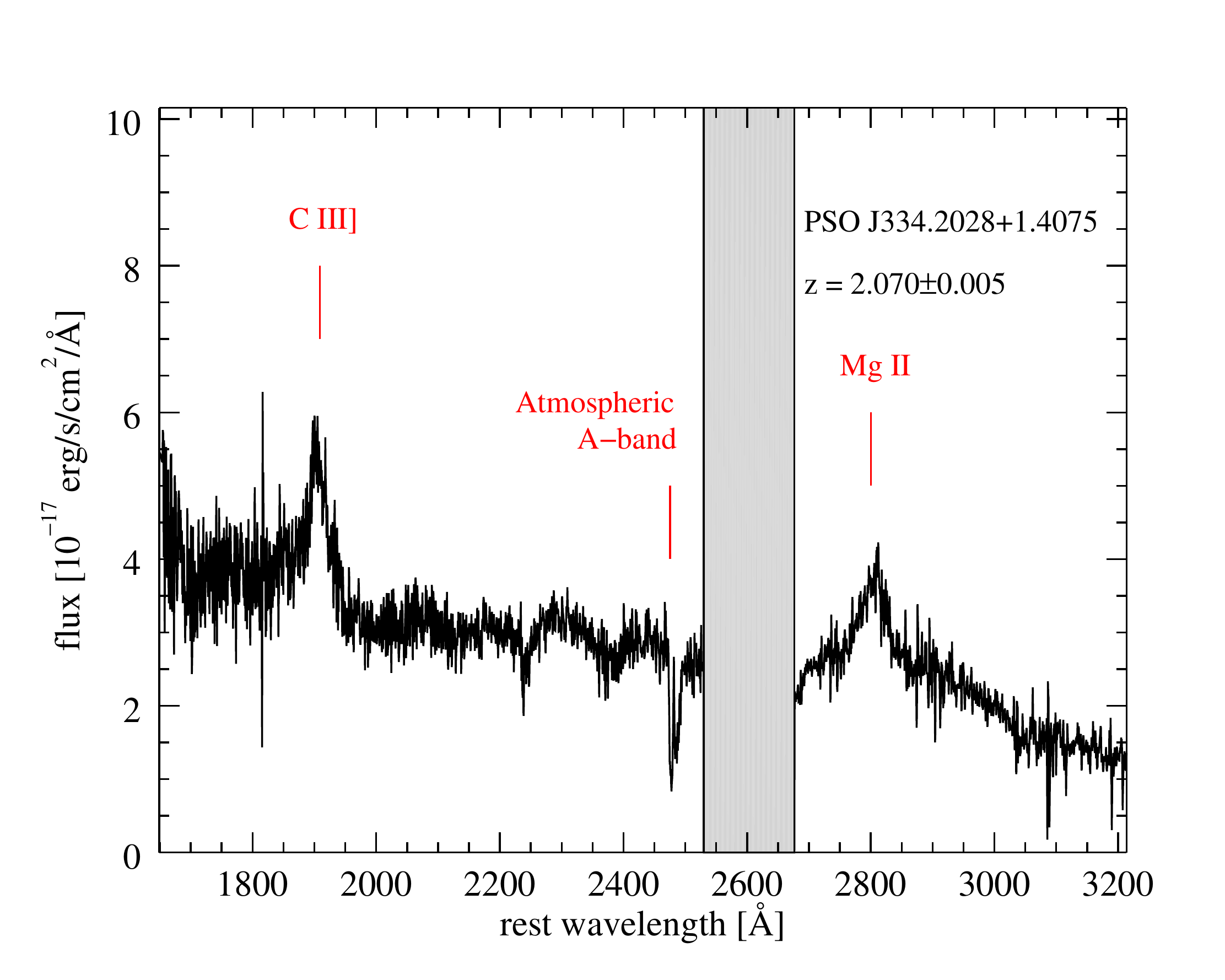,width=0.2\textwidth,clip=}
\epsfig{file=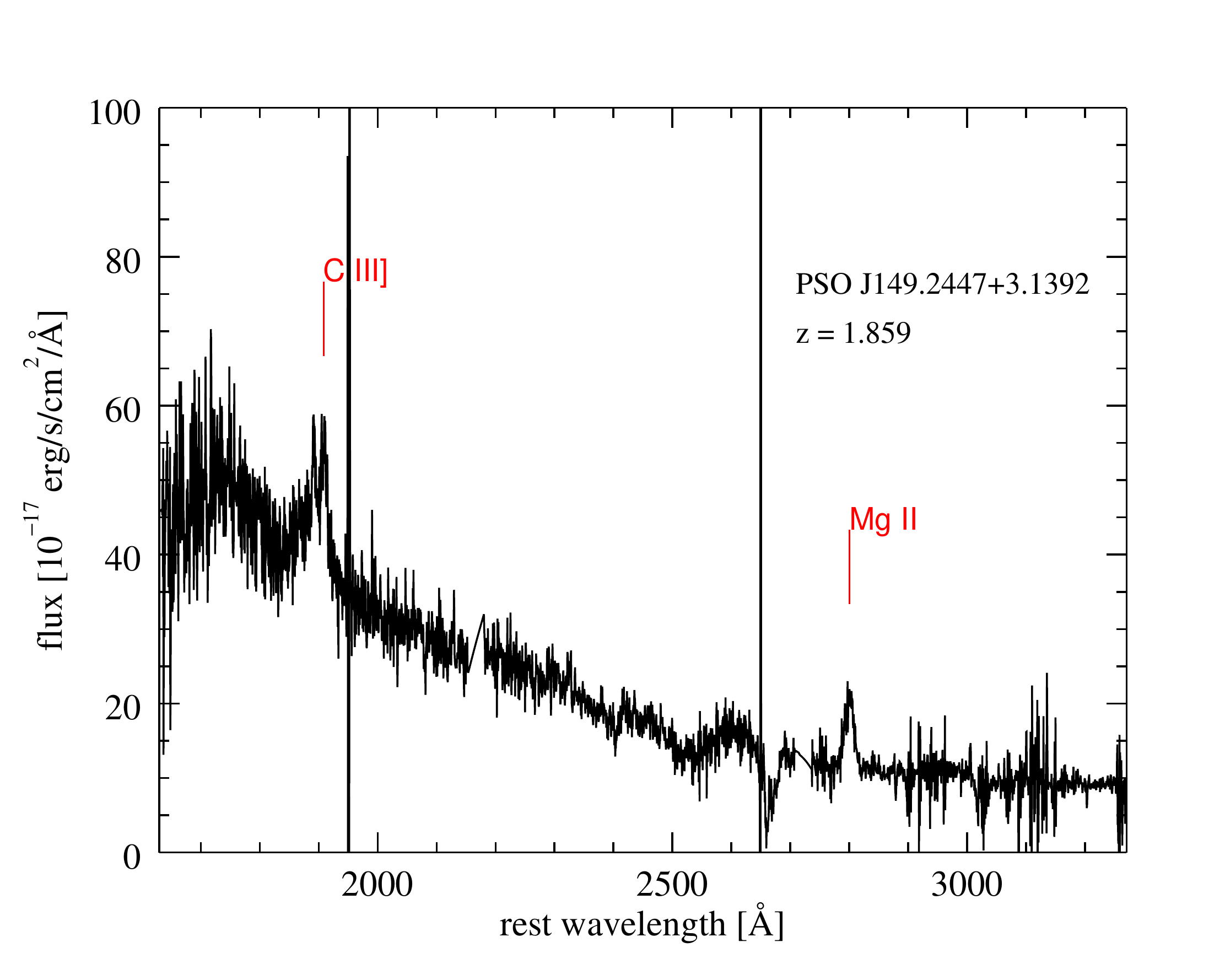,width=0.2\textwidth,clip=}
\epsfig{file=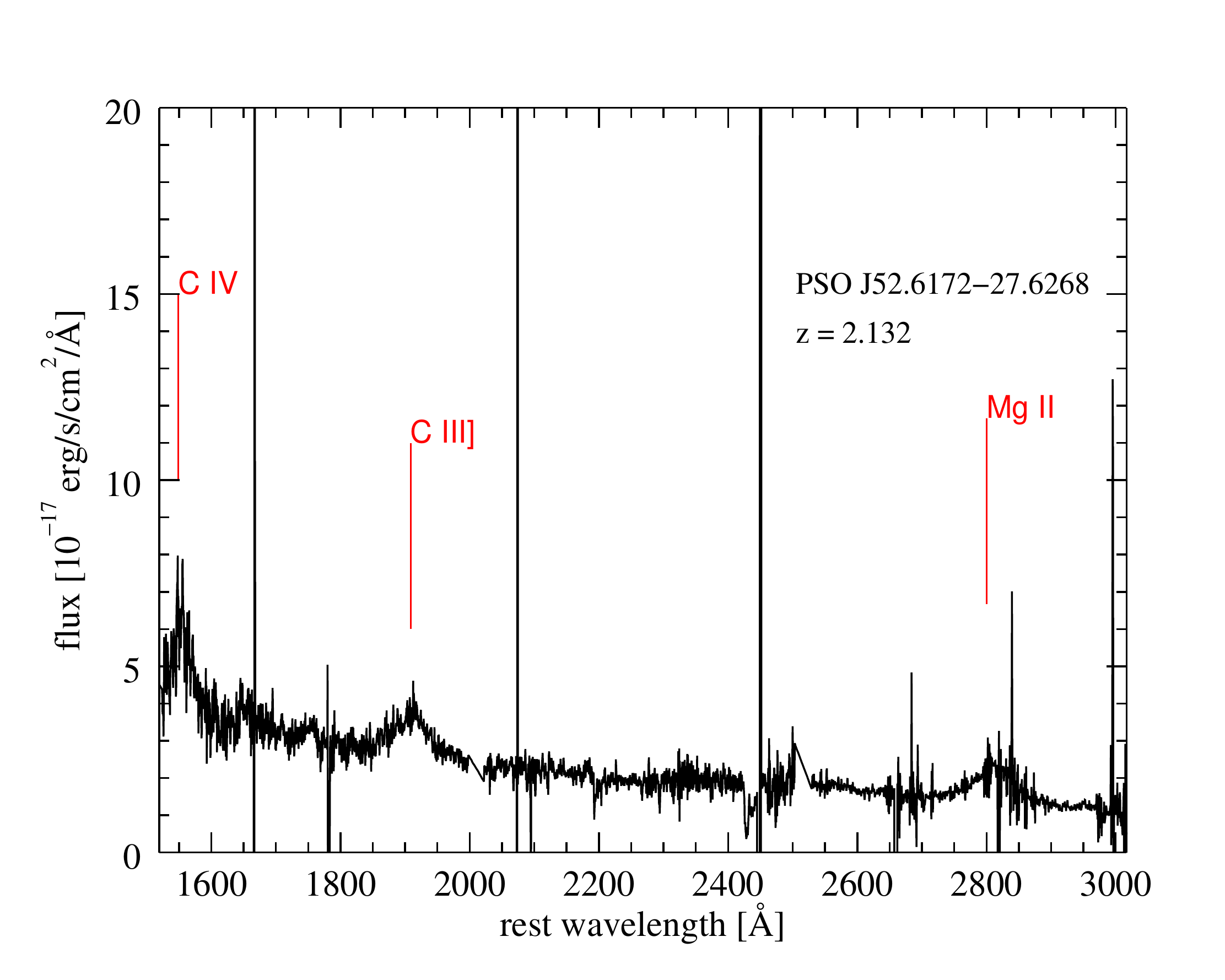,width=0.2\textwidth,clip=}
\epsfig{file=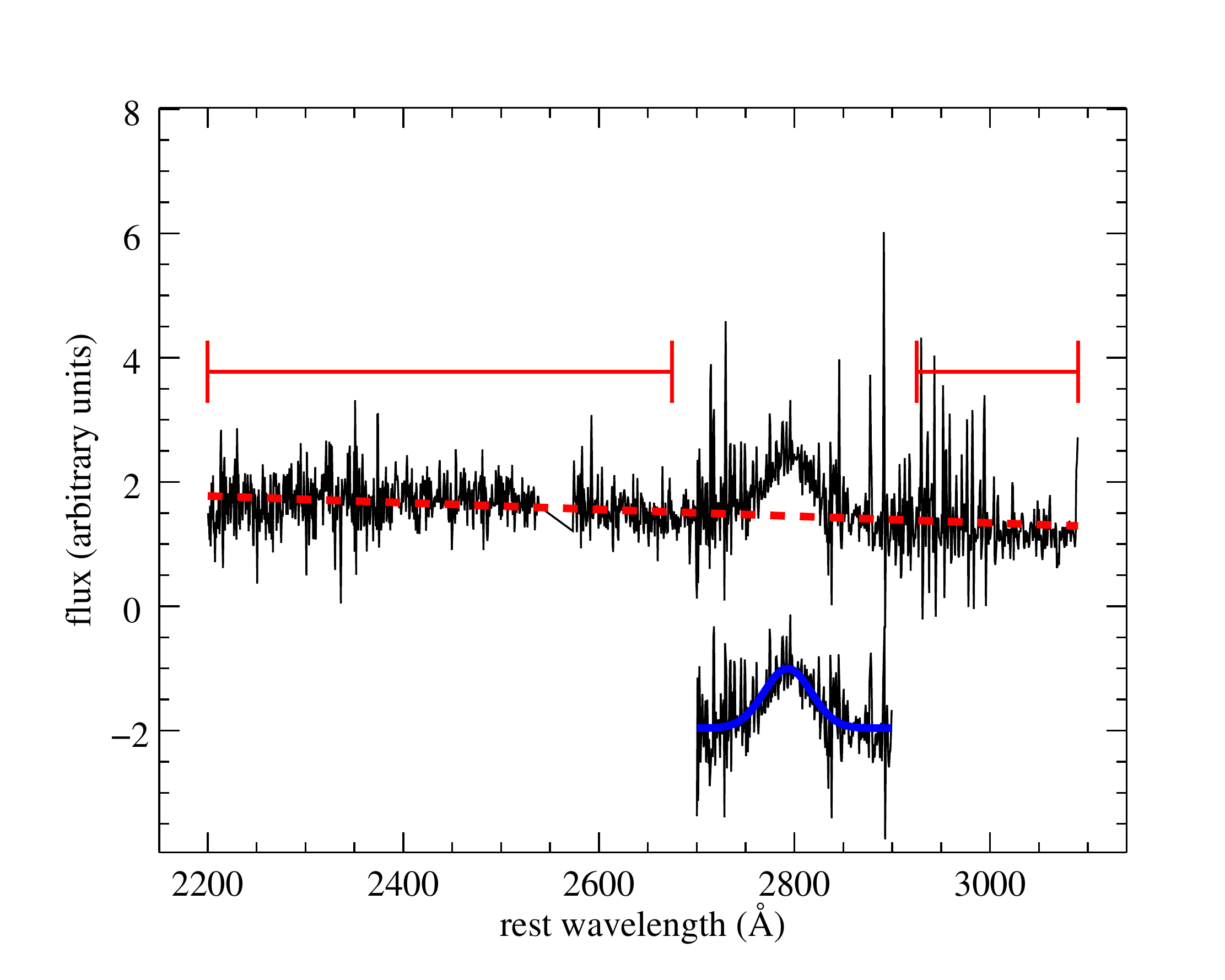,width=0.2\textwidth,clip=}
\epsfig{file=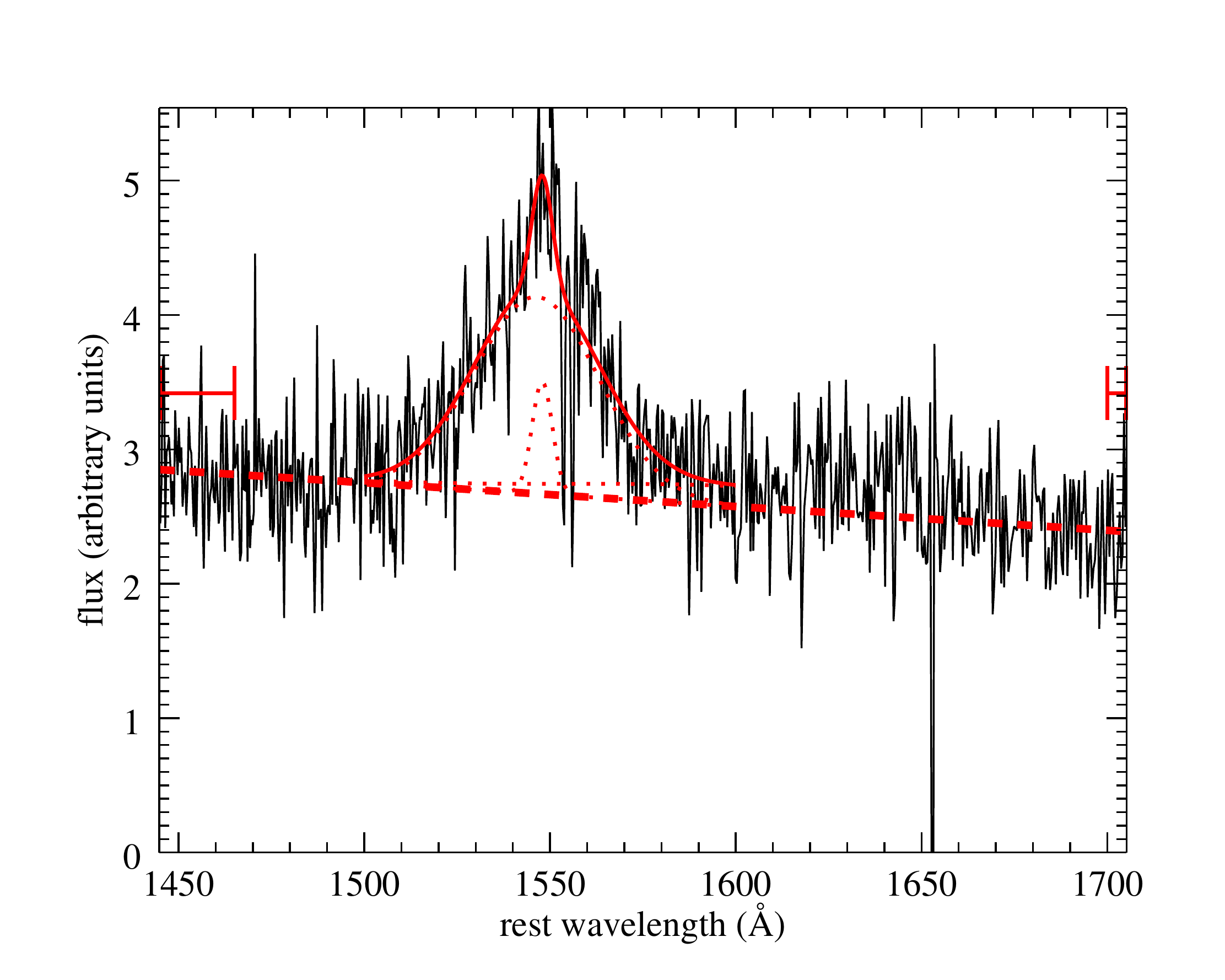,width=0.2\textwidth,clip=}
\label{fig:spec}
\caption{Archival and follow-up spectra of PS1 MDS candidates.}
\end{figure*} 


\end{document}